\documentclass[12pt,oneside,aps]{report}
\usepackage{graphicx, amsfonts,amssymb}
\def \f{\frac}

\let \pp=\partial

\newcommand{\be}{\begin{equation}}
\newcommand{\ee}{\end{equation}}
\newcommand{\bes}{\begin{eqnarray}}
\newcommand{\ees}{\end{eqnarray}}
\newcommand{\bean}{\begin{eqnarray*}}
\newcommand{\eean}{\end{eqnarray*}}
\def\sgn{\mathop{\rm sgn}\nolimits}
\def \ni{\noindent}

\def \tr{\textrm}
\newcommand{\N}{\mathrm{I\hspace{-.37ex}N}}
\newcommand{\R}{\mathrm{I\hspace{-.37ex}R}}

\def \w{\wedge}
\newcommand{\Ref}[1]{(\ref{#1})}
\def \tl{\tilde}
\def\g{{\mathfrak{g}}}
\newcommand{\su}[1]{| #1 \rangle}
\newcommand{\Rb}{{\rm \bf R}}
\newcommand{\Cb}{{\rm \bf C}}

\newcommand{\Z}{{\rm \bf Z}}
\def \w{\wedge}

\def \sl{SL(2,\Cb)}
\def \SA{{\cal A}}
\def \arr{\rightarrow}

\usepackage[centertags,sumlimits,intlimits,namelimits,reqno]{amsmath}

\def \tr{\textrm}

\def\N{{\mathbbm N}}
\def\R{{\mathbbm R}}
\def\Z{{\mathbbm Z}}
\def\sym#1{{\mathcal #1}}
\def\dim{\mathop{\rm dim}\nolimits}
\def\g{{\mathfrak{g}}}

\def\so{{\mathfrak{so}}}

\def\su{{\mathfrak{su}}}
\def\tr{\mathop{\rm tr}\nolimits}

\def\N{{\mathbb N}}
\def\R{{\mathbb R}}
\def\Z{{\mathbb Z}}

\begin{document}
\begin{titlepage}
\begin{center}

\vspace{2.0cm}

\begin{center}
\begin{tabular}{l}
\hline
   \\ \\
\Huge{{\bf Spin Foam Models}} \\ \\ \Huge{{\bf \qquad \,\,\,\,\,\,\, of}} \\ \\
\Huge{{\bf Quantum Spacetime}}\\
\\\\
\hline
\end{tabular}

\end{center}

\vspace{3.0cm}

{\Large \bf Daniele Oriti}\\ [2.0 cm]
{\Large  Girton College}\\ [1.5 cm]
{\large Dissertation submitted for the degree of Doctor of Philosophy} \\ [1.0 cm]
\centering
{\Large  University of Cambridge}\\[0.3cm]
{\Large  Faculty of Mathematics}\\[0.3cm]
{\Large  Department of Applied Mathematics \\ and Theoretical Physics}\\[0.5cm]
{\Large 2003}
\end{center}
\end{titlepage}

\vspace*{4 cm}
\hspace{10 cm}
\emph{{\large A Sandra}}

\vspace*{10 cm}
\hspace{10 cm}
\emph{{\large A Epifaniotto}}
\clearpage
\begin{quote}
\emph{(TALKING TO HIS BRAIN) Ok, brain. Let's face it. I don't like you, and you don't like me, but let's get through this 
thing, and then I can continue killing you with beer!}
\end{quote}
\hspace{5 cm} Homer J. Simpson
\vspace{1.5 cm}
\begin{quote}
\emph{....In formulating any philosophy, the first consideration must always be: what can we know? That is, what can we be 
sure we know, or sure that we know we knew it, if indeed it is at all knowable. Or have we simply forgotten it and are too 
embarassed to say anything? Descartes hinted at the problem when he wrote: \lq\lq My mind can never know my body, although 
it has become quite friendly with my legs." .....Can we actually \lq\lq know" the Universe? 
My God, it is hard enough to find your way around in Chinatown! The point, however, is: Is there anything out there? And why?
And must they be so noisy?....}
\end{quote}
\hspace{5 cm} Woody Allen 
\vspace{1.5 cm}
\begin{quote}
\emph{As far as the laws of mathematics refer to reality, they are not certain, and as far as they are certian, they do 
not refer to reality.}
\end{quote}
\hspace{5 cm} Albert Einstein
\vspace{1.5 cm}
\begin{quote}
\emph{Those are my principles, and if you don't like them..well, I have others.}
\end{quote}
\hspace{5 cm} Groucho Marx
\vspace{1.5 cm}
\begin{quote}
\emph{Whenever people agree with me, I feel I must be wrong...}
\end{quote}
\hspace{5 cm} Oscar Wilde

\clearpage

\tableofcontents

\chapter*{Declaration}
This dissertation is based on research  done at the Department of
Applied Mathematics and Theoretical Physics at the University of
Cambridge between January 2000 and June 2003.  No part of it or 
anything substantially the same has been previously submitted for a
degree or any other qualification at this or at any other University. 
It is the result of my own work and includes nothing
which is the outcome of work done in collaboration, except where
specifically indicated in the text. Section 4.4 contains work done in collaboration with R. M. Williams.
Section 4.3, section 4.7, and Chapter 6 contain work done in collaboration with E. R. Livine. 
Section 7.4 contains work done in collaboration with H. Pfeiffer. 
Section 5.2.4 and section 5.5 describe work in progress and to be published.
This thesis contains material which has appeared on the electronic
print archive {\sf http://arXiv.org}, and which has been published in the
following papers:
\begin{enumerate}
\item Daniele Oriti, Ruth M Williams, 
{\em Gluing 4-simplices: a derivation of the Barrett-Crane spin
foam model for Euclidean quantum gravity},
Phys.Rev. D 63 (2001) 024022; gr-qc/0010031;
\item Etera R. Livine, Daniele Oriti, {\em Barrett-Crane
spin foam model from a generalized BF-type action for gravity},
Phys. Rev. D65, 044025 (2002); gr-qc/0104043;
\item
D. Oriti, {\em Spacetime geometry from algebra: spin foam models for
non-perturbative quantum gravity}, Rep. Prog. Phys. 64, 1489 (2001);
gr-qc/0106091;
\item D. Oriti, {\em Boundary terms in the Barrett-Crane spin
foam model and consistent gluing}, Phys. Lett. B 532, 363 (2002);
gr-qc/0201077;
\item D. Oriti, H. Pfeiffer, {\em A spin foam model for pure
gauge theory coupled to quantum gravity}, Phys. Rev. D 66, 124010 (2002);
gr-qc/0207041;
\item E. R. Livine, D. Oriti, {\em Implementing causality in the spin foam quantum geometry},
Nucl. Phys. B 663, 231 (2003); gr-qc/0210064;
\item E. R. Livine, D. Oriti, {\em Causality in spin foam models for quantum gravity}, to appear in the
Proceedings of the XV SIGRAV Conference on General Relativity and
gravitational physics (2003); gr-qc/0302018;

\end{enumerate}

\chapter*{Acknowledgements}
\bigskip
There are so many people I would like to thank for having made this work possible. Acknowledging them all would take too 
long....and it will!

I am deeply grateful to my supervisor, Dr. Ruth M. Williams, for her unconditional support and help, both at the scientific 
and non-scientific level, for the many discussions, for her kindness, and for always being such a wonderful person with me. 
Thanks, Ruth!
  
A special thank goes to Etera Richard Livine, for being at the same time a great collaborator to work with and a wonderful
 friend to have fun with. It's amazing we managed somehow to do some work while having fun! Thanks!

Most of what I know about the subject of this thesis I've learnt from conversations and discussions with other people, 
that helped me to shape my ideas, to understand more, and to realize how much I still have to understand and to improve. 
To name but a few: John Baez, John Barrett, Olaf Dreyer, Laurent Freidel, Florian Girelli, Etera Livine, Renate Loll, 
Fotini Markopoulou, Robert Oeckl, Hendryk Pfeiffer, Mike Reisenberger, Carlo Rovelli, Hanno Sahlmann, Lee Smolin, Ruth 
Williams. To all of you (and the others I didn't mention), thanks for your help and for always being so nice! You reminded
me continously of how many great guys are out there!

Of course, the stimulating atmosphere of DAMTP has greatly helped me during the course of this work, so thanks to all the 
DAMTP people, faculty, staff, students and postdocs, for making it such a great place to work in.  

Also, the work reported in this thesis would have not been possibly carried out without the financial support coming from 
many different sources, covering my fees, my mantainance, and expenses for travel and conferences, that I gratefully 
acknowledge: Girton College, the Engineering and Physical Sciences Research Council, the Cambridge European Trust, the 
Cambridge Philosophical Society, Trinity College, the Isaac Newton Trust and the C.T. Taylor Fund.

Now it is time to thank all those who have indirectly improven this work, by improving my personal life, contributing 
greatly to my happiness, making me a better person, helping me when I needed it and teaching me so much.  

I would like to thank very much all my friends in Cambridge, who have made the last three years there so enjoyable, for 
the gift of their friendship; with no presumption of completeness, thanks to: Akihiro Ishibashi, 
Christian Stahn, David Mota and Elisabeth, Erica Zavala Carrasco, Francesca Marchetti, Francis Dolan, Giovanni Miniutti, 
Joao Baptista, Joao Lopez Dias, Martin Landriau, Matyas Karadi, Mike Pickles, Mohammed Akbar, Petra Scudo, Raf Guedens, 
Ruben Portugues, Sigbjorn Hervik, Susha Parameswaran, Tibra Ali, Toby Wiseman and Yves Gaspar. You are what I like most in 
Cambridge!   

I am grateful to my family, for their full support and encouragement, during the last three years just as much as during all
 my life. So, thanks to my father Sebastiano and my mother Rosalia, who have been always great in the difficult job of being 
my parents, vi voglio bene! Thanks to my sister Angela, because one could not hope for a better sister! Sono orgoglioso 
di te! And thanks to Epi, who has been, with his sweetness and love, the best non-human little brother one could have. 
Mi manchi, Epifaniotto!

I cannot thank enough all my friends in Rome and in Italy, for what they gave me by just being my friends and for the 
happiness this brings into my life; being sorry for surely neglecting someone: Alessandra, Andrea, Antonio, Carlo, Cecilia, 
Chiara, Claudia, Daniela, Daniele C., Daniele M., David, Davide, Federico, Gianluca M., Gianluca R., Giovanni, Luisa, Marco,
 Matteo, Nando, Oriele, Pierpaolo, Roberto, Sergio, Silvia, Simone E., Simone S., Stefano. I feel so lucky to have you all! 
Vi voglio bene, ragazzi!

Finally, Sandra; I thank you so much for being what you are, for making me happy just by looking at you and thinking of you, 
for being always so close and lovely, for making me feel lucky of having you, for sharing each moment of my life with me,
and for letting me sharing yours; for all this, and all that I am not able to say in words, thanks. Sei tutto, amore! 

\chapter*{Summary}
\bigskip
Spin foam models are a new approach to the construction of a quantum
theory of gravity. They aim to represent
a formulation of quantum gravity which is fully background independent and
non-perturbative, in that they do not rely on any
pre-existing fixed geometric structure, and covariant, in the spirit of
path integral formulations of quantum field theory.

Many different approaches converged recently to this new framework: loop
quantum gravity, topological field theory, lattice gauge theory,
path integral or sum-over-histories quantum gravity, causal sets, Regge
calculus, to name but a few.

On the one hand spin foam models may be seen as a covariant (no $3+1$
splitting) formulation of loop quantum gravity,
on the other hand they are a kind of translation in purely algebraic and
combinatorial terms of the path integral approach to quantum gravity.

States of the gravitational field, i.e. possible 3-geometries, are
represented by spin networks, graphs whose edges are labelled with
representations of some given group, while their histories, i.e. possible
(spacetime) 4-geometries, are spin foams, 2-complexes with faces labelled
by representations of the same group. The models are then characterized by
an assignment of probability amplitudes to each element of these
2-complexes and are defined by a sum over histories of the gravitational
field
interpolating between given states. This sum over spin foams defines the
partition function of the theory, by means of
which it is possible to compute transition amplitudes between states,
expectation values of observbles, and so on.

In this thesis we describe in details the general ideas and formalism of
spin foam models, and review many of the results obtained
recently in this approach. We concentrate, for the case of 3-dimensional
quantum gravity, on the Turaev-Viro model, and, in the 4-dimensional case,
which is our main concern,
on the Barrett-Crane model, based on a simplicial formulation of gravity,
and on the Lorentz group as symmetry group, both in its Riemannian and
Lorentzian formulations, which is currently the most promising model among
the proposed ones.

For the Turaev-Viro model, we show how it gives a complete formulation of
3-dimensional gravity, discuss its space of quantum states, showing also
the link with the loop quantum gravity approach, and that with
simplicial formulations of gravity, and finally give a
few example of the evaluation of the partition function for interesting
topologies.

For the Barrett-Crane model, we first describe the general ideas behind
its construction, and review what it is known up to date about this model,
 and then discuss in details its links with the classical formulations of
gravity as constrained topological field theory.

We show a derivation of the model from a lattice gauge theory perspective,
in the general case of manifold with
boundaries, discussing the issue
of boundary terms in the model, showing how all the different amplitudes
for the elements of the triangulations arise,
and presenting also a few possible variations of the procedure used,
discussing the problems they present.

We also describe how, from the same perspective, a spin foam model that
couples quantum gravity to any gauge theory may be
constructed, and its consequences for our understanding of both gravity
and gauge theory.

We analyse in details the classical and quantum geometry of the
Barrett-Crane model, the meaning of the variables it uses, the properties
of its space of quantum states,
how it defines physical transition amplitudes between these states, the
classical formualtions of simplicial gravity it corresponds to, all its
symmetry properties.

Then we deal with the issue of causality in spin foam models in general
and in the Barrett-Crane one in particular. We
describe a general scheme for
causal spin foam models, and the resulting link with the causal set
approach to quantum gravity. We show how the Barrett-Crane
model can be modified to implement causality and to fit in such a scheme,
and analyse the properties of the modified model.

\chapter{Introduction}
\section{The search for quantum gravity}
The construction of a quantum theory of gravity remains probably {\it
the} issue left unsolved in theoretical physics in the last century, in spite of a lot of
efforts and many important results obtained during an indeed long
history of works (for an account of this history see \cite{Rov0}). The
problem of a complete formulation of quantum gravity is still quite
far from being solved (for reviews of the present situation see
\cite{Hor,Rov00,Carlip}), but we can say that the last (roughly) fifteen years
have seen a considerable number of developments, principally
identifiable with the construction of superstring theory \cite{GSW,Polch}, in its
perturbative formulation and, more recently, with the understanding of
some of its non-perturbative features, with the birth of non-commutative geometry \cite{Connes, 
Majid, Majid2}, and with the renaissance of the
canonical quantization program in the form of loop quantum gravity \cite{Ashlect, RovSmo2},
based on the reformulation of classical General Relativity in terms of
the Ashtekar variables \cite{Ash0}. In recent years many different approaches on
the non-perturbative and background-independent side have
been converging to the formalism of the so-called spin foams
\cite{Baez,Baez2, Oriti}, and this class of models in the subject of this thesis.

Why do we want to quantize gravity? There are many reasons for this\cite{isham2}: the presence of singularities in 
classical General Relativity \cite{HawEll}, showing that something in this theory goes wrong when we try to use it 
to describe spacetime on very small scales; the fact that the most general form of mechanics we have 
at our disposal in describing Nature is quantum mechanics and that this is indeed the language that
proved itself correct in order to account for the features of the other three fundamental 
interactions, so that one is lead to think that the gravitational interaction too must be described 
within the same conceptual and semantic framework; the desire to unify all these interactions, i.e. 
to find a unified theory that explains all of them as different manifestation of the same physical 
entity; the ultraviolet divergences one encounters in quantum field theory, that may be suspected to 
arise because at high energies and small scales the gravitational interaction should be necessarily 
taken into account; the necessity for a theory describing the microscopic degrees of freedom 
responsible for the existence of an entropy associated with black
holes \cite{Bek, Sork} 
and more generally with any causal horizon \cite{JacobParent}; more generally, the need to explain the 
statistical mechanics behind black hole thermodynamics; the need to have a coherent 
theory describing the interaction of quantum gauge and matter fields with the gravitational field, 
beyond the approximation of quantum field theory in curved spaces; 
the unsolved issues in cosmology asking for a more fundamental theory to describe the origin of our 
universe, at energies and distances close to the Planck scale. 

The reader may add her own preferred 
reasons, from a physical or mathematical point of view, but what remains in our opinion the most 
compelling one is the purely conceptual (philosophical) one, well explained in \cite{carlo1, carlo2}, 
given by the need to bring together the two conceptual revolutions with which the last century in 
physics started, represented by General Relativity and Quantum Mechanics. The main elements of these
two paradigmatic shifts (in Kuhn's sense) of our picture of the natural world can be summarised, we 
think, as: 1) spacetime is nothing other than the gravitational field, and that it is a dynamical 
entity and not a fixed background stucture on which the other fields live; consequently, any 
determination of localization has to be fully relational, since no fixed structure exists that can be
 used to define it in an absolute way; 2) all dynamical objects, i.e. all dynamical fields, are 
quantum objects, in the sense of having to be described within the framework of quantum mechanics, 
whatever formalism for quantum mechanics one decides to use (operators on Hilbert spaces, consistent 
histories, etc.).

This requires a peculiar (and, admittedly, conceptually puzzling) relationship between observed 
systems and observers, and it is best characterized by a relational point of view that assigns
properties to physical systems only relative to a given observer, or that deals with the information 
that a given observer has about a given physical system, and that because of this favours an 
operational description of all physical systems.

The problem we face, from this perspective, is then to give a more fundamental description of 
spacetime and of its geometry than that obtained from General Relativity, to construct a more 
fundamental picture of what spacetime is, to understand it as a quantum entity, to grasp more of the
 nature of space and time.

Before getting to the particular approach to the construction of a quantum theory of gravity 
represented by spin foam models, we think it is useful to account for the main ideas and motivations,
that in one way or another are at the roots of this approach, in order to understand better what is,
in our opinion, the conceptual basis of the spin foam approach, as it is presently understood (by us).  

\section{Conceptual ingredients and motivations}

Let us list what we think are the conceptual ingredients that should enter in a complete formulation of 
a theory of quantum gravity, mostly coming from insights we obtain from the existing theory of the 
gravitational field, General Relativity, and from the framework of quantum mechanics, and partly resulting 
from independent thinking and intuitions of some of the leaders of quantum gravity research. 

The list is as follows: a theory of quantum gravity should be {\bf background independent} and 
fully {\bf relational}, it should reveal a fundamental {\bf discreteness} of spacetime, and be formulated 
in a {\bf covariant} manner, with the basic objects appearing in it having a clear {\bf operational} 
significance and a basic role played by {\bf symmetry} principles, and a notion of {\bf causality} built 
in at its basis.  

By background independence and relationality we mean that the physical significance of (active) 
diffeomorphism invariance of General Relativity has to be mantained at the quantum level, i.e there should 
be no room in the theory for any fixed, absolute background spacetime structure, for any non-dynamical 
object, that any physical quantity has to be defined with respect to one or more of the dynamical objects 
the theory deals with, and also the location of such objects has to be so defined.

A fundamental discreteness of spacetime at the Planck scale is to be expected for many reasons, included 
the way quantum mechanics describes states of systems in terms of a finite set of quantum numbers, the 
ultraviolet divergences in QFT and the singularities in GR, both possibly cured by a fundamental cut-off at
the Planck scale,  
and the finiteness of black hole entropy, also possibly explained by such a minimal length (the entropy 
of a black hole would be infinite if infinitesimal fluctuations of both graitational and matter degrees of
freedom were allowed at the horizon \cite{Sork}), not mentioning hints coming from 
successes and failures of several approaches to quantum gravity pursued in the long history of the subject, 
including the discrete spectrum for geometric operators obtained in the context of loop quantum gravity 
(showing also a minimal quantum of geometry of the order of the Planck length), and the difficulties in 
giving meaning to the path integral for quantum gravity in a continuum formulation.

A covariant formulation, treating space and time on equal footing, is also to be preferred both in the 
light of how space and time are treated in General Relativity and of the fact that such a covariant 
formulation can be shown to be the most general formulation for a mechanical theory, both at the classical
and quantum level, as we are going to discuss.

Operationality would imply the quantities (states and observables) appearing in the formulation of the 
theory to have as clear-cut a physical interpretation as possible, to be related easily to operations one 
can perform concretely and not to refer to abstract ontological entities away from experimental (although
 maybe idealized) verification; such operationality, however appealing on philosophical grounds, is even 
more necessary in a quantum mechanical setting, where the observables are not only needed to interpret the
 theory but also for its very definition.

Also, the use of symmetries to characterize physical systems has a long and glorious history, so does not 
need probably too much justification, but in this case it is doubly relevant: on the one hand, 
operationality applied to spacetime would suggest the use of symmetry transformations on spacetime 
objects to play a key role in defining the observables and the states of the theory (also recalling that 
it is the {\it parallel transport} of objects along suitable paths that indicates the presence of a 
non-trivial geometry of spacetime); on the other hand, the impossibilty of using geometric structures as 
starting point, since they have to emerge from more fundamental non-geometric ones, suggests that the 
proper language to be used in quantum gravity is algebraic, and coming from the algebraic context of the 
representation theory of some symmetry group.

As for causality, first of all it is a fundamental component of our understanding of the world, so its 
presence at some level of the theory is basically compulsory, second, it is very difficult to envisage a 
way in which it can emerge as the usual spacetime causal structure if not already present at least in some
primitive form at the fundamental level.
  
Let us analyse all these ideas in more detail.

\subsection{Background independence and relationality}
The gauge invariance of General Relativity, i.e. diffeomorphism invariance, requires the absence of any 
absolute, non-dynamical object in the theory \cite{GaulRov}; the spacetime background itself, spacetime geometry, has to 
be dynamical and not fixed a priori; therefore no statement in the theory can refer to a fixed background,
 but has to be fully relational in character, expressing a correlation between dynamical variables; this 
is what is meant by background independence, a necessary requirement for any complete theory of quantum 
gravity, also because it is already present in classical GR. 
In practice, this fact translates into the requirements that observables in gravity have to be 
diffeomorphism invariant, i.e. gauge invariant as in any gauge theory, so cannot depend on spacetime 
points\cite{carloObs}; in a canonical formulation they have to commute with the gravity constraints, both at the 
classical and quantum level; therefore they have to be either global, e.g. the volume of the universe, or
 local if locality is defined in terms of the dynamical quantities of the theory, e.g. some
 geometrical quantity at the event where some metric component has a given value, or, if some matter field
 is considered, some geometrical quantity where the matter field has a certain value\cite{carloObs}. In any case, any 
local observable will define some {\it correlation} between the dynamical fields (more generally, between the partial 
observables of the theory \cite{carlopartial}) which are present, and 
must therefore take into account the equaton of motion of these fields. This has to be true both at the 
classical and quantum level.

This has important consequences also for the interpretation of quantum gravity states.
The only physically meaningful definition of locality in GR is relational, as we said.
Once quantized, the metric field, spacetime, has to be described (in one formulation of quantum theory) in
 terms of states in a Hilbert space; however, because of the mentioned relationality of the concept of 
localization, the states of the gravitational field cannot
describe excitations of a field localized somewhere on some background space, but must describe excitations
 {\it of} spacetime itself, being thus defined regardless of any background manifold\cite{carlo1}.

One can push this idea even further and say that, at a fundamental level, because of diffeomorphism 
invariance, there should not be any spacetime at all, in the sense that no set of
spacetime points, no background manifold, should exist, with a spacetime point being {\it defined} by where a given 
field is, or, better, that we ought to reconstruct a notion of spacetime and of spacetime events from the 
background independent gravitational quantum states, from the quanta of spacetime, because of the 
identification between the gravitational field and spacetime geometry. 

The idea would then be
that they are (linear combinations of) eigenstates of
geometric operators, such as lengths, areas and volumes, although
they must be given in terms of non-geometric diffeomorphism invariant
informations, in order to make sense regardless of any spacetime
manifold and to be able to reconstruct such a manifold from them,
including its geometry; therefore they are non-local, in the sense of
not being localized anywhere in a background manifold, they are
instead the ``where'' with respect to which other fields may be
localized. If not geometric, the ingredients coming into their
definition have to be purely combinatorial and algebraic; they are like \lq\lq seeds" of space, 
\lq\lq producing" it by suitable translation of non-geometric into geometric information; in the same
sense their dynamics should also be given in purely algebraic and
combinatorial terms, expressed by the definition of transition
amplitudes between quantum gravity states only, either expressed as
canonical inner products or as sum-over-histories, and should not take place ``in time'' but rather
{\it define} what time is, for other gauge or matter fields;
therefore, these transition amplitudes will not depend on any time
variable at all, not a coordinate time, not a proper time, no time
variable at all. Of course this does not mean that they do not contain
any notion of time, but only that this notion is not of the familiar
classical form \cite{carlobook}.  We will
later see how these ideas are realized in practice in loop quantum
gravity and spin foam models.   

Let us briefly note that the relational point of view forced on us by GR is somehow reminescent of the 
ideas at the roots of category theory, where indeed objects are treated on the same footing as 
relationships between objects, and in a sense what really matters about 
objects is indeed only their relationships with other objects; the importance of category theory in 
quantum gravity is suggested by the way 3-d gravity can indeed be quantized by categorical methods and 
expressed nicely in categorical terms, as we will discuss. 
It has in fact been repeatedly argued that the general framework of topological field theory can be adapted 
to 4-dimensional quantum gravity, implying that the right formulation of 4-d quantum gravity has to be 
in categorical terms; also this is partly realized in spin foam models \cite{baezalgebra}. This is a way in which 
algebra, in this case, categorical algebra, can furnish the more fundamental language in which to describe
spacetime geometry \cite{crane95}. 

If taken seriously, background independence has important implications for the fomulations of the theory; 
for example, the absence of a fixed background structure implies that the theory must be formulated in a 
non-perturbative fashion, from the metric point of view, in order to be considered truly fundamental, and 
to incorporate satisfactorily the symmetries of the classical theory. 

This means that it cannot be 
formulated by fixing a metric, or its corresponding quantum state, and then only considering the 
perturbations around it, although of course this is an approximation that we should be able to use in 
some limited context, since it describes, for example, the situation we live in. Of course this does 
not mean that no perturbative expansion at all can be used in the formulation of the theory, but only that,
 if used, each term in such a perturbation expansion should be defined in a background independent way. 
We will see an explicit example of this in the group field theory
formulation of spin foam models.

A simple system to study, which nevertheless presents many analogies
with the case of General Relativity, is that represented by a
relativistic particle in Minkowski space, since it may be thought of
as something like ``General Relativity in 0 spatial dimensions''.
Its configuartion variables are the spacetime coordinates $x^\mu$ and
its classical action is:
\be
S(x)\,=\,\int_{\lambda_1}^{\lambda_2}\,(-m)
\sqrt{\frac{dx^\mu}{d\lambda}\frac{dx_\mu}{d\lambda}}\,d\lambda.
\ee
The system is invariant under reparametrization $\lambda\rightarrow
f(\lambda)$ of the trajectory of the particle, and this is analogous
to a time diffeomorphism in General Relativity (this is
the sense in which one may think
 of this elementary system as similar to General Relativity in 0 spatial
dimensions) that reduce to the identity
 on the boundary. It is often convenient to pass
to the Hamiltonian formalism. The action becomes: \be
S(x)\,=\,\int_{\lambda_1}^{\lambda_2}( p_\mu
\dot{x}^\mu\,-\,N\,\mathcal{H})\,d\lambda \ee where
$p_\mu$ is the momentum conjugate to $x^\mu$ and $\mathcal{H}=p_\mu
p^\mu + m^2 = 0$ (we are using here the signature $(- + + +)$) is the Hamiltonian constraint
that gives the dynamics of the system (it represents
 the equation of motion obtained by extremizing the action above with
respect to the variables $x$, $p$, and $N$).

\subsection{Operationality}
Operationality when applied to the construction of a quantum gravity theory would require the use of 
fundamental variables with as direct as possible an experimental meaning, constructed or obtained by 
performing possibly idealized operations with clocks, rods, gyroscopes or the like, or requires us to
 intepret such variables (and any object in the theory) as a convenient summary of some set of operations
we may have performed\cite{sorkinspeci}. In particular it is 
the set of observables that has to be given a clear operational significance, in order to clarify the 
physical meaning of the theory itself. Indeed we will see later how the most natural observables (physical 
predictions) of classical gravity, these being \lq\lq classical spin network", are of a clear operational 
significance and directly related to transformations we may perform on spacetime\cite{carlomech}. This can moreover be 
translated into the quantum domain more or less directly, as we shall see.

If the (operational) definition of the observables of the theory is important in classical mechanics, in 
order to provide it with physical meaning, it is absolutely crucial in quantum mechanics since it is part
 of the very definition of the theory. Indeed, a 
quantum mechanical theory is defined by a space of states {\it and} by a set of (hermitian) operators 
representing the observables of the theory \cite{Haag, DiracQM, Peres}. To 
stress even more the importance of observations and observables in quantum theory, just note that the states of a system 
(vectors in a Hilbert space) can be thought of either as results of a given measurement, if they are eigestates of a given 
operator, or as a summary of possible measurements to be performed on the system, when they are just linear combinations of 
eigenstates of a given observables (and this is the general case). Also, the observation (measurement) process is one of the two 
dynamical laws governing the evolution of the states of the system considered, formalized as the collapse 
of the wave function, with all the interesting and puzzling features this involves. From a more general 
perspective, quantum mechanics is about observations that some subsystem of the universe (observer) makes 
of some other (observed) system, and one can push this point of view further to state that the 
very notion of state of a given system is necessarily to be considered as relative to some observer in 
order to make sense, and as just a collection of possible results of her potential measurements 
(observations)\cite{carlo96}; This point of view has interesting connections with (quantum) information theory, in which
 context it is in some sense the natural point of view on quantum
 mechanics \cite{Peres, Fuchs}; also, it is 
natural in the context of topological field theory (of which 3-dimensional quantum gravity is an important 
example), and can be phrased, related to these, in categorical
language \cite{crane95,robertQM}. 

In a sense, thus, quantum mechanics is operational from the 
very onset, and can be made even more \lq\lq operational-looking" if one formulates it in such a way that it reflects more
closely what are the conditins and the outcomes of realistic
measurements and observations\cite{carlodon}. Such a reformulation \cite{carlomike} uses 
spacetime smeared states and a covariant formulation of both classical and quantum mechanics, where the basic idea is to 
treat space and time on equal footing. It is apparent that such a formulation is more naturally adapted to a quantum gravity
 context than the usual one based on a rigid splitting between a space manifold and a fixed time 
coordinate. We will describe such a formulation in more detail when discussing the issue of covariance.

\subsection{Discreteness}
As we said above, there are many reasons to question the correctness of the usual assumption that 
spacetime has to be represented by a smooth manifold, and to doubt the physical meaning of
 the very concept of a spacetime \lq\lq point"\cite{isham}. So, is spacetime a $C^\infty$-manifold? What is the 
meaning of a spacetime point? Is the very idea and use of points justified 
from the general relativistic and quantum mechanical perspective? If not, then what replaces it?

In both canonical loop quantum gravity and path integral approaches a continuum differentiable manifold is
 assumed as a starting point (even if in the path integral approach one then treats the topology of the 
manifold as a quantum variable to be summed over), but then in the loop approach the development of the 
theory itself leads beyond the continuum, since 
the states of the theory (at least in one (loop) representation) are non-local, combinatorial and algebraic
 only, and the spectrum of the geometric operators is discrete. 
One may instead take the point of view that the continuum is only a 
phenomenological approximation, holding away from the Planck scale, or in other words when the resolution
 of our measurements of spacetime is not good enough; therefore a continuum manifold is what we may use to represent
 spacetime only if we do not test it too closely, and only if we do not need any precise description of it.
One may also say that the continuum has no operational meaning, or that it is the result of an 
idealization in which our sensitivity is assumed to be infinite, in the same sense as classical mechanics 
results from quantum mechanics in the ideal limit in which our intervention does not influence the system
 under observation (Planck's constant going to zero), so that a continuum decsription cannot hold at a 
fundamental level. From the operational point of view, spacetime is modeled and should be modeled by what 
our observations can tell us about it, and because these are necessarily of finite resolution then 
spacetime cannot be modeled by a set of point with the cardinality of the continuum. 

What description should we look for, then? One may take the 
drastic attitude that the ultimate theory of spacetime should not be based on any concept of 
spacetime at all, and that only some coarse graining procedure will lead us, in several steps, to an 
approximation where the concept of a \lq\lq local region" does indeed make sense, but not as a subset 
of a \lq\lq spacetime", but as an emergent concept, and where a collection of such regions can be defined 
and then {\it interpreted} as the collection of open coverings of a continuum manifold. In this way it 
would be just like there exists some continuum manifold representing spacetime, to which our collection 
of local regions is an approximation, although no continuum manifold plays any role in the formulation 
of the fundamental theory, so that the collection of local regions, if not truly fundamental, 
is definitely \lq\lq more fundamental" than the continuum manifold itself. Of course, if the physical 
question one is posing allows for it, one may at this point use the emerging continuum manifold as a 
description of spacetime and use the theory in this approximation. It may also be, if one does not 
subscribe to any realistic point of view for scientific theories, that there is no real meaning to the 
idea of a \lq\lq fundamental theory", but also in this case, the description of spacetime in 
terms of collections of local regions as opposed to that using a continuum manifold seems to be more 
reasonable, also in light of the operational point of view mentioned above\cite{isham}\footnotemark. \footnotetext{It is maybe 
interesting to note that, even in the continuum case, a topology of a space $M$ is indeed defined as the 
family of open subsets of $M$.} 

If a continuum description of spacetime is not available at a fundamental level, either because we are 
forced by the theory itself to take into account the not infinite precision of our measurements, or 
because there exists a minimum length that 
works as a least bound for any geometric measurement, as implied by $SU(2)$ loop quantum gravity, and 
hinted at from many different 
points of view, or because simply we want to avoid the idealization of a spacetime {\it point}, then we 
are forced to 
consider what kind of discrete substratum may replace the usual smooth manifold. 
The continuum may be 
regarded as a fundamentally problematic idealization in that it involves an infinite number of elements 
also requiring an infinite amount of information to be distinguished
from each other, so we want to look for some substitute of it in the form of a topological space (in 
order to remain as close
as possible to the approximation we want to recover) in which any bounded region contains only a finite 
number of elements, 
i.e. we would work with a \lq\lq finitary" topological
space\cite{SorkinFinit, Porter}. The collection of local regions mentioned 
above would be in fact a space of this sort. Of course, a necessary condition for a finitary topological
 space to be sensibly used as a substitute for a continuum 
spacetime manifold is that the former does indeed approximate in a clear sense the latter, and that a 
limit can be defined in which this approximation does indeed become more and more precise.

Let us be a bit more precise.
Given a topological space $S$, from the operational point of view the closest thing to a point-event we 
can possibly measure is an open subset of $S$, and it is important to note that the definiton of the 
topology of $S$ is exactly given by a family of open subsets of $S$, so that to have access to only a 
finite number of open sets means to have access to a subtopology $U$ of $S$. Operationally then, our 
knowledge about the topology of $S$ can be summarized in the space $F(U)$ obtained by identifying with each
 other any point of $S$ that we cannot be distinguished by the open sets in $U$. Therefore, given 
the open cover $U$ of $S$, assumed to form a subtopology of $S$, i.e. to be finite and closed under union 
and intersection, 
we regard $x,y\in S$ as equivalent if and only if $\forall u\in U, x\in U \Leftrightarrow y\in U$.
 Then $F(U)$ is the quotient of $S$ with respect to this equivalence relation. Another way to 
characterize the space $F(U)$ is considering it as the $T_0$-quotient of $S$ with respect to the 
topology $U$\footnotemark. \footnotetext{A space is a $T_0$-space if for any pair of its points there is an open set 
containing one but not the other of the two} 
If one considers $F(U)$ as an \lq\lq approximation" to a topological space $S$, one can show that such approximation becomes
 exact
as more and more open sets are used, i.e. as the open covering is refined. Of course, without such a result, it would be 
impossible to claim that finitary topological 
spaces or posets represent a more fundamental substratum for a continuum spacetime\cite{SorkinFinit}. 

We note that a finitary topological space has an equivalent description as a poset, i.e. a partially 
ordered set; in fact the collection $C$ of subsets of the set $F$ is closed under union and intersection,
and consequently one can define for any $x\in F$ a smallest neighborhood 
$\Lambda(x)=\cap \{ A\in C | x\in A\}$. Because of this the natural ordering of subsets translates into a 
relation among the elements of $F$:
 $x\rightarrow y \Leftrightarrow \Lambda (x)\subset \Lambda(y) \Leftrightarrow x\in \Lambda (y)$.  
This relation is reflexive ($x\rightarrow x$) and transitive 
($x\rightarrow y \rightarrow z \Rightarrow x\rightarrow z$).
On the other hand, one can proceed the other way round, because any relation between elements of $F$ which
 satisfies these properties gives a topology, by defining for any 
$x\in F$, $\Lambda(x)\equiv \{ y\in F | y\rightarrow x \}$, and defining a subset  $A\subset F$ to be 
open if and only if $x\rightarrow y\in A \Rightarrow x\in A$.
Moreover, such a relation defines a topology that makes $F$ a $T_0$ space iff no circular order relation 
($x\rightarrow y \rightarrow x$) ever occurs. This means that $(F,C)$ is a $T_0$ space if and only if 
$F$ is a partially ordered set (poset). This characterization of a finitary space as a poset is useful to 
make a link to the causal set approach we will discuss in the following \cite{sorkinspeci, SorkinFinit}.

Another very important (although well known) equivalent characterization of a finitary space such as that
 represented by an open covering of a given manifold is in terms of simplicial complexes.
Indeed from any open covering one can construct a simplicial complex by considering the so-called 
\lq\lq nerve" of the covering, which is the simplicial complex having as
vertices the open sets of the open covering, and as simplices the
finite families of open sets of it whose intersection is non-empty. Therefore we clearly see how all 
models that one may construct for quantum gravity, which are based on the use of simplicial complexes as
substitutes for a continuum spacetime, are just applications of this idea of replacing a smooth manifold 
with a finitary substitute, and are thus not a technically useful approximation, but on the
contrary more close to a fundamental description of spacetime\cite{Porter}, at least if one accepts the point of view
 we are advocating here. In other words, simplicial
 gravity\cite{ReggeWill, Loll}, e.g. in the Regge calculus formulation\cite{Will}, 
represents such a finitary model for gravity with on top of it a certain operational flavor, with finitary
 patches of flat space replacing the spacetime continuum manifold, thus realising at the same time both 
this finitary philosophy and the equivalence principle. One can imagine each 4-simplex to represent 
something like an {\it imperfect} determination of localization\cite{sorkinspeci}, thus replacing the 
over-idealized concept of a spacetime point, so that the simplicial complex used in place of the continuum
 manifold has to be intepreted as the probing of spacetime by a number of imperfect thus real measurements,
 or in other words, the finitary topological space represented by the simplicial complex is the spacetime 
we reconstruct from our set of measurements. In this case, the simplicial complex we use is not the nerve 
of the open covering mentioned above, but equivalently the open covering itself; of course, one can then go
 to the simplicial complex dual to the original one, and this is precisely the {\it nerve} of the 
simplicial open covering; this is the dual complex used in spin foam models, which is then seen as also 
arising naturally from this operational perspective\footnotemark. \footnotetext{Note that the Regge Calculus action itself can be 
defined on a simplicial complex even if the simplicial complex is not a manifold, but only a pseudo-manifold.}  

It is very important to note that the structures of open coverings, nerves,
simplicial complexes associated to a space make perfect sense even
when such a space does not consist of points nor is it anything like a
continuum manifold, and are thus much more general than this. Both manifolds and non-manifold
spaces give rise to the same kind of open coverings and simplicial
complexes, and both are retrieved in the limit of finer coverings.
Therefore the difference between manifolds and non-manifold spaces
cannot be seen from the open coverings alone, but must lie in the
relationship between different open coverings, i.e. in the refining
maps. It is the refinement procedure that gives us either a manifold
or a non-manifold structure, depending on how it is defined and
performed, and these subtleties must be taken into account when considering a refinement of a simplicial or
finitary model for quantum gravity associated to a classical limit\cite{Porter}. 

In any case, from an operational perspective, the
triangulated structure, the open covering, the simplicial complex, all
these structure are not to be thought of as imposed on the manifold,
or as approximations of it, but are the result of the process of
``observing'' it, they are a representation of a set of observations
of spacetime. This change in perspective should affect the way certain
issues are faced, like for example the issue of ``triangulation
independence'' or the way a sum over triangulations or 2-complexes is
dealt with (should one take the number of vertices to infinity? what
are diffeomorphisms from this point of view? what if the underlying
space we try to recover is not a manifold?)\cite{Porter}.        

If a discrete substratum is indeed chosen and identified as a more fundamental description of what spacetime is, then
 in some sense the situation we would face in quantum gravity would closely resemble that of solid state
 physics where field theoretical methods are used adopting the {\it approximation}
of a smooth manifold in place of the crystal lattice, and where the cut-off enforced in the so constructed
 field theories has a clear explanation and justification in terms of the atomic structure of the space 
on which these fields live (the crystal lattice)\cite{sorkinspeci}. 

We note also that, depending on the attitude one takes towards these questions, one is then forced
to have different points of view on the role of the 
diffeomorphism group, on the way diffomorphism invariance has to be implemented and on its physical 
meaning, points of view that necessarily affect how one deals with the various quantum gravity models and 
their technical issues. 

\subsection{Covariance}
By covariance we mean an as symmetric as possible treatment for space
and time, as suggested, if not imposed, by the same ``covariance'' of
General Relativity, which is intrinsically 4-dimensional and where a
$3+1$ splitting of spacetime into a space-like hypersurface and a time
evolution is possible only if some additional restriction is imposed on
the spacetime manifold considered (global hyperbolicity).

There are two main ways in which one can look for a covariant
formulation of a quantum theory of gravity: one is by adopting a
conventional Hamiltonian approach but starting from a covariant
formulation of General Relativity as a Hamiltonial system, phrasing GR
into the general framework of covariant or presymplectic Hamiltonian
mechanics \cite{Arnold, carlomech}; the other is to use a
sum-over-histories formulation of quantum mechanics\cite{hartle}; these two
approaches are by no means alternative, and we will see in the
following how they actually co-exist.     

Let us first discuss the covariant formulation of classical mechanics.
The best starting point is the analysis done by Rovelli
\cite{carlomech, carlobook}, that we now
outline.

It can be shown\cite{Arnold,carlomech} that any physical system can be completely described in
terms of: 1) the configuration space $C$ of partial observables, 2) the
phase space of states $\Gamma$, 3) the evolution equation $f=0$, with
$f: \Gamma\times C\rightarrow V$, where $V$ is a linear space. The phase space can be taken to be the space of 
the orbits of the system, or as the space of solutions of the equations of motion, and each solution determines a surface in
$C$, so the phase space is a space of surfaces in $C$ (infinite-dimensional in the field theory case). However, in a field 
theory, because of this infinite-dimensionality, it is convenient to use a different space $\mathcal{G}$ defined as the space
 of boundary configuration data that specify a solution, which in turn are the possible boundaries of a portion of a motion 
in $C$, i.e. 3-dimensional surfaces in $C$; therefore we define the space $\mathcal{G}$ as the space of (oriented) 
3-d surfaces, without boundaries, in $C$. The function $f$ expresses a motion, i.e. an element of $\Gamma$, as a relation 
in the extended configuration space $C$, i.e. as a correlation of partial observables.

The main difference between a relativistic system and a
non-relativistic one is in the fact that the configuration space
admits a (unique) decomposition into $C=C_0\times\mathbb{R}$, with one
of the partial observables (configuration variables) named $t$ and
called ``time'', with a consequent splitting of the Hamiltonian
(evolution equation) as $f=H=p_t+H_0$, with $H_0$ being what is usually
named ``Hamiltonian'', giving rise to a Schroedinger equation in
quantum mechanics as opposed to a constraint or Wheeler-DeWitt
equation as in canonical quantum gravity.

In the case of a single massive relativistic particle the objects indicated
above are easy to identify. The configuration space $C$ of partial
observables is simply Minkowski space, with the partial observables
being given by the coordinates $X^{\mu}$ of the particle, and the
correlations of partial observables (physical predictions of the
theory) being the points-events  in Minkowski space. The space of
states or motions, identified in $C$ by means of the function $f$, is the space of timelike geodesics in Minkowski
space, the Hamiltonian is $f=H=p^\mu p_\mu  + m^2$, and everything is
Lorentz invariant.  

In the
case of Hamiltonian General Relativity \cite{carlomech, carlobook}, the configuration space $C$ is given by
the real $(18+4)$-dimensional space with coordinates given by the (space and Lorentz) components of
the gravitational connection $A_{a}^{IJ}$ and by the spacetime coordinates $x^\mu$; the space $\mathcal{G}$ is given by the 
space of parametrized 3-dimensional
surfaces without boundaries $A$ in $C$, with coordinates the spacetime coordinates $x^\mu$, $A_{a}^{IJ}$
and their conjugate momenta $E^{a}_{IJ}$ (dual to the 3d restriction
of the 2-form $B$ appearing in the Plebanski formulation of GR we will
deal with later); this space $\mathcal{G}$ is infinite dimensional as it should be, since we are dealing with a field 
theory; the evolution equations are given by the usual canonical
constraints (Gauss, 3-diffeomorphism, and Hamiltonian) imposing the
symmetries of the theory, expressed in terms of variations of the
action with respect to $A$. 

While this is basically all well known, the
crucial thing to note is that this is nothing but the presymplectic
structure that is possessed by any classical mechanical theory, if one
does not make use of any splitting of the configuration space leading
to a choice of a global time parameter; for more details see
\cite{carlomech, carlobook}.

Now, it is of course crucial to identify what are the observables of
the theory, by which we mean the physical predictions of it; we know
they are given by correlations of partial observables, but we still
have to identify them taking into account the symmetries of GR. These complete 
observables cannot be given simply by points in $C$, because of the non-trivial behaviour of $A$ under 
the symmetry transformations of the theory (gauge and diffeos), so they must be represented by extended objects in $C$.

These extended objects are what we may call \lq\lq classical spin
networks"\cite{carlomech, carlobook}: take a graph $\gamma$ (a set of points joined
 by links) embedded in $C$; a sensible question the theory has to answer is whether the set of correlations given by $\gamma$
is realised in a given state (point) in $\mathcal{G}$. However, it is easy to see that the theory does not distinguish 
between loops (closed unknotted curves) $\alpha$ in $C$ for which the holonomy of the connection $A$, given by
$T_\alpha = Tr U_\alpha = Tr P e^{\int_\alpha ds \frac{dx^\mu (s)}{ds}A^{IJ}_\mu(s)J_{IJ}}$, with $J_{IJ}$ being the 
generators of the Lorentz algebra, is the same, so the prediction depend on $T_\alpha$ only. Therefore the observables of 
the theory will be given by the knot class $[\gamma]$ to which the restriction of $\gamma$ to the spacetime manifold $M$
 belongs and by a set of holonomies $T_\alpha$ assigned to the all the possible loops $\alpha$ in $\gamma$. The object
$s=([\gamma],T_\alpha)$ is a classical spin network, and can be described as an invariant (with respect to the symmetries of
 the theory) set of correlations in the configuration space $C$. Its operational meaning is also clear, as it is its 
possible practical realization: it is given by the parallel transport of a physical reference frame (gyroscopes, etc.) along
 given paths in spacetime, so that the quantities $T_\alpha$ are basically given as collections of angles and relative 
velocities, with the invariant information provided by such an operation being thus given by the combinatorial topology of
 the paths (graph) chosen and by these relative angles and velocities. What the classical theory tells us is whether such 
a set of correlations is realised in a given motion, and what a quantum theory of gravity is expected to be able to tell us
 is the probability for a given quantum spin network (a set of quantum correlations) to be realized in some given quantum 
state\cite{carlomech, carlobook}.

Can the quantum theory of mechanical systems be formulated also
in a covariant way, being still based on the Hilbert space plus
hermitian operators formalism? the answer is yes \cite{carlomike} and we will now
discuss briefly what such a covariant formulation of quantum mechanics
looks like.

It is very important to note that, while a greater generality, and the
similarity with the structure of General Relativity was the motivation
for using classical mechanics in its pre-symplectic form, when coming
to the quantum theory we get additional motivations from the
operational point of view. Indeed the splitting of spacetime into a
3-dimensional manifold evolving in time is linked with a corresponding
idealization of physical measurements as able to achieve an exact time
determination of the time location of an event, i.e. as happening 
instantaneously. While maybe useful in practice, this is certainly an
approximation that may not be justified in most of the cases and it is
certainly detached from the reality of our interaction with the
world. Just as one is motivated by similar considerations in developing
quantum mechanics in terms of ``wave packets'' more or less spread in
space, and in using them as a basic tool, one can consider spacetime
smeared states or spacetime wave packets as more realistic objects to
deal with when describing physical systems.
Of course, the motivation for a covariant form of classical mechanics (greater
generality, bigger consonance with the 4-dimensional formalism of GR)
still holds in this quantum case.
  
Consider a single non-relativistic particle first; denoting the eigenstates of the unitarily evolving Heisenberg position 
operator $X(T)$ as $\mid X,T \rangle$, the spacetime 
smeared states are defined as 
\bes
\mid f \rangle=\int dXdT f(X,T)\mid X,T\rangle, \ees with $f$ being a suitable spacetime smearing 
function, and their scalar product is given by 
\bes \langle f\mid f'\rangle=\int dXdT \int dX'dT' f^{*}(X,T)W(X,T;X',T')f'(X',T'),\ees 
where $W(X,T;X',T')=\langle X,T\mid X',T'\rangle$
is the propagator of the Schroedinger equation. 

The states $\mid f\rangle$ are basically spacetime wave packets, and are the
realistic localized states that can be detected by a measurement apparatus with a spacetime resolution corresponding to the 
support of the function $f$. One can then show that for sufficiently small spacetime regions considered it is possible to 
define consistently the probability of detecting a system in such regions, having the usual quantum probabilistic formalism;
these states are the starting point for a covariant reformulation of quantum theory. Consider the space $\mathcal{E}$ of test
functions $f(X,T)$ and the linear map $P:\mathcal{E}\rightarrow \mathcal{H}$ sending each function $f$ into the state 
$\mid f\rangle$ defined as above, whose image is dense in $\mathcal{H}$, being the linear space of solutions of the 
Schroedinger equation; of course the scalar product for the states $\mid f\rangle$ can be pulled back to $\mathcal{E}$, by 
means of the propagator $W$; therefore the linear space $\mathcal{E}$, equipped with such a scalar product, quotiented by the 
zero norm states, and completed in this norm, can be identified with the Hilbert space $\mathcal{H}$ of the theory, and 
we see that the propagator $W(X,T;X',T')$ contains the full information needed to reconstruct $\mathcal{H}$ from $\mathcal{E}$.
Also, one can show that the propagator itself can be constructed directly from the Schroedinger operator 
$C=i \hbar \frac{\partial}{\partial T}+\frac{\hbar^2}{2m} \frac{\partial^2}{\partial X^2}$, for example using group averaging 
techniques, i.e. as the kernel of the bilinear form on $\mathcal{E}$:
\bes 
(f,f')_C=\int_{-\infty}^{+\infty}d\tau \int dXdT f^{*}(X,T)[e^{i \tau C}\,f'](X,T)
\ees    
defined in terms of $C$; note that the dynamics of the theory is expressed in terms of the constraint $C=0$ with no special 
role played by the time variable $T$; 
this framework can be generalised to any mechanical system, defined by an extended (in the sense of including any time 
variable, and of not being of the special form $\Sigma\times\mathbb{R}$) configuration space $\mathcal{M}$, with elements $x$,
 and by a dynamical (set of) constraint(s) $C=0$. Starting with the space $\mathcal{E}$ of test functions, we define the 
propagator $W(x,x')$ as the kernel of the bilinear form 
$(f,f')_C=\int_{-\infty}^{+\infty} d\tau\int dx f^{*}(x)[e^{i\tau C}f'](x)$;
this propagator defines the dynamics of the theory completely, and the Hilbert space of the theory is defined from 
$\mathcal{E}$ 
as outlined above, using the bilinear form written and the map $P$
  sometimes called the \lq\lq projector" onto physical states \cite{carlomike}.

A sketch of the generalization to the relativistic particle case can
also be given. The kinematical states $\mid s\rangle$ are vectors in a Hilbert space
$\mathcal{K}$, with basis given by complete eigenstates of suitable
self-adjoint operators representing the partial observables, e.g. the
spacetime coordinates $X^\mu$. In the same Hilbert space one can
construct spacetime smeared states just as in the non-relativistic
case described above, based on smearing functions $f$ on the extended
configuration space $C$, and the corresponding probability of detecting
the particle in some region of spacetime. The dynamics is defined by a
self-adjoint operator $H$ in $\mathcal{K}$, i.e. the quantized
relativistic Hamiltonian constraint, and one can from this define and compute
transition amplitudes again by means of the so-called ``projector operator''
from the kinematical space to the space of physical states (solutions
of the constraint equation), analogous to the one defined above, and
that may be thought of being defined by means of its matrix elements
given by the relativistic analogue of the scalar products $(f,f')_C$; this
relativistic formulation is described in \cite{carlomike}. 

Let us now turn to another approach to covariance in quantum
mechanics: the sum-over-histories approach\cite{hartle}.

First of all, let us give some motivations for taking such an
approach. 

In spite of the formal and conceptual differences, the
canonical framework may also be seen as a special case of a 
sum-over-histories formulation, arising when one allows for states $\Psi(t)$ associated with a given moment of time $t$ and 
realized as a particular sum over past histories; moreover, it can be taken to be just a way to \lq\lq define" the transition amplitudes (propagators) of the 
canonical theory in terms of path integrals, as we are going to
discuss in the following for the case of a relativistic
particle. Therefore, one may simply look for a sum-over-histories
formulation of canonical quantum mechanics because of its greater
generality\cite{hartle,sorkinforks}.

However, there is also hope that a sum-over-histories approach applied to gravity 
will make it easier to deal with time-related issues, which are instead particularly harsh to dealt with in the 
\lq\lq frozen" canonical formalism. Also, such an approach seems to be better suited for dealing with a closed system such as 
the universe as a whole, one of the \lq\lq objects" that a quantum theory of gravity should be able to deal with. Furthermore, 
because of diffeomorphism invariance, questions in gravity seem to be all of an unavoidable spacetime character, so that, 
again, a spacetime approach treating space and time on equal footing
seems preferrable.  Moreover, a sum-over-histories 
approach in quantum gravity is necessary if one wants to allow and study processes involving topology change, either spatial topology change or
change in spacetime topology\cite{sorkinforks}; if the topology itself has to be considered as a dynamical variable then the only way to do 
it is to include a sum over topologies in the path integral defining the quantum gravity theory. Also, the study of 
topological geons \cite{sorkingeons} seems to lead to the conclusion that a dynamical metric requires a dynamical topology, so, indirectly, 
requires a sum over histories formulation for quantum gravity. Because of Geroch's theorem, on the other hand, if one wants 
topology change in the theory and at the same time wants to avoid such pathologies as closed timelike curves, then one is 
forced to allow for mild degeneracies in the metric configurations, i.e. has to include, in the sum over metrics 
representing spacetime histories of geometry, configurations where the metric is degenerate at finite, isolated spacetime 
points\footnotemark \cite{sorkinforks} \footnotetext{It may also be that, even if one allows closed timelike curves in the formulation of the theory, then 
configurations where these are present do not correspond to any continuum configuration and result in being suppressed in the 
classical limit}. 

As we said,  a sum-over-histories formulation of quantum mechanics is not necessarily alternative to the usual canonical 
formulation, and can be used to define the transition amplitudes (propagators) of the 
canonical theory in terms of path integrals. Let us see how this is
done\cite{Teitelboim,Halliwell}.

Consider a single relativistic particle. Its path integral quantization is defined by:
\be
Z(x_1,x_2)\,=\,\int_{x_1=x(\lambda_1),x_2=x(\lambda_2)}\,(\prod_{\lambda\in
[\lambda_1,\lambda_2]} d^4 x)\,e^{i\,S(x)}. \ee 
We want to understand how different Green's functions and transition
amplitudes come out of this same
expression. 
After a gauge fixing such as, for example, $\dot{N}=0$, one may
proceed to quantization integrating the exponential of the action
with respect to the canonical variables, with a suitable choice of
measure. The integral over the \lq\lq lapse" $N$ requires a bit of
discussion. First of all we use as integration variable
$T=N(\lambda_2 -\lambda_1)$ (which may be interpreted as the
proper time elapsed between the initial and final state). Then
note that the monotonicity of $\lambda$, together with the
continuity of $N$ as a function of $\lambda$ imply that $N$ is
always positive or always negative, so
 that the integration over it may be divided into two disjoint classes
$N>0$ and $N<0$. Now one can show that the integral
over both classes yields the Hadamard Green function:
\bes
G_H(x_1,x_2)\,=\,\langle x_2\mid x_1\rangle &=&
\int_{-\infty}^{+\infty} dT\,
\int (\prod_\lambda d^4 x\, d^4 p)\, e^{i\,\int d\lambda (p x - T
\mathcal{H})} \nonumber \\
&=& \int (\prod_\lambda d^4 x\, d^4 p)\, \delta(p^2\, +\,
m^2)\,e^{i\,\int d\lambda\, (p x)} \ees which is related to the
Wightman functions $G^{\pm}$, in turn obtained from the previous
expression by inserting a $\theta(p^0)$ in the integrand and a
$\pm$ in the exponent, by: \be G_H(x_1,x_2) \, = \,
G^+(x_1,x_2)\,+\,G^-(x_1,x_2)
\,=\,G^+(x_1,x_2)\,+\,G^{+}(x_2,x_1). \label{eq:had} \ee This
function is a solution of the Klein-Gordon equation in both its
arguments, and does not register any order between them, in fact
$G_H (x,y)=G_H (y,x)$. Putting it differently, it is an acausal
transition amplitude between physical states, or a physical inner
product between them, and the path integral above can be seen as a
definition of the generalized projector operator that projects
kinematical states onto solutions of the Hamiltonian constraint:

\be
G_H(x_1,x_2)\,=\,\langle x_2\mid x_1\rangle_{phys}\,=\,_{kin}\langle x_2 \mid
\mathcal{P}_{\mathcal{H}=0}\mid x_1\rangle_{kin} .
\ee

On the other hand, one may choose to integrate over only one of the two
classes corresponding to each given sign of $N$, say $N>0$.
This corresponds to an integration over all and only the histories for
which the state $\mid x_2\rangle$ lies in the future
of the state $\mid x_1\rangle$, with respect to the proper time $T$, and
yields the Feynman propagator or causal amplitude:
\be
G_F(x_1,x_2)\,=\,\langle x_2\mid x_1\rangle_C\,=\,\int_{0}^{+\infty} dT\,
\int (\prod_\lambda d^4 x\, d^4 p)\, e^{i\,\int d\lambda (p x - T
\mathcal{H})},
\ee
which is related to the Wightman functions by:
\bes
G_F(x_1,x_2) = \langle x_2\mid
x_1\rangle_C &=&
\theta(x_1^0 - x_2^0) G^+(x_1,x_2) +
 \theta(x_2^0 - x_1^0) G^-(x_1,x_2)
\nonumber \\
&=&
\theta(x_1^0 - x_2^0) G^+(x_1,x_2)
 + \theta(x_2^0 - x_1^0) G^+(x_2,x_1)\,\,\,.
\ees

Note that all trajectories of interest, for both orientations of $x_0$
with respect to $\lambda$, are included in the above
integral, if we consider both positive and negative energies, so no
physical limitation is implied by the choice made above.
The resulting function is not a solution of the Klein-Gordon equation, it
is not a realization of a projection onto solutions
 of the Hamiltonian constraint, but it is the physical transition
amplitude between states which takes into account causality
requirements (it corresponds, in field theory, to the time-ordered
2-point function).

In which sense is this sum-over-histories formulation more general
 that the canonical one? On the one hand, in fact, it gives a more
 restrictive definition of the transition
amplitudes, since it seems to favour configuration space variables over momentum space variables, while the usual 
Schroedinger-Heisenberg formulation uses the whole set of momentum
 states given by Fourier transformation. On the other hand, however, 
it allows easy generalizations of the spacetime configurations to be considered other than sequences of alternative regions
of configuration space at definite successive moments of time, i.e. those only which are assigned probabilities by the 
Schroedinger-Heisenberg formalism, by which we mean configurations that cannot be represented by products of 
measurements-projections and unitary evolutions in a time sequence, e.g. it can be generalised to include spacetime configurations 
in which the system moves back and forth in a given coordinate
 time. Let us stress that the reason for this impossibility does not reside in 
the use of path integrals versus operators {\it per se}, but in the fundamental spacetime nature of the variables considered 
(the histories). In a sum-over-histories formulation of quantum gravity, based on 4-dimensional manifolds and metrics 
interpolating between given 3-dimensional ones, weighted by the (exponential of $i$ times the) gravitational action, such generalized alternatives  
would correspond to 4-dimensional metrics and manifolds that do not admit a slicing structure, i.e. that are not globally 
hyperbolic. Of course a sum-over-histories formulation should be completed by a rule for partitioning the whole set of 
spacetime histories into a set of exhaustive alternatives to which the theory can consistently assign probabilities, using 
some coarse-graining procedure, and this is where the mechanism of
 decoherence is supposed to be necessary; such a mechanism looks 
particularly handy in a quantum gravity context, at least to some
 people \cite{hartle}, since it provides a definition of physical 
measurements that does not involve directly any notion of observer, and can thus apply to closed system as the universe as 
a whole. Whether such notions of observer-independent systems, closed systems, or universe-as-a-whole are of any significance
is of course questionable, expecially if one adopts the \lq\lq operational" point of view advocated here. In any case, 
the sum-over-histories formulation of quantum mechanics does not need these motivations to be preferred to the usual one, 
as its greater generality, as we said, makes it a preferrable framework if spacetime, as opposed to space, is the system under investigation\cite{hartle,sorkinforks}.

\subsection{Symmetry}
We anticipated above the expectation that algebra, together with
combinatorics, should represent the main language to describe a
quantum spacetime. We have said that categorical algebra may be one
way in which this may happen. The discussion regarding general
relativistic observables in the context of covariant Hamiltonian
General Relativity lead naturally to the understanding of why and how the
algebra of the representation theory of the Lorentz group is going to
play a crucial role in the definition of observables, states and
histories at the quantum level. Another line of thought leading to the
same conclusion is that motivated by considerations about
operationality.

Geometric quantities, and all the geometry of spacetime, are to be characterized by operations 
we can perform on objects (maybe idealized) in spacetime. In particular, we have seen already at the classical level that the natural
observables for pure GR are classical spin networks, i.e. sets of
correlations 
obtained by suitable operations defined 
in terms of the gravitational connection, which is a Lorentz
connection. 
It is then natural to expect that it is the group of local transformations of spacetime 
objects, i.e. the Lorentz group, that furnishes the more fundamental characterization of such observables, also at the 
quantum level. Indeed a straightforward quantization of these
gravitational 
observables and states would then amount to a replacement of the
classical holonomies living on the links of the graph $\gamma$ with
quantum holonomies, i.e. elements of the Lorentz group living in some
representation of it, in other words with quantum operators acting on
a representation space associated with the link itself. Of course, the
quantum spin networks so constructed (with representations of the
Lorentz group on the links), being diffeomorphism invariant, i.e. not
embedded in any spacetime manifold, should also be gauge invariant,
and this has to be achieved by associating Lorentz intertwiners to the
nodes of the graph, mapping the tensor product of the representations
on the adjacent links to the invariant representation. The resulting
quantum objects are indeed quantum spin networks, as used in loop
quantum gravity and spin foam models. These characterize both
observables and states of the quantum theory, in those approaches

Therefore we see that, as anticipated, quantum gravity would live in a space
characterized by the possible representations of the Lorentz group,
i.e. of the fundamental transformations we can perform on spacetime objects. The geometry of spacetime will be described in terms of the algebra 
coming from the representation theory of the Lorentz group, from the
symmetry group of spacetime objects.

\subsection{Causality}
If one is wondering what ingredients a \lq\lq more fundamental" theory of spacetime should contain, in order to be able 
to determine its geometry in a quantum domain and then, with a suitable limiting or approximating procedure, in a classical
one, then a crucial observation is that, in a classical setting, spacetime causal structure (meaning the relation between 
points specifying whether a given point is causally related (in the future or in the past) of which other) is almost sufficient to 
determine its entire geometrical properties\cite{sorkincausal}. Usually one starts with a differentiable manifold, a metric field defining a 
light cone structure, and then reads out from this light-cone structure the causal structure of the manifold. However, also 
the opposite procedure actually works: given a set of elements (points) endowed with a partial ordering relation, then the 
embedding of such a set into a manifold of suitable smoothness allows the reconstruction of the topology of the manifold, 
of its differentiable structure and of its metric structure as well, up to a conformal factor. So one can say that a metric 
field and a poset structure are almost equivalent objects, but on the other hand it is difficult not to think that a partial
ordering relation, because of its simplicity, has to be regarded as somehow more fundamental. Of course a poset structure 
has a spacetime causal structure intepretation if and only if the signature of such a spacetime is the Lorentzian one, so 
that one may consider taking a poset as fundamental as a (maybe un-necessary) restriction of the class of theories we 
construct, or as an explanation of why our spacetime has to be of
Lorentzian signature\cite{sorkincausal}.

We can then say that, if a theory or model for quantum gravity has
encoded in it a clear notion of causal ordering, so that it allows for
the identification of a partially ordered set of events, then it
basically contains all the information that is needed to reconstruct a
metric field in some continuum limit/approximation. Also the
identification of a poset structure brings immediately the
identification of a finitary topology then leading to a reconstruction
of a spacetime manifold via suitable refinement procedure outlined
above. We see here a nice interplay between the notions of causality
and that of discreteness\cite{sorkinspeci}. 

The issue of causality and the previous considerations offer a new
perspective on the crucial and problematic issue of time in quantum
gravity\cite{ishamtime}.

Let us discuss this briefly.
From the point of view of causality as partial order, and from the
discussed implications of diffeomorphism invariance, we expect only some of the many features of
time and time variables in the continuum classical theory to be present
in the fundamental quantum gravity theory. These characteristics,
limiting ourself to those present in a general relativistic context,
are \cite{carlobook, Fraser}: directionality, i.e. the possibility to distinguish a ``before''
and an ``after''; metricity, i.e. the possibility to assign a length
to a time interval; 1-dimensionality, i.e. the possibility of
arranging the time intervals ina  1-dimensional manifold, e.g. the
real line. These are the only properties of time which make sense in
GR, since already in this classical context other properties, like
uniqueness of a time variable, or a global definition of time, or the
possibility of time being external to the dynamical fields appearing
in the theory, are all lost. 

In a more fundamental quantum gravity context, we may expect also the metricity to be dropped, or at least to be of a
much more subtle realization, as much more subtle should turn out to
be any arrangement of time instants or intervals in a 1-dimensional
array, if meaningful at all. Therefore, coherently with the point of view
outlined above, the only remaining property of time that we may want to
build in the fundamental theory is the notion of an ordering, although
its identification with a ``time ordering'' would certainly turn out
to be problematic. 

Of course it may well be that the theory we are looking for is such that it admits an acausal formulation and a causal 
one, that come into use depending on whether causality is an element entering the physical {\it question} we are interested
in getting an answer for from our theory. Even in this case, the possibility of a causal formulation, and its realization, is 
necessary for our theory to be considered satisfactory, since one can definitely come up with definite, meaningful questions
involving some sort of causal order.
In a sense, we may expect to be able to distinguish
in the theory two types of quantities, ordered or oriented ones, in
which one may recognise the seed of a time ordering, a kind of
thermodynamical time, where a future direction but no time variable is
specified, and un-ordered/un-oriented ones, where no time at all can
be identified. Both types of objects may cohexist because they may
correspond to different types of physical questions being posed, one
involving time and the other not, so that no harm comes from the fact
that the theory includes both. The example would be having two sorts
of transition amplitudes between two states A and B, one corresponding
to a question like ``what is the correlation between A and B?'', and
the other to a more temporal question like ``what is the probability
of observing B after having observed A?''.      
An example of a situation where this happens is again given by the
relativistic particle; we can define acausal transition
amplitudes, the Hadamard or Schwinger functions we defined above, naturally intepreted as real inner products of canonical
states, not distinguishing ``past'' and ``future'' states, not
registering any ordering among them, and
encoding correlations between states. Also we can define properly
causal transition amplitudes, like the Wightman functions or the
Feynman propagator, distinguishing an ordering of their arguments,
although not involving any explicit external temporal parameter (of
course an external temporal parameter in the relativistic particle
case would be the affine parameter along the particle trajectory, not
the coordinate time $x^0$ which is part of the definition of the states). 

The crucial point is thus the idea that the notion of time can be reduced
to a notion of ``ordering'' among elements of a given set, and that
this, being more fundamental that any continuum, classical notion of
a time variable (whatever this is taken to be), is what survives as
``time'' in a fundamental, quantum regime, as a seed from which
classical time emerges; the results of
discrete ordered calculus\cite{kauffman}, with the reconstruction of the notions of
derivatives, connections and metric, of canonical commutation
relations, of gauge fields, non-commutative versions of
electromagnetism and Dirac equation, from the basic notion of a
discrete ordered set of elements, are of clear support of this
possibility.

\section{Approaches to quantum gravity}
We want now to review briefly the main features of several approaches to quantum gravity, which are related in one way or 
another to spin foam models, either historically or conceptually or both. We will also highlight how and to which extent
the ideas discussed above play a role in all of them. The lines of research we shall discuss are: the path integral approach, 
topological field theory, loop quantum gravity, simplicial gravity in both the Regge calculus and the dynamical 
triangulations versions, causal sets and quantum causal histories.

\subsection{The path integral approach to quantum gravity}
Spin foam models can be naturally interpreted, as we shall see, as
a new form of the path integral approach to quantum gravity, so we give here the basic ideas of 
this line of research, referring to the literature \cite{Misn, Haw, HarHaw}
for more details. 

The key idea is that one can define a partition function (path integral) for quantum gravity as a sum
over histories of the gravitational field between two given states, or from the vacuum to the vacuum, 
if one just wants to define the partition function of the theory. The states are given by 3-geometries
relative to 3d hypersurfaces, not necessarily connected, and the histories are 4-geometries
inducing the given 3-geometries on the hypersurfaces (see figure 1.1). The weight for
each history, in the Lorentzian case, is the exponential of $i$ times the classical action, the
Einstein-Hilbert action (modulo boundary terms), as usual.

More precisely, the action, in
Hamiltonian terms \cite{Thiemann}, is: \be S\,=\,\int_\mathcal{M}\,( \pi^{ij}
\dot{g}_{ij}\,-\,N^i\,\mathcal{H}_i\,-\,N\,\mathcal{H} )d^4 x, \ee
where the variables are $g_{ij}$, the 3-metric induced on a
spacelike slice of the manifold $\mathcal{M}$, $\pi^{ij}$, its
conjugate momentum, the shift $N^i$, a Lagrange multiplier that
enforces the (space) diffeomorphism constraint $\mathcal{H}^i=0$,
and the lapse $N$ that enforces the Hamiltonian constraint
$\mathcal{H}=0$, again encoding the dynamics of the theory and the
symmetry under time diffeomorphisms, which are required to reduce
to the identity on the boundary, while the space diffeomorphisms
are unrestricted.

We are interested in the transition amplitude between
3-geometries. After a proper gauge fixing (such as $N^i=0$ and
$\dot{N}=0$, \lq\lq proper time gauge"), and the addition of
suitable ghost terms (see \cite{Teitelboim}), one may write a
formal path integral, integrating over the variables with some
given measure the exponential of ($i$ times) the action written
above. The integration over $N^i$ is now absent because of the
gauge condition chosen, but it could be anyway carried out without
problems, with $N^i(x)$ having integration range $(-\infty,
+\infty)$ at each point in space, projecting the states $\mid g_1\rangle$, $\mid
g_2\rangle$ onto diffeomorphism invariant states, while, again, as
in the particle case, the integral over $N(x)$ can be split into two
disjoint classes according to its sign, again at each point in space. Again, different choices
of the integration range yield different types of transition
 amplitudes.
Thus,
\be
\langle h_{2}, S_{2}\mid h_{1},S_{1}\rangle\,=\,\int_{g/
g(S_{1})=h_{1}, g(S_{2})=h_{2}}\mathcal{D}g\,\exp{(i\,I_{EH}(g))}
\ee 
where, as we said, we have to specify only the given 3-geometries,
meaning only the equivalence class of 3-metrics on the boundary
$\partial M$ of the 4-manifold $M$, under diffeomorphisms mapping the
boundary into itself.

\begin{figure}
\begin{center}
\includegraphics[width=7cm]{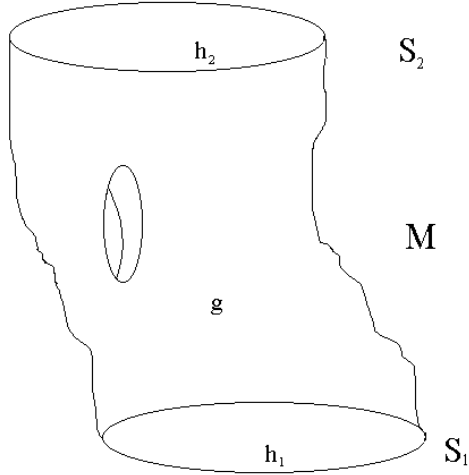}
\caption{Schematic representation of a 4-manifold (of genus one) with two disconnected boundary components}
\end{center}
\end{figure}

Of course one then asks for the transition from a 3-geometry $h_{1}$
to another $h_{2}$, and then to a third $h_{3}$ to be independent of
the intermediate state:
\be
\langle h_{3}, S_{3}\mid h_{1},S_{1}\rangle\,=\,\sum_{h_{2}}\langle
h_{3}, S_{3}\mid h_{2},S_{2}\rangle\langle h_{2}, S_{2}\mid
h_{1},S_{1}\rangle
\ee
corresponding to the gluing of two manifolds $M_{1}$, with boundaries
$S_{1}$ and $S_{2}$, and $M_{2}$, with boundaries $S_{2}$ and $S_{3}$,
along the common boundary $S_{2}$.
This is of course nothing more than the usual composition law for quantum mechanical amplitudes,
which however can be tricky in a sum-over-histories formulation \cite{Halliwell}

Also, the state themselves can be defined through a path integral of
the same kind, considering now manifolds with only one boundary, assuming the 
vacuum or ground state to be given by the manifold consisting of no points (the so-called Hartle-Hawking state), 
and consequently no metric \cite{HarHaw}:

\be
\langle 0\mid h_{1},S_{1}\rangle\,=\,\int_{g/
g(S_{1})=h_{1}}\mathcal{D}g\,\exp{(i\,I_{EH}(g))}
\ee

Topology itself is also allowed to change, and the path integral can
be completed by a sum over different topologies, i.e. over all possible 
4-manifolds having the given fixed boundaries, giving rise to a
foam-like structure of spacetime \cite{wheeler}, with quantum
fluctuation from one 4-metric to another for a given topology, but
also from one topology to another.
 
As we said, different choices of the integration range yield different types of transition amplitudes, 
as in the relativistic particle case \cite{Teitelboim}.  
The integration over both classes implements the projection onto
physical states, solutions of the Hamiltonian constraint, so that
we may formally write: \bes \langle g_2 \mid g_1\rangle_{phys}
&=&{}_{kin}\langle g_2\mid P_{\mathcal{H}=0}\mid
g_1\rangle_{kin}\,=\,_{kin}\langle g_2\mid
\int_{-\infty}^{+\infty}\mathcal{D}N(x)\,e^{i\,N\,\mathcal{H}}\mid
g_1\rangle_{kin}\nonumber \\
&=&\int_{-\infty}^{+\infty}\,\mathcal{D}N(x)\,\int_{g_1,g_2}
(\prod_{x}\mathcal{D}g_{ij}(x)\mathcal{D}\pi^{ij}(x))\,e^{i\,S},
\ees
where we have omitted the ghost terms. The symbol $\mathcal{D}$
indicates a formal product of ordinary integrals over $N(x)$ one for
each point in space, each with the integration range shown.

As in the particle case, this amplitude satisfies all the constraints, i.e.
$\mathcal{H}^\mu \langle g_2 \mid g_1\rangle = 0$,
and does not register any ordering of the two arguments. In this sense it
can be thus identified with the analogue of the
Hadamard function for the gravitational field. It gives the physical inner
product between quantum gravity states.

If we are interested in a physical, causal, transition amplitude between
these states, on the other hand, then we must take
into account the causality requirement that the second 3-geometry lies in
the future of the first, i.e. that the proper time
elapsed between the two is positive. This translates into the restriction
of the integration range of $N$ to positive values
only, or to only half of the possible locations of the final hypersurface
with respect to the first (again, for each point in space).

Then we define a causal tansition amplitude as:
\bes
\langle g_2 \mid g_1\rangle_C
&=&{}_{kin}\langle g_2\mid \mathcal{E}\mid
g_1\rangle_{kin}\,=\,_{kin}\langle g_2\mid
\int_{0}^{+\infty}\mathcal{D}N\,e^{i\,N\,\mathcal{H}}\mid
g_1\rangle_{kin} \nonumber \\
&=&\int_{0}^{+\infty}\,\mathcal{D}N\,\int_{g_1,g_2}
(\prod_{x}\mathcal{D}g_{ij}(x)\mathcal{D}\pi^{ij}(x))\,e^{i\,S},
\ees where we have formally defined the path integral with the
given boundary states as the action of an evolution operator
$\mathcal{E}$ on kinematical states, and the integral over $n(x)$ must
be understood as stated above.

The causal amplitude is not a solution of the Hamiltonian constraint, as a
result of the restriction of the average over only
half of the possible deformations of the initial hypersurface, generated
by it; on the other hand, in this way causality is
incorporated directly at the level of the sum-over-histories formulation
of the quantum gravity transition amplitude.

This approach faces many technical problems, most notably in the definition 
of the integration measure on the space of 4-metrics modulo spacetime diffeomorphisms, i.e. in the 
space of 4-geometries, and the definition of the partition function and of the amplitudes remains 
purely formal. This may be also thought to be due to the problems, both technical and conceptual, of 
basing the theory on a continuum spacetime. However, we see at work in it some of the ideas mentioned
 above as key features of a future quantum theory of gravity: covariance, as the model is a 
sum-over-histories formulation, background independence, although being realized just as in classical 
General Relativity and at the quantum level only formally, and causality, with the possibility of 
constructing (again, formally) both causal and acausal transition amplitudes.

\subsection{Topological quantum field theories}
A second approach that converged recently to the spin foam formalism
is represented by topological quantum field theories, as axiomatized
by Atiyah\cite{Atiyah} and realized in several different
forms, including spin foam methods\cite{Witten, Witten2, RT,PR,TV,CY,CKY}. In
fact, 3-d gravity is a topological field theory with no local degrees
of freedom, as we shall discuss, and fits in the framework of Atiyah's
axioms, that in turn can be realized by path integral methods, either
in the continuum \cite{Witten} or in the discrete \cite{PR,TV}
setting, in close similarity with the path integral approach to
quantum gravity summarized in the previous paragraph. Let us explain
the general structure of a topological field theory\cite{Atiyah,Picken}.

The most succint (but nevertheless complete) definition of a
topological quantum field theory is: a TQFT $Z$ is a functor from the
category of n-dimensional cobordisms (nCob) to the category of Hilbert
spaces (Hilb): $Z: nCob\rightarrow Hilb$.
More explicitely, a TQFT $Z$ assigns a Hilbert space $Z(S)=V_S$ (object in
$Hilb$), over a field $K$, to each
$(n-1)$-dimensional manifold $S$ (object in $nCob$), with vectors in
$Z(S)$ representing states ``of the universe'' given that the (spatial) universe is the manifold $S$, and a linear operator $Z(M): Z(S)\rightarrow
Z(S')$  (morphism in $Hilb$) to each n-cobordism ($n$-dimensional
oriented manifold with boundary) $M$
(morphism in $nCob$) interpolating between two $(n-1)$-dimensional
manifolds $S$ and $S'$. Equivalently, $Z(M)$ can be thought of as an
element of $Z(\partial M)=V_{\partial M}$. 
  
This assignment satisfies the following axioms: 1) the assignment is
functorial, i.e. holds up to isomorphisms in the relevant categories;
2) $V_{-S}=V^{*}_{S}$, where $-S$ is the same manifold $S$ but with
opposite orientation, and $V_{S}^{*}$ is the dual vector space to
$V_{S}$; 3) with $\cup$ denoting disjoint union, $V_{S_1\cup
  S_2}=V_{S_1}\otimes V_{S_2}$ and $Z_{M_1\cup M_2}=Z_{M_1}\otimes
Z_{M_2}$; 3) (gluing axiom) $Z_{M_1\cup_{S} M_2}=\langle Z_{M_1} \mid
Z_{M_2}\rangle$, where $\cup_S$ indicates the gluing of $M_1$ and
$M_2$ along a common (portion of the) boundary $S$, and
$\langle\cdot\mid\cdot\rangle$ is in this case given by the evaluation
of $V_S^{*}$ on $V_{S}$; 4) appropriate non-triviality conditions,
including $V_{\empty}= K$, so that the TQFT associates a numerical
invariant to any closed $n$-dimensional manifold. 
 
 Also, to fulfil the axioms of a category, we must have a
 notion of composition of cobordisms, which is non-commutative, and
 the gluing axiom can be understood as the corresponding usual composition of linear
 maps, and an identity cobordism $I_S:S\rightarrow S$, mapped to the
 identity on the Hilbert space, such that $I_S M= M I_S =M$. 

The overall analogy with the path integral approach to quantum
 gravity should be clear, as it should be clear that the structure
 given by these axioms furnishes a very abstract algebraic
 characterization of the notion of time evolution, that can be
 suitable for quantum gravity as well, with the adjoint operation on
 operators acting on the Hilbert spaces being the quantum translation
 of the ordering reversal on the cobordism, exchanging the role of
 past and future. 

In the original formulation, the manifolds considered are smooth
manifolds, and the isomorphisms in their category are diffeomorphisms,
and the Hilbert spaces considered are finite dimensional,
but the axioms are more general in that they allow for the use of the
piecewise linear category, with simplicial manifolds, instead, for example, or of infinite
dimensional vector spaces, while the data needed for the actual
construction of the TQFT functor may be taken from the algebra of the
representation theory of a given group, as it is indeed done in spin
foam models, as we shall see.       

Indeed, this general formulation of a TQFT has been convincingly
argued by many \cite{barrett, crane95, baezalgebra,carlomikeproj,robertQM} to
be general enough to furnish a mathematical and conceptual framework
for a 4-dimensional theory of quantum gravity, if one uses infinite
dimensional Hilbert spaces as spaces of states for the theory.  

\subsection{Loop quantum gravity}
Another approach closely related to spin foam models is that of loop
quantum gravity; in fact, the first spin foams models ever constructed \cite{mike, mike2,carlomikeproj} were indeed directly
inspired or derived from the formalism of loop quantum gravity, and
also the very first spin foam model for 3d gravity \cite{PR}, recognised as
such in retrospective, is closely linked with it \cite{carloPR}, as we shall discuss. 

Let us then give an outline of the main features of this approach,
referring to the existing reviews \cite{Thiemann, carloreview} for a more complete account of these.

We start with an ADM formulation of classical gravity in terms of
local triads $e^{i}_{a}$ (related to the 3-metric by $h_{ab}=e^{i}_{a}e^{i}_{b}$), on a 4-dimensional manifold
$\mathcal{M}$ with topology $\mathbb{R}\times M$, $M$ compact. Thus there is an
additional SU(2) local gauge symmetry (coming from the reduction
(partial gauge fixing) of
the 4-dimensional Lorentz invariance to the 3-dimensional local
symmetry group of $M$), given by arbitrary local frame
rotations. We have $\{
E^{a}_{i},K^{i}_{a}\}$ as canonical pair on the phase space of the
theory, where $E^{i}_{a}=ee^{i}_{a}$ ($e$ is the determinant of
$e^{i}_{a}$) and $K_{a}^{i}$ is related to the extrinsic curvature by
$K^{i}_{a}=K_{ab}E^{bi}/\sqrt{h}$ with $h$ the determinant of the 3-metric
$h_{ab}$. Then given the canonical transformation
$A^{i}_{a}(x)=\omega^{i}_{a}+\beta K^{i}_{a}$, with $\omega^{i}_{a}$
being the SU(2) spin connection compatible with the triad, we arrive
at the new canonical pair of variables in the phase space,
$(A^{i}_{a}(x),E^{a}_{i})$. The new configuration variable is now the
$su(2)$-Lie algebra-valued connection 1-form on $M$, $A^{i}_{a}$, and $E^{a}_{i}$ is
the conjugate momentum. 
At the quantum level the canonical variables
will be replaced by operators acting on the state space of the theory
and the full dynamical content of General Relativity will be encoded
in the action of the first-class constraints on the physical states:
the SU(2) Gauss constraint, imposing the local gauge invariance on the
states, the diffeomorphism constraint, generating 3-dimensional
diffeomorphisms on $M$, and the Hamiltonian constraint, representing
the evolution of $M$ in the (unphysical) coordinate time. The issue of
quantization is then the realization of the space of the physical
states of the theory (in general, functionals of the connection) with the correct action of the constraints on
them, i.e. they should lie in the kernel of the quantum constraint
operators. Consequently let us focus on these states.
 
Consider a graph $\Gamma_{n}$, given by a collection of n {\it links}
$\gamma_{i}$, piecewise smooth curves embedded in $M$ and meeting only
at their endpoints, called {\it vertices}, if at all. Now we can
assign group elements to each link $\gamma_{i}$ taking the holonomy or parallel transport $g_{i}=\mathcal{P}exp\int_{\gamma_{i}}A$, where the
notation means that the exponential is defined by the path ordered
series, and consequently assigning an element of $SU(2)^{n}$ to the
graph. Given now a complex-valued function $f$ of $SU(2)^{n}$, we define
the state $\Psi_{\Gamma_{n},f}(A)=f(g_{1},...,g_{n})$. The set of
states so defined (called {\it cylindrical functions}) forms a subset of the space of
smooth functions on the space of connections, on which it is possible to
define consistently an inner product \cite{Ba,AshLew}, and then
complete the space of linear combinations of cylindrical functions
in the norm induced by this inner product. In this way we obtain a
Hilbert space of states $\mathcal{H}_{aux}$. This is not the physical
space of states, $\mathcal{H}_{phys}$, 
which is instead given by the subspace of it annihilated by all the
three quantum constraints of the theory. 

An orthonormal basis in
$\mathcal{H}_{aux}$ is constructed in the following way. Consider the
graph $\Gamma_{n}$ and assign {\it irreducible} representations to the
links. Then consider the tensor product of the $k$ Hilbert spaces of
the representations associated to the $k$ links intersecting at a
{\it vertex} $v$ of the graph, fix an orthonormal basis in this space,
and assign to the vertex an element $\iota$ of the basis. The corresponding state is then
defined to be:
\be 
\Psi_{\Gamma_{n}}(A)=\otimes_{i}\rho^{j_{i}}(g_{i})\otimes_{v}\iota_{v},
\ee where the products are over all the links and all the vertices of
the graph, $\rho$ indicates the representation matrix of the group
element $g$ in the irreducible representation $j$, and the indices of
$\rho$ and $\iota$ (seen as a tensor) are suitably contracted. It is
possible to show that the states so defined for all the possible
graphs and all the possible colorings of links and vertices are
orthonormal in the previously defined inner product. Moreover, if we
define $\iota_{v}$ to be an {\it invariant} tensor, i.e. to be in the
singlet subspace in the decomposition of the tensor product of the $k$ Hilbert
spaces into irreducible parts, then the resulting quantum state
$\Psi_{\Gamma}$ is {\it invariant} under SU(2). As a result, the set
of all these states, with all the possible choices of graphs and
colorings gives an orthonormal basis for the subspace of
$\mathcal{H}_{aux}$ which is annihilated by the first of the quantum
constraint operators of the theory, namely the Gauss constraint. The
invariant tensors $\iota_{v}$ are called {\it intertwiners}, they make it
possible to couple the representations assigned to the links
intersecting at the vertex, and are given by the standard recoupling
theory of SU(2) (in this case). The colored graph above is a {\it spin
network} $S$ and defines a quantum state $\mid S\rangle$, represented in
terms of the connection by a functional $\Psi_{S}(A)=\langle A\mid
S\rangle$ of the type just described. 

We have still to consider the
diffeomorphism constraint. But this is (quite) easily taken into
account by considering the equivalence class of embedded spin networks
$S$ under the action of $Diff(M)$, called {\it s-knots}. It is important to note that modding out by diffeomorphisms does 
not leave us without any informations about the relationship between the 
manifold and our graphs. There is still (homotopy-theoretic) information about how loops in the graph wind around holes in 
the manifold, for example, or about how the edges intersect each other. What is no longer defined is \lq\lq where" the graph 
sits inside the manifold, its location, and its metric properties (e.g. length of its links,...). Each s-knot defines an 
element
$\mid s\rangle$
of $\mathcal{H}_{Diff}$, that is the Hilbert space of both gauge
invariant and diffeomorphism invariant states of the gravitational
field, and it can be proven that the states $(1/\sqrt{is(s)})\mid
s\rangle$, where $is(s)$ is the number of isomorphisms of the s-knot
into itself, preserving the coloring and generated by a diffeomorphism
of $M$, form an orthonormal basis for that space. 

Now we can say that we have found which structure gives the quantum
states of the gravitational field, at least at the kinematical level,
since the dynamics is encoded in the action of the Hamiltonian
constraint, that we have not yet considered. They are spin networks,
purely algebraic and combinatorial objects, defined regardless of any
embedding in the following way (here we also generalize the previous
definition to any compact group $G$)

- a spin network is a triple $(\Gamma, \rho, \iota)$ given by: a
1-dimensional oriented (with a suitable orientation of the edges or
links) complex $\Gamma$, a labelling $\rho$ of each edge $e$ of
$\Gamma$ by an
irreducible representation $\rho_{e}$ of $G$, and a labelling $\iota$
of each vertex $v$ of $\Gamma$ by an intertwiner $\iota_{v}$ mapping
(the tensor product of) the irreducible representations of the edges
incoming to $v$ to (the tensor product of) the irreducible
representations of the edges outgoing from $v$. 

Let us note that this definition is not completely precise unless we specify an ordering of the edges; without this, the 
assignment of intertwiners to the vertices of the graph and the diagrammatic representation of the spin network would be 
ambiguous (think for example at a spin network were three edges are all labelled by the vector $j=1$ representation of 
$SU(2)$, so that the value of the intertwiner, being the completely antisymmetric symbol $\epsilon^{IJK}$, is just $+1$ or 
$-1$ depending on the ordering only).

A simple example of a spin network is given in the picture, where we indicate
also the corresponding state in the connection representation.

\begin{figure}

\begin{center}

\includegraphics[width=6cm]{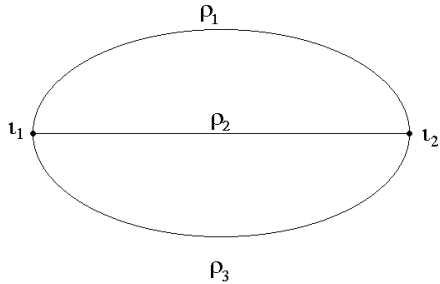}

\caption{A spin network whose corresponding functional is: 
$\psi_{S}(A)=\rho_{e_{1}}(\mathcal{P}\,e^{\int_{e_{1}}A})^{a}_{b}\,\rho_{e_{2}}(
\mathcal{P}\,e^{\int_{e_{2}}A})^{c}_{d}\,\rho_{e_{3}}(
\mathcal{P}\,e^{\int_{e_{3}}A})^{e}_{f}\,(\iota_{v_{1}})_{ace}(\iota_{v_{2}})^{bdf}$}

\end{center}

\end{figure}

These are, as we stress again,  the 3d diffeomorphism invariant
quantum states
of the gravitational field (provided we add to them homotopy-theoretic information of the kind mentioned above), and, accordingly, they do not live anywhere
in the space, but {\it define} \lq\lq the where'' itself (they can be seen
as elementary excitations of the space itself). Moreover,
we will now see that they carry the geometrical information necessary
to construct the geometry of the space in which we {\it decide} to
embed them.

Having at hand the kinematical states of our quantum theory, we want
now to construct gauge invariant operators acting on them. The
simplest example is the trace of the holonomy around a loop $\alpha$,
$\hat{T}(\alpha)=TrU(A,\alpha)$. This is a multiplicative operator,
whose action on spin network states is simply given by:
\be 
\hat{T}(\alpha)\Psi_{S}(A)\,=\,TrU(A,\alpha)\Psi_{S}(A). \ee
Since our configuration variable is the connection $A$, we then look
for a conjugate momentum operator in the form of a derivative with
respect to it, so replacing the $E^{i}_{a}$ field by the operator
$-i\delta/\delta A$. This is an operator-valued distribution so we
have to suitably smear it in order to have a well-posed operator. It
is convenient to contract it with the Levi-Civita density and to
integrate it over a surface $\Sigma$, with embedding in $M$ given by
$(\sigma^{1},\sigma^{2})\rightarrow x^{a}(\vec{\sigma})$, where
$\vec{\sigma}$ are coordinates on $\Sigma$. We then define the
operator 
\be 
\hat{E}^{i}(\Sigma)=-i\int_{\Sigma}d\vec{\sigma}\,n_{a}(\vec{\sigma})\frac{\delta}{\delta
A^{i}_{a}} \ee
where $n(\sigma)$ is the normal to $\Sigma$. This operator is well
defined but not gauge invariant. We then consider its square given by
$\hat{E}^{i}(\Sigma)\hat{E}^{i}(\Sigma)$, but we discover that its
action on a spin network state $\Psi_{S}(A)$ is gauge invariant only
if the spin network $S$, when embedded in $M$, has only one point of
intersection with the surface $\Sigma$. But now we can take a
partition $p$ of $\Sigma$ in $N(p)$ surfaces $\Sigma_{n}$ so that
$\cup\Sigma_{n}=\Sigma$ and such that each $\Sigma_{n}$ has only one point
of intersection with $S$, if any. Then the operator 
\be
\hat{\mathcal{A}}(\Sigma)\,=\,\lim_{p}\sum_{n}\sqrt{\hat{E}^{i}(\Sigma_{n})\hat{E}^{i}(\Sigma_{n})},
\ee where the limit is an infinite refinement of the partition $p$, is
well-defined, after a proper regularization, independent on the
partition chosen, and gauge invariant. Given a spin network $S$ having
(again, when embedded in $M$)
a finite number of intersections with $\Sigma$ and no nodes lying on
it (only for simplicity, since it is possible to consider any other
more general situation)(see figure), the action of $\hat{\mathcal{A}}$ on the
corresponding state $\Psi_{S}(A)$ is 
\be
\hat{\mathcal{A}}(\Sigma)\Psi_{S}(A)=\sum_{i}\sqrt{j_{i}(j_{i}+1)}\Psi_{S}(A),
\ee
where the sum is over the intersection between $\Sigma$ and $S$ and
$j_{i}$ is the irreducible representation of the edge of $S$
intersecting $\Sigma$. 

\begin{figure}
\begin{center}
\includegraphics[width=7.5cm]{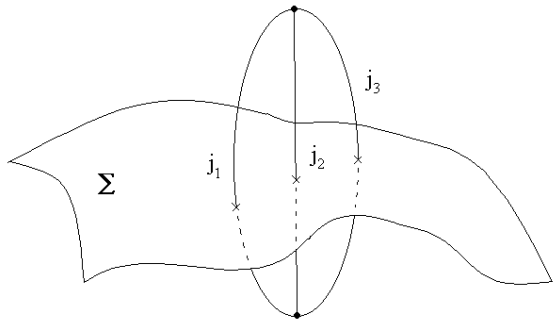}
\caption{A spin network embedded in a manifold and having three points of intersection with a surface $\Sigma$}
\end{center}
\end{figure}

So the spin network states are eigenstates of
the operator with discrete eigenvalues. The crucial point is that the
operator we considered has the classical interpretation of the
area of the surface $\Sigma$. In fact its classical counterpart is
 \be
\mathcal{A}(\Sigma)\,=\,\int_{\Sigma}d^{2}\sigma\sqrt{n_{a}(\vec{\sigma})E^{ai}(\vec{x}(\vec{\sigma}))n_{b}(\vec{\sigma})
E^{bi}(\vec{x}(\vec{\sigma}))}\,=\,\int_{\Sigma}d^{2}\sigma\sqrt{det(^{2}h)},\ee

which describes exactly the area of $\Sigma$ ($^{2}h$ is the 2-metric
induced by $h_{ab}$ on $\Sigma$).

This means that the area is quantized and has a discrete spectrum of
eigenvalues! Moreover we see that the \lq\lq carriers'' of this area
at the fundamental level are the edges of the spin network $S$ that
we choose to embed in the manifold. 

The same kind of procedure can be applied also to construct a
quantum operator corresponding to the volume of a 3-hypersurface and
to find that the spin network states diagonalize it as well and that
the eigenspectrum is again discrete. In this case, however, the volume is given by the vertices of the spin network
inside the hypersurface.

So we can say that a spin network is a kinematical quantum state of the
gravitational field in which the vertices  give volume and the edges
give areas to the space in which we embed it.

So far we have considered only smooth embeddings of s-knots, but we stress that spin networks can be 
defined and have been studied in the context of piecewise flat embeddings, and used in a similar 
way for studying simplicial quantum geometry \cite{barbieri}. 

What about the evolution of such quantum states? Several definitions
of the quantum Hamiltonian constraint exist in loop quantum gravity
\cite{thiemHam, gaulcarlo}, and the basic action of such an operator on spin netowork
states is also known. However, both doubts about the correctness of
such definitions \cite{smolinham, nelson} for the recovering of a
classical limit, and technical and conceptual difficulties in
understanding time evolution in a canonical formalism, have led to
investigating ways of defining such evolution in terms of the
``projector operator'' \cite{carloprojector} discussed above, constructing directly
transition amplitudes between spin networks and using them to {\it
  define} the action of the Hamiltonian constraint and the physical
inner product of the theory. Interestingly, this way of approaching
the problem leads naturally to the idea of spin foams.

   In the canonical approach to quantum gravity, as we said, the
(coordinate) time evolution of the gravitational field is generated by the Hamiltonian 
$H_{N,\vec{N}}(t)=C[N(t)]+C[\vec{N}(t)]=\int d^3 x [N(t,x) C(x) +
     N^a(t,x) C_a(x)]$, i.e. it is given by the sum of the Hamiltonian constraint (function of the lapse 
function $N(t,x)$)  and the diffeomorphism constraint (function of the shift vector $\vec{N}(t,x)$). The quantum operator giving 
evolution from one hypersurface $\Sigma_{i}$ ($t=0$) to another $\Sigma_{f}$ ($t=1$) is given by 
\bes
U_{N,\vec{N}}\,=\,e^{-i\int_{0}^{1}dt  H_{N,\vec{N}}(t)}.
\ees
The proper time evolution operator is then defined
\cite{carlomikeproj} as:
\be U(T)\,=\,\int_{T}dN d\vec{N}\,U_{N,\vec{N}} \ee
where the integral is over all the lapses and shifts satisfying
$N(x,t)=N(t)$ and $N^a(x,t) = N(t)$, and
 $\int_{0}^{1}dt N(t)=T$, and $T$ is the proper time separation
between $\Sigma_{i}$ and $\Sigma_{f}$ (this construction can be generalized to a multifingered proper time \cite{carlomikeproj}); it is important to stress that
these conditions on the Lapse and Shift functions involve a particular
choice for the slicing of the 4-manifold into equal proper time
hypersurfaces and a gauge fixing that should be compensated by
appropriate ghost terms as we have mentioned also when discussing the
path integral approach to quantum gravity in the traditional metric
variables. 
Now we want to calculate the 
matrix elements of this operator between two spin network states (two s-knots). These are to be interpreted as transition 
amplitudes between quantum states of the gravitational field, and computing them is equivalent to having solved the theory 
imposing both the Hamiltonian and diffeomorphism constraints.   

Now \cite{carlomikeproj} we take a large number of hypersurfaces separated by
small intervals of coordinate time, and write
$U_{N,\vec{N}}=D[g]U_{N_{\vec{N}},0}$, where $g$ is the finite
diffeomorphism generated by the shift between the slices at $t=0$ and
$t$ and $D(g)$ is the corresponding diffeomorphism operator acting on
states (in fact we can always rearrange the coordinates to put the shift equal to zero, then compensating with a finite 
change of space coordinates (a diffeomorphism) at the end).
Considering $N_{\vec{N}}=N=const$ (partial gauge fixing) we can expand $U_{N,0}$ as:
\bes
U_{N,0}\,=\,1\,+\,(-i)\int_{0}^{\tau}dt\,C[N(t)]\,+\,(-i)^{2}\int_{0}^{\tau}dt\int_{0}^{\tau}dt'\,C[N(t')]C[N(t)]\,+....
\ees
and compute its matrix elements as:
\bes
\langle S_{f}\mid U_{N,0}(T)\mid S_{i}\rangle = \langle S_{f}\mid S_{i}\rangle + (-i)\int_{0}^{\tau}dt\langle S_{f}\mid 
C[N(t)]\mid S_{i}\rangle +\nonumber \\+ (-i)^{2}\int_{0}^{\tau}dt\int_{0}^{\tau}dt'\langle S_{f}\mid C[N(t')]\mid S_{1}
\rangle\langle S_{1}\mid C[N(t)]\mid S_{i}\rangle +.....
\ees
having inserted the resolution of the identity in terms of a complete set of states $\mid S_{1}\rangle$ (with the summation 
implicit).

Using the explicit form of the Hamiltonian constraint operator \cite{carlomikeproj,thiemHam}: 
\bes
C[N]\mid S\rangle\,=\,\sum_{v,e,e',\epsilon,\epsilon'}A_{vee'\epsilon\epsilon'}(S)D_{vee'\epsilon\epsilon'}\mid S\rangle\,=\,
\sum_{\alpha}A_{\alpha}(S)D_{\alpha}\mid S\rangle\,+\,h.c.
\ees
where the operator $D_{vee'\epsilon\epsilon'}$ acts on the spin network $S$ creating two new trivalent vertices $v'$ and 
$v''$ on the two edges $e$ and $e'$ (intersecting at the vertex $v$), connecting them with a new edge with label 1, and 
adding $\epsilon=\pm 1$ (and $\epsilon'=\pm 1$) to the color of the edge connecting $v$ and $v'$ (and $v$ and $v''$); the 
$A$s are coefficients that can be explicitely computed. Doing this and carrying out the integrals we find:
\bes 
  \lefteqn{\langle S_{f}\mid U_{N,0}(T)\mid S_{i}\rangle = \langle S_{f}\mid S_{i}\rangle +} \nonumber \\ &+&\,(-iT)\left( 
\sum_{\alpha\in S_{i}} A_{\alpha}(S_{i})\langle S_{f}\mid D_{\alpha}\mid S_{i}\rangle+ \sum_{\alpha\in S_{f}} A_{\alpha}
(S_{f})\langle S_{f}\mid D^{\dagger}_{\alpha}\mid S_{i}\rangle\right)+\nonumber \\ &+&\frac{(-iT)^{2}}{2!}\sum_{\alpha\in 
S_{i}}\sum_{\alpha'\in S_{1}}A_{\alpha}(S_{i})A_{\alpha'}(S_{1})\langle S_{f}\mid D_{\alpha'}\mid S_{1}\rangle\langle S_{1}
D_{\alpha}\mid S_{i}\rangle + .....
\ees
So at each order $n$ we have the operator $D$ acting $n$ times, there are $n$ factors $A$ and a finite number of terms 
coming from a sum over vertices, edges, and $\epsilon=\pm 1$. Moreover since the sum over intermediate states $S_{1}$ is 
finite, the expansion above is finite order by order.
Now we should integrate over $N$ and $\vec{N}$ to obtain the matrix
elements of $U(T)$. It happens that the integration over the lapse is
trivial, since there is no dependence on it in the integral, while the
integration over the shift is just the imposition of the
diffeomorphism constraint, so its effect is to replace the spin
networks in the matrix elements with their diffeomorphism equivalence classes. At the end we have:
    \bes 
  && \langle s_{f}\mid U(T)\mid s_{i}\rangle = \nonumber \\ &=&\langle s_{f}\mid s_{i}\rangle + (-iT)\left( \sum_{\alpha\in s_{i}} A_{\alpha}
(s_{i})\langle s_{f}\mid D_{\alpha}\mid s_{i}\rangle+ \sum_{\alpha\in s_{f}} A_{\alpha}(s_{f})\langle s_{f}\mid D^{\dagger}
_{\alpha}\mid s_{i}\rangle\right)+\nonumber \\ &+&\frac{(-iT)^{2}}{2!}\sum_{\alpha\in s_{i}}\sum_{\alpha'\in s_{1}}A_{\alpha}
(s_{i})A_{\alpha'}(s_{1})\langle s_{f}\mid D_{\alpha'}\mid s_{1}\rangle\langle s_{1}D_{\alpha}\mid s_{i}\rangle + .....
\label{eq:exp}
\ees
This is the transition amplitude between a 3-geometry $\mid s_{i}\rangle$ and a 3-geometry $\mid s_{f}\rangle$.
The crucial observation now is that we can associate to each term in the expansion above a 2-dimensional colored surface 
$\sigma$ in the manifold defined up to a 4-diffeomorphism. The idea is the following. Consider the initial hypersurfaces 
$\Sigma_{i}$ and $\Sigma_{f}$ and draw $s_{i}$ in the first and $s_{f}$ in the second; of course location is chosen 
arbitrarily, again up to diffeomorphisms, since there is no information in $s_{i}$ and $s_{f}$ about their location in 
spacetime. Now let $s_{i}$ slide across the manifold $\mathcal{M}$ from $\Sigma_{i}$ towards $s_{f}$ in $\Sigma_{f}$ (let 
it \lq\lq evolve in time"). The edges of $s_{i}$ will describe 2-surfaces, while the vertices will describe lines. Each 
\lq\lq spatial" slice of the 2-complex so created will be a spin network in the same s-knot (with the same combinatorial 
and algebraic structure), unless an \lq\lq interaction" occurs, i.e. unless the Hamiltonian constraint acts on one of these 
spin networks. When this happens the Hamiltonian constraint creates a spin network with an additional edge (or with one 
edge less) and two new vertices; this means that the action on the 2-complex described by the evolving spin network is 
given by a creation of a vertex in the 2-complex connected by two edges to the new vertices of the new spin network, 
originating from the old one by the action of the Hamiltonian constraint. So at each event in which the Hamiltonian 
constraint acts the 2-complex \lq\lq branches" and this branching is the elementary interaction vertex of the theory. So an
 $n$-th order term in the expansion (~\ref{eq:exp}) corresponds to a 2-complex with $n$ interaction vertices, so with $n$ 
actions of the operator $D_{\alpha}$ on the s-knot giving the 3-geometry at the \lq\lq moment" at which the action occurs. 
Moreover, each surface in the full 2-complex connecting in this way $s_{i}$ to $s_{f}$ can be coloured, assigning to each 
2-surface (face) in it the irrep of the spin network edge that has swept it out, and to each edge in it the intertwiner of 
the corresponding spin network vertex. We give a picture of a second order term.

\begin{figure}
\begin{center}
\includegraphics[width=8cm]{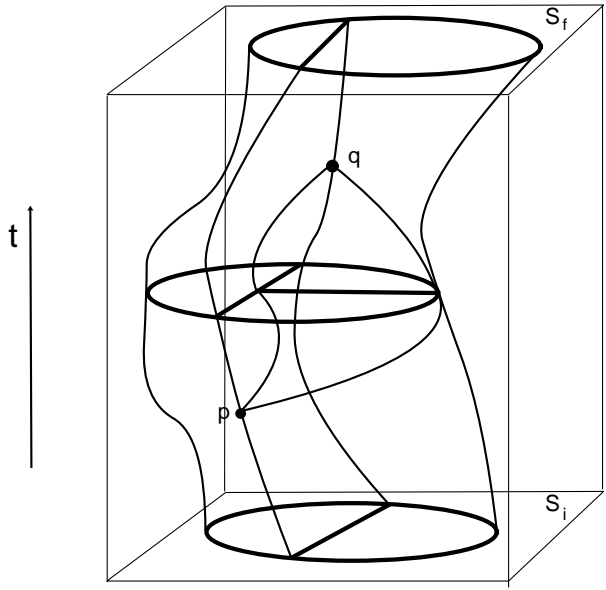}
\caption{The 2-complex correspondent to a second order term in the expansion of the amplitude} 
\end{center}
\end{figure}

If we also fix an ordering of the vertices of the 2-complex, then each term in (~\ref{eq:exp}) corresponds uniquely to a 
2-complex of this kind, in a diffeomorphism invariant way, in the sense that two 2-complexes correspond to the same term if 
and only if they are related by a 4-diffeomorphism.

Given this, we can rewrite the transition amplitudes as sums over topologically inequivalent (ordered) 2-complexes $\sigma$ 
bounded by $s_{i}$ and $s_{f}$, and with the weight for each 2-complex being a product over the $n(\sigma)$ vertices of the 
2-complex $\sigma$ with a contribution from each vertex given by the coefficients of the Hamiltonian constraint:
\bes
\langle s_{f}\mid U(T)\mid s_{i}\rangle\,=\,\sum_{\sigma}\mathcal{A}[\sigma](T)\,=\,\sum_{\sigma}
\frac{(-iT)^{n(\sigma)}}{n(\sigma)!}\prod_{v}A_{\alpha}
\ees

These colored 2-complexes arising from the evolution in time of
spin networks are precisely spin foams. This derivation is not entirely rigorous, but it shows very clearly what should be 
expected to be the 4-dimensional history of a spin network, and how the same algebraic and combinatorial characterization of
quantum gravity states will extend to the description of their evolution. 

How does the loop approach incorporate the ideas we discussed above? As far as background independence and relationality are
concerned, loop quantum gravity is basically the paradigm for a background independent theory, and for a theory where the 
states and the observables are defined in a purely relational and background independent way; indeed, they turn out to be 
a quantization of the classical spin network we defined before, that were also characterized by a straightforward 
operational
intepretation. Also, although derived from a $3+1$ splitting of spacetime, the end result of the quantization process does 
not reflect too much this splitting, and the basic structure of spin network states and observables with the same 
characteristics can easily be seen in the context of a covariant formulation of both classical and quantum mechanics, since 
coordinates do not play any role at all. Causality is not an easy issue to deal with in this approach, in light of the 
troubles
in defining the evolution of spin network states, so not much can be said on this at present. Moroever, although we started 
from a smooth manifold and the usual action for 
gravity, we have reached, at the end of the quantization process, a realm where there is no spacetime, no points, no 
geometry,
but all of them can be reconstructed from the more fundamental, discrete (because of the spectum of the geometric 
observables), algebraic structures that represent the quantum spacetime. It is in other words a marvellous example of a 
theory 
that leads us naturally beyond itself, pointing to more fundamental ideas and structures than those it was built from.

\subsection{Simplicial quantum gravity}
Simplicial approaches to quantum gravity have also a quite long history behind them, motivated by the conceptual 
considerations we discussed, and by the idea that lattice methods that proved so useful in quantizing non-Abelian 
gauge theory could be important also in quantum gravity, but also by the practical advantages of dealing with discrete 
structures, and achieved important results, both analytical and numerical. We want to give here an outline of the two main 
simplicial approaches to quantum gravity, namely quantum Regge calculus and dynamical triangulations, since they both turn 
out to be very closely related to the spin foam approach. For more extensive review of both we refer to the literature 
\cite{ruth1, ruth2, ruth3, Loll, amblol1, amblol2}.
\subsubsection{Quantum Regge calculus}
Consider a Riemannian simplicial manifold $S$, that may be thought of as an approximation of a continuum manifold $M$.
More precisely, one may consider the simplicial complex to represent a piecewise flat manifold made up out of patches of 
flat 4-dimensional space, the 4-simplices, glued together along the common tetrahedra.
Classical Regge calculus\cite{Regge} is a straightforward discretization of General Relativity based on the Einstein-Hilbert
action, resulting in the action (with cosmological constant $\lambda$): 
\bes
S_R(l_i) &=& \frac{1}{8\pi G}\sum_{t}\,A_t(l)\,\epsilon_t(l)\,-\,\lambda\,\sum_\sigma\,V_\sigma =\, \nonumber 
\\ &=& \frac{1}{8\pi G}\sum_t\,
A_t(l)\,\left( 2\pi\,-\,\sum_{M\ni\sigma\supset t}\theta_\sigma^t(l)\right)\,-\,\lambda\,\sum_\sigma\,V_\sigma
\ees
where one sums over all the triangles (2-simplices) $t$ in the simplicial complex, which is where the curvature is 
distributionally located, $A_t$ is the area of the triangle $t$, $\epsilon_t$ is the deficit angle (simplicial measure of 
the
intrinsic curvature associated to the triangle), which in turn is expressed as a sum over all the 4-simplices $\sigma$ 
sharing the triangle $t$ of the dihedral angles associated with it in each 4-simplex, the dihedral angle being the angle 
between the normals to 
the two tetrahedra (3-simplices) sharing the triangle; $V_\sigma$ is the 4-volume of the 4-simplex $\sigma$. The fundamental
 variables are the edge (1-simplices) lengths $l_i$, so 
all the other geometrical quantities appearing in the action are functions of them, and the equations of motion are obtained 
by variations with respect to them.

The quantization of the theory based on such an action is via Euclidean path integral methods. One then defines the 
partition 
function of the theory by:
\bes
Z(G)\,=\,\int \mathcal{D}l\;e^{-\,S_R(l)}
\ees 
and the main problem is the definition of the integration measure for the edge lengths, since it has to satisfy the 
discrete analogue of the diffeomorphism invariance of the continuum theory; the most used choices are the $l dl$ and the
$dl/l$ measures; one usually imposes also a cut-off both in the 
infrared and in the ultraviolet limits, to make the integral converge. Also, the integration over $l$ may be reduced to a 
summmation over (half-)integers $j$, by imposing a quantization condition on the edge lengths as $l= l_p j$, where $l_p$ is 
taken to represent the Planck length.

A Lorentzian version of the classical Regge calculus is formulated much in a similar way, with the only difference  being in
the definition of the deficit angle that is now given just by the sum of the dihedral angles (more on the Lorentzian 
simplicial geometry will be explained in the rest of ths work), and one may define a Lorentzian path integral based on this 
action.
Extensive analytical and numerical investigations have been done on this model, often using an hypercubic lattice instead of
the simplicial one, for practical convenience, and sometimes adding higher derivative terms. The results may be shown to 
agree with the continuum calculations in the weak field limit, and indications were found on the existence of a transition 
between a smooth and a rough phase of spacetime geometry, maybe of second order; also, 2-point correlation functions were 
studied in the proximity of the transition point.   
Matter and gauge fields can also be coupled to the gravity action, using techniques similar to those used in lattice gauge 
theory, and also important issues like gauge (diffeomorphism) invariance have been investigated. Other results concerned 
applications to quantum cosmology, connections with loop quantum gravity and Ashtekar variables, using gauge-theoretic 
variables (analogues of a gauge connection) instead of metric ones (the edge lengths).

From the conceptual point of view, so neglecting for a moment the way Regge calculus enters in the formulation of spin foam
models, and the results obtained in this approach, the main interest for us is in the concrete implementation in a quantum 
gravity model of the ideas about fundamental discreteness we outlined above. As a simplicial approach quantum Regge calculus 
implements the finitary approach to spacetime discretenss we discussed and, with a suitable intepretation of the physical 
meaning of the simplicial complex, also the operational flavor we would like a quantum theory of gravity to possess, in 
addition of being a fully covariant approach being based on a sum-over-histories formulation of quantum theory. As for 
background independence, the approach is implementing it, although this can be really said to be so only when 
a continuum limit is taken, with all the difficulties it implies, since any triangulation represents a truncation of the 
degrees of freedom of the gravitational field.

\subsubsection{Dynamical triangulations}
In Regge calculus, as we have said, the lengths of the edges of the simplicial complex are the dynamical variables, while 
the simplicial complex itself (and consequently the topology of the manifold) is held fixed, with the full dynamical content 
of gravity recoved when a refinement of it is performed going to the continuum limit (and thus recovering the full topology 
of the continuum manifold from the finitary one represented by the simplicial complex).
Dynamical triangulations are based on the opposite, and in a way complementary, approach. One still describes spacetime as a
simplicial manifold and uses the Regge action as a discretization of the gravity action, but now one fixed the edges lengths 
to a fixed value, say the Planck length $l_p$, and treats as variable the connectivity of the triangulations, for fixed 
topology.
In other words, now in constructing a path integral for gravity, one sums over all the possible equilateral triangulations 
for a given topology (usually one deals with the $S^4$ topology).

All the triangles being equilateral, the action takes a very simple form: $S_R=-k_2 N_2(T)+k_4 N_4(T)$, where $k_2$ and 
$k_4$ 
are coupling constants related to the gravitational and cosmological constat respectively, and $N_2$ and $N_4$ are the 
number of triangles and 4-simplices in the triangulation $T$.

The path integral is then given, again in the Euclidean form, by:
\bes
Z(k_2,k_4)\,=\,\sum_T \,\frac{1}{C(T)}\,e^{-\,S(T)},
\ees
where $C(T)$ is the number of automorphisms (order of the automorphism group) of the triangulation $T$. We see that the 
problem of constructing any quantum gravity quantity, including the partition function for the theory itself, is reduced to 
a combinatorial (counting) problem, since as we said the metric information is completely encoded in the connectivity of the 
triangulations considered, in a diffeomorphism invariant way, the only remnants of diffeomorphisms at this simplicial level
being the automorphisms of the triangulation. 

Also this system is extremely suitable for both analytical and especially numerical calculations, and indeed one can obtain 
quite a number of results. These include the full solution of quantum gravity in 2d (due to the fact that an explicit 
formula
exists in this case for counting all the triangulations for fixed topology), the coupling of matter and gauge fields, an 
extensive exploration of the quantum geometric properties, and a detailed analysis of the phase 
structure of the theory in 4 dimensions, with a crumpled phase  with small negative curvature, a large Hausdorff dimension 
and a high connectivity, and an elongated phase with large positive curvature and Hausdorff dimension $d\simeq 2$, with a 
possible second order phase transition. So no smooth phase has been found. However, interesting new features appear in the
 Lorentzian (causal) version of this approach, where the construction of the triangulations to be summed over in the path 
integral is done using a causal evolution algorithm building up the spacetime triangulation from triangulated hypersurfaces,
 forbidding spatial topology to change so that only the cylindrical topology is allowed for spacetime. In this causal 
version of the approach the phase structure of the theory appears to be entirely different, with a transition between a 
crumpled phase and a smooth classical-looking one. 

We see that this type of models involves again a finitary approach to the issue of modelling spacetime using more 
fundamental
structures, but in a complementary way with respect to the Regge calculus approach, because the variables used (or in other 
words the choice of which degrees of freedom of the gravitational field should be held fixed and which should be not) are 
opposite in the two, 
edge lengths versus connectivity of the simpicial complex, and consequently the way in which the full theory (with the same 
dynamical content of the classical gravitational theory) is approached is complementary as well, choosing either a refining 
of the triangulation or sending the edge length to zero (i.e. in each case, making dynamical the set of degrees of freedom 
that were first held fixed), in both cases obtaining in this way a background independent theory (no fixed set of background 
degrees of freedom). As far as the topological properties of the underlying spacetime manifold are considered, in one 
case one chooses to refine the finitary open covering giving the subtopology of the manifold, while in the other these are 
recovered by summing over all the possible subtopologies (in line with what the actual mathematical definition of a topology
 is). Being simplicial, dynamical triangulations are also operational, again assuming the interpretation of the 4-simplices 
as not idealized determinations of location, or as operational substitutes for continuum spacetime points. Symmetry considerations,
on the other hand, and the Lorentz group do not play any particularly fundamental role here. As for causality, the issue is 
very tricky in this context, but the mentioned results on Lorentzian dynamical triangulations show that the correct 
implementation of a dynamical causal structure may well turn out to be crucial for having a sensible theory.

\subsection{Causal sets and quantum causal histories}

Let us now briefly describe an approach to quantum gravity which {\it starts} from the idea that a correct implementation of
causality is {\bf the} crucial ingredient for any fundamental description of spacetime at the classical and especially 
quantum level. The causal set approach \cite{sorkincausal} is indeed a radical attempt to define a quantum gravity theory 
basically using causal 
structures alone.

Motivated by the recognition that the causal structure is almost sufficient for reconstructing a full metric field, as 
discussed 
above, the starting point for this approach is just a poset, a set of points endowed with a (causal) partial ordering 
relation 
among them.
More precisely, consider a discrete set of events $\{p, q, r, s, ...\}$, endowed
with an ordering relation $\leq$, where the equal sign implies
that the two events coincide. The ordering relation is assumed to
be reflexive ($\forall q, q\leq q$), antisymmetric ($q\leq s,
s\leq q \Rightarrow q=s$) and transitive ($ q\leq r, r\leq s\,
\Rightarrow q\leq s$). These properties characterize the set as a
partially ordered set, or poset, or, interpreting the ordering
relation as a causal relation between two events, as a causal set (see figure 1.5). In this last case, the 
antisymmetry of the ordering
relation, together with transitivity, implies the absence of
closed timelike loops. 

 \begin{figure}
\begin{center}
\includegraphics[width=9cm]{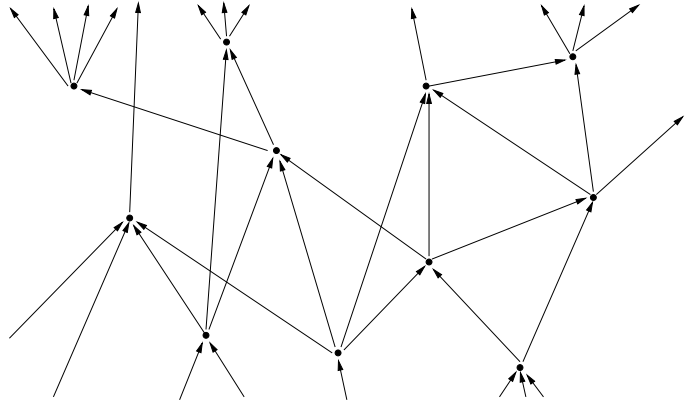}
\caption{A causal set}
\end{center}
\end{figure}

In addition, the poset is required to be \lq\lq locally finite", i.e. the requirement is that the 
Alexandroff set $A(x,y)$ of any two points $x,y$ is made out of only a finite number of elements, this
 being defined as $A(x,y)=\{ z : x\leq z \leq y \}$.

However, we said before that there is one (only one) degree of freedom of the gravitational field, 
i.e. of spacetime geometry, which is not specified by the causal structure and this is a length scale, 
or the conformal structure, or the volume element, depending on how one is actually reconstructing 
the metric from the causal structure. In the causal set approach one usually takes the attitude that 
the volume of a given region of spacetime where the causal set is embedded is obtained by simply 
{\it counting} the number of elements of the causal set contained in that region.

As far as the topology of spacetime is concerned, the crucial observation is that, as we noted above, 
a poset is an equivalent description of an open covering of a manifold, i.e. of a subtopology, just 
like a simplicial complex, so giving a poset is the same as giving a finitary description of spacetime 
topology and a refinement procedure will give us the full characterization of it.
In any case, the causal set is the only structure from which one has to reconstruct the full 
spacetime.

At the classical level, this procedure, although not straightforward nor easy, is fairly 
uncontentious, and many results on this have been obtained\cite{sorkincausal}, giving support to the basic idea, 
including the embedding procedure for a causal set into a continuum manifold and the 
reconstruction of a metric field. Important results have been obtained also in the definition 
of background independent cosmological observables \cite{causalobs} and in the definition of interesting laws 
of evolution for causal sets as (classical) stochastic motion \cite{sorkinstoc}.

The difficult part is however at the quantum level.
Here one should specify a quantum amplitude for each causal set, define a suitable quantum measure, 
construct a suitable path integral, and compute, after having defined them, appropriate observables.
This has not been done yet.

The most recent and interesting attempt to define a qauntum framework on these lines, or to define 
{\it quantum causal sets} resulted in the construction of the Quantum Causal Histories framework.

Starting from the causal set as just defined, we recognise a few other structures within it: 1) the causal past of an event 
$p$ is the set $P(p)=\{ r | r\leq p\}$;
2) the causal future of an event $p$ is the set $F(p)=\{ r | p\leq r\}$;
3) an acausal set is a set of events unrelated to each other;
4) an acausal set $\alpha$ is a complete past for the event $p$ if
$\forall r\in P(p), \exists s\in\alpha \,|\, r\leq s\,\,or\,\,s\leq r$;
5) an acausal set $\beta$ is a complete future for the event $p$ if
$\forall r\in F(p), \exists s\in\beta \, |\, r\leq s\,\,or\,\,s\leq r$;
6) the causal past of an acausal set $\alpha$ is
$P(\alpha)=\cup_i\,P(p_i)$ for all $p_i\in\alpha$, while its causal
future is $F(\alpha)=\cup_i\,F(p_i)$ for all $p_i\in\alpha$;
7) an acausal set $\alpha$ is a complete past (future) of the
acausal set $\beta$ if $\forall p\in P(\beta) (F(\beta))\,\,\exists
q\in\alpha \,\,|\,\,p\leq q\,\,or\,\,q\leq p$;
8) two acausal sets $\alpha$ and $\beta$ are a complete pair when
$\alpha$ is a complete past of $\beta$ and $\beta$ is a complete future
for $\alpha$.

From a given causal set, one may construct another causal set
given by the set of acausal sets within it endowed with the
ordering relation $\rightarrow$, so that $\alpha\rightarrow\beta$
if $\alpha$ and $\beta$ form a complete pair, also required to be
reflexive, antisymmetric and transitive. This poset of acausal
sets is actually the basis of the quantum histories model.

The quantization of the causal set is as follows. It can be seen as a functor
between the causal set and the categories of Hilbert spaces.

We attach an Hilbert space $\mathcal{H}$ to each node-event and
tensor together the Hilbert spaces of the events which are
spacelike separated, i.e. causally unrelated to each other; in
particular this gives for a given acausal set
$\alpha=\{p_i,...,p_i,...\}$ the Hilbert space
$\mathcal{H}_\alpha=\otimes_i \mathcal{H}(p_i)$.

Then given two acausal sets $\alpha$ and $\beta$ such that
$\alpha\rightarrow\beta$, we assign an evolution operator  between
their Hilbert spaces: \be
E_{\alpha\beta}\,:\,\mathcal{H}_{\alpha}\,\rightarrow\,\mathcal{H}_{\beta}.
\ee

In the original Markopoulou scheme, the Hilbert spaces considered
are always of the same (finite) dimension, and the evolution
operator is supposed to be unitary, and fully reflecting the
properties of the underlying causal set, i.e. being reflexive:
$E_{\alpha\alpha}=Id_\alpha$, antisymmetric:
$E_{\alpha\beta}E_{\beta\alpha}=Id_\alpha
\Leftrightarrow\,E_{\alpha\beta}=E_{\beta\alpha}=Id_\alpha$, and
transitive: $E_{\alpha\beta}E_{\beta\gamma}=E_{\alpha\gamma}$.

The evolution operators mapping complete pairs that are causally
unrelated to each other, and can be thus tensored together, may also
be tensored together, so that there are cases when given evolution
operators may be decomposed into elementary components. More on the
dynamics defined by this type of models can be found in \cite{fot}\cite{fot2}.

Another possibility is to assign Hilbert spaces to the causal
relations (arrows) and evolution operators to the nodes-events.
This matches the intuition that an event in a causal set (in
spacetime) is an elementary change. This second possibility gives
rise to an evolution that respects local causality, while the
first one does not.   

In any case, given a complete pair $ij$ of acausal sets, we define the transition amplitude between them as a sum over all 
the possible causal sets connecting weighted by a quantum amplitude $A_C$ for each causal set $C$ constructed out of the 
evolution operators $E$ associated to any intermediate pair of acausal sets between them, with suitable tensoring of 
operators when needed:
\be
\mathcal{A}_{i\rightarrow j}\,=\,\sum_C\,A_C(E)\,=\,\sum_C\,\prod\,E(C)
\ee.

A refinement of such a framework was defined recently \cite{HMS}, and uses directly the 
initial causal set instead of working with the poset of acausal sets, assigning $C^*$-algebras 
of operators to the events, instead of Hilbert spaces, and completely positive maps to each 
pair of events, instead of unitary operators to complete pairs of acausal sets. This new 
implementation is at the same time more general and respects more closely the causal structure 
of the underlying causal set. Also, the general logical and mathematical structure of a theory of 
cosmology defined in terms of causal histories was investigated \cite{fotlogic}, showing the role 
that relationality and intuitionistic logic play in such a description. 

The main advantage of quantum causal histories is their generality, but the same property makes
 it difficult to come up with a specific non-trivial example of Hilbert spaces and,
above all, evolution operators with a clear link with gravity and geometry. 

The most relevant example of quantum causal histories in this respect is that of causal evolution of 
($SU(2)$) spin networks \cite{marsmo, fotdual, marsmo2}, where the Hilbert spaces associated with the
events of the causal set are given by the spaces of intertwiners between representations of $SU(2)$, 
so that an $SU(2)$ spin network is associated to each acausal set, and several properties of their 
quantum evolution can be identified and studied; however, also in this case it is not easy to define 
a non-trivial evolution operator associated to complete pairs of acausal sets. We will see in the 
following how a non-trivial example of a quantum evolution amplitude can be actually constructed in 
the context of spin foam models.

This is very good, since this kind of approach seems to encode all the fundamental ideas we outlined 
above: it is background independent, easily related to spin networks, with their relational and 
operational description of quantum geometry states, it furnishes a discrete and finitary description 
of spacetime, and it has causality built in at the most fundamental level. What is missing is in fact a specific, 
concretely defined, highly non-trivial model with a precise and clear way of encoding spacetime geometry, and a connection 
with classical gravity.

\section{General structure of spin foam models}
So, many roads led to spin foams. For reviews on spin foams see \cite{Baez2,Oriti,AlejandroReview}. 
A spin foam picture emerges when considering the evolution
 in time of spin networks \cite{carlomikeproj,carlosum}, which were discovered to represent states of 
quantum General Relativity at the kinematical level
\cite{RovSmolin,Baez3,Ash1,Ash3,Ash4}, so from the canonical side.
 Spin foam models were developed also for topological field
theories in different dimensions, including 3d quantum gravity \cite{PR,TV,CY,CKY,BaWe},
and this represents a completely different line of research coming to
 the same formalism. In these models category theory plays a major
 role, since their whole construction can be rephrased in terms 
 of operations in the category of a Lie group (or quantum group),
 making the algebraic nature of the models manifest. A spin foam model
 can also be seen, as we will discuss in the following, as a state sum
 of the type used in statistical mechanics, giving another bridge
 between different areas. Many of these models, moreover, are based on simplicial formulations of 
classical gravity, representing a different approach to its quantization.
 A spin foam formulation can also be given
 \cite{Reis1,OePf} for lattice Yang-Mills theory; here the partition function of the spin foam model is equal to that of the lattice gauge theory and can be used to calculate the strong coupling expansion. Finally, and most important for our present
 interest, pioneered by the works of Reisenberger \cite{Reis1}, many different spin foam models have 
been developed for gravity in 4d
\cite{BC,Reis1,Reis2, Reis3, Iwa1,Iwa2,BC2}, opening a new path towards a
 formulation of a non-perturbative quantum gravity theory; some of
 these models make use of methods and ideas from category theory as well \cite{crane95}, and 
moreover can be derived using a generalization of the matrix models developed for 2d gravity and 
string theory, namely the group field theory formalism. Also, the evolution of spin network can be 
recast in the form of quantum causal histories, and again, leads to a spin foam formalism.

Thus, we can say that, strikingly, spin foam models lie at the point
of convergence of research areas as different as canonical quantum
gravity, topological quantum field theories, category theory, statistical mechanics
 and field theory, lattice gauge theory, path integral formulation of
quantum gravity, matrix models, simplicial gravity, causal sets, to name but a few, with consequent
 and remarkable
cross-fertilization of ideas and techniques.

Let us now give the abstract definition of spin foams and spin foam models.

Let us start with the definition of what is a spin foam \cite{Baez}. It is very analogous to the definition of a spin network, but everything is one dimension higher. More precisely:

- given a spin network $\Psi=(\Gamma,\rho,\iota)$, a spin foam $F:0\rightarrow\Psi$ is a triple   $(\kappa,\tilde{\rho},\tilde{\iota})$ where: $\kappa$ is a 2-dimensional oriented complex such that $\Gamma$ borders $\kappa$, $\tilde{\rho}$ is a labeling of each face $f$ of $\kappa$ by an irreducible representation $\tilde{\rho}_{f}$ of a group $G$, and $\tilde{\iota}$ is a labelling of each edge $e$ not lying in $\Gamma$ by an intertwiner mapping
(the tensor product of) the irreducible representations of the faces
incoming to $e$ to (the tensor product of) the irreducible
representations of the faces outgoing from $e$, such that: for any edge $e'$ of $\Gamma$ $\tilde{\rho}_{f}=\rho_{e'}$ if 
$f$ is incoming to $e'$ and $\tilde{\rho}_{f}=\rho^{*}_{e'}$ if it is outgoing from it, and for any vertex $v$ in $\Gamma$ 
$\tilde{\iota}_{e}=\iota_{v}$ with appropriate dualization.

Similarly, given two disjoint spin networks $\Psi$ and $\Psi'$, a spin foam $F:\Psi\rightarrow\Psi'$ is defined to be the 
spin foam $F:0\rightarrow\Psi^{*}\otimes\Psi'$, where the dual of a spin network is defined to be a spin network with the 
same underlying 1-dimensional oriented complex with each edge labeled by dual representations $\rho^{*}_{e}$ and 
consequently the intertwiners appropriately dualized, while the tensor product of two spin networks is the spin network whose underlying oriented 1-complex is the disjoint union of the complexes of the original spin networks and with the labelling coming from them. 

Now some properties. A spin foam is {\it non-degenerate} if every vertex is the endpoint of at least one edge, every edge of
 at least one face, and every face is labelled with a non trivial irrep of $G$. We can define an equivalence relation for 
spin foams  so that two spin foams are equivalent if one can be obtained from the other by a sequence of the following 
moves: {\it affine transformation} - a spin foam $F=(\kappa, \rho, \iota)$ is obtained from $F'=(\kappa', \rho', \iota')$ by
 an affine transformation if there is a 1-1 affine map between cells of $\kappa$ and of $\kappa'$, preserving their 
orientations, and $\rho_{f}=\rho'_{\phi(f)}$ and $\iota_{e}=\iota'_{\phi(e)}$; {\it subdivision} - $F'$ is obtained from 
$F$ by subdivision if: the oriented complex $\kappa'$ is obtained from $\kappa$ by subdivision, when $f'$ is contained in 
$f$ then $\rho'_{f'}=\rho_{f}$, when $e'$ is contained in $e$ then $\iota'_{e'}=\iota_{e}$, while if $e'$ is the edge of two faces of $\kappa'$ contained in the same face $f$ of $\kappa$ then $\iota'_{e'}=1_{\rho_{f}}$; {\it orientation reversal} - $F'$ is obtained from $F$ by orientation reversal if: $\kappa$ and $\kappa'$ have the same cells but possibly different orientations, $\rho'_{f}=\rho_{f}$ or $\rho'_{f}=\rho_{f}^{*}$ if the complexes $\kappa'$ and $\kappa$ give equal or opposite orientations to the face $f$ respectively, and $\iota'_{e}=\iota_{e}$ after appropriate dualization.

We can then define {\it composition} between two (equivalence classes of) spin foams $F:\Psi\rightarrow\Psi'$ and 
$F':\Psi'\rightarrow\Psi''$ as follows. We choose representatives of $F$ and $F'$ in the same space $\mathbb{R}^{n}$ such 
that $\Psi'$ is the same in both and the affine maps $c,c':\Gamma'\times[0,1]\rightarrow\mathbb{R}^{n}$, by which 
$\Gamma'$ borders both $\kappa$ and $\kappa'$, fit together to a single affine map 
$C:\Gamma'\times[-1,1]\rightarrow\mathbb{R}^{n}$. Then the composite $FF'$ is defined to be the spin foam whose complex is 
$\kappa\cup\kappa'$, with label given by the original labels, while the edges of $\Gamma'$ are now labelled by 
(appropriately dualized) identity intertwiners.

It is important to note that we can define, for any given group $G$, a category
$\mathcal{F}$ in which the objects are non-degenerate spin networks
and the (non-degenerate) spin foams represent morphisms between them \cite{Baez,Baez2}. In order to have associativity of composition and unit laws to hold, in addition to the given definitions of equivalence and composition, we impose the extra equivalence relations $F(GH)\sim (FG)H$, for any spin foams $F, G, H$, and $1_{\Psi}F\sim F\sim F 1_{\Psi}$, where, for any spin network $\Psi$, $1_{\Psi}:\Psi\rightarrow\Psi$ is a left and right unit.    

We note also that a generic slice of a spin foam is a spin network so that every spin foam can be considered as a composition of \lq\lq smaller" spin foams composed along common spin networks.

Moreover, with this formulation the close analogy is apparent between the category of spin foams and the category of cobordisms used in topological quantum field theory. In fact, if we choose $G$ to be the trivial group, then a spin network is just a 1-dimensional complex, thought to represent space, and a spin foam is just a 2-dimensional complex representing spacetime. The use of a non trivial group corresponds to the addition of extra labels, representing fields, in our present case the gravitational field, i.e. the geometry.

This analogy leads us to consider spin foams as a tool for calculating transition amplitudes for the gravitational field and for formulating a quantum theory of gravity.

Consider an n-dimensional compact oriented cobordism $\mathcal{M}:S\rightarrow S'$ (spacetime), with $S$ and $S'$ compact 
oriented $(n-1)$-dimensional manifolds (space). A triangulation of $\mathcal{M}$ induces triangulations on $S$ and $S'$ with 
dual skeletons $\Gamma$ and $\Gamma'$ respectively (a dual skeleton has one vertex at the center of each $(n-1)$-simplex and 
an edge intersecting each $(n-2)$-simplex). We can thus consider spin networks whose underlying graph is this dual 
1-skeleton (so we work with embedded spin networks). We know that the space of all the possible spin networks embedded in 
$S$ (or $S'$) (so for all the possible triangulations) defines the gauge invariant state space $\mathcal{H}$ (or 
$\mathcal{H}'$) on $S$ (or $S'$), so that time evolution is naturally given by an operator $Z(\mathcal{M}):\mathcal{H}\rightarrow\mathcal{H'}$. Since as we said spin networks are a complete basis for the state space, to characterize this operator it is enough to have the transition amplitudes $\langle\Psi'\mid Z(\mathcal{M})\mid\Psi\rangle$ between two spin networks. The idea is then to write this amplitude as a sum over all the possible spin foams going from $\Psi$ to $\Psi'$:
\bes
  Z(\Psi',\Psi)\,=\,\langle\Psi'\mid\Psi\rangle\,=\,\sum_{F:\Psi\rightarrow\Psi'}\,Z(F) \label{eq:ampli}
\ees
where the sum over spin foams implies both a sum over 2-dimensional complexes (or, in the dual picture, over all the triangulations of $\mathcal{M}$ matching the graphs of $\Psi$ and $\Psi'$ on the boundary), and  a sum over all the possible labelling of the elements of them by irreps of $G$. We see that in this picture spin networks are states while spin foams represent histories of these states. So the problem is now to find the right form of the amplitude $Z(F)$ for a given spin foam $F$. Moreover, this amplitude should satisfy
\bes 
Z(F')Z(F)\,=\,Z(F'F)      
\ees
where $FF':\Psi\rightarrow\Psi''$ is the spin foam obtained gluing together $F:\Psi\rightarrow\Psi'$ and $F':\Psi'\rightarrow\Psi''$ along $\Psi'$. When this happens, and the sum in (~\ref{eq:ampli}) converges in a sufficiently nice way, we have:
\bes 
Z(\mathcal{M}')Z(\mathcal{M})\,=\,Z(\mathcal{M}'\mathcal{M})      
\ees
for composable cobordisms $\mathcal{M}:S\rightarrow S'$ and $\mathcal{M}':S'\rightarrow S''$.

We want to stress that even though we used a picture of spin networks and spin foams living in a 
triangulated manifold, this is a priori not necessary and one can instead work with abstract (non 
embedded) objects.

The expression for the amplitudes of the propagation operator in loop quantum gravity can be seen as an example of a spin 
foam model for 4-dimensional quantum gravity. More generally, a spin foam model is given by a partition function $Z$ 
expressed as a sum over topologically inequivalent spin foams, this meaning a sum over all the possible 2-complexes and a 
sum over all the possible representation of the group $G$ labelling the faces of it, with a weight for each spin foam given,
 in general, by a product of amplitudes for the elements of the spin foam, i.e. faces, edges, vertices:
\be
Z\,=\,\sum_{\sigma:\partial\sigma=\Psi\cup\Psi'}w(\sigma)\sum_{J}\prod_{f}A_{f}\prod_{e}A_{e}\prod_{v}A_{v} \label{eq:spinfoam} \ee
including the case in which $\Psi$ and/or $\Psi'$ are null spin networks, i.e. the vacuum.

We stress that, although we have labelled the representations of the group used with the 
\lq\lq $SU(2)$-like" symbol $J$, and the sum over the representations labelling the spin foam faces
 as a discrete sum, these representations can be, and indeed are in the known cases, continuous real 
parameters when dealing with non-compact group, such as the Lorentz group, in both 3 and 4 dimensions, and 
consequently in these cases the sum over representations is actually an integral over all of them.

Also, there is no {\it a priori} reason to expect only one set of algebraic data labelling the spin 
foam faces, although this is the original definition; one could envisage a situation where the 
geometry of the (simplicial) manifold is decribed in terms of two sets of variables, as in a first 
order formulation of gravity (where we have both a triad/tetrad field and a spin connection), both
expressed of course in a purely algebraic form, and both appearing in the formula for the amplitudes 
to be assigned to the different elements of the 2-complex. Therefore we would also have two sets of 
sums/integrals over them, representing the sum over gravitational histories. Of course we would 
consequently have two sets of variables labelling the states of quantum gravity, i.e. the spin 
networks associated with the spin foams. We will see examples of this situation in the following.

There are several ways to look at a spin foam model. One is to
 consider it as giving the transition amplitudes for the gravitational
 field, consequently giving a precise formulation of the dynamics of
 the quantum theory (a problem still not fully resolved in the
 canonical loop quantum gravity approach), in the form of a sum over
 histories of its states, so we have a formulation which seems a
 precise implementation of the idea of a gravitational path integral
 \cite{HarHaw, Haw}. In this sense spin foam models represent a new
 form of the covariant approach to quantum gravity. Correspondingly we
 would like to interpret them as realizing the sum over geometries
 proposed in this approach as the way to realize a quantum gravity
 theory. And indeed this can be made more precise, as we will see in
 more detail in the following, since one can look at a spin foam as
 giving a quantum 4-geometry just like a spin network gives a quantum
 3-geometry. In fact one can think of the 2-complex of a spin foam as
 being the dual 2-skeleton of a 4-dimensional triangulation, so that
 it has a vertex at the centre of each 4-simplex, an edge intersecting
 each tetrahedron, and a face intersecting each triangle. Then we can think that the representation of $G$ labelling a face 
gives an area to the triangle this face intersects, and that the intertwiner labelling an edge gives a volume to the 
tetrahedron the edge intersects. This idea will be made precise and proven in the following. 
We note also that, seeing the 2-complex as the dual 2-skeleton of a triangulation, the sum over 2-complexes in 
(~\ref{eq:spinfoam}) is interpretable as a sum over triangulations, a crucial step, necessary to recover triangulation 
independence and the full degrees of freedom of the gravitational field in 4-dimensions when the theory itself is not 
topological.
Another way to look at a spin foam model, suitable in a simplicial context, is to regard it as a state sum model \cite{B} 
for a (triangulated) manifold of the kind used in statistical mechanics. Consider a 4-dimensional triangulated manifold $M$,
 and assign a set of states $S$ to each $n$-simplex $s_{n}$, with vertices $0, 1,...,n$, for each $n\leq 4$ and the same set
 for any simplex of the same dimension. There are then maps $\partial_{i}:S(s_{n})\rightarrow S(s_{n-1})$, for 
$i=0,...,n$, by means of which a state on a $n$-simplex can specify a state on anyone of its faces, the $i$-th map 
corresponding
 to the $i$-th ($n-1$)-face. Of course the states on intersecting simplices are related, because the boundary data must 
match. Then we specify a weight ($\mathbb{C}$-number) giving an amplitude to each state $w:S(s_{n})\rightarrow\mathbb{C}$. 
Finally we use all this information, which together specifies a {\it configuration} $c$, to construct a partition function 
on the triangulated manifold of the form: 
\be
Z(M)\,=\,\sum_{c}\prod_{s} w(s(c))
\ee where the sum is over all the possible configurations and the product is over all the simplices (of every dimension) of 
the triangulation.

We recognize that, again with the identification of the 2-complex of a spin foam with the 2-skeleton of a triangulation, and
 so of 4-simplices with vertices, tetrahedra with edges and triangles with faces, this state sum is exactly analogous to a 
spin foam model (~\ref{eq:spinfoam}) if we think of the sum over representations in this as the sum over configurations 
above. Of course if we want a state sum model for quantum gravity we should at the end get rid of the dependence on the 
particular triangulation chosen, and there are several ways in which this can be done, as we will see.

The crucial point is that a spin foam model aims to be a non-perturbative (since it is not based
on any kind of perturbative expansion, with the exception of the
formulation based on the field theory over a group, where, as we will
see, the spin foam model is given by the perturbative expansion of the
group field theory\footnotemark \footnotetext{However, even in this case it is a
perturbative expansion of a very unusual kind since it is not about a fixed
background geometry, but about ``nothingness'', as we shall see.}) and background independent (with the geometry and the 
metric being emerging concepts and not a priori fixed structures) formulation of quantum gravity.  

With respect to the other approaches to quantum gravity we described
above, we can easily see that elements of all of them are involved in
the definition of spin foam models. The very definition of a spin foam
model is exactly of the same kind as that defining a topological
quantum field theory, and indeed, as we have noted already, it can be
argued that quantum gravity should have the same general structure of a
topological quantum field theory, but with an infinite number of
degrees of freedom, to allow for local (but not localized!) geometric
excitations propagating causally (i.e. quantum gravitational waves,
or gravitons); the Hilbert spaces of states used and the
language by which the amplitudes are constructed are very close to
those of loop quantum gravity, with the models themselves being
defined by a sum over histories of spin networks, thus realising in a
background independent and purely combinatorial and algebraic way a
path integral approach to quantum gravity; in addition, the same states
and histories have a clear-cut intepretation in terms of simplicial
geometry, as we shalll see, and the models can be seen as a dynamical
triangulation approach, with a sum over all the triangulations and all
the topologies as well, but with additional dynamical geometric data
summed over, as in quantum Regge calculus; the spin foam amplitudes of
the models we will deal with,
moreover, will have a manifest connection with the Regge calculus
action for gravity, as we are going to see, making the connection with
simplicial gravity even more apparent. Also the connections with
causal sets and quantum causal histories will become manifest in the
following.  

Already from the general abstract structure of spin foam models we see
at play all the fundamental ideas we discussed. The theory is
background independent and fully relational in the same sense as loop
quantum gravity (with which they share the form of quantum states and
histories) or dynamical triangulations are, finitary and discrete just
as simplicial quantum gravity is, the Lorentz group and representation
theory algebras play a central role since they furnish the variables
with which to describe the geometry, with an operational flavor in
both the definition of the observables and states of the quantum
theory (quantum spin networks) and in the way the topology of
spacetime is described. The approach is also fully covariant, both in
the definition of the states and in the formulation of their evolution
as a sum-over-histories. As for causality, a suitable definition of
the spin foam amplitudes would turn the models into the quantum causal
histories form, if also the dual 2-complex is suitably put in
correspondence with a poset structure, and consequently causality
would have full implementation, as we shall discuss.   

\chapter{Spin foam models for 3-dimensional quantum gravity}
We are now going to show how the ideas discussed in the previous section are implemented in the case of 3-dimensional 
gravity, where the quantization of classical general relativity can be carried out to a great extent and in several 
alternative ways\cite{Carlip2}. The main reason why 3-d gravity is so much easier to quantize than the 4-dimensional case is the fact that 
pure general relativity (i.e. in the absence of matter) is a topological field theory, i.e. a theory with no local degrees 
of freedom, no real dynamics, and posseses only global or topological degrees of freedom, at least when the manifold 
considered is closed. In the presence of boundaries, due to the non-trivial interplay between gauge symmetry and boundary 
conditions, the theory may acquire dynamical boundary degrees of freedom, even in the absence of matter fields. The typical 
example of this situation is given by Chern-Simons theory, which gives rise to a Wess-Zumino theory on the boundary, if 
particular boundary conditions are chosen.
The topological nature of pure gravity in $3d$ will affect several
aspects of the spin foam models we are going to present in the
following. In particular, although all using a triangulation of the
spacetime manifold to define the partition function, they result in being
invariant under any topology preserving change of the triangulation
itself, and their partition function defines a topological invariant
of the manifold, i.e. a complex number characterizing its topology. 

\section{3-dimensional gravity as a topological field theory: continuum and discrete cases}
Consider the classical action for general relativity in the first
order (Palatini) formalism:

\bes
S(e,\omega)\,=\,
-\frac{1}{2}\int_{\mathcal{M}} 
tr \left( e\wedge
F(\omega)\right)\,=
\,-\,\int_{\mathcal{M}}\epsilon_{ijk}\,
e^i\,\wedge\,F^{jk}(\omega).\ees

The variables are a 1-form $e^{i}$, the triad field, with values in the Lie algebra of
$SU(2)$ in the Riemannian case and of $SL(2,\mathbb{R})$ in the
Lorentzian case, and a 1-form $\omega$, the spin connection, also with value in the Lie
algebra of the same group, and with curvature
$F(\omega)=d\omega+\omega\wedge\omega=d_\omega \omega$, where $d$ is
the exterior derivative of forms and $d_\omega$ is the covariant
derivative with respect to the connection $\omega$. The metric field is
given in terms of the triad field as $g_{\mu\nu}=\eta_{ij} e^{i}_\mu
e^{j}_\nu$, with $\eta= diag(+++)$ in the Riemannian case and
$\eta=diag(-++)$ in the Lorentzian. The generators of the Lie algebra
are denoted by $J^{i}$.

The equations of motion obtained by variation of this action
(Einstein's equations) are:

\bes
d_\omega\, e\,=\,0 \;\;\;\;\; F(\omega)\,=\,0,
\ees
which enforce the metricity of the connection on the one hand,
i.e. the compatibility between the triad (metric field) $e$ and the connection
$\omega$, and on the other hand force the curvature $F(\omega)$ to be
vanishing everywhere, thus excluding any non-flat
configuration. Therefore no local excitations (gravitational waves)
are possible.  
 
In fact, the action above can be seen as the 3-dimensional case of a
general $BF$ topological field theory action\cite{horowitz,birblau}:

\bes
S[B,A]\,=\,-\frac{1}{2}\int_{\mathcal{M}} tr\left(B\,\wedge\,
F(A)\right)\,=\,-\,\int_{\mathcal{M}}\epsilon_{ijk}\,B^{i}\,\wedge\,F^{jk}(A)
\ees
based on an $(n-2)$-form $B$, with values in the Lie algebra of a given
group (here the 3-dimensional Lorentz group), and a 1-form connection
$A$. This action is topological in any dimension, since the equations
of motion always constrain the connection to be flat.  

It is interesting to analyse the symmetries of this action
\cite{lldiff}.
The theory is first of all invariant under local Lorentz gauge
symmetry ($SU(2)$ or $SL(2,\mathbb{R})$ gauge transformations,

\bes
\delta_X^L\,\omega\,=\,d_\omega\,X\;\;\;\;\;\;\delta_X^L\,e\,=\,\left[\,e,\,X\,\right],
\ees 
where $X$ is an arbitrary Lie algebra element.

Then we have a translational symmetry:

\bes 
\delta_\phi^T\,\omega\,=\,0\;\;\;\;\;\;\delta_\phi^T\,e\,=\,d_\omega\,\phi,
\ees again for a Lie algebra element $\phi$, and it is easy to see that
this symmetry is due to the Bianchi identity
$d_\omega\,F(\omega)\,=\,0$.

Finally we have invariance under diffeomorphisms, i.e. under
transformations of the field variables as:

\bes
\delta_\xi^D\,\omega\,=\,d\,(\iota_\xi\,\omega)\,+\,\iota_\xi\,(\,d\,\omega)\;\;\;\;\;
\delta_\xi^D\,e\,=\,d\,(\iota_\xi\,e)\,+\,\iota_\xi\,(\,d\,e),
\ees
 
where $\iota_\xi$ is the interior product on forms.

These symmetries are not independent as one can show that\cite{lldiff}:

\bes
\delta_\xi^D\,e\,=\,\delta_{(\iota_\xi\omega)}^L\,e\,+\,\delta_{(\iota_\xi
  e)}^T\,e\,+\,\iota_\xi\,(d_\omega\,e) \\  \delta_\xi^D\,\omega\,=\,\delta_{(\iota_\xi\omega)}^L\,\omega\,+\,\delta_{(\iota_\xi
  e)}^T\,\omega\,+\,\iota_\xi\,(d_\omega\,\omega).
\ees

Therefore on shell we can simply write:
$\delta_\xi^D\,=\,\delta_{(\iota_\xi \omega)}^L\,+\,\delta{(\iota_\xi
  e)}^T$, so the diffeomorphism symmetry is a combination of Lorentz
and translation symmetries with field dependent parameters
$X=\iota_\xi \omega$ and $\phi=\iota_\xi e$. This is just a local
Poincare symmetry, as can be seen by introducing a Poincare connection
$A=\omega^i J_i + e^i P_i$, with $P_i$ the generators of local translations\cite{Witten}.  

This analysis of the symmetry helps to understand \cite{lldiff} the
regularization that will be later needed to make the Ponzano-Regge
model finite.

Let us now describe how the discretization of this action is perfomed.
Consider an oriented triangulation $T$ of the manifold
$\mathcal{M}$. The triad field, being a 1-form, is naturally integrated
along 1-dimensional objects, i.e along the edges of the triangulation,
and when we do so, we obtain a collection of Lie algebra elements
$E^{i}=\int_e e^{i}(x)$ associated to the edges of the simplicial
complex.
We want now to discretize the connection field, in such a way that it is
also associated to 1-dimensional objects and that defines a simplicial
measure of the curvature located on the edges, as we know is the case
in simplicial gravity. The way to do it is to introduce the simplicial
complex $T*$ dual to the triangulation $T$, having a $(n-d)$- simplex for any
$d$-simplex of $T$. Then the connection $\omega$ can be integrated
along the links $e*$ of $T*$ thus giving holonomies associated with
them. These holonomies are group elements $g_{e*}$. The curvature is
then naturally defined as the product $g_{f*}=\prod_{e*\subset
  \partial f*} g_{e*}$ of the group elements $g_{e*}$ associated with
the links of the boundary of the dual face $f*$, and it is thus
associated with the dual face itself. This is in turn dual to the
edges of the triangulation $T$, so we have the simplicial curvature
associated to them, as we wanted.       
The logarithm of the group element $g_{f*}$ gives a Lie algebra
element $\Omega_e$, that we may think as the actual discretization of
the curvature field on the edges of $T$.
The discretized set of variables is thus $(E_e,\Omega_e)$, all
associated to the edges of the triangulation.

Now the corresponding discretized action is taken to be:

\bes
S(E.\Omega)\,=\,\sum_{e\in T}\,tr\left(E_e\,\Omega_e\right).\label{eq:bfdiscrete}
\ees

Although clearly plausible, we have to check that this action has all
the (discrete analogues of the) symmetries of the continuum action. It
can be shown \cite{lldiff} that it is indeed invariant under:
1) local Lorentz transformations: $\Omega_e\rightarrow
g_{v*}^{-1}\Omega_e g_{v*}\;\;\;E_e\rightarrow g_{v*}^{-1}E_e g_{v*}$,
where $g_{v*}$ is a group element associated to the vertex $v*$ of the
dual complex; 2) discrete translations: $\delta E_e = \Phi_v -
[\omega_e^v , \Phi_v]$, where $\Phi_v$ is the discrete analogue of the
field $\phi$ associated to the vertices of $T$, and $\omega_e$ may be
thought of as the object obtained by integrating the 1-form $\omega$
over the edge $e$ of $T$ starting from the vertex $v$, and can be expressed entirely in terms of
the $\Omega_e$ of the edges starting from that vertex\cite{lldiff}.
We do not expect full invariance under what were the continuum diffeomorphisms, since the very
choice of a triangulation breaks such a symmetry, and this has
important consequences for the spin foam models as well
(\cite{lldiff}.

\section{Quantum simplicial 3-geometry} 
We now want to translate the above simplicial geometry, with edges and
edge lengths, triangles and areas, tetrahedra and 3-volumes, into a
picture of quantum 3-geometry.

The idea is to quantize the basic variables of the theory and to obtain
a state associated to each 2-dimensional surface in the simplicial
manifold, given by a collection of triangles glued together along
common edges, and an amplitude for each 3-dimensional manifold, given
by a collection of 3-simplices (tetrahedra) glued together along
common triangles, and then to use these amplitudes to construct a
suitable path integral expressions for the transition amplitudes of
the theory. 

Here we first show how all the elements of the triangulation are given
a quantum description, and quantum states are constructed and the
fundamental building blocks for the transition amplitudes can be
identified. In the next section we give an explicit derivation of the
partition function and transition amplitudes for the theory based on
these building blocks.

While the discussion up to now did not refer to any choice of
signature for spacetime explicitly, here we consider explicitly the
Riemannian case first, and we will later discuss the
Lorentzian case.

Also, we concentrate on a single tetrahedron (an atom of spacetime)
and then discuss how to consider a collection of many tetrahedra,
i.e. the whole Riemannian simplicial manifold.

Recall \cite{barbieri} that a classical tetrahedron is the convex
envelope of four points in $\mathbb{R}^3$ and it is thus completely
characterized by the 12 cartesian coordinates for the 4 points, so
that the intrinsic geometry of the tetrahedron,  i.e. a tetrahedron
modulo translations (3 parameters) and rotations (3 parameters) in
$\mathbb{R}^3$, i.e. modulo its location and orientation in
$\mathbb{R}^3$, is characterised by only 6 parameters ($12-3-3=6$). If
we have a triad of independent vectors $E_a^i$ associated to three of its
edges, we can take these 6 parameters to be the scalars:
$E_1\cdot E_1$, $E_2\cdot E_2$, $E_3\cdot E_3$, $E_1\cdot E_2$,
$E_2\cdot E_3$, $E_1\cdot E_3$, i.e. three edge lengths (squared) and three
angles.
Alternatively, we can use a full set of six edge vectors $E^i_a$, one
for each edge of the tetrahedron, and use as parameters the six edge
lengths (square): $E_1\cdot E_1$, $E_2\cdot E_2$, $E_3\cdot E_3$,
$E_4\cdot E_4$,
$E_5\cdot E_5$, $E_6\cdot E_6$.

Let us also note (since it will be important later) that from these six
numerical parameters alone it is not posible to distinguish a
tetrahedron from its mirror image, i.e. from its parity-transform,
since all these parameters are invaraint under the spacetime reversal (parity)
transformation $E^i\rightarrow -E^i$, so any model for quantum
geometry (and any transition amplitude) constructed out of these only
will not distinguish a tetrahedron from its parity transform.  

The fundamental metric variable to be used to express geometric quantities is the
discrete variable corresponding to the triad field $E^i$. It is
associated to the edges of the triangulation, as we have seen,
therefore it gives exactly the edge vectors whose lengths characterize
completely the tetraedron geometry, and it
is given by an $SU(2)$ Lie algebra element. 

There is one direct and natural way to quantize such a variable, and
it is to choose a representation $j$ of $SU(2)$ and turn this Lie
algebra element into an operator acting on the corresponding
representation space $V^j$. In this way we have a Hilbert space and a
set of operators acting on it associated to each edge of the
triangulation. In particular, we can associate to each edge an element
of the canonical basis of the $SU(2)$ Lie algebra, $J^i$, in some
representation $j$. Doing this, the operator corresponding to the
square of the edge length $E_a\cdot E_a$ for the edge $a$ is given by
the $SU(2)$ Casimir $C=L^2=J_a \cdot J_a$, which is diagonal on the
representation space $V^{j_a}$ with eigenvalue
$L^2_a=j_a(j_a+1)$\footnotetext{One obtains again the Casimir of the
  representation when one quantizes directly the continuum length
  operator for a 1-dimensional object, in this case the edge $a$, but
  in this case one choice of ordering of the operators entering in the definition of the
  length gives the above result, while another gives the eigenvalues
  as $L_a^2= (j+1/2)^2$\cite{alekseev}, which, as we will see, fits
  better with the asymptotics results for the amplitudes of the
  Ponzano-Regge spin foam model and with the Regge calculus
  interpretation of these.}. Therefore we see that the representation
label $j$ gives the quantum length of the edge to which the
corresponding representation is assigned, thus the Hilbert space $V^j$
can be interpreted as the Hilbert space for a 3-vector with squared length
$j(j+1)$.
One can thus define the Hilbert space of a quantum vector, or, in this
simplicial context, of a quantum edge, as $\mathcal{H}_e=\oplus_{j_e}
V^j_e$, i.e. as a direct sum of all the Hilbert spaces corresponding
to a possible fixed edge length. Note that, of course, in
characterizing an edge by its length only, we are neglecting some
geometric information; the point is that however the limited
information provided by the six edge lengths is enough to characterize the full
geometry of the tetrahedron built out of them, as we said, and that
this is the spacetime geometry we are trying to quantize in the end.

This is the fundamental Hilbert space out of which the other Hilbert
spaces the theory may deal with, including the full state space, are
constructed. 

So let us go one step further and construct the Hilbert space of a
{\bf quantum triangle}.
Consider a classical triangle made out of the three edges with edge
vectors $E^i_a\;\;a=1,2,3$; it is specified by any pair of these edge
vectors alone, but we can have a more symmetric characterization of it
using all three vectors subject to the constraint $E_1+E_2+E_3=0$,
that we may call a {\it closure constraint}, enforcing the requirement
that the three vectors $E_a^i$ come indeed from a closed oriented (a
change in orientation of any edge changes $E_a$ to $-E_a$) geometric
triangle in $\mathbb{R}^3$. Also, in order to be a well-defined
geometric triangle, the edge lengths have to fullfill the Riemannian triangle inequalities.

Using the Hilbert spaces for edges defined above, it is clear that a
state for a quantum triangle with given edge lengths is an element
$\psi$ in the Hilbert space
$V^{j_1j_2j_3}=V^{j_1}\otimes V^{j_2}\otimes V^{j_3}$; however, we
have to impose the quantum analogue of the closure constraint; this is
easily seen to be just the $SU(2)$ invariance of the state $\psi$; in
other words, the state is an element of the space of invariant tensors
for the given three tensored representation spaces, $\psi\in
Inv\left(V^{j_1}\otimes V^{j_2}\otimes V^{j_3}\right)$, so
$\psi:V^{j_1}\otimes V^{j_2}\otimes V^{j_3}\rightarrow \mathbb{C}$.
If one takes into account the triangle inequalities, now translated
into the same inequalities for the quantum representation parameters
$j_a$, i.e. being given by $\mid j_1-j_2\mid \leq j_3 \leq j_1+j_2$
and the like, there is only one possible choice for such a state for
the quantum triangle, up to a constant factor; there exist only one
quantum triangle with give quantum edges. This reflects the fact that
the classical Riemannian geometry of a triangle is specified, up to its
embedding in $\mathbb{R}^3$, entirely by its edge lengths. The same
uniqueness can be explained more rigorously using geometric
quantization \cite{bb}.         

This invariant tensor is a well-known object in the representation
theory of $SU(2)$: the intertwiner between the three representations
$j_1,j_2,j_3$, i.e. the $3j$-symbol:

\bes
\psi\,=\,C^{j_1j_2j_3}_{m_1m_2m_3}\,=\,\begin{pmatrix} 
j_1 &j_2 &j_3
\\ m_1 &m_2 &m_3 
\end{pmatrix}
\ees

We can call this unique quantum state ``a vertex'' referring to the
representation of the quantum triangle as a spin network (see figure 2.1).

\begin{figure}
\begin{center}
\includegraphics[width=7cm]{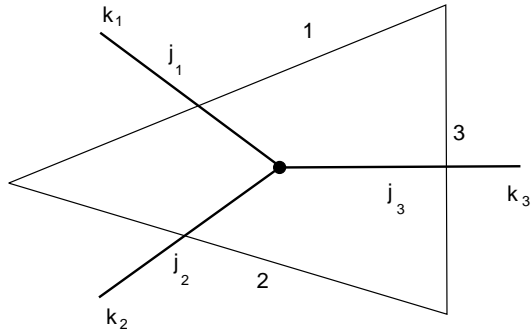}
\caption{A spin network (trivalent vertex) representing a state of the quantum triangle, with the representations of SU(2) and the state labels
associated to its three edges}
\end{center}
\end{figure}

The full Hilbert space for the quantum triangle $abc$ is then given
by:
\bes
\mathcal{H}_{abc}\,=\,\oplus_{j_a,j_b,j_c} Inv\left(V^{j_1}\otimes
V^{j_2}\otimes V^{j_3}\right).
\ees

A generic 2-surface (to which we would like to associate a quantum
gravity state) is built out of many triangles glued together along
common edges, as we said, so we construct the associated state in
a similar way, i.e. by gluing together the states associated to the
individual triangles. These states being tensors as explained, the
gluing is just the trace over the common representation space of all
the tensors for the triangles.

There is one more step further we should go. We have to construct
suitable amplitudes to be associated to the tetrahedra in the
triangulation, as building blocks for quantum gravity transition
amplitudes in a path integral-like formulation of the theory.

As the basic formalism of topological quantum field theory teaches us,
these should be linear maps between the Hilbert spaces associated to
the boundaries of the manifold we are considering. We now know what
these Hilbert spaces are, and we can easily construct the amplitude
for a single tetrahedron since we have to use only the data on its
edges. Consider first a tetrahedron with fixed edge lengths (so fixed
representations $j$ at the quantum level). A tetrahedon has the topology of $B^3$, i.e. it has a single
connected boundary component with the topology of $S^3$, in this
discrete case given by the 4 boundary triangles. It can be seen as a
cobordism $tetra: S^3\rightarrow \mathbb{C}$, therefore a TQFT should
associate to it the quantum 
amplitude 
\bes
tetra (j) : \otimes_i \psi_i\,=\,\otimes_i Inv\left(V^{j_{1_i}}\otimes
V^{j_{2_i}}\otimes V^{j_{3_i}}\right)\,\rightarrow\,\mathbb{C}.
\ees
    
The natural and simplest choice for an amplitude of this kind, that
uses the representations $j$ associated to the six edges of the
tetrahedron only, that satisfies the triangle inequalities and is
invariant under the group of rotations $SU(2)$ (our spacetime symmetry
group) is obtained by fully constracting the four invariant tensors
($3j$-symbols) for the four triangles, an operation that mimics the
closure of the four triangles to build up the boundary sphere $S^3$,
thus getting a scalar as a result.

The amplitude is thus given, up to a phase, by:
\bes
tetra (j)\,=\,C^{j_1j_2j_3}_{m_1m_2m_3}\,C^{j_3j_4j_5}_{m_3m_4m_5}
C^{j_5j_1j_6}_{m_5m_1m_6} C^{j_6j_2j_4}_{m_6m_2m_4}\,=\,\{\, 6j\,\},
\ees
 
i.e. by the $6j$-symbol:

\bes
tetra (j)\,=\,\{ \,6j\, \}\,=\,\begin{bmatrix} 
j_1 &j_2 &j_3
\\ j_4 &j_5 &j_6 
\end{bmatrix},
\ees
another well-known object in the representation theory of $SU(2)$.

Therefore, to conclude, we have labelled each edge of the
triangulation by a representation $j$ of $SU(2)$, determined a state
for each triangle, graphically represented by  a tri-valent spin
network vertex, and obtained a quantum amplitude associated to each
tetrahedron, given by a $6j$-symbol. This can also be given a
graphical representation, and an evaluation, as a spin network (the
``tetrahedral'' spin network) by: 

\begin{figure}
\begin{center}
\includegraphics[width=7cm]{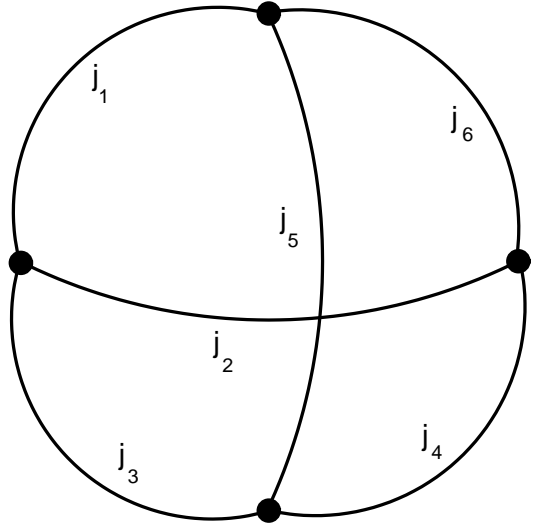}
\caption{The tetrahedral spin network, obtained by connecting the four trivalent 
vertices corresponding to the four triangles in the boundary of a tetrahedron.}
\end{center}
\end{figure}

Up to now we have worked with fixed edge lengths, but it is easy to lift
this restriction by defining the amplitude associated to a
tetrahedron, for any choice of edge lengths, as:

\bes
tetra\,=\,\left(\prod_i\sum_{j_i} \Delta_{j_i}\,\right)\,\{\,6j\,\}
\ees
where we have included a yet undetermined measure factor for each
representation $j$. 

For a generic simplicial complex we then expect
the amplitude to be given by a product of $6j$-symbols, one for each
tetrahedron, and by a similar sum over representations for all the
edges in it.

This procedure allowed a complete translation of the simplicial
geometry  into purely combinatorial and algebraic terms, i.e. into a
spin foam language.

\section{The Ponzano-Regge spin foam model: a lattice gauge theory derivation}
We are now going to show how the construction we have just outlined is
indeed the result of a lattice gauge theory type of quantization \cite{Ooguri,F-K,OLough} of
the simplicial action for 3d gravity, i.e. for 3d BF theory, and how a
complete and explicit model using the representations $j$ as variables
and the $6j$-symbols as quantum
amplitudes  can be obtained. This is the so-called Ponzano-Regge
model.

We want to realize explicitely, in a discrete context, the path
integral 

\bes
Z\,=\,\int\,\mathcal{D} e\,\mathcal{D} \omega\,e^{i\,\int_{\mathcal{M}}tr\left(
  e\,\wedge\,F(\omega)\right)},
\ees
 which becomes, using the discretization procedure outlined above\cite{lldiff}:
\bes
Z(T)\,=\,\int_{\g^E}\prod_{e\in E}
dE_e\,\int_{G^{E*}}\prod_{e*\in E*}d g_{e*} \,e^{i\,\sum_{e\in T}\,tr\left(E_e\,\Omega_e\right)},
\ees
where $\g$ is the Lie algebra of $G=SU(2)$, and $E$ and $E*$
are the sets of edges of $T$ and dual edges (links of $T*$)
respectively, and the measures used are the invariant measures on the
Lie algebra and group, such that the volume of the group is normalized
to $1$. We are going to make active use of the dual complex $T*$
in what follows, so we are basically constructing quantum 3d gravity
or quantum BF theory as a lattice gauge theory on the simplicial
complex dual to the spacetime triangulation; note that no metric
information about $T*$, however, enters in the action nor in the
partition function, and that we are only using the combinatorial
information provided to us by $T*$, as is to be expected, since we
are dealing with a theory which is supposed to {\it define} this
metric information through the field $E^i_e$.  
Here and in the following the triangulation $T$ is supposed to be closed (no boundaries).
Note also that the partition function above is invariant under change
of global orientation, given in the continuum by a change in the orientation
of the triad field $e$, since in the discrete case both $E_e$ and
$\Omega_e$ change sign under such a transformation. This means that
the quantum model we are going to derive will not distinguish the
different possible orientation of the simplicial complex on which it
is based, as was to be expected from the discussion about quantum
simplicial geometry given above.    

Now, the integral over the $E_e$ variables is easily perfomed, since
it basically plays the role of a Lagrange multiplier enforcing the
flatness constraint on the discretized connection variable $g_{f*}$:
\bes
\int_{\g_e} dE_e\,e^{i\,tr\left(E_e\,\Omega_e\right)}\,=\,\delta\left(
e^{\Omega_e}\right)\,=\,\delta\left(g_{f*}\right)
\ees

Therefore the partition function has the expression:
\bes
Z(T)\,=\,\int_{G^{E*}}\prod_{e*\in E*}d g_{e*} \,\prod_{f*}\delta\left(g_{f*}\right),
\ees
because of the one to one correspondence between edges and dual faces.
Now we have to introduce the representations of $SU(2)$ into the expression, and this is simply done by the unique 
character decomposition of the delta function (Plancherel formula):
\bes
\delta\left(g_{f*}\right)\,=\,\sum_{j_{f*}}\Delta_{j_{f*}}\,\chi^{j_{f*}}\left(g_{f*}\right),
\ees
where the sum is over the irreducible representations of $SU(2)$ (all half-integers $j$), $\Delta_j=(2j+1)$, and 
$\chi^j(g)$ is the character of the group element $g$ in the representation $j$, so the partition functions acquires
 the expression:
\bes
  Z(T)\,=\,\int_{G^{E*}}\prod_{e*\in E*}d g_{e*} \,\prod_{f*}\,\sum_{j_{f*}}\Delta_{j_{f*}}\,\chi^{j_{f*}}\left(g_{f*}\right)\, 
=\, \nonumber \\ =\,\left(\prod_{f*}\,\sum_{j_{f*}} \Delta_{j_{f*}}\right)\left( \prod_{e*}\int_{SU(2)} d g_{e*} \right)\prod_{f*}\, \chi^{j_{f*}}\left(g_{f*}\right)\,= 
\nonumber \\ =\,\left(\prod_{f*}\,\sum_{j_{f*}} \Delta_{j_{f*}}\right)\left( \prod_{e*}\int_{SU(2)} d g_{e*} \right)\prod_{f*}\, \chi^{j_{f*}}
(\prod_{e*\in\partial f*}g_{e*})\label{eq:ZPR}.
\ees    

Comparing this expression with the one we started with it is clear that what we have done up to now is to replace the 
variables $E_e$ associated with the edges of $T$ with the representation labels $j_{F*}$, associated with the faces of the 
dual simplicial complex $T*$, which are in one to one correspondence with the edges of $T$, with a corresponding switch of 
the integrals with sums and the introduction of the measure $\Delta_j$; note also that all the variables are now on the first 
two layers only of the dual simplicial complex, i.e. the model has data on the dual 2-complex only. This dual 2-complex 
will in fact in the end be the 2-complex underlying the spin foams representing the histories of the gravitational field in 
the final model. 

Now we expand the characters explicitely in terms of the Wigner representation functions, in order to be able to carry out 
the integrals: $\chi^j(\prod g)=\sum_m\prod D^j_{mm'}(g)$.
It is easy to see that, because there are three dual faces sharing each dual edge, just as there are three edges bounding a
 triangle in $T$, we obtain an integral of three representation functions with the same argument for each dual edge, thus 
obtaining: 
\bes 
Z(T)\,=\,Z(T*)\,=\,\left(\prod_{f*}\,\sum_{j_{f*}} \Delta_{j_{f*}}\right)\prod_{e*} \int_{SU(2)} d g_{e*} D^{j_{f_1^{*}\ni
 e*}} _{k_1 k'_1}(g_{e*})D^{j_{f_2^{*}\ni e*}}_{k_2 k'_2}(g_{e*})D^{j_{f_3^{*}\ni e*}}_{k_3 k'_3}(g_{e*})\;\;\;\;\; 
\ees
The spin foam formulation is obtained simply by performing these integrals over the group, using the formula expressing these integrals in terms of $3j$-symbols:
\bes
 \int_{SU(2)} d g_{e*} D^{j_1}_{k_1 k'_1}(g_{e*})D^{j_2}_{k_2 k'_2}(g_{e*})D^{j_3}_{k_3 k'_3}(g_{e*})\,=\,C^{j_1\,j_2\,j_3}_{k_1\, k_2\,k_3}\,C^{j_1\,j_2\,j_3}_{k'_1\, k'_2\,k'_3}  
\ees

We thus get two $3j$-symbols for each dual edge of $T*$, i.e. for each triangle in $T$, one with the (angular momentum) 
indices $k$'s referring to one of the two vertices (tetrahedra) sharing the edge (triangle), one to the other. The indices 
referring to the same tetrahedron are fully contracted pairwise, so that we get a full contraction of four $3j$-symbols for 
each tetrahedron, i.e. a $6j$-symbol for each tetrahedron (vertex $v*$ of $T*$). 
Therefore, the partition function assumes the 
form:
\bes
Z(T)\,=\,Z(T*)\,=\,\left(\prod_{f*}\,\sum_{j_{f*}}\right)\,\prod_{f*}\Delta_{j_{f*}}\prod_{v*}\,(-1)^{c(j)}\,\begin{bmatrix} 
j_1 &j_2 &j_3
\\ j_4 &j_5 &j_6 
\end{bmatrix}_{v*}, 
\ees
where $c(j)$ is a simple linear combination of the six representations labels in each $6j$-symbol.

Now, as it stands, the partition function is badly divergent and needs a regularization; a simple regularization can be 
given adding suitable factors to the expression above \cite{PR,lldiff}, and it can be shown that the rationale for this 
regularization is the need to take into account the translation invariance of the discrete classical action \cite{lldiff}.

In any case, with such a regularization taken into account, we have obtained the explicit expression for the Ponzano-Regge 
spin foam model. The model has in fact as configurations spin foams, i.e. 2-complexes labelled by representations of $SU(2)$ 
on their faces, and suitable intertwiners on their edges. The final expression we got gives the partition function for 
quantum gravity in purely combinatorial and algebraic terms, as desired.

Of course, suitable boundary terms should be added for open manifolds, as when computing transition amplitudes between spin
network states \cite{OLough}.

\section{A spin foam model for Lorentzian 3d quantum gravity}
\label{sec:PRLor}
The lattice gauge theory derivation we have just described can be adapted in an
almost straightforward manner to the Lorentzian case, i.e. when the
group considered is the 3-dimensional Lorentz group $SO(2,1)\sim
SL(2,\mathbb{R})/\mathbb{Z}_2$, as done in\cite{freidellorentz}.
The main difficulty in extending the previous derivation to the
Lorentzian case is not conceptual, but lies in the non-compactness of
the Lorentz group, which requires a carefully done gauge fixing to
avoid divergences due to this non-compactness, and a different
definition of the (analogue of the) $6j$-symbols that cannot be
defined just as a contraction of $3j$-symbols for the same reason (it
would just give an infinite result).

However, the main technical ingredient needed in the $SU(2)$
derivation, namely the Plancherel formula for the decomposition of the
delta function on the group, can be generalized to the Lorentzian case
\cite{Ruhl,gelfand} and we may thus use the same machinery we used in
the previous case.

The discretized simplicial action and partition function are the same
as above, and again the integral over the $E_e$ variables can be
performed treating them as Lagrange multipliers, so we obtain again:

\bes
Z(T)\,=\,\int_{G^{E*}}\prod_{e*\in E*}d g_{e*} \,\prod_{f*}\delta\left(g_{f*}\right),
\ees
where of course now $G=SL(2,\mathbb{R})$, and where all the variables
live in the dual complex $T*$ (actually, only on the dual 2-complex).
Now we can use the $SL(2,\mathbb{R})$ generalization of the Plancherel
formula \cite{Ruhl,gelfand}, which expresses the delta function on the
  group in terms of irreducible unitary representations of it:
\bes
\delta(g)\,=\,\sum_j\,\left(2\,j\,+\,1\right)\left[
  \chi^j_+(g)\,+\,\chi^j_-(g)\right]\,+\,\sum_{\epsilon = 0,1}\int_0^\infty\,d\rho\,\mu(\epsilon,\rho)\,\chi^\rho_\epsilon(g) 
\ees
with the measure
$\mu(\rho,\epsilon)=2\rho\tanh(\pi\rho+i\epsilon\,\frac{\pi}{2})$.

The irreducible unitary representations appearing in this
decomposition and on which the spin foam model will be based, belong
to three series: 1) the principal series $T_{\rho,\epsilon}$ labelled
by a parameter $\epsilon=0,1$ and a continuous positive real number
$\rho>0$, and with Casimir $C_\rho\,=\,\rho^2\,+\,\frac{1}{4}>0$; 2) the
holomorphic discrete series $T_j^+$, labelled by a discrete parameter
(indeed a half-integer number) $J$, with Casimir $C_j=-j(j+1)<0$; and
the anti-holomorphic discrete series $T_j^-$, also labelled by the
same kind of parameter and with the same expression for the Casimir.
We will say more about the geometric interpretation of these three
series of representations in the following.

Using this formula in the expression above, we again introduce a sum
(or integral) over representations as a representation-theoretic substitute for the
integration over the $E_e$ variables. The rest of the derivation is
analogous to the $SU(2)$ case, but we now have to choose a suitable
gauge fixing to avoid divergences in the partition function coming
from the non-compactness of the Lorentz group.

This is done as follows: we consider a {\it maximal tree} in the dual
2-complex, i.e. a set of dual links which does not contain any closed loop and which
 cannot be extended without creating a loop, then we fix all the group elements associated to the 
links in the maximal tree to the identity, dropping the corresponding integrals from the expression
for the partition function.

The rest of the integrals are computed as before, expressing the characters in terms of matrix 
representation functions for $SL(2,\mathbb{R})$ in the canonical basis, and using the formulae 
for the recoupling theory of $SL(2,\mathbb{R})$. Each edge gets a representation of the group 
assigned to it, and again each tetrahedron $tet$ has an amplitude given by a ($SL(2,\mathbb{R})$ 
analogue of the) $6j$-symbol $T(j^+,j^-,\rho)$, defined as the matrix element of a unitary transformation on the space 
of invariant operators acting on the tensor product of four unitary representations of $SL(2,\mathbb{R})$
(which by the way is an alternative definition also for the $SU(2)$ $6j$-symbol).

To each edge we can assign a continuous representation $\rho$, and we then label it $e^0$, or a 
holomorphic discrete one $j^+$, and we label it $e^+$, or an anti-holomorphic discrete one, and we 
label it $e^-$. We can express the partition function as a sum over possible assignments $c$ of 
different types of representations to the different edges with a partition function defined for each
given assignment:

\bes
\lefteqn{Z(T)\,=\,Z(T*)\,=\,\sum_c\,Z(T,c)} \\ &Z(T,c)\,=\,\left( \prod_{e^+}\sum_{j^+}(2\,j^+\,+\,1)\right)
\left( \prod_{e^-}\sum_{j^-}(2\,j^-\,+\,1)\right) \\ 
&\left( \prod_{e^0} \sum_{\epsilon=0,1}\int d\rho_{e^0}\mu(\rho_{e^0},\epsilon)\right)
\,\prod_{tet}\,T(j^+,j^-,\rho).
\ees

If we consider an oriented triangulation, then we can relate the assignment of representations to it
 saying that a reversed orientation for the triangulation corresponds to a reversed assignment of 
 discrete representations (holomorphic $\rightarrow$ anti-holomorphic).

The $6j$-symbols have to satisfy a number of conditions in order to be non-vanishing, involving the 
three representations labelling each of the triangles in the tetrahedra. If we define the orientation
 of a triangle to be given by an ordering of its vertices modulo even permutations, these conditions
 restrict the possible assignments of representations to the three edges of a triangle to:
1) $(j^+_1,j^+_2,j^+_3)$ with $j_3 > j_1+j_2$ and $j_1+j_2+j_3$ an integer; 2) $(j^-_1,j^-_2,j^-_3)$
 with $j_3 > j_1+j_2$ and $j_1+j_2+j_3$ an integer; 3) $((\rho_1,\epsilon_1),j^+_2,j^-_3)$, with 
$j_2+j_3+\epsilon_1/2$ an integer; 4) $((\rho_1,\epsilon_1),(\rho_2,\epsilon_2),j^-_3)$, with 
$\epsilon_1/2+\epsilon_2+j_3$ an integer; $((\rho_1,\epsilon_1),(\rho_2,\epsilon_2),(\rho_3,\epsilon_3))$, with 
$\epsilon_1/2+\epsilon_2+\epsilon_3$ an integer.

The definition of the model is completed by a regularization of the same type as that used for the 
$SU(2)$ case, and motivated analogously.

The partition function $Z(T)$ can be shown to be then invariant under change of triangulation for the same 
topology, i.e. it evaluates to the same number for any simplicial complex of given topology. 
This is not true for the functions $Z(T,c)$ depending on a given assignment of 
representations, with the only exceptions being if all the edges of the triangulation are labelled by
holomorphic or all by anti-holomorphic discrete representations.

Let us conclude by discussing the geometric intepretation of the different types of representations.
The expression for the Casimir of the different representations, and its intepretation as length of 
the edge to which a given representation is assigned, leads to the natural interpretation of the 
continuous representations as describing spacelike edges, and of the discrete representations as 
describing timelike ones. Moreover, we characterize an edge labelled by an holomorphic representation 
as a future-pointing timelike one, and one labelled by a anti-holomorphic representation as a 
past-pointing timelike one. Several justifications can be given for this geometric, thus physical, 
interpretation, the most immediate being that, if one adopts it, then the admissibility conditions 
listed above for the edges in a triangle are just a representation-theoretic expression for the 
well-known rules of vector addition in $\mathbb{R}^{2,1}$. Also, given this interpretation, the 
assignments $c$ of representations are nothing but choices of possible {\it causal structures}.

\section{Quantum deformation and the Turaev-Viro model} \label{sec:TV}
There exists another spin foam model for gravity, which is related as
we will see to 3-dimensional Riemannian gravity with (positive)
cosmological constant\cite{MT}. This is the Turaev-Viro model \cite{TV,ArcWil}, which is defined
in a similar way to the Ponzano-Regge model we derived and discussed
in the previous sections, but it uses the representation theory of the
quantum $SU(2)$ algebra to define the ingredients of the spin foam
model, i.e. assigns representations of the q-deformed $SU(2)$,
$SU(2)_q$, to the edges of the triangulation $T$ (faces of the dual
2-complex), with suitable measure factors in the sum over them, and
the q-deformed analogue of $SU(2)$ $6j$-symbols as  amplitudes for the
tetrahedra of $T$ (vertices of $T*$). 

Let us see how it is defined in more details (see \cite{TV, ArcWil, MT}. We work again with a
closed manifold $\mathcal{M}$, and choose a triangulation $T$ of
it. Again we could re-phrase the whole construction in terms of the dual
complex to $T$.  

We fix an integer $r\geq 3$, and define $q=e^{\frac{2\pi i}{r}}$ as
the quantum deformation parameter in $SU(2)_q$; consider a set $I=\{0,
\frac{1}{2}, 1, ..., \frac{1}{2}(r-2)\}\subset\mathbb{Z}$; for any
$0\neq n\in\mathbb{N}$, we define a corresponding ``q-integer'' $[
n]$ given by:
\bes 
[n]\,=\,\frac{\left( q^{\frac{n}{2}}\,-\,q^{-\frac{n}{2}}\right)}{\left( q^{\frac{1}{2}}\,-\,q^{-\frac{1}{2}}\right)}    
\ees with $[n]!=[n][n-1]...[1][0]$; we call a triple $(j,k,l)$ of elements of
$I$ {\it admissible} if: $r-2 \geq j+k+l \in \mathbb{N}_{+}$ and $j
\leq k+l$, $k \leq j+l$, and $l \leq j+k$; correspondingly, we call
admissible a six-tuple $(i,j,k,l,m,n)$ if each of the four possible triples
in it are admissible; for an admissible six-tuple
$(j_1,j_2,j_3,j_4,j_5,j_6)$ we define:
\bes
\mid 6j\mid\,=\,\begin{vmatrix}
j_1 &j_2 &j_3
\\ j_4 &j_5 &j_6 
\end{vmatrix}
\,=\, (-1)^{\sum_{i=1}^6 j_i}\,\begin{bmatrix}
j_1 &j_2 &j_3
\\ j_4 &j_5 &j_6 
\end{bmatrix}_q 
\ees
where $\begin{bmatrix}
j_1 &j_2 &j_3
\\ j_4 &j_5 &j_6 
\end{bmatrix}_q$ is the q-deformed $6j$-symbol of the $SU(2)$
representation theory, whose explicit expression can be found in \cite{TV}.
The elements of $I$ label the representations of $SU(2)_q$ for $q$ a
root of unity, and we see that it is a finite set, thus providing an
immediate and elegant regularization cut-off for the sum over representations that will define
the spin foam partition function. It is clear that the admissibility conditions are just the usual
triangle inequalities if we assume the interpretation of the elements
of $I$ as possible edge lengths, as we have done in the Ponzano-Regge case.
 
These are the ``initial data'' for the definition of the partition
function of the Turaev-Viro model. We assign an element of $I$ to each
and every edge of $T$. We can any of such assignments $\phi$, and call
it a ``coloring'' of the simplicial manifold. 

The partition function is given by:

\bes
Z(T)\,=\,w^{-2\, a}\,\sum_\phi\,\prod_e\, w_e^2\,\prod_{tet}\,\mid
6j\mid_{tet}, 
\ees 
where $a$ is the number of vertices in the triangulation, $w^2_e=[2
    j_e + 1]_q$, and $w^2=-\frac{2r}{\left( q-q^{-1}\right)^2}$ is a
factor that can be thought of the q-analogue of the regularitation
factor needed in the Ponzano-Regge case to make the partition function
finite. We see that in this case, on the contrary, the partition
function is finite by definition, i.e. because the set of
allowed representations if finite, and this is a consequence of having
chosen the quantum deformation parameter to be a root of unity.   

It is apparent that the partition function above is basically just a
q-deformed version of the Ponzano-Regge one, defined from the same
type of elements but using data from the representation theory of
$SU(2)_q$ instead of $SU(2)$. 
Moreover, it can be shown that:
\bean 
& \begin{bmatrix}
j_1 &j_2 &j_3
\\ j_4 &j_5 &j_6 
\end{bmatrix}_q \,=\,\begin{bmatrix}
j_1 &j_2 &j_3
\\ j_4 &j_5 &j_6 
\end{bmatrix}\,+\,O(r^{-2}) \\ & [2\,j_e\,+\,1]_q\,=\,(2\,j_e\,+\,1)\,+\,O(r^{-2}) \\ 
& -\,\frac{2r}{\left(
  q\,-\,q^{-1}\right)^2}\,=\,\frac{r^3}{2\pi^2}\left( 1\,+\,O(r^{-2})\right),
\eean
so that the various elements entering in the Turaev-Viro partition function reduce to the analog un-deformed ones appearing 
in the Ponzano-Regge model; this suggest that the partition function itself of the Turaev-Viro model reduces to
 the one of the Ponzano-Regge model in the limit $r\rightarrow\infty$
 or $q\rightarrow 1$, i.e. when the $SU(2)_q$ quantum Lie algebra goes
 to the un-deformed $SU(2)$ case. However, the facts mentioned are not enough to prove such a reduction, since also the range 
of the sum over representations changes as a function of $r$, and therefore the limit for $r\rightarrow\infty$ of the 
partition function can not be trivially deduced from the limit of the amplitudes.

The partition function above defines a topological invariant, since it
can be proven \cite{TV} to be independent of the particular
triangulation $T$ chosen for the manifold $\mathcal{M}$. Also, one can
extend the definition to the case of a manifold  with boundary
\cite{TV, ArcWil, Ionicioiu}, and prove \cite{TV} the functorial nature of
the invariant. The Turaev-Viro spin foam model represents then a
rigorous construction of a topological field theory satisfying Atiyah's
axioms.

\section{Asymptotic values of the $6j$-symbols and the connection with
  simplicial gravity}

We have seen how the Ponzano-Regge model can be obtained by a
discretization and quantization of a classical gravity
action. However, one would also like to be able to ``obtain'' a
gravity action {\it from} the quantum theory, showing that a sector of
the quantum theory really describes a classical geometry, at least in
some limit.

In particular, we have seen that the quantum amplitudes of the spin foam
models we described are given by $6j$-symbols from either $SU(2)$,
$Sl(2,\mathbb{R})$ or $SU(2)_q$ representation theory, associated to
the tetrahedra in the triangulated manifold, so one would expect these
$6j$-symbols to behave in some limit as the exponential of some
simplicial gravity action.  

In this section, we are going to show that this is indeed the case\footnotemark\footnotetext{Actually, the idea that a model for quantum gravity in 
3d could be obtained by a product of $6j$-symbol for $SU(2)$ treated as amplitudes in an algebraic path integral formulation 
was {\it motivated} in the first place \cite{PR} by the observation that their asymptotics contain terms of the form  of the
 exponential of a simplicial action for gravity.} This was first shown in \cite{PR} heuristically and then rigorously in 
\cite{Roberts}, while a simpler derivation of the same result was proven in \cite{llasym}, both for the $SU(2)$ and the 
$SL(2,\mathbb{R})$ case, while the q-deformed case was dealt with in \cite{MT}.

Let us analyse the $SU(2)$ case first. Consider the $6j$-symbol:   
$\begin{bmatrix}
l_{01} &l_{02} &l_{03}
\\ l_{23} &l_{13} &l_{12} 
\end{bmatrix}$ where we have labelled the vertices of the tetrahedron by $I=0,...,3$ and the edges accordingly, and each edge 
is assigned an integer $l_{IJ}=j_{IJ}$ with $j$ being a representation of $SU(2)$. Its square can be expressed in terms of 
the integral \cite{barrettintegral}:

\bes
I(l_{IJ})\,=\,\begin{bmatrix}
l_{01} &l_{02} &l_{03}
\\ l_{23} &l_{13} &l_{12} 
\end{bmatrix}^2\,=\,\int_{G^4}[dg_{I}]\,\prod_{I<J}\chi^{l_{IJ}}(g_I
g_J^{-1}) \label{eq:6jint},
\ees i.e. as an integral over four copies of the group $G=SU(2)$, with the $l$'s for any three edges meeting at the same 
vertex satisfying the usual admissibility conditions. The symmetry under the transformation $g_I\rightarrow h k g_I k^{-1}$, 
for $h, k \in SU(2)$ makes the variables $g_I$ redundant, while a set of gauge invariant variables is given by the six 
angles $\theta_{IJ}\in [0,\pi]$ defined by: $\cos\theta_{IJ}=\frac{1}{2}\chi^{1}(g_I g_J^{-1})$, which can also be 
interpreted as spherical edge lengths of a tetrahedron in $S^3$. In terms of these variables,
the integral has the expression:
\bes
I(l_{IJ})\,=\,\frac{2}{\pi^4}\int_{\mathcal{D}_\pi}\,\left[ \prod d\theta_{IJ}\right]\,\frac{\prod_{I < J}\sin( l_{IJ}+1)\theta_{IJ}}{\sqrt{det[\cos\theta_{IJ}]}}                 
\ees
where $\mathcal{D}_\pi = \{ \theta_{IJ}, I\neq J \in (0,1,2,3) \mid \theta_{IJ}\leq \theta_{IK}+\theta_{JK}, \theta_{IJ}+\theta_{IK}+\theta_{JK}\leq 2\pi\}\subset [0,\pi]^6$, 
i.e. it is the set of all possible spherical tetrahedra. The boundary $\partial\mathcal{D}_\pi$ is the set of degenerate 
tetrahedra having volume equal to zero, and the denominator in the above formula vanishes if and only if $\theta$ belongs 
to this boundary.
We consider now a global rescaling of all the edge representations $l_{IJ}\rightarrow N l_{IJ}$ and we look for the 
behaviour of the $6j$-symbol in the limit $N\rightarrow\infty$. 

The rationale for considering such a large $N$ limit as a
semi-classical limit is to consider the length of the edges of the
tetrahedron in the simplicial manifold as given by $L_i=\hbar (2
j_i+1)$ so that the semi-classical limit for fixed length can be taken
to be the combined limit $\hbar\rightarrow 0$ and $j_i\rightarrow
\infty$. In this limit/approximation one would not see anymore any
discreteness or quantization for the edges, since the relative spacing
between two successive levels, $\frac{\Delta j}{j}=\frac{1}{2j}$ would
go to zero.   

The asymptotic expression is found by stationary phase methods, but care is to be taken in taking into account the singular 
behaviour of the integrand on $\partial\mathcal{D}_\pi$. 
Therefore, one splits the integration domain into: $\mathcal{D}_\pi=\mathcal{D}_{\pi,\epsilon}^{<}\cup\mathcal{D}_{\pi,\epsilon}^{>}$, with
 $\mathcal{D}_{\pi,\epsilon}^{>}=[\epsilon,\pi-\epsilon]^6 \cap \mathcal{D}_{\pi}$ and $\mathcal{D}_{\pi,\epsilon}^{<}=\mathcal{D}_{\pi}-\mathcal{D}_{\pi,\epsilon}^{>}$, 
for $1>>\epsilon >0$.     
The integral above gets consequently split into two separate integrals $I^{>}(l)$ and $I^{<}$.  Basically what we are doing 
is to separate the degenerate sector (where the integrand is singular)  from the non-degenerate one, not only because this 
makes it easier to deal with the singularity in the integrand, but also because this allows a direct comparison of the relative amplitudes that the model assigns to the two sectors, to see which one dominates.

The analysis is then carried out by recasting the integral into an exponential form and then looking for solution of the stationarity equations, expanding the phases of the exponential around such solutions \cite{llasym}. The results are as follows: for a flat tetrahedron with volume $V$ and edge lengths $l_{IJ}$ and dihedral angles $\Theta_{IJ}$, the integral corresponding to non-degenerate configurations has, for $N\rightarrow\infty$, the asymptotic form:
\bes
I^{>}(N\,l_{IJ})\,\sim\,-\,\frac{\sin\left(\,\sum_{I<J} ( N\,l_{IJ}\,+\,1 )\,\Theta_{IJ}\right)}{3\pi\,N^{3}\,V}
\ees
while the one for the degenerate configurations is:
\bes
I^{<}(N\,l_{IJ})\,\sim\,\frac{1}{3\pi\,N^{3}\,V}.
\ees

The degenerate configurations correspond to tetrahedra which are
flattered to a 2-dimensional surface in $\mathbb{R}^3$, and indeed can
be uderstood in terms of the representation theory of $\mathbb{E}^2$,
i.e. the symmetry group of a flat 2-surface in 3d \cite{baezasym}.
We see that the two contributions have similar weight.
Thus the overall asymptotic expression for the $6j$-symbols \cite{PR,Roberts,llasym} is:

\bes
\begin{bmatrix}
l_{01} &l_{02} &l_{03}
\\ l_{23} &l_{13} &l_{12} 
\end{bmatrix}\,\sim\,\frac{\sqrt{2}\,\cos\left( \sum_{I<J} ( N\,l_{IJ}\,+\,1)\,\Theta_{IJ}/2\,+\,\frac{\pi}{4}\right)}{\sqrt{3\pi\,N^3\,V}}.
\ees

The function of the edge lengths $\sum_{I<J}( N\,l_{IJ}\,+\,1)\,\Theta_{IJ}(l)/2$ is nothing but the Regge calculus action for the manifold given by a single tetrahedron. Therefore, splitting the cosine into a sum of two exponentials, we see that the full partition function for the manifold $\mathcal{M}$ with triangulation $T$, made out of a product of tetrahedral amplitudes, contains a term like: $e^{i S_{R}}$ with a well defined measure, where $S_{R}$ is the Regge calculus action for the full triangulation $T$, that is the amplitude we would expect in a simplicial path integral quantization of gravity so that the sum over representations $j$ of $SU(2)$ is the algebraic equivalent of the integral over edge lengths in the simplicial path integral approach.  This gives support to the 
interpretation of the spin foam model as a sum-over-histories
formulation of quantum gravity.

The other terms in the asymptotic expansion can be interpreted as
corresponding to other choices of orientations for the edges of the
triangulation, and the reason for their presence (which changes
drastically the amplitudes for the model, since different terms in the
amplitudes may interfere) can be understood as the result of the model representing a definition of the projector operator onto physical quantum states, as we shall discuss.

In order to obtain the asymptotic expression above, we have assumed
the triangle inequalities to be satisfied, in order to be able to use
the expression ~\ref{eq:6jint}. This set of conditions on the spins
$j_i$ can be shown \cite{PR} to be equivalent to the condition $V^2\geq 0$, where
$V$ is the volume of the tetrahedron having the edge lengths given by
$2j_i+1$. In other words, the admissibility conditions constrain the
spins to characterize a {\it real} tetrahedron embedded in a Riemannian space
$\mathbb{R}^3$. However, the $6j$-symbol, although very small (as can
be seen from numerical tebles), is actually not zero for a
six-tuple of spins not satisfying those conditions, and thus
corresponding to a tetrahedron having $V^2 < 0$ (an {\it imaginary}
tetrahedron). 

The above formula for the asymptotics of the $6j$-symbol
may be analytically continued to this sector, and the result
is \cite{PR,barrettfoxon}:
\bes
\begin{bmatrix}
l_{01} &l_{02} &l_{03}
\\ l_{23} &l_{13} &l_{12} 
\end{bmatrix}\,\sim\,\frac{1}{2\sqrt{12\pi\mid
    V\mid}}\,(\cos\phi)\;e^{-\,\mid\,\sum_{I<J} (2 \,l_{IJ}\,+\,1)\,Im(\theta_{IJ})\mid},
\ees 
where the $\theta$'s are functions of the $l$'s, being the dihedral
angles of the tetrahedron given now by: $\theta_{IJ}=m_{IJ}\pi+
i\,Im \theta_{IJ}$, for an integer $m_{IJ}$, and $\cos\phi=(-1)^{(\sum_{I<J}(l_{IJ}-\frac{1}{2})m_{IJ})}$.  

We see that such combinations of spins, giving $V^2<0$ are
exponentially suppressed in the asymptotic limit.

How do we interpret such configurations, which are in principle present
in the partition function for the Ponzano-Regge model? Amazingly, it
can be shown \cite{barrettfoxon} that these can be naturally
intepreted as corresponding to tetrahedra embedded in Lorentzian space
$\mathbb{R}^{2,1}$, with the imaginary part of the dihedral angles
corresponding to the Lorentzian angles.

The situation we face is then as follows. The assignment of six
$SU(2)$ spins
corresponds to defining a quantum tetrahedron with amplitude being
given by a $6j$-symbol. In a semi-classical limit, this amplitude has
the form of a path integral amplitude for simplicial gravity, but with
a sum over the two possible choices of orientation (for each
tetrahedron), and describe a well-defined geometrical tetrahedron
embedded in Riemannian space. However, we cannot forbid fluctuations
in the geometry in our partition function and we will then sum also
over configurations which ``cross the signature barrier'' (think of a
particle confined in a potential well) and,
although we set the model in a Riemannian signature, correspond to
Lorentzian geometries, i.e. to tetrahedra embedded in Lorentzian
space. Of course these classically forbidden configurations are
strongly suppressed in a semiclassical limit.

Up to now we have considered only the Riemannian $SU(2)$ case. What
happens in the Lorentzian?
We have seen that in the Lorentzian case the amplitude for a
tetrahedron is given by an $SL(2,\mathbb{R})$ $6j$-symbol, and that we
can have different types of unitary representations associated to the
edges with the interpretation of spacelike or timelike edges,
depending on whether we assign continuous or discrete parameters to
them. Consequently we can have different types of $6j$-symbols for
different types of tetrahedra, and we can distinguish a purely
spacelike case, when all the edges are assigned continuous
representations and are thus spacelike, purely timelike, with all the
edges timelike (either future or past directed) and only discrete
representations used, and the mixed cases. 
Also in this case we can write the (square of the) $6j$-symbol in
integral form, valid for all these cases\cite{llasym}:
\bes
\begin{bmatrix}
\lambda_{01} &\lambda_{02} &\lambda_{03}
\\ \lambda_{23} &\lambda_{13} &\lambda_{12} 
\end{bmatrix}^2\,\int_{SL(2,\mathbb{R})^4}\prod_{I<J}\chi_{\lambda_{IJ}}(g_j\,
g_I^{-1})\,\prod_I dg_I,
\ees
with the representations $\lambda_{IJ}$ satisfying the admissibility
conditions discussed before.
Also in this case we can pass to gauge invariant variables, and again
it is convenient to split the domain of integration and consequently
the integral into two terms, one corresponding to the contributions
from degenerate geometries and the other to well-defined ones.
We consider only the non-degenerate sector. 
For spacelike tetrahedra we have all the edges spacelike and the
associated Lorentzian dihedral angles $T_{IJ}$ corresponding to group
elements $g_I g_J^{-1}$ being pure boosts.
The asymptotic expression for the (square of the) $6j$-symbol is given
by \cite{llasym}:
\bes
I^>\,(N\,\rho_{IJ})\,\sim\,\frac{\pi^4\,\sin\left(\sum_{I<J}\,\rho_{IJ}\,T_{IJ}\,+\,\sigma\right)\,\pi/2}{2\,N^2\,3\pi^3\,V(\rho)},
\ees    
an expression similar to the Riemannian case, where $V$ is
again the 3-volume of the tetrahedron having the $\rho$'s as edge
lengths and where $\sigma$ is a $\rho$-independent integer.

For timelike tetrahedra, i.e. with all the edges being timelike and
labelled by discrete representations, and the dihedral angles $\Theta$
now being  conjugated to group elements that are pure rotations.
Modulo permutations of the vertices and changes in orientation, we can
always consider all the edges being future-pointing timelike vectors,
so labelled by representations $l_{IJ}^+$. The asymptotic analysis, again using the stationary point method, gives
\cite{llasym}:

\bes
I^>\,(i
N\,l_{IJ}^+)\,\sim\,\frac{e^{i\,\sum_{I<J}\,N\,l_{IJ}^+\,\Theta_{IJ}}}{N^3\,\sqrt{V(l)}},
\ees
where now the path integral-like expression is already apparent.

Therefore we see that the same intepretation of the spin foam model as
an algebraic sum-over-histories formulation of the quantum gravity
theory, with the usual simplicial path integral expression holding in
the semi-classical limit, is valid in the Lorentzian case as well as
in the Riemannian.

It remains to discuss the q-deformed case of $SU(2)_q$, i.e. the
Turaev-Viro model. The crucial point is to understand what is the
physical and geometrical input coming from the quantum deformation
parameter $q$. 
The asymptotic expression for the $SU(2)_q$
$6j$-symbol was analysed in \cite{MT}.
Of course, in this case the quantum deformation parameter should be
sent towards $1$ as $N$ is sent to infinity, since we have to send to
infinity the cut-off on the representations coming from the quantum deformation. 
The result is:

\bes
\begin{bmatrix}
l_{01} &l_{02} &l_{03}
\\ l_{23} &l_{13} &l_{12} 
\end{bmatrix}\,\sim\,\frac{\sqrt{2}\,\cos\left( \sum_{I<J} ( N\,l_{IJ}\,+\,1)\,\Theta_{IJ}/2\,-\,\frac{8\pi^2}{k^2}\,+\,\frac{\pi}{4}\right)}{\sqrt{3\pi\,N^3\,V}},
\ees
where $k$ is related to $q$ by: $q=e^{\frac{i\pi}{k}}$, so that we are
in the limit $k>>1$.

the extra term appearing in this asymptotic expression, compared to
the Ponzano-Regge $SU(2)$ case, is of the form and it has the
interpretation of a cosmological constant term
$\Lambda=\frac{8\pi^2}{k^2}$. This is also consistent with the
algebraic fact that a finite $k$ introduces a cut-off or upper bound
on the representations just as a cosmological constant itroduces an
infra-red cut-off on the physical lengths.
The same intepretation can be supported by a reformulation of BF
theory with cosmological constant in terms of a Chern-Simons theory,
and by the relation between the Turaev-Viro topological invariant and
the Chern-Simons invariant, as we shall see in the following.
In any case, the intepretation of spin foam models as algebraic path
integrals for quantum gravity is again confirmed.      

\section{Boundary states and connection with loop quantum gravity: the
Ponzano-Regge model as a generalized projector operator}
We want now to identify the states of the Ponzano-Regge model, which
are located at the boundaries of the triangulated manifold we use to
construct the model itself, and then characterize better the
amplitudes between these states that the spin foam defines.

Consider the triangulation with boundary $T$, and call its boundary
$\partial T$. This is a collection of triangles joined at common
edges. Consider now the 1-complex (simply, a graph) dual to this
boundary triangulation, having a vertex in each triangle and a link
for each boundary link. This dual graph is just the boundary of the
2-complex dual to the triangulation $T$. One can define the
Ponzano-Regge partition function in the same way as we did above,
starting form the discretized $BF$ action, in the presence of boundaries,
and the partition function may or may not acquire additional boundary
terms depending respectively on whether we fix the $B$ or metric field or the
connection $A$ on the boundaries \cite{OLough}.
In any case the boundary links $e$ are assigned a group element $g_e$
(representing the boundary connection),
and a corresponding representation function $D^{j_e}(g_e)$ (an holonomy),
in a given representation $j_e$ of $SU(2)$. These boundary links are
the boundary of the dual faces to which the model assigns a
representation, so the $j_e$ are just the representations $j_f$ we
considered in the definition of the partition function. Then, each
boundary vertex, i.e. each vertex of the dual graph, is assigned an
intertwiner or $3j$-symbol contracting the three representation
functions of the incoming edges, and mapping the product of the three
representations to the invariant representation. There is thus one
$3j$-symbol for each triangle on the boundary. If the connection is
fixed on the boundary, the collection of all these $3j$-symbols and
holonomies is what we have as boundary terms.  

Now, the contraction of
a $3j$-symbol for each vertex in the dual graph  with a representation function for
each connection on the dual edges is precisely a spin network
function, which is a gauge invariant function, corresponding to a spin network with
the underlying graph being indeed the graph dual to the boundary
triangulation and a representation of $SU(2)$ assigned to each of its edges. 

These functions $\Phi_{\gamma,j}(g)$, depending on a graph $\gamma$, a set
of representations $j$ and a set of connections (group elements) $g$, may be interpreted as the spin network
states $\mid \gamma, j\rangle$ written in the ``connection
representation'' $\mid g\rangle$, i.e. $\Phi_{\gamma,j}(g)=\langle g
\mid \gamma, j\rangle$. 

These spin networks label the kinematical states of the theory, which
are then the same kinematical states of $3d$ loop quantum gravity,
with a restriction on the allowed valence of the vertices coming from
their origin as dual to triangles in a triangulation \cite{Ooguri, carloPR}. This restriction
does not limit the generality of these states, again because of the
topological invariance of the theory, and also because any spin
network with higher valence may be decomposed into a 3-valent one
(much in the same way as any 2-d polygon may be chopped into triangles).    

Just as in loop quantum gravity, there are two possible bases for the states in the theory: the \lq\lq loop basis" (or spin
network or representation basis) and the \lq\lq connection basis", and the spin network functions above may be thought of as 
as the coefficients of the change of basis: a generic state of the theory $\mid \Psi\rangle$ can be written in the connection
basis as $\langle g\mid\Psi\rangle=\Psi(g)=\sum_j \langle g\mid\gamma, j\rangle\langle \gamma, j\mid\Psi\rangle=\sum_j \Phi_{\gamma,j}(g) \Psi_\gamma(j)$.  
While in $4d$ we would need to sum over the possible graphs as well as over their colorings, this is not needed here, again, 
because of the triangulation invariance\cite{carloPR, Ooguri}. The kinematical Hilbert space can then be taken to be the 
space of (square integrable) invariant functions of an $SU(2)$ connection. 

The physical states are obtained by restricting the support of these
functions to flat connections only, as required by the constraints of
BF theory. Their inner product would then formally be: 
\bes
\langle \Psi_1\mid \Psi_2\rangle\,=\,\int \mathcal{D} g\,\,\,\delta(F(g))\,\,\Psi^{*}_1(g)\,\Psi_2(g)
\ees
How these physical states are to be constructed in practice
is what we will discuss in the following. The idea is that the Ponzano-Regge model defines a generalized projector 
operator \cite{carloprojector}
from kinematical states to physical states, that can also be used to define the physical inner product between them, i.e. the 
transition amplitudes of the theory. 

Let us first note, however, that this link with loop quantum gravity also
allows another justification for the intepretation of the spins $j$
assigned to each edge as giving its length. In fact, one can just use
the definition of the length operator in a canonical formulation of
first order gravity and show that its spectrum, when applied to an
edge with a single intersection with a spin network link (as it is in
this case) has spectrum given by the $SU(2)$ Casimir, so expressed in
terms of the representation $j$.

That the Ponzano-Regge model is a realization of the projector
operator (which is how we have been using it here) is well-known and was shown, for example, in \cite{Ooguri}
and in \cite{ansdorf}. Let us recall the argument in \cite{Ooguri}.
Consider a 3-manifold $M$ and decompose it into three parts $M_1$,
$M_2$ and $N$, with $N$ having the topology of a cylinder
$\Sigma\times[0,1]$ ($\Sigma$ compact), and the boundaries $\partial M_1$ and $\partial
M_2$ being isomorphic to $\Sigma$. The Ponzano-Regge partition
function may then be written as:

\bes
Z_M\,=\,\mathcal{N}_T\,\sum_{j_e\in\Delta_1,j_{\tilde{e}}\in\Delta_2}\,Z_{M_1,\Delta_1}(j_e)\,P_{\Delta_1,\Delta_2}(j_e,j_{\tilde{e}})\,Z_{M_2,\Delta_2}(j_{\tilde{e}})
\ees where we have chosen a triangulation in which no tetrahedron
is shared by any two of the three parts in which we have
partitioned the manifold, and $\Delta_i$ are triangulations of the
boundaries of these parts. We have included also the sum over
spins assigned to edges internal to $M_1$, $M_2$ and $N$ in the
definition of the functions $ Z_{M_i,\Delta_i}$ and
$P_{\Delta_1,\Delta_2}$.

Because of the topological invariance of the model, and of the
consequent invariance under change of triangulation, the following
relation holds:

\bes
\mathcal{L}_{\Delta_2}\sum_{j_{\tilde{e}}\in\Delta_2}\,P_{\Delta_1,\Delta_2}(j_e,j_{\tilde{e}})\,P_{\Delta_2,\Delta_3}(j_{\tilde{e}},j'_e)\,=\,P_{\Delta_1,\Delta_3}(j_e,j'_e)
\ees
where $\mathcal{L}$ is another constant depending only on the
triangulation $\Delta_2$.

Because of this property, the following operator acting on spin network
states $\phi_\Delta$ living on the boundary triangulation is a
projector:

\bes
\mathcal{P}[\phi_\Delta](j)\,=\,\mathcal{L}_\Delta\,\sum_{j'}\,P_{\Delta,\Delta}(j,j')\,\phi_\Delta(j')
\ees
i.e. it satisfies: $\mathcal{P}\cdot\mathcal{P}=\mathcal{P}$, and we
can re-write the partition function as:
\bes
Z_M\,=\,\mathcal{N}_T\,\sum_{j_e\in\Delta_1,j_{\tilde{e}}\in\Delta_2}\,\mathcal{P}[Z_{M_1,\Delta_1}](j_e)\,P_{\Delta_1,\Delta_2}(j_e,j_{\tilde{e}})\,\mathcal{P}[Z_{M_2,\Delta_2}](j_{\tilde{e}}).
\ees
This allows us to define the physical quantum states of the theory
as those satisfying:

\bes
\phi_\Delta\,=\,\mathcal{P}[\phi_\Delta] \label{eq:proj}
\ees
being the anlogue of the Wheeler-DeWitt equation,
and the inner product between them as:
\bes
\langle \phi_\Delta \mid
\phi'_\Delta\rangle_{phys}\,=\,\sum_{j,j'\in\Delta}\,\phi_\Delta(j)\,P_{\Delta,\Delta}(j,j')\,\phi'_\Delta(j'),
\ees
so that the functions $Z_{M_1}$ and $Z_{M_2}$ are solutions to the
equation \Ref{eq:proj} and the partition function for $M$ basically
gives their inner product.

In this argument a crucial role is played by the triangulation
invariance of the model, so that it is not possible to repeat it
for the case of 4-dimensional gravity (i.e. Barrett-Crane model),
where a sum over triangulations or a refining procedure is
necessary to avoid dependence on the given triangulation. For a generalization to
 the 4-dimensional case a better starting point is represented by the group field theory formulation
of the Ponzano-Regge model, that we are going to discuss later in the following. However,
it is nevertheless possible to identify in the Ponzano-Regge model a
distinguishing feature of the projector operator as realized in
path integral terms, and that can be used to characterize the projector also in the 4-dimensional 
case, for a fixed triangulation. This is the analogue of the $Z_2$
symmetry we have seen in the relativistic particle case and in the
formal path integral quantization of gravity in the metric
formalism. This is the symmetry that ``kills causality'' by
integrating over both signs of the proper time, and is realized in
the present case as a symmetry under change of orientation for the
simplicial manifold (we will give more details on this link
between orientation and causality when discussing the
Barrett-Crane model). This is clear in the Lorentzian context, where
future oriented $(d-1)$-simplices are changed into past oriented
ones and vice-versa. This symmetry is actually evident at the level
of $(d-2)$-simplices, writing the model in terms of characters as we
did in \ref{eq:ZPR}. In fact, under a change of orientation of the edges in the
manifold, the plaquettes of the dual complex also
change their orientation and this is reflected by substituting the
group elements assigned to the boundary links of each plaquette
(wedge) with their inverses. Clearly, the partition function, and
each amplitude in it, is not affected by this change due to the
equality between the characters of group elements which are
inverse of each other. Indeed, in the case of $SU(2)$, a group
element is conjugate to its inverse (the Weyl group is $Z_2$) and
they both have the same (real) character. It is the identification
of an analogous symmetry in the 4-dimensional case that will show
how the Barrett-Crane model realizes a projection onto physical
states. Indeed, once again, $Spin(4)$ and $SL(2,\mathbb{C})$ group
elements are conjugate to their inverse (the Weyl group is still
$Z_2$) and the model is invariant under change of orientation.

We note that the same invariance under change of orientation was noted in \cite{fkvol} and pinpointed 
as marking the difference between $BF$ theory and gravity, while here we are suggesting that
it marks the difference between an orientation dependent and an orientation-independent transition 
amplitude, this last one being the inner product between physical states provided by the spin foam 
models considered so far.   

The same construction can be formally carred out for the Lorentzian model discussed in section 
\ref{sec:PRLor}. The boundary states are again given by spin networks, based on $Sl(2,\mathbb{R})$ 
representations, with an appropriate gauge fixing that makes them well-defined objects \cite{el}, 
and the intepretation of these representations as edge lengths (continuous for spacelike edges and 
discrete for timelike ones) can again be supported by the corresponding canonical analysis of $2+1$ 
gravity in a first order formalism \cite{elc}. 

Also, the same intepretation of the partition function as defining the generalized projector holds, 
and again the characteristic feature is the $\mathbb{Z}_2$ symmetry under change of orientation. 
However, the situation is more subtle and thus more interesting than in the Riemannian case. While 
nothing changes for the spacelike continuous representations (the Weyl group is again $\mathbb{Z}_2$) 
and is reflected in the symmetry of the characters under change of orientation of the corresponding 
edges, the characters for discrete representations are not; in other words, for these representations, it turns
out that the $Z_2$ symmetry is killed (the Weyl group is trivial)
and that the character are simple exponentials; $Sl(2,\mathbb{R})$ distinguishes automatically between two time arrows, 
the past and the future as reflected by the existence of two separate classes of discrete 
representations.
This model is therefore is a good candidate for Lorentzian $2+1$
general relativity and represents a motivation to deal with the
$3+1$ case in a similar fashion.

\section{Group field theory formulation and the sum over topologies}
We have seen that both the (Riemannian and Lorentzian) Ponzano-Regge model and 
the Turaev-Viro model are invariant under change of the triangulation
by means of which they are defined, and this is a consequence of the
fact that they define a topological field theory. Their degrees of
freedom are purely global and topological, related to the manifold
underlying the model and not to the metric field living on it, which
is constrained to be locally flat everywhere. 

As long as quantizing the metric, or the spacetime geometry, is what
we want to achieve, this is all we need from a model of 3d
quantum gravity, and it is provided by the spin foam models we
discussed. However, we have argued that also the topology of spacetime
is expected to be a dynamical variable in a full quantum theory of
gravity, i.e. also the global degrees of freedom related to the choice
of a spacetime manifold (or its algebraic and combinatorial
equivalent) have to be dynamical ones.

 In other words, if one accepts
the definition of a quantum gravity theory as a description of a
{\bf quantum spacetime} and of its dynamics, then there is one more
ingredient which is necessary for the definition of the full theory,
having as a starting point the spin foam models given above. This is a
suitably defined sum over topologies or manifolds interpolating
between given boundary ones. In a spin foam language, and accordingly
to the definition of a full spin foam model given in the introduction,
what we need is a definition and a construction of a sum over (labelled)
2-complexes (or spin foams) interpolating between given boundary
(labelled) graphs (or spin networks).

This is what is achieved using the formalism of field theories over
group manifolds \cite{boulatov, ooguri, DP-F-K-R, P-R, RR, RR2}, which
can be also seen as a generalization of the matrix models used to
construct quantum gravity models in 2d.

For the case of interest here, the Riemannian Ponzano-Regge model,
whose group field theory formulation was obtained in \cite{boulatov} the
group used is of course $SU(2)$ and the field is defined as a real function
of three group elements:

\bes
\phi(g_1,g_2,g_3)\,=\,\bar{\phi}(g_1,g_2,g_3)
\ees
on which we impose the following symmetries: invariance under even
permutation $\pi$ and invariance under simultaneous right translations
of all its arguments:
\bes
\phi(g_1,g_2,g_3)\,=\,\phi(g_{\pi(1)},g_{\pi(2)},g_{\pi(3)})\,
\hspace{1cm} \phi(g_1,g_2,g_3)\,=\,\phi(g_1 g, g_2 g, g_3
g)\;\;\;\forall g\in SU(2)\;\;\;\;\;\;.
\ees 

For this field we can define the action to be:

\bes
S[\phi] &=& \frac{1}{2}\int
dg_1..dg_3 [P_g\phi(g_1,g_2,g_3)]^2 \,- \nonumber \\&-&\frac{\lambda}{4!}\int
dg_1..dg_6 [P_{h_1}\phi(g_1,g_2,g_3)][P_{h_2}\phi(g_3,g_5,g_4)][P_{h_3}\phi(g_4,g_2,g_6)][P_{h_4}\phi(g_6,g_5,g_1)]\;\;\;\;\;\;\;\;,
\ees
where the projectors $P_g$ are used to enforce the gauge invariance of
the field: $P_g \phi(g_1,g_2,g_3)=\int_{SU(2)}dg \phi(g_1 g, g_2 g,
  g_3 g)$.

One may understand the form of this action by associating the field to
a triangle in a simplicial manifold, and its arguments to the edges of the
triangle, and thus the kinetic term describes the gluing of two
triangles while the interaction term describes four triangles building
a tetrahedron (and this explains the way we wrote the arguments of the
fields). The integrals are group integrals performed with the
normalized Haar measure.

We can re-write the action in a form more easily related to Feynman
graphs, i.e. identifying clearly a kinetic operator and a potential
term, using the shorthand notation $\phi(g_1,g_2,g_3,g_4)=\phi(g_i)$:

\bes
S[\phi]\,=\,\frac{1}{2}\int
dg_i\,d\tilde{g}_j\,\phi(g_i)\,\mathcal{K}(g_i,\tilde{g}_j)\,\phi(\tilde{g}_j)\,-\,\frac{\lambda}{4!}\int
dg_{ij}\,\mathcal{V}(g_{ij})\,\phi(g_{1j})\,\phi(g_{2j})\,\phi(g_{3j})\,\phi(g_{4j})\;\;\;\;\;\;,
\ees 
with $\phi(g_{1j})=\phi(g_{12},..,g_{14})$ etc.

The kinetic operator is then given by:
\bes
\mathcal{K}(g_i,\tilde{g}_j)\,=\,\sum_\pi\,\int
dg\,\prod_{i=1}^3\,\delta\left(g_i\,g\,\tilde{g}_{\pi(j)}^{-1}\right)
\ees
having the graphical structure: 

\begin{figure}
\begin{center}
\includegraphics[width=7cm]{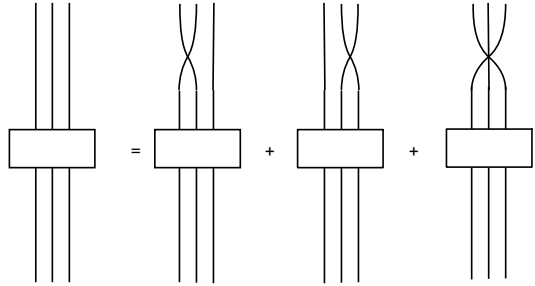}
\caption{The propagator of the 3-dimensional group field theory.}
\end{center}
\end{figure}
 and we note that, being a
projector, it is
identical to the propagator of the theory since its inverse in the
space of gauge invariant fields (the propagator) is itself. 

The potential term is given by:

\bes 
\mathcal{V}\,=\,\int dh_i\,\prod_{i<j}\,\delta\left( h_j\,g_{ji}^{-1}\,g_{ij}\,h_i^{-1}\right),
\ees
with its structure represented as: 
\begin{figure}
\begin{center}
\includegraphics[width=7cm]{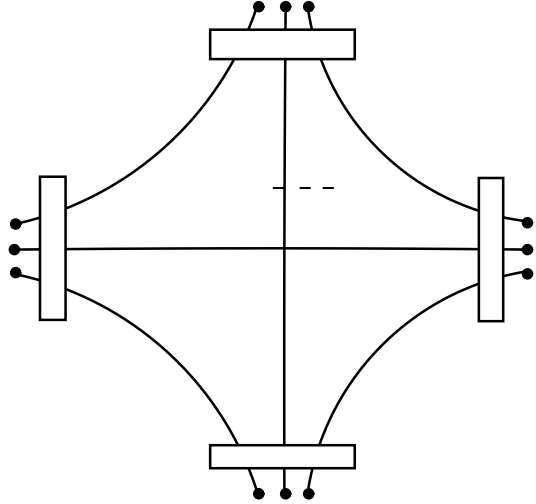}
\caption{The vertex of the 3-dimensional group field theory.}
\end{center}
\end{figure}

In these pictures each line represent a delta function, and we have represented each group integration (imposing gauge 
invariance) as a little box, while drawing explicitely the sum over odd permutations of the arguments of the field only in 
the picture for the propagator, leaving it implicit in that for the vertex.
It is
apparent that the vertex has the structure of a tetrahedron, while
each set of incoming lines corresponds to a triangle.

Having at hand the basic building blocks of Feynman graphs, we can
thus expand the partition function for the group field theory as a
perturbative expansion given by a sum over all the possible
vacuum-to-vacuum Feynman graphs $\Gamma$, each given by a certain number of
interaction vertices glued together by means of propagators.

\bes
Z\,=\,\int d\phi\,e^{-S[\phi]}\,=\,\sum_{\Gamma}\,\frac{\lambda^N}{sym[\Gamma]}\,Z(\Gamma)
\ees
where $N$ is the number of vertices in the Feynman graph $\Gamma$
while $sym[\Gamma]$ is the number of automorphisms of the graph
onto itself, and we have called $Z(\Gamma)$ the amplitude for each graph. 

 The group
variables which are the original arguments of the field are associated
to the open ends of lines in both propagators and vertices, so that one can
follow their propagation along the whole Feynman graph through
propagators and vertices; for a closed graph, as all the graphs
in the expansion of the partition function are closed, the line corresponding to
each group variable will close, so that we can associate a surface to
each of these closed lines. The set of interaction vertices, links
corresponding to propagators and these surfaces build up a
2-complex for each Feynman graph of the theory and for each choice of
permutations in the propagators. The sum over Feynman graphs is thus a
sum over 2-complexes. Moreover, because of the permutation symmetry we imposed on the field, the 2-complexes so generated
are all oriented \cite{DP-F-K-R}. These 2-complexes can also be put in 1-1
correspondence with 3-dimensional triangulations, again because of the
intepretation of propagators as gluing of triangles and of vertices as
tetrahedra, with a corresponding association of surfaces of the
2-complexes with edges of the triangulations. They are then the
2-complexes dual to 3-dimensional triangulations. The different ways
of pairing vertices with propagators then give rise to different
triangulations not only of the same topology but also of different
topologies for the same number of tetrahedra, so we get a sum over
topologies from the perturbative expansion of the partition function. 
A crucial question is
whether these triangulations are manifolds or not, i.e. whether it is
true that each point of them has a closed neighbourhood homeomorphic to
an n-disc $D^{n}$ ($n=3$ in this case). Precise conditions for the
answer to be positive can be found\cite{DPP}, and translated into conditions on
the graphs and it turns out \cite{DPP} that in this case all the
resulting triangulations are indeed manifolds.   

Now let us see what are the amplitudes for each Feynman graph $\Gamma$,
i.e. for each 2-complex. If one
uses the expressions we gave for propagators and interaction vertices,
it is easy to see that, by integrating the group variables $g$ at the
end of each open line in vertices and propagators (i.e. the original
arguments of the field), thus realizing
their gluing, the resulting amplitudes are (we treat the different
graphs obtained from the different terms in the propagators
corresponding to different permutations as different graphs):

\bes
Z(\Gamma)\,=\, \left(\prod_{e\in \Gamma} \int d g_{e}\right)
\,\prod_{f}\,\delta (\prod_{e\in\partial f} g_{e} ),
\ees
i.e. we get a delta function for each face of the dual 2-complex, with
argument a product of $SU(2)$ group elements one for each edge of the
2-complex in the boundary of the face. This is exaclty the expression
for the Ponzano-Regge model as a lattice gauge theory on the
2-complex!

As in the lattice gauge theory derivation, we could now use this
expression to derive the full spin foam expression for the partition
function, expanding each delta function. However, the spin foam
formulation can be derived also directly from the group field theory,
as a ``momentum space'' expression for the same Feynman graph amplitudes. 

Consider in fact the Fourier decomposition of the field $\phi$:

\bes
\phi(g_1,g_2,g_3)\,=\,\sum_{j_1,j_2,j_3}\Delta_{j_1}\Delta_{j_2}\Delta_{j_3}\,\phi^{j_1j_2j_3}_{m_1n_1
  m_2n_2 m_3n_3}\,D^{j_1}_{m_1n_1}(g_1)D^{j_2}_{m_2n_2}(g_2)D^{j_3}_{m_3n_3}(g_3),\ees 
with $\Delta_j=2j+1$, giving then for the gauge invariant field (performing the group integral):
\bes
\phi(g_1,g_2,g_3)\,&=&\,P_g\phi(g_1,g_2,g_3)\,= \nonumber \\ 
&=&\sum_{j_1,j_2,j_3}\Delta_{j_1}\Delta_{j_2}\Delta_{j_3} \phi^{j_1j_2j_3}_{m_1k_1 m_2k_2
  m_3k_3}\,D^{j_1}_{m_1n_1}(g_1)D^{j_2}_{m_2n_2}(g_2)D^{j_3}_{m_3n_3}(g_3)\,C^{j_1j_2j_3}_{n_1 n_2 n_3}\,
C^{j_1j_2j_3}_{k_1 k_2 k_3} = \nonumber \\ 
&=&\,\sum_{j_1,j_2,j_3}\sqrt{\Delta_{j_1}\Delta_{j_2}\Delta_{j_3}}\,\Phi^{j_1j_2j_3}_{m_1m_2m_3}\,D^{j_1}_{m_1n_1}(g_1)D^{j_2}_{m_2n_2}(g_2)D^{j_3}_{m_3n_3}(g_3)\,C^{j_1j_2j_3}_{n_1 n_2 n_3}, 
\ees
where the $C$ are the $3j$-symbols and we have defined the Fourier components of the field as $\Phi=\phi^{123}\sqrt{\Delta_1\Delta_2\Delta_3}C^{123}$.
Of course the symmetries we imposed on the field give corresponding symmetries on the Fourier modes.

In terms of this decomposition, the action takes the form:
\bes
S[\phi] = \frac{1}{2}\sum_{j_1,j_2,j_3} \mid \Phi^{j_1j_2j_3}_{m_1m_2m_3}\mid^2 - \frac{\lambda}{4!}\sum_{j_1,..,j_6}
\Phi^{j_1j_2j_3}_{m_1m_2m_3}\Phi^{j_3j_5j_4}_{m_3m_5m_4}\Phi^{j_4j_2j_6}_{m_4m_2m_6}\Phi^{j_6j_5j_1}_{m_6m_5m_1} 
\begin{bmatrix}
j_{1} &j_{2} &j_{3}
\\ j_{4} &j_{5} &j_{6} ,
\end{bmatrix}\,\,\,\,
\ees
from which we easily read out the expression for propagator and potential to be used in the Feynman expansion. 
The propagator is just a product of Kronecker deltas for the representation $j$'s and the projections $m$'s, with all 
the even permutations of the arguments considered, and the vertex is just given by a $6j$-symbol, again with additional 
Kronecker delta matching those of the propagators, and responsible for the correct gluing.

The resulting amplitude for the 2-complex (Feynman graph) $\Gamma$ is:
\bes
Z(\Gamma)\,=\,\left(\prod_{f}\,\sum_{j_{f}}\right)\,\prod_{f}\Delta_{j_{f}}\prod_{v}\,\begin{bmatrix} 
j_1 &j_2 &j_3
\\ j_4 &j_5 &j_6 
\end{bmatrix}_{v}, 
\ees
exactly the Ponzano-Regge amplitude for the 2-complex $\Gamma$!

In the end the partition function for the field theory is therefore:

\bes
Z\,=\,\int d\phi\,e^{-S[\phi]}\,=\,\sum_{\Gamma}\,\frac{\lambda^N}{sym[\Gamma]}\,\left(\prod_{f}\,\sum_{j_{f}}\right)\,\prod_{f}\Delta_{j_{f}}\prod_{v}\,\begin{bmatrix} 
j_1 &j_2 &j_3
\\ j_4 &j_5 &j_6 
\end{bmatrix}_{v} 
\ees
and it has the full structure of a spin foam model, including both a sum over representations and a sum over 2-complexes. 
These 2-complexes as we said correspond to different topologies as well as to different triangulations of the same topology, 
and thus the theory defines a full sum-over-histories quantum theory for a 3-dimensional {\bf quantum spacetime}. 
Also, being a spin foam model, it encompasses both the dynamical triangulations and quantum Regge calculus approaches to 
quantum gravity, incorporating both a sum over triangulations and a sum over metric variables (think of the asymptotic results). 

Of course, the first objection one may have is that we have indeed too much; in particular, a sum over topologies, although 
well-defined here as a sum over 2-complexes and in terms of their fundamental building blocks, can be considered to be 
doomed 
to be badly divergent, being thus physically meaningless and making mathematically flawed the perturbation expansion.
Amazingly, this is not so, and the sum over 2-complexes can be non-perturbatively defined rigorously by a simple modification,
and this is obtained by {\it adding} terms to the sum itself \cite{llsum}! Indeed, it can be shown \cite{llsum} that simply 
by adding a term like 
\bes
\alpha \;\;[P_{h_1}\phi(g_1,g_2,g_3)][P_{h_2}\phi(g_3,g_5,g_4)][P_{h_3}\phi(g_4,g_5,g_6)][P_{h_4}\phi(g_6,g_2,g_1)]
\ees
as an additional interaction term in the action for the field, the corresponding partition function is Borel summable when 
given in terms of a perturbative expansion of Feynman graphs. The modified model amounts just to a simple generalization of 
the Ponzano-Regge model, with the same propagators but with additional possible vertices of a different combinatorial structure 
(not having anymore the structure of tetrahedra) with potential $\frac{\delta_{j_3,j_6}}{2j_3+1}$.

The new term has the structure of a \lq\lq pillow", i.e. of a set of 4 triangles glued together in such a way that two pairs 
of them share a single edge each, while two other pairs share two edges; it can also be thought of as two tetrahedra glued 
together along two common triangles, so that the sum over regular triangulations (made only with tetrahedra) and 
irregular traingulations (made using also pillows) can be re-arranged as a sum over regular triangulations only, but with
 different amplitudes.  

Not too much is understood of this modified model, but the crucial point is only that also the sum over topologies that one
 obtains naturally by means of the group field theory, and that completes the definition of the spin foam model, can be given
a rigorous meaning with a non-perturbative definition.

\section{Quantum gravity observables in 3d: transition amplitudes}
The field theory over a group gives the most complete definition of the spin foam models and thus permits us to define and 
compute a natural set of observables, namely the n-point functions representing transition
amplitudes between eigenstates of geometry, i.e. spin networks (better
s-knots), meaning states with fixed number of quanta of geometry \cite{Mikov,P-R-Obs}. 
We start the discussion of these observables, following
\cite{P-R-Obs}, from their definition from a canonical loop quantum
gravity point of view, and then turn to their realization in the field theory
over a group framework. We will give a bit more details about these observables when dealing with the 4-dimensional case.

The kinematical state space of the canonical theory is given by a Hilbert space of s-knot states, solutions of the
gauge and diffeomorphism constraints, $\mid s\rangle$, including the
vacuum s-knot $\mid 0\rangle$. We can formally define a projection $P:\mathcal{H}_{diff}\rightarrow\mathcal{H}_{phys}$
from this space to the physical state space of the solutions of the
Hamiltonian constraint, $\mid s\rangle_{phys}=P\mid s\rangle$, as we have already discussed. The
operator $P$ is assumed to be real, meaning that $\langle s_{1}\cup
s_{3}\mid P\mid s_{2}\rangle=\langle s_{1}\mid P\mid s_{2}\cup
s_{3}\rangle$, and thus giving invariance under exchange in the order of the arguments. The $\cup$ stands for the disjoint
 union of two s-knots, which is another s-knot. The quantities
\bes
W(s,s')\equiv \,\,_{phys}\langle s\mid s'\rangle_{phys}\,=\,\langle s\mid P\mid s'\rangle
\ees
are fully gauge invariant (invariant under the action of all the
constraints) objects and represent transition amplitudes between
physical states. We have seen how a spin foam formalism allows a
precise definition of these transition amplitudes, indirectly defining
the projector operator $P$, without an explicit construction of the
operator $P$ itself. We are about to see how this definition can be
phrased in terms of the group field theory formalism. 
 
We can introduce in $\mathcal{H}_{phys}$ the operator
\bes
\phi_{s}\mid s'\rangle_{phys}\,=\,\mid s\cup s'\rangle_{phys}
\ees
with the properties of being self-adjoint (because of the reality of
$P$) and of satisfying $[\phi_{s},\phi_{s'}]=0$, so that we can define
\bes
W(s)\,=\,\,_{phys}\langle 0\mid \phi_{s}\mid 0\rangle_{phys}
\ees
and
\bes
W(s,s')\,=\,\,_{phys}\langle 0\mid \phi_{s}\phi_{s'}\mid
0\rangle_{phys}\,=\,W(s\cup s').
\ees
In this way we have a field-theoretic definition of the $W$'s as
n-point functions for the field $\phi$. 

Using the field theory described above, its n-point functions are
given, as usual, by:

\bes
W(g_{1}^{i_{1}},...,g_{n}^{i_{1}})=\int\mathcal{D}\phi\,\,\phi(g^{i_{1}}_{1})...\phi(g^{i_{n}}_{n})\,e^{-S[\phi]},
\ees
where we have used a shortened notation for the three arguments of the
fields $\phi$ (each of the indices $i$ runs over the three arguments of the field).

Expanding the fields $\phi$ in \lq\lq momentum space'', we have \cite{P-R-Obs} the following explicit (up to a
rescaling depending on the representations $J_{i}$)
expression in terms of the ``field components''
$\Phi_{j_{1}j_{2}j_{3}}^{\alpha_{1}\alpha_{2}\alpha_{3}}$:
\bes
W_{j^{1}_{1}j^{1}_{2}j^{1}_{3}}^{\alpha^{1}_{1}\alpha^{1}_{2}\alpha^{1}_{3}}.....
_{j^{n}_{1}j^{n}_{2}j^{n}_{3}}^{\alpha^{n}_{1}\alpha^{n}_{2}\alpha^{n}_{3}}\,=
\,\int\mathcal{D}\phi\,\,\phi_{{j}^{1}_{1}j^{1}_{2}j^{1}_{3}}^{\alpha^{1}_{1}\alpha^{1}_{2}\alpha^{1}_{3}}...
\phi_{j^{n}_{1}j^{n}_{2}j^{n}_{3}}^{\alpha^{n}_{1}\alpha^{n}_{2}\alpha^{n}_{3}}\,e^{-S[\phi]}.
\ees
However, the W functions have to be invariant under the gauge group
$G$ to which the $g$'s belong, and this requires all the indices $\alpha$ to be 
suitably paired (with the same representations for the paired indices)
and summed over. Each independent choice of indices and of their pairing defines an
independent W function. If we associate a 3-valent vertex to each
$\phi_{j^{i}_{1}j^{i}_{2}j^{i}_{3}}^{\alpha^{i}_{1}\alpha^{i}_{2}\alpha^{i}_{3}}$
 with $j^{i}_{h}$ at the i-th edge,
and connect all the vertices as in the chosen pairing, we see that we
obtain a 3-valent spin network, so that independent  n-point functions
$W$ are labelled by spin networks with n vertices.

To put it differently, to each spin network $s$ we can associate a
gauge invariant product of field operators $\phi_{s}$

\bes
\phi_{s}\,=\,\sum_{\alpha}\,\prod_{n}\,\phi_{j^{1}_{1}j^{1}_{2}j^{1}_{3}}^{\alpha^{1}_{1}\alpha^{1}_{2}\alpha^{1}_{3}}.
\ees
This provides us with a functional on the space of spin networks
\bes
W(s)\,=\,\int\mathcal{D}\phi\,\phi_{s}\,e^{-S[\phi]}
\ees
that we can use, if positive definite, to reconstruct the full Hilbert space
of the theory, using only the field theory over the group, via the GNS
construction. The transition functions between spin networks can be easily computed using a perturbative
expansion in Feynman diagrams. As we have seen above, this turns out
to be given by a sum over spin foams $\sigma$ interpolating between the n spin
networks representing their boundaries, for example:

\bes
W(s,s')\,=\,W(s\cup s')\,=\,\sum_{\sigma/\partial\sigma=s\cup
s'}A(\sigma).
\ees 

Let us note that the 2-point functions we discussed are naturally associated to the Hadamard or Schwinger functions of usual 
quantum field theory, or of the relativistic particle, since they do not distinguish between different ordering of the two arguments 
and are real functions. 
They represent a-causal amplitudes between quantum geometry states, or correlations among them, i.e. they provide their 
inner product. They are nevertheless fundamental, fully gauge-invariant, non-perturbative observables of 3d quantum gravity.

For a different way of using the field theoretical techniques
in this context, giving rise to a Fock space of spin networks
 on which creation and annihilation operators constructed 
from the field act, see \cite{Mikov}. 

\chapter{The Turaev-Viro invariant: some properties}
In this section we intend to give more details about the Turaev-Viro invariant, the most extablished and rigorously defined
 of the spin foam models discussed so far, and to show some example of explicit calculations that can be performed with it,
making clear how it involves only combinatorial and algebraic manipulations, while describing quantum gravity (i.e. geometric 
and topological) transition amplitudes.

\section{Classical and quantum 3d gravity, Chern-Simons theory and BF theory}
Let us first recall how different theories can be defined to describe 3d geometry (gravity)\cite{Carlip2}, and how they are 
related to each other at the classical and quantum level.

Consider first the action for Riemannian first order gravity with positive cosmological constant in 3d:

\bes
S_{gr}\,=\,-\int_{\mathcal{M}}\left( e\wedge F(A)\,-\,\frac{\Lambda}{6}\,e\,\wedge\,e\,\wedge\,e\right)
\ees

The variables are a 1-form $e$ with values in the Lie algebra of $SU(2)$ and a connection 1-form $A$, with curvature 
$F(A)=d_A A$ again with values in the same Lie algebra. The gauge group is thus $SU(2)$ (the trace on the Lie algebra
is understood in the action above). Let us recall that in turn this action is a particular instance of a BF theory action,
a topological theory that can be defined in any spacetime dimension. The solutions of the equations of motion are spaces of 
constant positive curvature $F=\Lambda\,e\wedge e$. The group of isometries of the covering space (de Sitter space) is thus 
$SO(4)$. We can now define a new gauge connection, with values in the Lie algebra of $Spin(4)\simeq SU(2)\times SU(2)$, by:

\bes
A\,=\,(A^+,A^-)\,\,\,\,\,\,\,\,\,\,\,A^{\pm}\,=\,A\,\pm\,\sqrt{\Lambda}\,e.
\ees 

The above action can now be re-written as:
\bes
S_{gr}\,=\,-\int_{\mathcal{M}}\left( e\wedge F(A)\,-\,\frac{\Lambda}{6}\,e\,\wedge\,e\,\wedge\,e\right)\,=\,S_{CS}[A]\,=
\, S_{CS}[A^+]\,-\,S_{CS}[A^-],
\ees

where

\bes 
S_{CS}[A]\,=\,\frac{k}{4\pi}\,\int_{\mathcal{M}}\left( A\,\wedge\,dA\,+\,\frac{2}{3}\,A\,\wedge\,A\,\wedge\,A\right)
\ees
is the action for {\it Chern-Simons theory}, which is another topological field theory in 3d, that can be defined for
 any gauge group $G$; here, as we said, $G=SO(4)$ and $k=\frac{4\pi}{\sqrt{\Lambda}}$. We thus see that 3d gravity with 
cosmological constant is classically equivalent to two copies of Chern-Simons theory, with connections each being the complex 
conjugate of the other, so any field configuration of the gravitational theory (in particular any solution of the equation 
of motion) is in 1-1 correspondence with a field configuration of the Chern-Simons theory. 
There is a perfect equivalence between the two theories at the classical level.

Most important from our point of view is that Witten \cite{Witten} has shown that a path integral quantization of the 
Chern-Simons theory can be obtained, thus using non-perturbative methods, when the theory is, just as 3d gravity is, 
perturbatively non-renormalizable. In other words, the quantum Chern-Simons theory can be defined by means of the partition 
function:
\bes
Z_{CS}\,=\,\int \mathcal{D}A\,e^{i\,S_{CS}[A]}
\ees
and this partition function defines by quantum field theoretic means a topological invariant of 3-manifolds, related to the 
Jones polynomial \cite{Witten2,Kauffman}, called the {\it Chern-Simons invariant}.

At the quantum level, however, the correspondence between gravity and Chern-Simons theory is more subtle, although one would 
expect from the considerations just made that the partition function for 3d gravity with cosmological constant and gauge 
group $SU(2)$ would be equal to the modulus squared of the Chern-Simons partition function for the same gauge group, i.e.:
\bes
Z_{gr}[\mathcal{M}]\,=\,\int\mathcal{D}A\,\mathcal{D}e\,e^{i\,S_{gr}[e,A]}\,=\,\int\mathcal{D}A^+\,\mathcal{D}A^-\,e^{i\,
S_{CS}[A^+]\,-\,i\,S_{CS}[A^-]}\,=\,\mid Z_{CS}[\mathcal{M}]\mid^2\,\,\,\,\,
\ees 
thus giving a different but strictly related topological invariant. The difficulties in making this correspondence precise are
of course in making sense of the path integral expressions.

Luckily it is indeed possible to make sense of the expressions above by adopting purely combinatorial and algebraic methods 
in constructing the above-mentioned invariants. The rigorously defined topological invariant corresponding to the Chern-Simons 
classical theory has been obtained by Reshetikhin and Turaev in \cite{RT}, using quantum group techniques and represent an 
alternative realization of the Chern-Simons invariant, while the Turaev-Viro invariant we are going to discuss in the 
following is the counterpart for the gravity invariant. In fact, while there is no rigorous construction of the path 
integral expression for $Z_{gr}$, and the most direct connection between the Turaev-Viro spin foam model and 3d gravity is 
the asymptotic calculation for the quantum $6j$-symbol we already presented, Roberts \cite{roberts} has proven that the 
Turaev-Viro invariant is given exactly by the modulus square of the Reshetikhin-Turaev invariant, i.e. the Chern-Simons 
invariant, and thus realises precisely the correspondence guessed above. This correspondence can be shown to be valid also by 
direct calculations in several instances, with a perfect matching of continuum calculations based on Chern-Simons theory and 
combinatorial and algebraic calculations within the Turaev-Viro model, as we are going to show in the following.

\section{The Turaev-Viro invariant for closed manifolds}
Let us recall the basic elements of the Turaev-Viro invariant, as exposed already in section \ref{sec:TV} (for more details 
see \cite{TV, ArcWil}).
The topological invariant for the 3-dimensional manifold is defined by 
the partition function:
\bes
Z_{TV}\,=\,Z(T)\,=\,w^{-2\, a}\,\sum_\phi\,\prod_e\, w_e^2\,\prod_{tet}\,\mid
6j\mid_{tet}, 
\ees 
where again $\phi$ indicate a coloring of the edges $e$ of the triangulation (faces of the dual 2-complex) by elements of the 
set $I$ (representations of $SU(2)_q$), and $w$ and the $w_e$'s are as defined in section \ref{sec:TV}.
The basic amplitudes are the quantum $6j$-symbols $\mid 6j\mid=\begin{vmatrix}
j_1 &j_2 &j_3
\\ j_4 &j_5 &j_6 
\end{vmatrix}$, that we denote for simplicity of notation $\begin{vmatrix}
1 &2 &3
\\ 4 &5 &6 
\end{vmatrix}$ here and in the following.

These are assumed to possess the following symmetries:
\bes
\begin{vmatrix}
1 &2 &3
\\ 4 &5 &6 
\end{vmatrix}\,=\,\begin{vmatrix}
2 &1 &3
\\ 5 &4 &6 
\end{vmatrix}\,=\,\begin{vmatrix}
1 &3 &2
\\ 4 &6 &5 
\end{vmatrix}\,=\,\begin{vmatrix}
1 &5 &6
\\ 4 &2 &3 
\end{vmatrix}\,=\,\begin{vmatrix}
4 &5 &3
\\ 1 &2 &6 
\end{vmatrix}\,=\,\begin{vmatrix}
4 &2 &6
\\ 1 &5 &3 
\end{vmatrix}. 
\ees
In adddition, we assume the following conditions to be verified:

\begin{enumerate}
\item (orthogonality) \bes \sum_j\,w_j^2\,w_{4}^2\,\begin{vmatrix}
2 &1 &j
\\ 3 &5 &4 
\end{vmatrix}\,\begin{vmatrix}
3 &1 &6
\\ 2 &5 &j 
\end{vmatrix}\,=\,\delta_{4,6}\ees for admissible six-tuples;
\item (BE) \bes \sum_j\,w_j^2\,\begin{vmatrix}
2 &a &j
\\ 1 &c &b 
\end{vmatrix}\,\begin{vmatrix}
3 &j &e
\\ 1 &f &c 
\end{vmatrix}\,\begin{vmatrix}
3 &2 &4
\\ a &e &j 
\end{vmatrix}\,=\,\begin{vmatrix}
4 &a &e
\\ 1 &f &b 
\end{vmatrix}\,\begin{vmatrix}
3 &2 &4
\\ b &f &c 
\end{vmatrix}
\ees 
for admissible six-tuples $(4,a,e,1,f,b)$ and $(3,2,4,b,f,c)$;
\item \bes \forall j\in I\;\;\;\;\;\;w^2\,=\,w_j^{-2}\,\sum_{k,l : (j,k,l)\,adm}w_k^2\,w_l^2\,\,. \ees
\end{enumerate}
While the third condition is a purely algebraic one in nature, the other two have a geometric 
intepretation as operations on tetrahedra in the triangulation (as shown in the picture for the Biedenharn-Elliott identity) 
\begin{figure}
\begin{center}
\includegraphics[width=8cm]{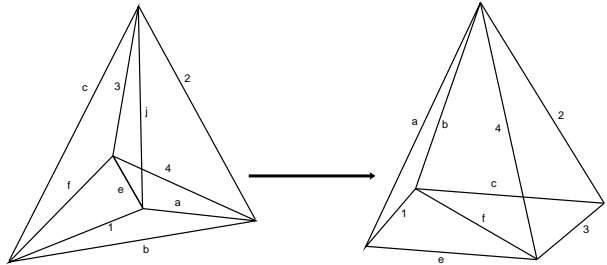}
\caption{The geometric interpretation of the Biedenharn-Elliott identity, with three tetrahedra sharing an edge replaced by two tetrahedra.}
\end{center}
\end{figure}
and axiomatize the 
orthogonality and Biedenharn-Elliott identities for quantum $6j$-symbols. Also, the symmetries imposed on the symbol above 
correspond to the invariance of the amplitude for a tetrahedron under the action of the permutation group $S_4$ on the 
its 4 vertices. As a result, the amplitude does not register the orientation of the tetraedron itself.

The partition function above can be interpreted, in the context of axiomatic topological quantum field theories, as a 
vacuum-to-vacuum transition amplitude, i.e. as the probability amplitude for a 3-dimensional universe with topology given by 
the chosen topology for $\mathcal{M}$. In the case of 3-manifolds with boundaries, instead, the model provides a function of 
the boundary triangulations that can be intepreted as a transition amplitude between quantum states (function of the 
colorings of the boundary edges) associated to the boundaries. It is important to stress that the model does not register 
the orientation of the manifold considered, and therefore does not register any order in the states which are the arguments 
of the transition amplitude \cite{TV}.

The above mentioned conditions on the quantum $6j$-symbols and on the amplitudes $w$ and $w_j$'s make the partition function
independent of the particular triangulation chosen to define it, i.e. imply the fact that $Z_{TV}(\mathcal{M})$ is indeed a 
topological invariant characterizing $\mathcal{M}$. In fact, any two triangulations of a given polyhedron (so of a given 
topology) can be transformed into one another by a finite sequence of \lq\lq Alexander moves" and their inverses 
\cite{Alexander, TV}, and these in turn can be factorised into three \lq\lq elementary moves", one replacing a tetrahedron 
in the triangulation with three of them, its inverse, and one changing a pair of tetrahedra into a different pair. The 
Alexander moves can also be formulated in the dual picture, i.e. as moves changing the 2-skeletons (2-complexes) dual to the
triangulations into one another. It can then be shown \cite{TV} that the conditions 1-3) listed above, and satisfied by the 
building blocks of our partition function $Z_{TV}$, imply its invariance under the Alexander moves. In other words, the 
function $Z_{TV}(\mathcal{M})$ evaluates to the same number for fixed $\mathcal{M}$ for any triangulation of it we choose.
We can also use this invariance, and the properties 1-3) to our advantage when evaluating the invariant, since appropriate 
sequences of application of the conditions lead to important simplification of the expression we have to evaluate. This will      
become apparent in the examples to be shown below.

\section{Evaluation of the Turaev-Viro invariant for different topologies}
Before showing in details how a non-trivial evaluation of the invariant is performed, we give a few examples of its values 
in simple but significant cases.
 
The simplest case to consider is that of the 3-sphere $S^3$, and the invariant does not involve any $6j$-symbol and evaluates 
to \cite{TV}: \bes
Z_{TV}(S^3)\,=\,w^{-4}\,\sum_j\,w_j^4. \ees
 
Another simple case is for $\mathcal{M}=S^1\times S^2$ for which again we do not need any $6j$-symbol and we get \cite{TV} 
simply $Z_{TV}=1$.

More interesting, although still simple, is the evaluation of the invariant for the 3-ball $B^3$. In this case, one  gets as 
we said a function of the boundary states, i.e. of the triangulation of the boundary $S^2$ and of its coloring by 
representations; therefore the result can be labelled by using the number of vertices, edges and triangles in the 
triangulation of $S^2$ one used. Consider the following three possible triangulations of $B^3$ \cite{Ionicioiu}. 

One can triangulate $B^3$ by taking the cone with vertex $(4)$ over two triangles $(123)$ and $(123)$ joined along the edge 
$(23)$ and with the opposite vertex $(1)$ identified (see figure 3.2).

\begin{figure}
\begin{center}
\includegraphics[width=7cm]{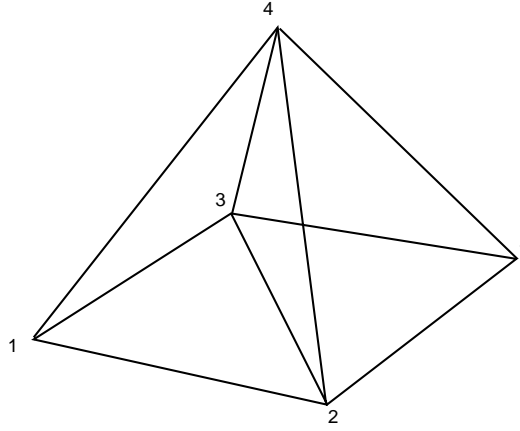}
\caption{A triangulation of $B^3$ using two tetrahedra with the base triangles identified.}
\end{center}
\end{figure}

 In this case the boundary triangulation has 3 vertices, 
3 edges and 2 triangles. The triangulation of $B^3$ is instead given by 4 vertices, 6 edges and 2 tetrahedra. The amplitude for 
the transition $\emptyset \rightarrow S^2$ (or equivalently $S^2 \rightarrow\emptyset$) is then given by \cite{Ionicioiu}:
\bes
Z_{TV}(B^3,(3,3,2),j)\,=\,w^{-5}\,w_1\,w_2\,w_3\,\sum_{4,5,6}\,w^2_4\,w^2_5\,w^2_6\,\begin{vmatrix}
1 &2 &3
\\ 6 &5 &4 
\end{vmatrix}^2\,= \nonumber \\ =\,w^{-5}\,w_1\,w_2\,w_3\,\sum_{5,6}\,\frac{w^2_5\,w^2_6}{w^2_3}\,=\,w^{-3}\,w_1\,w_2\,w_3,
\ees 
where we have used orthogonality first and property 3) in the last step.
 
The most common and simple triangulation of $B^3$ is however given by a single tetrahedron, whose 4 boundary triangles 
constitute then a triangulation of $S^2$. The boundary triangulation has then 4 vertices, 6 edges and 4 triangles.
 In this case there are no internal edges, and thus no summation has to be performed, with the invariant being just \cite{Ionicioiu}:
\bes
Z_{TV}(B^3,(4,6,4),j)\,=\,w^{-4}\,w_1\,...\,w_6\,\begin{vmatrix}
1 &2 &3
\\ 6 &4 &5 
\end{vmatrix}.
\ees

Another possible triangulation is that using two tetrahedra glued along a common face (just think of the previous picture without
the identification of the opposite vertices 1 in the base triangles), resulting in 5 vertices, 
9 edges and 6 triangles on the boundary, with again no internal edges, so that \cite{Ionicioiu}:
\bes 
Z_{TV}(B^3,(5,9,6),j)\,=\,w^{-5}\,w_1\,...\,w_9\,\begin{vmatrix}
4 &5 &6
\\ 2 &1 &3 
\end{vmatrix}\,\begin{vmatrix}
4 &5 &6
\\ 9 &7 &8 
\end{vmatrix}.
\ees

A nice check of the above results is to note that one can obtain a 3-sphere $S^3$ by gluing together two 3-balls $B^3$ along 
their common $S^2$ boundary, so that:
\bes
Z(S^3)\,=\,\sum_{j\in\partial B^3\sim S^2}\,Z^2(B^3,(v,e,t),j),
\ees
using the general properties of topological quantum field theories, and the results above.

Another application of the axioms of topological quantum field
theories leads to the calculation of the TV invariant for the cylinder
$S^2\times I\sim B^3 \cup B^3$; since the general rule for
disconnected sums is: $Z(A\cup B)=\frac{Z(A)Z(B)}{Z(S^3)}$, we have
for the cylinder: 
\bes
Z(S^2\times I,j,k)\,=\,\frac{Z(B^3,j)Z(B^3,k)}{Z(S^3)}, \ees
and the result of
course depends on the particular triangulation we choose for the
$S^2$ boundaries of the two 3-balls.

It is also interesting to note (and to check) that, whatever this triangulation is
chosen to be, the partition function for $Z(S^2\times I, j,k)$ acts
like a Kronecker delta $\delta_{j,k}$, expressing the topological
statement that attaching a cylinder $S^2\times I$ to the $S^2$
boundary of an arbitrary triangulated manifold $\mathcal{M}$ does not change its topology, but
only the coloring of the boundary triangulation:

\bes
\sum_j\,Z(\mathcal{M},S^2,j)\,Z(S^2\times I, j,
k)\,=\,\sum_j\,\frac{Z(\mathcal{M})\,Z(B^3,j)}{Z(S^3)}\,\frac{Z(B^3,j)\,Z(B^3,k)}{Z(S^3)}\,
\\ =\,\frac{Z(\mathcal{M})\,Z(B^3,k)}{Z(S^3)}\,\sum_j\,\frac{Z^2(B^3,j)}{Z(S^3)}\,=\,Z(\mathcal{M},S^2,k),
\ees 
 where we have used the previous results and the fact that an $S^2$ boundary can be obtained in a manifold 
$\mathcal{M}$ by simply taking the disjoint union of it with a 3-ball. This is just an expression
 of the fact that $Z(S^2\times I)$ is a projection operator on the
 vector space associated to the surface $S^2$ \cite{Atiyah,barrett}.

We can also compute the partition function for $S^2\times S^1$
starting from this result by gluing together the two boundaries of a
cylinder, recovering the result anticipated above. 

Let us now show a non-trivial evaluation of the Turaev-Viro invariant
in full details. We compute the Turaev-Viro invariant for the 3-torus
$T^3\sim S^1\times S^1\times S^1$. On the one hand this should make it clear how
evaluations of the invariant are performed in general, on the other
hand it provides a check of the correspondence between the Turaev-Viro
and the Chern-Simons invariant, showing how the continuum and
sophisticated calculations of the latter can be reproduced by purely
combinatorial and algebraic methods using
the former. The calculation may be tedious but it is nevertheless
absolutely straightforward.

We obtain a triangulation of $T^3$ by taking a cube, chopping
each face into two triangles, considering one body diagonal, and
identifying all the opposite pairs of edges in each of its faces (see in the figure 3.3).

\begin{figure}
\begin{center}
\includegraphics[width=8cm]{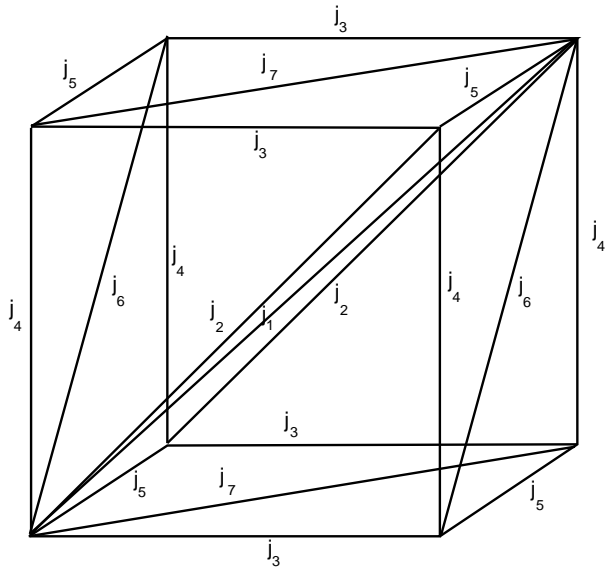}
\caption{A triangulation of the 3-torus $T^3$, using six tetrahedra.}
\end{center}
\end{figure} 

The resulting triangulation is given by one vertex, seven
edges and six tetrahedra.

The partition function is thus given by:
\bes
Z(T^3)\,=\,w^{-2}\,\sum_{1,...,7}\,w_1^2\,...\,w_7^2\,\begin{vmatrix}
3 &5 &7
\\ 4 &1 &6 
\end{vmatrix}\,\begin{vmatrix}
3 &5 &7
\\ 1 &4 &2 
\end{vmatrix}\,\begin{vmatrix}
4 &5 &6
\\ 1 &3 &2 
\end{vmatrix}\, \begin{vmatrix}
4 &5 &6
\\ 3 &1 &7 
\end{vmatrix}\,\begin{vmatrix}
2 &3 &4
\\ 7 &1 &5 
\end{vmatrix}\,\begin{vmatrix}
2 &3 &4
\\ 6 &5 &1 
\end{vmatrix}\,\,\,\,.
\ees
  
Now we use the inverse of the B-E identity to substitute the 3rd and
last tetrahedron with three tetrahedra, insering an additional edge
labelled by $j_8$:

\bes
\begin{vmatrix}
4 &5 &6
\\ 1 &3 &2 
\end{vmatrix}\begin{vmatrix}
2 &3 &4
\\ 6 &5 &1 
\end{vmatrix}\,=\,\sum_8\,w_8^2\,\begin{vmatrix}
5 &5 &8
\\ 2 &2 &1 
\end{vmatrix}\begin{vmatrix}
4 &8 &4
\\ 2 &3 &2 
\end{vmatrix}\begin{vmatrix}
4 &5 &6
\\ 5 &4 &8 
\end{vmatrix}.
\ees

Next we repeat this operation with what were the 2nd and fifth
tetrahedra inserting a new edge labelled by $j_9$, then we use the
proper B-E identity summing over the spin $j_7$ and eliminating the
corresponding edge, obtaining after
these operations:

\bes
Z(T^3)\,=\,w^{-2}\sum_{1,..,6,8,9}\,w_1^2\,...\,w^2_6\,w^2_8\,w^2_9\,\begin{vmatrix}
6 &5 &4
\\4  &9 &6 
\end{vmatrix}\begin{vmatrix}
1 &1 &9
\\ 2 &2 &5 
\end{vmatrix}\begin{vmatrix}
1 &3 &6
\\ 6 &9 &1 
\end{vmatrix} \nonumber \\ \begin{vmatrix}
4 &9 &4
\\ 2 &3 &2 
\end{vmatrix}\begin{vmatrix}
5 &5 &8
\\ 2 &2 &1 
\end{vmatrix}\begin{vmatrix}
4 &8 &4
\\ 2 &3 &2 
\end{vmatrix}\begin{vmatrix}
4 &5 &6
\\ 5 &4 &8 
\end{vmatrix}\,\,\,\,\,.
\ees 

We now apply again the inverse B-E condition to the 4th and 6th
tetrahedra inserting a new edge with spin $j_{10}$.   
Recall that all
these moves are basically combinations of Alexander moves changing a
triangulation into another for the same topology, so we are always
exploiting the topological invariance of the model.

We get:

\bes
Z(T^3)\,=\,w^{-2}\sum_{1,..,6,8,..,10}\,w^2_1\,..\,w^2_6\,w^2_8\,w^2_9\,w^2_{10}\,\begin{vmatrix}
6 &5 &4
\\ 4 &9 &6 
\end{vmatrix}\begin{vmatrix}
1 &1 &9
\\ 2 &2 &5 
\end{vmatrix}\begin{vmatrix}
1 &3 &6
\\ 6 &9 &1 
\end{vmatrix}\begin{vmatrix}
8 &9 &10
\\2  &2 &2 
\end{vmatrix} \nonumber \\ \begin{vmatrix}
4 &10 &4
\\ 2 &3 &2 
\end{vmatrix}\begin{vmatrix}
9 &8 &10
\\ 4 &4 &4 
\end{vmatrix}\begin{vmatrix}
5 &5 &8
\\ 2 &2 &1 
\end{vmatrix}\begin{vmatrix}
4 &5 &6
\\ 5 &4 &8 
\end{vmatrix}\,\,\,\,\,\,.
\ees
  
And we do the same with the 4th and 6th tetrahedra inserting a new
edge with spin $j_{11}$, obtaining:

\bes
Z(T^3)\,=\,w^{-2}\sum_{1,..,6,8,..,11}\,w^2_1\,..\,w^2_6\,w^2_8\,..\,w^2_{11}\,\begin{vmatrix}
 6 &5 &4
\\ 4 &9 &6 
\end{vmatrix}\begin{vmatrix}
1 &1 &9
\\ 2 &2 &5 
\end{vmatrix}\begin{vmatrix}
1 &3 &6
\\ 6 &9 &1 
\end{vmatrix}\begin{vmatrix}
4 &2 &11
\\ 2 &4 &10 
\end{vmatrix}\nonumber \\ \begin{vmatrix}
4 &4 &8
\\ 2 &2 &11 
\end{vmatrix}\begin{vmatrix}
4 &10 &4
\\ 2 &9 &2 
\end{vmatrix}\begin{vmatrix}
5 &5 &8
\\ 2 &2 &1 
\end{vmatrix}\begin{vmatrix}
4 &5 &6
\\ 5 &4 &8 
\end{vmatrix}\,\,\,\,\,\,.
\ees

Up to now the impression may be that we have not simplified the
expression for the invariant at all, and it is indeed the case, but
now the partition function has a form that allows for the
simplifications we look for.

We can use orthogonality of the $6j$-symbols, by summing over
$j_{10}$, thus eliminating two tetrahedra:
\bes
\sum_{j_{10}}\,w_{10}^2\,w_{11}^2\,\begin{vmatrix}
4 &10 &4
\\ 2 &9 &2 
\end{vmatrix}\begin{vmatrix}
4 &2 &11
\\ 2 &4 &10 
\end{vmatrix}\,=\,\delta_{9,11}.
\ees 
After this, we eliminate the edge labelled by $j_8$ summing over the
corresponding spin, and we get:

\bes
Z(T^3)\,=\,w^{-2}\sum_{1,..,6,9}\,w^2_1\,..\,w^2_6\,w^2_9\,\begin{vmatrix}
6 &5 &4
\\ 4 &9 &6 
\end{vmatrix}\begin{vmatrix}
1 &1 &9
\\ 2 &2 &5 
\end{vmatrix}\begin{vmatrix}
1 &3 &6
\\ 6 &9 &1 
\end{vmatrix}\begin{vmatrix}
4 &9 &4
\\ 2 &9 &2 
\end{vmatrix}\begin{vmatrix}
6 &4 &5
\\ 2 &1 &9 
\end{vmatrix}\begin{vmatrix}
5 &4 &6
\\ 9 &1 &2 
\end{vmatrix}\,\,\,\,\,\,.
\ees
At this stage the partition function has the same complexity as the
one we started with, having again seven edges and six tetrahedra;
however, the configuration of edges is such that two sequences of
an inverse B-E move and two orthogonality moves lead immediately to
the final result.

This is what we do. We apply the inverse B-E move to the 1st and 3rd
tetrahedra, inserting a new edge with label $j_7$, and then we use
orthogonality twice to eliminate the edges labelled by $j_3$ and
$j_6$.

The result is:
\bes
Z(T^3)\,=\,w^{-2}\sum_{1,2,4,5,7,9}\,w^2_1\,w^2_5\,w^2_7\,w^2_9\,\begin{vmatrix}
1 &7 &4
\\ 4 &9 &1 
\end{vmatrix}\begin{vmatrix}
1 &1 &9
\\ 2 &2 &5 
\end{vmatrix}\begin{vmatrix}
4 &9 &4
\\ 2 &9 &2 
\end{vmatrix}\,\,\,\,\,\,\,\,\,.
\ees 

The simplification is apparent. We then repeat the same sequence of
moves, by first replacing the 1st and 3rd tetrahedra with three new
ones with a new edge inserted labelled by $j_3$, and then use
orthogonality twice to eliminate the spins $j_7$ and $j_9$.

The result do not involve $6j$-symbols anymore, and reads:

\bes
Z(T^3)\,=\,w^{-2}\sum_{1,2,4,5}\,w^2_1\,w^{-2}_2\,w^2_5.
\ees

We finally make use of the condition 3):

\bes
w_2^{-2}\sum_{1,5}\,w^2_1\,w^2_5\,=\,w^2.
\ees

The final expression we obtain is:

\bes
Z(T^3)\,=\,\sum_{2,4}\,1.
\ees

On top of being extremely simple, this result is in agreement with
the continuum calculation perfomed using Chern-Simons theory in
\cite{Jeffrey}, confirming the Turaev-Viro invariant as the square of
the Chern-Simons one.

\section{The Turaev-Viro invariant for lens spaces}
In this section we shall compute the Turaev-Viro invariant for more interesting manifolds than the ones considered so far, 
showing how it encodes their topological properties. The manifolds we consider are lens spaces, denoted $L_{p,q}$, where $p$ 
and $q$ are integers.

Lens spaces \cite{Ionicioiu,Rofsen} are 3-manifolds with genus 1 Heegaard splitting, i.e. they are constructed out of two solid 
tori $H_1=D^2\times S^1$ by gluing them along the common boundary $T^2=S^1\times S^1$ with an appropriate homeomorphism 
$h_{p,q}:T^2\rightarrow T^2$: $L_{p,q}=H_1\cup_{h_{p,q}}H_1$. The homeomorphism is defined as follows: calling $m$ and $l$ 
the meridian and longitudinal circles of $T^2$, $h_{p,q}$ is defined as the identification of the meridian $m$ of one torus 
with the curve winding $p$ times long the longitude and $q$ times along the meridian of the second torus, with $p,q\in\mathbb{Z}$:
\bes
h_{p,q}\,:\,m\,\rightarrow\,p\,l\,+\,q\,m.
\ees

In addition to the interest in lens spaces from a purely topological point of view, that will be our concern in the following,
we note also that, being prime manifolds \cite{hempel}, lens spaces have an interpretation as {\it geons} \cite{sorkgeon}, 
i.e. topologically non-trivial excitation of the metric.

Let us give a few more details on the topology of lens spaces \cite{guadagninipilo}.
Lens spaces $L_{p,q}$ have fundamental group $\mathbb{Z}_p$ (so it depends on $p$ only). Two lens spaces $L_{p,q}$ and 
$L_{p',q'}$ are homeomorphic if and only if $\mid p\mid = \mid p'\mid$ and $q =\pm q'$ (mod p) or $q q' = \pm 1$ (mod p). 
Because of this, we must consider only the cases when $p > 1$ and $p > q >0$, with $p$ and $q$ relatively prime. Moreover, 
two lens spaces are homotopic if and only if $\mid p\mid =\mid p'\mid$ and $q q'$ is, up to a sign, a quadratic residue 
(mod p).

Therefore, we can have lens spaces that are homotopic but not homeomorphic, such as $L_{13,2}$ and $L_{13,5}$, or that are 
not homotopic and not homeomorphic but have the same fundamental group, such as $L_{13,2}$ and $L_{13,3}$.

Because of these subtleties, lens spaces provide a very interesting class of manifold on which to test topological 
invariants of 3-manifold. In particular they shed light on the properties of the Chern-Simons invariant and thus on the 
Turaev-Viro one. 

 It was conjectured \cite{guadagninipilo} that, when the Chern-Simons invariant $Z_{CS}(\mathcal{M})$ is non-vanishing, its
 absolute value $\mid Z_{CS}(\mathcal{M}) \mid = \sqrt{Z_{TV}(\mathcal{M})}$ depends only on the fundamental group of the 
manifold $\mathcal{M}$, $\pi_1(\mathcal{M})$. 

To the best of our knowledge, there is not yet any proof or disproof of this conjecture in the general case, but it can 
indeed be proven for the particular class of manifolds given by lens spaces, and for the choice of $SU(2)$ as gauge group 
in the Chern-Simons invariant \cite{guadagninipilo}, which is the case we are most interested in from a quantum gravity 
perspective. For the case of $G=SU(3)$ only numerical evidence for the conjecture can be provided \cite{guadagninipilo}. 
The reason for testing such a conjecture using lens spaces is of course the mentioned fact that there are examples of lens 
spaces which are not homeomorphic (they are not, topologically speaking, the \lq\lq same manifold"), but have nevertheless 
the same fundamental group. In these cases one would then expect, if the conjecture is true, that the evaluation of the 
modulus of the Chern-Simons invariant, and thus of the Turaev-Viro invariant, would give the same result.

The proof in the case of $SU(2)$ and for a lens space $L_{p,q}$ is by direct computation of the invariant \cite{guadagninipilo,Jeffrey}.
The modulus square of the Chern-Simons invariant, i.e. the Turaev-Viro invariant, in this case, for level $k\geq 2$ is given 
by the formulae:
\bes
Z_{TV}(L_{2,1})\,=\,\mid Z_{CS}(L_{2,1})\mid^2\,=\,[1\,+\,(-1)^k]\,\frac{\sin^2(\frac{\pi}{2 k})}{\sin^2(\frac{\pi}{k})}
\ees
for $p=2$; 
\bes
Z_{TV}(L_{p,q})\,=\,\frac{1}{2}\,[1\,-\,(-1)^p]\,\frac{\sin^2[\frac{\pi\,(k^{\phi(p)}\,-\,1)}{k\,p}]}{\sin^2(\frac{\pi}{k})}\,
+\,\frac{1}{2}\,[1\,+\,(-1)^p]\,[1\,+\,(-1)^\frac{p}{2}]\frac{\sin^2[\frac{\pi\,(k^{\phi(p/2)}\,-\,1)}{k\,p}]}{\sin^2(\frac{\pi}{k})},
\nonumber \ees
with $p>2$ and where $p$ and $k$ are coprime integers;
\bes
Z_{TV}(L_{p,q})\,=\,\frac{g}{4\sin^2(\pi/k)}\,[\delta_g(q\,-\,1)\,+\,\delta_g(q\,+\,1)], \ees
with $p>2$ and where the greatest common divisor of $p$ and $k$, $g$, is greater than one, i.e. $g>1$, and $p/g$ is odd;
\bes
Z_{TV}(L_{p,q})\,=\,\frac{g}{4\sin^2(\pi/k)}\,\left\{ \delta_g(r\,+\,1)\,[1\,+\,(-1)^\frac{k p}{2 g^2}\,(-1)^\frac{r+1}{g}]\,
+\, \delta_g(r\,-\,1)\,[1\,+\,(-1)^\frac{k p}{2 g^2}\,(-1)^\frac{r-1}{g}]\, \right\} \nonumber ,
\ees
with $p>2$ and for $g>1$ and $p/g$ even, where we have indicated by $\phi(n)$ the Euler function and by $\delta_p(x)$ the 
modulo-p Kronecker delta being $0$ for $x\neq 0$ (mod p) and $1$ for $x=0$ (mod p).

The crucial point is the dependence of $Z_{TV}(L_{p,q})$ on $q$, and it can be shown that, when $Z_{CS}(L_{p,q})\neq 0$, we have:
\bes
Z_{TV}(L_{p,q})\,=\,\frac{1}{\sin^2(\pi/k)}
\ees for $g = 2 $ and $p/g$ odd,
\bes
Z_{TV}(L_{p,q})\,=\,\frac{g}{4\,\sin^2(\pi/k)}
\ees
for $g > 2$ and $p/g$ odd,
\bes
Z_{TV}(L_{p,q})\,=\,\frac{g}{2\,\sin^2(\pi/k)}.
\ees

We see that indeed $Z_{TV}(L_{p,q})=\mid Z_{CS}(L_{p,q})\mid^2$ does not depend on the integer $q$ and is thus a function of 
the fundamental group $\mathbb{Z}_p$ only.

The calculations leading to these results are based on surgery
presentation of the lens spaces, and on the expression of the
Chern-Simons invariant as a polynomial invariant (the Jones
polynomial) for oriented framed, coloured links embedded in the
continuum manifold. Beautiful as it may be this approach to the
evaluation, it is nevertheless both conceptually and technically quite
complex. 

The evaluation of the Turaev-Viro invariant for the same manifolds is
considerably simpler and really straightforward and it involves only the three basic
combinatorial and algebraic conditions we discussed and used for the
evaluation of the invariant for the 3-torus.    

To write down the TV partition function we have to choose a
triangulation of the lens spaces. This is done as follows
\cite{Rofsen, Ionicioiu}.
Consider a $p$-polygon and a double cone over it, thus obtaining a
double solid pyramid glued along the common polygonal base and thus
forming a 3-ball $B^3$. Now identify each point in the upper half of
$\partial B^3$ with a point in the lower half after a rotation by
$\frac{2\pi q}{p}$ in the base plane and a reflection with respect to
it. The resulting simplicial manifold is indeed a lens space
$L_{p,q}$, for relatively prime integers $p$ and $q$ with $q < p$. 

\begin{figure}
\begin{center}
\includegraphics[width=8cm]{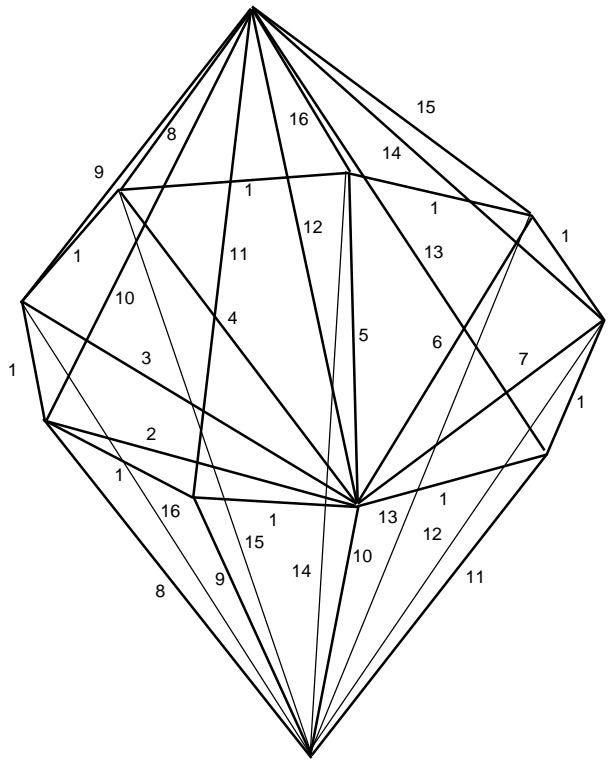}
\caption{A triangulation of the lens space $L_{9,2}$, obtained with the procedure described, starting with a 9-polygon, 
taking a double cone over it, and identifying upper and lower half points after a rotation of $4\pi/9$ in the base plane and 
a final reflection with respect to it.}
\end{center}
\end{figure}

We now give the evaluation of the Turaev-Viro invariant for all the
lens spaces with $p$ up to $9$. The calculations for $p$ up to $8$ were
reported in \cite{TV} and \cite{Ionicioiu}. Those for $p=9$ are new.

We have:
\bes
&& Z_{TV}(L_{2,1}\sim \mathbb{R}P^3)\,=\,w^{-2}\sum_1\,w^2_1 \nonumber
\\
&& Z_{TV}(L_{3,1})\,=\,w^{-2}\sum_{1 : (1,1,1) \nonumber
  adm}\,w^2_1
\\
&& Z_{TV}(L_{4,1})\,=\,w^{-2}\sum_{1,2 : (1,1,2) adm}\,w^2_1\,w^2_2\,\begin{vmatrix}
1 &1 &2
\\ 1 &1 &2 
\end{vmatrix} \nonumber
\\
&& Z_{TV}(L_{5,1})\,=\,w^{-2}\sum_{1,2,3}\,w^2_1\,w^2_2\,w^2_3\,\begin{vmatrix}
1 &2 &3
\\ 1 &2 &1 
\end{vmatrix}\begin{vmatrix}
1 &2 &3
\\ 1 &1 &3 
\end{vmatrix} \nonumber
\\
&& Z_{TV}(L_{5,2})\,=\,w^{-2}\sum_{1,2}\,w^2_1\,w^2_2\,\begin{vmatrix}
1 &1 &2
\\ 1 &2 &2 
\end{vmatrix} \nonumber
\\
&& Z_{TV}(L_{6,1})\,=\,w^{-2}\sum_{1,..,4}\,w^2_1\,...\,w^2_4\,\begin{vmatrix}
1 &2 &3
\\ 1 &2 &1 
\end{vmatrix}\begin{vmatrix}
1 &2 &3
\\ 1 &4 &3 
\end{vmatrix}\begin{vmatrix}
1 &3 &4
\\ 1 &1 &4 
\end{vmatrix} \nonumber
\\
&& Z_{TV}(L_{7,1})\,=\,w^{-2}\sum_{1,..,5}\,w^2_1\,...\,w^2_5\,\begin{vmatrix}
1 &4 &5
\\ 1 &1 &5 
\end{vmatrix}\begin{vmatrix}
1 &4 &5
\\ 1 &4 &3 
\end{vmatrix}\begin{vmatrix}
1 &2 &3
\\ 1 &2 &1 
\end{vmatrix}\begin{vmatrix}
1 &2 &3
\\ 1 &4 &3 
\end{vmatrix} \nonumber
\\
&& Z_{TV}(L_{7,2})\,=\,Z_{TV}(L_{7,3})\,=\,w^{-2}\sum_{1,2,3}\,w^2_1\,w^2_2\,w^2_3\,\begin{vmatrix}
1 &2 &3
\\ 1 &2 &1 
\end{vmatrix}\begin{vmatrix}
1 &2 &3
\\ 1 &3 &3 
\end{vmatrix} \nonumber
\\
&& Z_{TV}(L_{8,1})\,=\,w^{-2}\sum_{1,..,6}\,w^2_1\,...\,w^2_6\,\begin{vmatrix}
1 &2 &3
\\ 1 &2 &1 
\end{vmatrix}\begin{vmatrix}
1 &3 &4
\\ 1 &3 &2 
\end{vmatrix}\begin{vmatrix}
1 &4 &5
\\ 1 &4 &3 
\end{vmatrix}\begin{vmatrix}
1 &4 &5
\\ 1 &6 &5 
\end{vmatrix}\begin{vmatrix}
1 &5 &6
\\ 1 &1 &6 
\end{vmatrix} \nonumber
\\
&& Z_{TV}(L_{8,3})\,=\,w^{-2}\sum_{1,2,3}\,w^2_1\,w^2_2\,w^2_3\,\begin{vmatrix}
1 &2 &3
\\ 1 &2 &1 
\end{vmatrix}\begin{vmatrix}
1 &2 &3
\\ 3 &2 &3 
\end{vmatrix} \nonumber
\\
&& Z_{TV}(L_{9,1})\,=\,w^{-2}\sum_{1,..,7}w^2_1\,...\,w^2_7\,\begin{vmatrix}
7 &1 &1
\\ 7 &1 &6 
\end{vmatrix}\begin{vmatrix}
6 &5 &1
\\ 6 &7 &1 
\end{vmatrix}\begin{vmatrix}
1 &3 &4
\\ 1 &5 &4 
\end{vmatrix}\begin{vmatrix}
1 &2 &3
\\ 2 &4 &3 
\end{vmatrix}\begin{vmatrix}
1 &2 &3
\\ 2 &2 &1 
\end{vmatrix}\begin{vmatrix}
4 &5 &1
\\ 6 &5 &1 
\end{vmatrix} \nonumber 
\\
&& Z_{TV}(L_{9,2})\,=\,Z_{TV}(L_{9,4})\,=\,w^{-2}\sum_{1,..,4}\,w^2_1\,..\,w^2_4\,\begin{vmatrix}
3 &2 &4
\\ 3 &2 &1 
\end{vmatrix}\begin{vmatrix}
1 &2 &3
\\ 1 &2 &1 
\end{vmatrix}\begin{vmatrix}
5 &2 &2
\\ 5 &2 &3 
\end{vmatrix} \nonumber
\ees

These results are in agreement with those based on Chern-Simons
theory \cite{guadagninipilo,Jeffrey}, of course, and in particular we see the
Guadagnini-Pilo result confirmed explicitely for $p=7$ and $q=2,3$,
and $p=9$ and $q=2,4$. It may also be possible to simplify further
some of the
expressions above by using properties of the quantum $6j$-symbols other
than the orthogonality and the B-E identity that have been used here.

\chapter{Spin foam models for 4-dimensional quantum gravity}
We turn now to the 4-dimensional case. General Relativity  in four spacetime dimensions is a highly non-trivial theory, not
topological anymore in that it possesses local (although \lq\lq not localizable", because of diffeomorphism invariance) degrees 
of freedom and is as a consequence much more difficult to quantize. However, it is close enough to a topological theory that 
one can make use of many of the techniques we used to quantize BF theory, and that one can even fit the yet-to-be-constructed 
quantum gravity theory in the general formalism of topological quantum field theories \cite{barrett,crane95}, as we shall see.
Most important for our concerns, spin foam models for a 4-dimensional spacetime can be obtained along similar lines as in the 3-dimensional 
case, and this is indeed the object of the following chapters.

\section{4-dimensional gravity as a constrained topological field theory: continuum and discrete cases} \label{sec:pleb}
The action for General Relativity in the first order formalism is the so-called Palatini action:
\bes
S(e,\omega)\,=\,\int_{\mathcal{M}}*\,e\,\wedge\,e\,\wedge\,F(\omega)\,=\,\int_{\mathcal{M}}\epsilon_{IJKL}\,e^I\,\wedge\,e^J\,\wedge\,F^{KL}(\omega),
\ees
where the field variables coming into the action are: a 1-form tetrad field $e^I=e^I_\mu dx^\mu$ with internal $\mathbb{R}^{3,1}$ index $I$ 
(in $\mathbb{R}^4$ in the Riemannian case), giving the spacetime metric as $g_{\mu\nu}=\eta_{IJ}e^I_\mu\otimes e^J_\nu$, and a 
1-form Lorentz connection $\omega^{IJ}=\omega^{IJ}_\mu dx^\mu$ with values in the Lie algebra of the Lorentz group $so(3,1)$ 
($so(4)$ in the Riemannian case) in the adjoint representation, so that we use the fact $\wedge^2(\mathbb{R}^{3,1})\sim so(3,1)$, 
 with a 2-form curvature $F^{IJ}(\omega)=\mathcal{D}\omega^{IJ}$. The corresponding equations of motion are: $\mathcal{D}e=0$, 
expressing  the compatibility of the tetrad field and the connection, i.e. the fact that the connection we have is a 
{\it metric} connection leaving the metric (tetrad) field covariantly constant, and $e \wedge F(\omega) = 0$, being the 
Einstein equations describing the dynamics of the spacetime geometry. The symmetries of the action are, just as in the
3-dimensional case, the usual diffeomorphism invariance and the invariance under the internal gauge group, i.e. the Lorentz
 group.

The starting point for the spin foam quantization we are going to discuss is the fact that this action is classically a 
subsector of a more general classical action for the gravitational field, namely the so-called Plebanski action \cite{plebanski, CDJ}, 
which describes gravity as a constrained topological theory.

The Plebanski action is a a
BF-type action, in the sense that it gives gravity as a constrained BF
theory, with quadratic constraints on the B field. More precisely the action is given by:
\bes 
S\,=\,S(\omega,B,\phi)\,=\,\int_{\mathcal{M}}\left[
B^{IJ}\,\wedge\,F_{IJ}(\omega)\,-\frac{1}{2}\phi_{IJKL}\,B^{KL}\,\wedge\,B^{IJ}\right]
\ees
where $\omega$ is a connection 1-form valued in $so(3,1)$ ($so(4)$),
$\omega=\omega_{a}^{IJ}J_{IJ}dx^{a}$, $J_{IJ}$ are the generators of
$so(3,1)$ ($so(4)$), $F=\mathcal{D}\omega$ is the corresponding two-form curvature,
$B$ is a 2-form also valued in $so(4)$ ($so(3,1)$),
$B=B_{ab}^{IJ}J_{IJ}dx^{a}\wedge dx^{b}$, and $\phi_{IJKL}$ is a
Lagrange multiplier, with symmetries $\phi_{IJKL}=\phi_{[IJ][KL]}=\phi_{[KL][IJ]}$, satisfying $\phi_{IJKL}\epsilon^{IJKL}=0$.
 
The equations of motion are:
\bes
d B\,+\,[\omega,B]=0 \;\;\;\;\;\;\;\;
F^{IJ}(\omega)\,=\,\phi^{IJKL}B_{KL}\;\;\;\;\;\;\;\; 
B^{IJ}\,\wedge\,B^{KL}\,=\,e\,\epsilon^{IJKL} \label{eq:constrB}
\ees
where $e=\frac{1}{4!}\epsilon_{IJKL}B^{IJ}\wedge B^{KL}$.

When $e\ne0$, i.e. for non-degenerate metric configurations, the constraint \Ref{eq:constrB}
is equivalent to
$\epsilon_{IJKL}B^{IJ}_{ab}B^{KL}_{cd}=\epsilon_{abcd}e$
\cite{DP-F,reisenbergerareas}, which implies that
$\epsilon_{IJKL}B^{IJ}_{ab}B^{KLab}=0$
i.e. $B_{ab}$ is a simple bivector. This is the form of the constraint that we are going to use and that leads to the 
Barrett-Crane spin foam model. In other words, \Ref{eq:constrB}
is satisfied if and only if there exists a real tetrad field 
$e^{I}=e^{I}_{a}dx^{a}$ so that one of the following equations holds:
\bes &I&\;\;\;\;\;\;\;\;\;\;B^{IJ}\,=\,\pm\,e^{I}\,\wedge\,e^{J} \\ 
&II&\;\;\;\;\;\;\;\;\;\;B^{IJ}\,=\,\pm\,\frac{1}{2}\,\epsilon^{IJ}\,_{KL}e^{K}\,\wedge\,e^{L}
.  
\ees
Restricting the field B to be always in the sector $II_{+}$ (which is
always possible classically), the action becomes:
\bes 
S\,=\,\int_{\mathcal{M}}\,\epsilon_{IJKL}\,e^{I}\,\wedge\,e^{J}\,\wedge\,F^{KL}
\ees
which is the action for General Relativity in the first order Palatini
formalism. 

Also, the other sector, differing by a global change of sign only,
is classically equivalent to this, while the other two, related by
Hodge duality to the \lq\lq geometric" ones, corresponds to \lq\lq
pathological geometries" with no physical interpretation (see the
cited literature for more details). 

The possibility of this restriction at the classical level was shown in \cite{reisenbergerareas}, where it is proven that 
initial data in the gravity sector do not evolve into any of the others provided that the tetrad field remains 
non-degenerate. Therefore we see that this action defines a diffeomorphism and Lorentz invariant theory, a subsector of which 
describes classical General Relativity in the 1st order formalism, as we had anticipated. This is basically all, as long as 
the classical level is concerned. 

The two theories, however, are different at the quantum level, since in the quantum theory one cannot
 avoid interference between different sectors. In fact in a partition function for the Plebanski action we have to integrate
 over all the possible values of the $B$ field, so considering all the 4 sectors. Another way to see it is the existence in
 the Plebanski action of a $\mathbb{Z}_{2}\times\mathbb{Z}_{2}$ symmetry $B\rightarrow -B$, $B\rightarrow *B$ responsible 
for this interference. This is discussed in \cite{DP-F,simplicity}.
This adds to the usual subtlety in dealing with a 1st order action, where degenerate configurations of the metric field cannot 
be easily excluded, instead of a 2nd order action (metric or ADM formalism) where one integrates over non-degenerate configurations 
only.

In particular, the $\mathbb{Z}_2$ symmetry between the two geometric
sectors of the theory may affect the path integral quantization of
the theory, and the very meaning of the path integral; remember in
fact that a $\mathbb{Z}_2$ symmetry on the lapse function $N\rightarrow
-N$ makes the difference between a path integral realization of
the projector onto solutions of the Hamiltonian constraint and a
path integral representing the Feynman propagator between states,
or causal transition amplitude, both in the relativistic particle
case and in quantum gravity in metric formalism, as we have shown
above. A simple argument suggests that this may happen also in our
spin foam context; in fact, a $3+1$ splitting of the Plebanski
action (see \cite{Peldan}), after the imposition of the
constraints on the $B$ field (we are then analysing the $3+1$
splitting of the Palatini action for gravity), shows that a change
of sign in the $B$ field is equivalent to a change of sign in the
lapse function, so that both sectors of solutions are taken into
account in a path integral realization of the projector operator.
The $B$ field has in fact the role of metric field in this $BF$
type formulation of gravity, and a canonical 3+1 splitting
basically splits its independent components into the triple
($h_{ab}$, $N^a$, $N$) (with all these expressed in terms of the
tetrad field) as in the usual metric formulation. The reason for
and effect of this will be explained below. Of course, more
worrying would be the presence, in the quantum theory, of the two
\lq\lq non geometric" sectors. We will see that luckily these pathological sectors do not appear in the quantum theory we
 are going to discuss in the following.

Before turning to the quantum theory based on the Plebanski action, we have to discuss in detail the description of 
simplicial geometry based on it, since it is the real starting point of the spin foam quantization.
In fact, a quantization of gravity along such lines should start by identifying
suitable variables corresponding
to the $B$ and $\omega$ variables of the Plebanski action, and then the correct
translation at the quantum
level of the above constraints on the $B$ field, leading to a realization
of the path integral
\bes
\mathcal{Z}\,=\,\sum_\mathcal{M}\int\mathcal{D}B\,\mathcal{D}\omega\,\mathcal{D}\phi\,\,e^{i\,S(B,A,\phi)}
\ees
(we have included a sum over spacetime manifolds),
possibly in a not only formal way.
However, both in light of the \lq\lq finitary" philosophy mentioned above
and hoping to make sense of the path integral by a lattice type of
regularization, we pass to a simplicial setting in which the
continuum manifold is replaced by a simplicial complex, and the
continuum fields by variables assigned to the various elements of
this complex. 

Just in the same way as in the 3-dimensional 
case, the connection field is naturally discretized along the links of the dual complex, by integrating the 1-form 
connection along them $\omega^{IJ}=\int_{e*}\omega^{IJ}(x)$, so that we obtain a holonomy $g_{e*}$ (with values in $SO(3,1)$ or $SO(4)$) associated to each link; 
in this way, again, the curvature is obtained by choosing a closed path of links, in particular for each dual face we have a
product of group elements $g_{f*}=\prod_{e*\subset
  \partial f*} g_{e*}$ of the group elements $g_{e*}$ associated with
the links of the boundary of the dual face $f*$, and it is thus
associated with the dual face itself. This is in turn dual to the
triangles of the triangulation $T$, so we have the simplicial curvature
associated to them, as it is common in 4-dimensional simplicial gravity. The logarithm of $g_{f*}$ gives a Lie algebra
 element $\Omega_t$, the proper discretization of the curvature field of the Plebanski action.

The crucial point is however the discretization of the $B$ field, since this is what marks the difference between gravity 
and BF theory. Being a 2-form, the $B$ field is naturally discretized along the triangles in the triangulation obtaining a Lie
algebra element associated to each triangle, thus to each face of the dual complex, via $B^{IJ}(t)=\int_t
B^{IJ}_{\mu\nu}(x) dx^\mu \wedge dx^\nu$.It is crucial to note that in this discretization, the sign of the bivector reflects the orientation of the triangle to which 
it is associated.

With this discretization, the constraints on the $B$ field become
constraints on the bivectors
$B^{IJ}\in so(3,1)\simeq \wedge^2(\mathbb{R}^{3,1})$ associated to the various
triangles.        

The constraint term in the action is discretized analogously, by integrating over pairs of triangles, to get: 
$\phi_{IJKL}B^{IJ}(t)B^{KL}(t')$.

Therefore, the discrete action we get is:
\bes
S(B,\omega)\,=\,\sum_t \tr B_t\,\Omega_t\,+\,\sum_{t,t'}\,\phi_{IJKL}\,B^{IJ}(t)\,B^{KL}(t') \label{eq:Pleb}.
\ees
 
Let us now analyse the constraints on the bivectors more closely. 
Consider first a tetrahedron $T$, whose boundary is made of four triangles $t$, we have: 
$0 = \int_T dB^{IJ}(x) = \int_{\partial T} B^{IJ}(x) = \sum_t \int_t B^{IJ}(x) = \sum_t B^{IJ}(t) = 0$.
In other words, the four bivectors associated to the same tetrahedron sum to zero, as a result of the tetrahedron being 
enclosed by the corresponding four triangles.

Consider then the constraints expressed as: $\epsilon_{IJKL}B^{IJ}_{ab}B^{KL}_{cd}=\epsilon_{abcd} e$, where 
$e = \frac{1}{4!}\epsilon_{IJKL}\epsilon^{abcd} B^{IJ}_{ab} B^{KL}_{cd}\neq 0$ is the spacetime volume element after the 
imposition of the constraints, so its being non zero implies that the bivector field is non degenerate.

Integrating this expression of the constraints over pairs of surfaces (triangles) we obtain \cite{DP-F}:
\bes
V(t,t')\,=\,\int_{x\in t, y\in t'} e\,\epsilon_{abcd}\,dx^a \wedge dx^b \wedge x^c \wedge dx^d\,=\,\epsilon_{IJKL}\,B^{IJ}(t)\,B^{KL}(t'),
\ees
where $V(t,t')$ is obviously the 4-volume spanned by the two triangles $t$ and $t'$.

This formula clearly implies two constraints on the bivectors associated to the triangles of the simplicial manifold, 
corresponding to the two cases in which the 4-volume spanned is zero: each bivector associated to a triangle $t$ satisfies 

$\epsilon_{IJKL}\,B^{IJ}(t)\,B^{KL}(t) = 0$, which corresponds geometrically to the requirement of the bivector being formed 
as a vector product of two (edge) vectors, i.e. of being a \lq\lq simple" bivector; if we decompose the bivector into its
selfdual and anti-selfdual part (this implies in the Lorentzian case a complexification of the bivector itself), this 
constraint also imposes the equality of these two parts; also, for two triangles sharing an edge, thus belonging to the same 
tetrahedron, the corresponding bivectors must be such that $\epsilon_{IJKL}\,B^{IJ}(t)\,B^{KL}(t') = 0$, which in turn 
implies, together with the previous constraint, that $\epsilon_{IJKL}\,B^{IJ}(t+t')\,B^{KL}(t+t') = 0$, i.e. that the 
bivector $B^{KL}(t+t')= B^{KL}(t) + B^{KL}(t')$ obtained as a sum of the two bivectors corresponding to the two triangles $t$ and $t'$ sharing an
 edge is also a simple bivector. The geometric meaning of this constraint is to impose that the 
triangles intersect pairwise in lines in $\mathbb{R}^{3,1}$, i.e. that they pairwise span 
3-dimensional subspaces of $\mathbb{R}^{3,1}$ \cite{BC,bb,BC2} (this is a rather strong condition in 4 dimensions, since the 
generic case is for two surfaces to intersect in a point only).
 
At the continuum level, as we said, the Plebanski constraints on the bivector field $B$ make it a geometric field, i.e. put 
it in correspondence with a tetrad field and thus with a spacetime metric, so allow for a description of spacetime geometry 
in terms of this bivector field; at the discrete level, when the variables we have are bivectors assigned to the triangles, 
the above constraints on these bivectors allow for a description of simplicial geometry in terms of them. 

However, the same ambiguity that we have seen at the continuum level for the solutions of the 
Plebanski constraints exists at this discretized level, since there are again four sectors of 
solutions to these constraints corresponding to the bivectors 1) $B^{IJ}$, 2) $- B^{IJ}$, 
3) $*B^{IJ}=\epsilon^{IJ} _{KL} B^{KL}$, 4) $- * B^{IJ}$. 
Again, the cases 1) and 2) correspond to well-defined simplicial geometries, differing only by a 
change in orientation, while the cases 3) and 4) are pathological cases with no geometric 
interpretation. Again, we do not consider degenerate configurations.
It is possible to distinguish between these
four cases by defining the quantitites:
\bes
U^{\pm}\,=\,\pm\,B^{\pm}(t)\,\cdot\,B^{\pm}(t')\,\times\,B^{\pm}(t'')
\ees 
for any triple of triangles $t$, $t'$ and $t''$ sharing an edge, i.e. belonging to the same 
tetrahedron, where $B^{\pm}(t)$ is the selfdual (respectively, anti-selfdual) part of the bivector 
associated to the triangle $t$. In terms of them, the four cases are characterized by \cite{bb}: 
1) $U^{+} > 0$, $U^{-} < 0$, 2) $U^{+} < 0$, $U^{-} > 0$, 3) $U^{+} > 0$, $U^{-} > 0$, 4) $U^{+} < 0$, $U^{-} < 0$,    
and in all these cases we have $\mid U^{+}\mid = \mid U^{-}\mid$. Therefore we can restrict the 
configurations considered to the geometric ones by imposing a further constraint of the form: 
$U^{+} + U^{-} = 0$. This constraint would not come directly from the discretization of the Plebanski 
action.

Resuming, the geometry of a 4-simplex, and thus the geometry of a full simplicial complex (where one glues 4-simplices along 
common tetrahedra imposing that the bivector data on the common tetrahedra match), is determined by a set of 
bivectors associated to the triangles in the complex and satisfying the following requirements:

\begin{itemize}
\item the bivectors change sign when the orientation of the triangles is
changed (orientation constraint);
\item the bivectors are \lq\lq simple", i.e. they satisfy $B(t)\cdot \ast
B(t)=0$ (simplicity constraint);
\item the bivectors associated to neighbouring triangles sum to simple
bivectors,
i.e. $B(t)\cdot \ast B(t')=0$ if $t$ and $t'$ share an edge (decomposition constraint);
\item the four bivectors associated to the faces of a tetrahedron sum to
zero (closure constraint).
\item $U^{+} + U^{-} = 0$ (chirality constraint). 
\end{itemize}
The reason why we call the third constraint \lq\lq decomposition constraint" will become apparent 
upon quantization, while the relation of the fifth with issues of chirality will be clarified in the
 following. These constraints, together with their quantum counterparts, were given in 
\cite{BC,BC2,bb,Baez}, while their relation with the Plebanski formulation of gravity was shown in \cite{DP-F, reisenbergerareas}.
We stress again that this description of simplicial geometry in terms of bivectors, with the 
associated constraints, holds in both the Lorentzian and Riemannian case, although in the Lorentzian 
case the definition and use of selfdual and anti-selfdual components for the bivectors is less 
straightforward since it implies a complexification procedure.

Interestingly, this description of gravity as a constrained BF theory, with quadratic constraints on the $B$ field, can be 
generalized to any dimension, at least in the Riemannian case \cite{FKP}. We will indeed see in the following how the quantum
version of this description can be generalized as well.

In this description, all the geometric quantities such as areas, 3-volumes and 4-volumes have to be 
expressed in terms of the bivectors. Let us see briefly how this is done. 
The easiest objects to consider are the areas of the triangles to which the bivectors are attached, 
as they are given simply by the square root of the modulus of the bivectors themselves:
\bes
A(t)\,=\,\sqrt{B(t)\,\cdot\,B(t)}\,=\,\sqrt{B^{IJ}(t)\,B_{IJ}(t')}.
\ees
The 3-volume $V$ of a tetrahedron can also be expressed in terms of the bivectors associated to its 
faces, by taking three of them (which one is left out does not change the result) and writing:
\bes
V^2\,=\,U^{+}\,+\,U^{-}. 
\ees
This is indeed the most immediate definition of the (square of the) 3-volume \cite{bb, alejandro}. However, this 
definition does not reflects the orientation (up or down in $\mathbb{R}^{3,1}$ with respect to the
direction normal to the hyperplane in which it is embedded) of the tetrahedron itself, basically as a consequence 
of the scalar product (which is a trace in the Lie algebra if we use the isomorphism 
$\wedge^2 \mathbb{R}^{3,1}\sim so(3,1)$), while the 3-volume is indeed a chiral object (it is made out of three bivectors) and 
should reflect the orientation of its faces, i.e. whether we have $B$ or $-B$ associated to them. On
the other hand, we have seen that the quantitities $U^{\pm}$ do reflect this distinction. As 
a consequence, with this definition, the 3-volume $V$ would be always zero \cite{alejandro}, so a better definition of 
the 3-volume operator $V$ would be:
\bes
V\,=\,\sqrt{\mid U^{\pm}\mid}\,=\,\frac{1}{2}\,\left( U^{+}\,-\,U^{-}\right).
\ees
The simplicity constraint, as we said, can be interepreted as imposing the equality \lq\lq in norm" 
of the geometries defined by selfdual and anti-selfdual parts of the bivectors, but the two 
descriptions are still different when it comes to orientation properties of the simplicial structures.
This problem is not present for the areas of the triangles, which are defined by two bivectors 
and give the same result when computed using selfdual or anti-selfdual structures. 

Let us discuss finally the 4-volume operator for a 4-simplex $\sigma$ \cite{Re97b,danhend}. 
This can be given in terms of triangle bivectors as:

\begin{equation}
\label{eq_volume3}
  V_\sigma\,=\,\frac{1}{30}\sum_{t,t^\prime}\frac{1}{4!}
    \epsilon_{IJMN}\,\sgn(t,t^\prime)\,B^{IJ}(t)\,B^{MN}(t^\prime),
\end{equation}
where the sum is over all pairs
of triangles $(t,t^\prime)$ in $\sigma$ that do not share a common
edge, with a sign factor $\sgn(t,t^\prime)$
depending on their combinatorial orientations. Let $(12345)$ denote
the oriented combinatorial four-simplex $\sigma$ and $(PQRST)$ be a
permutation $\pi$ of $(12345)$ so that $t=(PQR)$ and $t^\prime=(PST)$
(two triangles $t,t^\prime$ in $\sigma$ that do not share a common
edge have one and only one vertex in common). Then the sign factor is
defined by $\sgn(t,t^\prime)=\sgn\pi$~\cite{Re97b}. The sum over all
 pairs of triangles $(t,t^\prime)$
provides us with a particular symmetrization which can be thought of
as an averaging over the angles that would be involved in an exact calculation of the volume of a
four-simplex.
An alternative expression for
the four-volume from the context of a first order formulation of Regge
calculus~\cite{CaAd89} is given by,
\begin{equation}
  (V_\sigma)^{3} =
  \frac{1}{4!}\epsilon^{abcd}N_{a}\wedge N_{b}\wedge
  N_{c}\wedge N_{d}, \label{4sop}.
\end{equation}
where the indices $a,b,c,d$ run over four out of the five tetrahedra
of the four-simplex $\sigma$ (the result is independent of the
tetrahedron which is left out), and the $N_a$ are vectors normal to
the hyperplanes spanned by the tetrahedra, whose lengths are
proportional to the three-volumes of the tetrahedra. These vectors can of course be expressed in 
terms of the bivectors as well \cite{CaAd89}. This formula would then involve the dihedral 
angles of the 4-simplex, i.e. the angles between the normals to the tetrahedra sharing a triangle,
but we did not introduce them in the simplicial description of the manifold yet. However, they will
play a crucial role in the following since they appear naturally as additional variables corresponding 
to the Lorentz connection in the spin foam models we will discuss shortly. 

We conclude by noting that, since the basic variables of the theory are associated to triangles, and thus in some sense one 
is building the simplicial complex from this level upwards, a definition of the edge lengths in terms of bivectors or triangle 
areas is much less natural.

\section{Quantum 4-dimensional simplicial geometry} \label{sec:qg4d}
We turn now to the quantization of the simplicial geometry described in terms of bivectors as given in the previous section.
The strategy is just as in the 3-dimensional case, we obtain a definition of a quantum version of the basic geometric 
variables, i.e. of \lq\lq quantum bivectors" and thus of \lq\lq quantum triangles", and then use this to define quantum states 
associated to 3-dimensional hypersurfaces in spacetime, obtained by gluing together tetrahedra along common triangles, thus 
passing through the definition of \lq\lq quantum tetrahedra". Finally we try to define appropriate quantum amplitudes to 
be associated to the 4-simplices, and to be used as building blocks for the partition function and transition amplitudes
between quantum states in the models that we will derive later on.

The first step is to turn the bivectors associated to the triangles into operators. To this end, we make use of the 
isomorphism between the space of bivectors $\wedge^2 \mathbb{R}^{3,1}$ and the Lie algebra of the Lorentz group $so(3,1)$ (or,
 in the Riemannian case, between $\wedge^2\mathbb{R}^4$ and $so(4)$ identifying the bivectors associated with the triangles
with the generators of the algebra: $B^{IJ}(t)\rightarrow * J^{IJ}(t)=\epsilon^{IJ}\,_{KL}J^{KL}$. After this preliminary step we are ready to turn these 
variables into operators by associating to the different triangles $t$ an irreducible representation $\rho_t$ of the group and the corresponding 
representation space, so that the the generators of the algebra act on it (as derivative operators). In other words, we use a 
quantization map of the kind: $B^{IJ}(t) \rightarrow \rho_t$, so that for each triangle we have a Hilbert space and 
operators acting on it. The representation we choose to use are the unitary and (in the Lorentzian case) belonging to the 
principal series. While at this stage this choice is rather arbitrary, We will see in the following how this class of 
representations is the appropriate choice, basically because it is that entering in the harmonic analysis for functions
 on the group \cite{VK, Ruhl, gelfand}, which is the basic tool for both the lattice-gauge theory and the group field theory
 derivation of the Barrett-Crane model.
The groups we use are in the Lorentzian case $SL(2,\mathbb{C})$ (the double cover of the proper orthocronous Lorentz group 
$SO(3,1)$) and in the Riemannian $Spin(4)$ (the double cover of $SO(4)$). 
In the Lorentzian case, the irreducible unitary representations in the principal series are characterized by a pair 
$(n,\rho)$ of a natural number $n$ and a real number $\rho$, while in the Riemannian case the unitary representations may be 
characterized by two half-integers $(j_1,j_2)$, reflecting the splitting $Spin(4)\sim SU(2)\times SU(2)$.

Therefore we can define the Hilbert space of a quantum bivector to be $\mathcal{H}=\oplus_{n,\rho}\mathcal{H}^{(n,\rho)}$ in 
the Lorentzian case, and $\mathcal{H} = \oplus_{j_1,j_2}\mathcal{H}^{(j_1,j_2)}$ in the Riemannian. 
However, this is not yet the Hilbert space of a quantum geometric triangle, simply because a set of bivectors does not 
describe a simplicial geometry unless it satisfies the constraints we have given in the previous section. The task is then to 
translate the above geometric constraints in the language of group representation theory into the quantum domain.

Of the constraints given for the bivectors, only the first two refer to a triangle alone, and these are then enough to 
characterize a quantum triangle. 
The change in orientation of the triangle, and the consequent change in sign of the
associated bivector are translated naturally into a map from a given representation $\rho_t$ and representation space
 $\mathcal{H}^{\rho_t}$ to the dual representation $\rho_t^{*}$ and complex conjugate Hilbert space $\mathcal{H}^{\rho_t^*}$. 
The most important constraint is however the simplicity constraint, that forces the bivectors to be simple, i..e formed as 
wedge product of two edge vectors. Recall that classically this was expressed by: $B(t)\cdot * B(t) = 0$. Now we use the 
quantization map given above and substitute the bivectors with the (Hodge dual of the) Lie algebra elements in the given 
representation $\rho_t$ for the triangle $t$, obtaining the condition:
\bes
J(\rho_t)\,\cdot\,* J(\rho_t)\,=\,\epsilon_{IJKL}\,J^{IJ}(\rho_t)\,J^{KL}(\rho_t)\,=\,C_2(\rho_t)\,=\,0,
\ees
i.e. it translates into the condition of having vanishing second Casimir of the group in the given representation.

For $Spin(4)$ the second Casimir in the representation $(j_1,j_2)$ is given by: $C_2 = j_1 (j_1 + 1) - j_2 (j_2 + 1)$, so 
that the simplicity constraint forces the selfdual and anti-selfdual parts to be equal, and we are restricted to considering 
only representations of the type $(j,j)$.

For $SL(2,\mathbb{C})$ in the representation $(n,\rho)$ we have for the second Casimir: $C_2 = 2 n \rho$, so that the only 
representations satisfying the constraint are those of the form: $(n,0)$ and $(0,\rho)$. 

In both cases, the type of representations singled out by the simplicity constraint are the so-called \lq\lq class 1" 
representations, realized on the space of functions on suitable homogeneous spaces, obtained as cosets of the group manifolds
with respect to given subgroups, as we are going to see.

Therefore the Hilbert space of a {\bf quantum triangle} is given by:
\bes
\mathcal{H}_t\,=\,\oplus_n\mathcal{H}^{(n,0)}\cup\oplus_\rho\mathcal{H}^{(0,\rho)}\;\;\;\;\;\;\;\;\;\mathcal{H}_t = \oplus_{j}\mathcal{H}^{(j,j)}, 
\ees 
in the Lorentzian and Riemannian cases respectively.

Before considering how this Hilbert space is used to construct that of tetrahedra, let us discuss briefly the geometric 
interpretation of the parameters labelling the representations. 
Recall that the area of a triangle is given in terms of the associated bivector by: $A^2 = B(t) \cdot B(t)$, using the 
quantization map above this translates into: $A^2 = J^{IJ}(\rho_t) J_{IJ}(\rho_t) = C_1(\rho_t)$, i.e. the area operator is
just the first Casimir of the group and it is diagonal on each representation space associated to the triangle. Its 
eigenvalues are, in the Riemannian case, $A^2 = C_1((j_1,j_2)) = j_1 (j_1 + 1) + j_2 (j_2 + 1)$, and in the Lorentzian, 
$A^2 = C_1((n,\rho)) = n^2 - \rho^2 - 1$. Therefore, for simple representations we have $A^2 = C_1 ((j,j)) = 2 j (j + 1)$ 
in the Riemannian, and $A^2 = C_1 ((0,\rho)) = - \rho^2 - 1$ or $A^2 = C_1 ((n,0)) = n^2 -1$ in the Lorentzian case. 
In each case we see that the representation label characterizes the quantum area of the triangle to which that representation 
is associated, and also that representations of the form $(0,\rho)$ correspond to spacelike triangles (we have chosen the 
signature $(+ - - -)$), while $(n,0)$ labels timelike triangles. This timelike/spacelike distinction is also in perfect 
agreement with the geometric properties of the corresponding bivectors, in the sense that the way spacelike and timelike 
bivectors are summed in $\mathbb{R}^{3,1}$ is mirrored exactly by the decomposition properties of tensor products of 
representations $(n,0)$ and $(0,\rho)$.

States of the quantum theory are assigned to 3-dimensional hypersurfaces embedded in the 4-dimensional spacetime, and are 
thus formed by tetrahedra glued along common triangles. We then have to define a state associated to each tetrahedron in the 
triangulation, and then define a tensor product of these states to obtain any given state of the theory. In turn, the Hilbert 
space of quantum states for the tetrahedron has to be obtained from the Hilbert space of its triangles, since these are the 
basic building blocks at our disposal corresponding to the basic variables of the theory. 

Each tetrahedron is formed by gluing 4 triangles along common edges, and this gluing is naturally represented by the tensor
product of the corresponding representation spaces; the tensor product of two representations of $SL(2,\mathbb{C})$ or 
$Spin(4)$ decomposes into irreducible representations which do not necessarily satisfy the simplicity constraint. The third
of the constraints on bivectors translates at the quantum level (using the same procedure as for the simplicity constraint)
 exactly as the requirement that only simple (class 1) representations appear in this decomposition. 
Therefore we are considering for each tetrahedron with given representations assigned to its triangles a tensor in the tensor product:
$\mathcal{H}^1\otimes\mathcal{H}^2\otimes\mathcal{H}^3\otimes\mathcal{H}^4$ of the four representation spaces for its faces, 
with the condition that the tensored spaces decompose pairwise into vector spaces for simple representations only.
 We have still to impose the closure constraint; this is an expression of the invariance under the gauge group of the tensor 
we assign to the tetrahedron, thus in order to fulfill the closure cosntraint we have to associate to the tetrahedron an 
invariant tensor, i.e. an intertwiner between the four simple representations associated to its faces 
$B^{\rho_1 \rho_2 \rho_3 \rho_4}: \mathcal{H}^1\otimes\mathcal{H}^2\otimes\mathcal{H}^3\otimes\mathcal{H}^4\rightarrow \mathbb{C}$.

It turns out that the last constraint, the chirality constraint, that forbids the presence of pathological configurations 
which are however solutions of the other constraints, is automatically implemented at the quantum level, i.e. that the 
operator $\hat{U}^+ + \hat{U}^-$ is identically zero on the space of solutions of the other constraints \cite{bb}.

Therefore the Hilbert space of a {\bf quantum tetrahedron} is given by:
\bes
\mathcal{H}_{tet}\,=\,Inv\left( \mathcal{H}_1\otimes\mathcal{H}_2\otimes\mathcal{H}_3\otimes\mathcal{H}_4\right),
\ees
with $\mathcal{H}_i$ being the Hilbert space for the $i$-th triangle defined above. Each state in this Hilbert space is 
then  a group intertwiner called \lq\lq the Barrett-Crane intertwiner" \cite{BC,BC2}. This construction was rigorously 
perfomed in \cite{bb} in the Riemannian case, using geometric quantization methods, while the uniqueness of the 
Barrett-Crane intertwiner as solution of the geometric constraints was shown in \cite{MikeIntert}.

The Barrett-Crane intertwiner can also be given an integral representation, which is very useful for both the asymptotic 
analysis and the geometric interpretation of the spin foam models we are going to discuss in the following. Moreover, it is 
this expression that appears at first in all the derivations of the same models.

In the Riemannian case the simplicity of the representation is equivalent to the requirement that the representation admits 
an invariant vector under an $SU(2)$ subgroup of $Spin(4)$ \cite{VK}, and the formula is the following:
\bes
B^{J_1 J_2 J_3 J_4}_{k_1 k_2 k_3 k_4}\,=\,w^{J_1}_{l_1}w^{J_2}_{l_2}w^{J_3}_{l_3}w^{J_4}_{l_4}\int_{Spin(4)}dg\,D^{J_1}_{l_1 k_1}(g)\,D^{J_2}_{l_2 k_2}(g)\,D^{J_3}_{l_3 k_3}(g)\,D^{J_4}_{l_4 k_4}(g)\,= \nonumber \\
=\, \int_{Spin(4)/SU(2) \sim S^3}dx\,D^{J_1}_{0 k_1}(x)\,D^{J_2}_{0 k_2}(x)\,D^{J_3}_{0 k_3}(x)\,D^{J_4}_{0 k_4}(x) \label{eq:BCintRie}, 
\ees
where the $D^{J_i}_{lk}(g)$ are matrix elements of the representation function for the group element $g$ in the 
representation
$J_i$, one for each triangle, and the $w^{J_i}$ are the $SU(2)$ invariant vectors in the same representation, with which we 
construct the 
representation functions to ensure that the resulting object is invariant under this $SU(2)$ subgroup; as a consequence, 
the integral runs only over the coset space $Spin(4)/SU(2)$ and this is in turn isomorphic to the 3-sphere $S^3$, so that 
the representation functions themselves can then be considered as functions on the 3-sphere; using the additional 
isomorphism $S^3\sim SU(2)$, one can write the Barrett-Crane intertwiner using $SU(2)$ representation functions, with the 
$j$ labelling the simple representations of $Spin(4)$ $(j,j)$ now interpreted as a label of an $SU(2)$ representation 
\cite{barrettintegral}.

In the Lorentzian case, there are several realizations of the constraints \cite{BC2,P-R2, P-R3}, and we discuss here in 
more detail only
that involving only representations of $SL(2,\mathbb{C})$ labelled by continuous parameters, i.e. that in which all the 
triangles and consequently all the tetrahedra are spacelike.
In this case, the simplicity constraint is realized as invariance under an $SU(2)$ subgroup of $SL(2,\mathbb{C})$, so there 
exists a timelike invariant vector in the representation \cite{VK,Ruhl}, and the formula is analogous:
\bes
B^{\rho_1 \rho_2 \rho_3 \rho_4}_{j_1k_1 j_2k_2 j_3k_3 j_4k_4} &=& w^{\rho_1}_{j'_1l_1}w^{\rho_2}_{j'_2l_2}w^{\rho_3}_{j'_3l_3}w^{\rho_4}_{j'_4l_4}\int_{SL(2,\mathbb{C})}dg\,D^{\rho_1}_{j'_1l_1 j_1k_1}(g) 
D^{\rho_2}_{j'_2l_2 j_2k_2}(g) D^{\rho_3}_{j'_3l_3 j_3k_3}(g) D^{\rho_4}_{j'_4l_4 j_4k_4}(g) = \nonumber \\
&=& \int_{SL(2,\mathbb{C})/SU(2) \sim H^3}dx\,D^{\rho_1}_{00 j_1k_1}(x)\,D^{\rho_2}_{00 j_2k_2}(x)\,D^{\rho_3}
_{00 j_3k_3}(x)\,D^{\rho_4}_{00 j_4k_4}(x),\;\;\; 
\label{eq:BCintLor} \ees
where again the $D^\rho(g)$ are matrix elements of the representation $\rho$ of the group element $g$, and again the $w$'s 
are invariant vectors under the $SU(2)$ subgroup; here, the invariance under $SU(2)$ turns the integral over $SL(2,\mathbb{C})$
into an integral over the homogeneous space $H^3$, i.e. the upper hyperboloid in Minkowski space. Of course, in the Lorentzian
 case, where the integration domain is non compact, we are not at all guaranteed that the integral makes sense at all. We 
leave aside this problem for now, but will comment again on this issue when dealing with spin networks and with the full spin 
foam model based on it.

In any case, this invariant tensor represents as we said the state of a tetrahedron whose faces are labelled by the given
representations; it can be represented graphically as in figure 4.1. 
\begin{figure}
\begin{center}
\includegraphics[width=10cm]{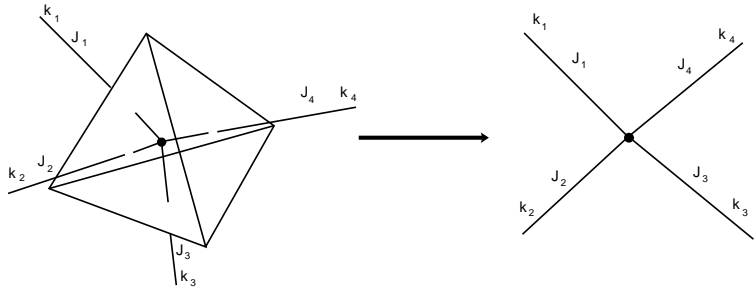}
\caption{A vertex corresponding to a quantum tetrahedron in 4d, with links labelled by the representations (and state labels)
associated to its four boundary triangles.}
\end{center} 
\end{figure}
.

The result is that a quantum tetrahedron
is characterized uniquely by 4 parameters, i.e. the 4 irreducible
simple representations of $so(4)$ assigned to the 4 triangles in it,
which in turn are interpretable as the (oriented) areas of the
triangles. 
Even if rigorous, this result is geometrically rather puzzling, since the geometry of a tetrahedron
 is classically determined by its 6 edge lengths, so imposing only the values of the 4 triangle areas
 should leave 2 degrees of freedom, i.e. a 2-dimensional moduli space of tetrahedra with given 
triangle areas. For example, this is what would happen in 3 dimensions, where we have to specify 6 parameters also at the 
quantum level. So why does a tetrahedron have fewer degrees of freedom in 4 dimensions than in 3 dimensions, at the quantum level, so that its quantum geometry is
 characterized by only 4 parameters? The answer was given in \cite{bb}, to which we refer for 
more details. The essential difference between the 3-dimensional and 
4-dimensional cases is represented by the simplicity constraints that have to be imposed on the bivectors
 in 4 dimensions. At the quantum level these additional constraints reduce 
significantly the number of degrees of freedom for the tetrahedron, as can be shown using geometric quantization 
\cite{bb}, leaving us at the end with a 1-dimensional state space for each assignment of simple irreps to the faces 
of the tetrahedron, i.e. with a unique quantum state up to normalization, as we have just seen above. 
Then the question is: what is the classical geometry corresponding to this state?   
In addition to the four triangle areas operators, there are two other operators that can be 
characterized just in terms of the representations assigned to the triangles: one can consider the parallelograms with
 vertices at the midpoints of the edges of a tetrahedron and their areas, and these are given by the
 representations entering in the decomposition of the tensor product of representations labelling the neighbouring triangles.
Analyzing the commutation relations of the quantum operators
corresponding to the triangle and parallelogram area operators, it
turns out (see \cite{Baez, bb}) that while the 4 triangle area
operators commute with each other and with the parallelogram areas
operators (among which only two are independent), the last ones  have
non-vanishing commutators among themselves. This implies that we are
free to specify 4 labels for the 4 faces of the tetrahedron, giving 4
triangle areas, and then only one additional parameter, corresponding
to one of the parallelogram areas, so
that only 5 parameters determine the state of the tetrahedron itself,
the other one being completely randomized, because of the uncertainty
principle. 
Consequently, we can say that a quantum tetrahedron does not have a unique metric geometry, since
 there are geometrical quantities whose value cannot be determined even if the system is in a
 well-defined quantum state. In the context of the Barrett-Crane spin foam model, this means that a
 complete characterization of two glued 4-simplices at the quantum level does not imply that we
 can have all the informations about the geometry of the tetrahedron they share. This is a very
 interesting example of the kind of quantum uncertainty relations that we can expect to find in a
 quantum gravity theory, i.e. in a theory of quantum geometry.

A generic quantum gravity state is to be associated to a 3-dimensional hypersurface in spacetime, and this 
will be triangulated by several tetrahedra glued along common triangles; therefore a generic state will live in the tensor
product of the Hilbert spaces of the tetrahedra of the hypersurface, and in terms of the Barrett-Crane intertwiners it will 
be given by a product of one intertwiner $B^{J_1 J_2 J_3 J_4}_{k_1 k_2 k_3 k_4}$ for each tetrahedron with a sum over the parameters (\lq\lq angular momentum projections") 
labelling the particular triangle state for the common triangles (the triangles along which the tetrahedra are glued).
The resulting object will be a function of the simple representations labelling the triangles, intetwined 
by the Barrett-Crane intertwiner to ensure gauge invariance, and it will be given by a graph which has representations of 
$Spin(4)$ or $SL(2,\mathbb{C})$ labelling its links and the Barrett-crane intertwiner at its nodes; in other words, it will 
be given by a (simple) spin network (see figure 4.2).

\begin{figure}
\begin{center}
\includegraphics[width=7cm]{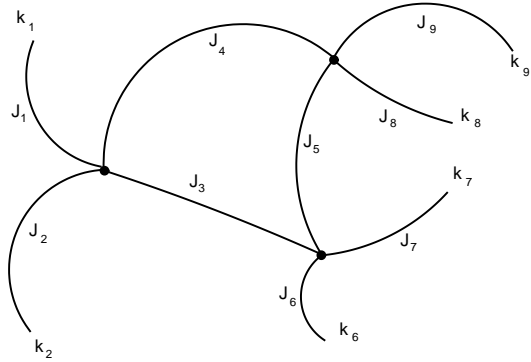}.
\caption{An example of a (simple) spin network in 4d, with three vertices (to which the Barrett-Crane intertwiners are 
associated), and both open and closed links.}
\end{center}
\end{figure}
 We stress that this spin network should not be thought of as embedded anywhere in the 
spacetime manifold, but it is just given by the combinatorial data necessary to determine its graph (and corresponding to the
combinatorial structure of a triangulated hypersurface), and by the algebraic data that are its labels on links and nodes.
 We will say a bit more on these quantum states later on.

We are left with a last ingredient of a quantum geometry still to be determined: the quantum amplitude for a 4-simplex $\sigma$, 
interpreted as an elementary change in the quantum geometry and thus encoding the dynamics of the theory, and representing
the fundamental building block for the partition function and the transition amplitudes of the theory. As in the 
3-dimensional case, this amplitude is to be constructed out of the tensors associated to the tetrahedra in the 4-simplex, so 
that it immediately fulfills the conditions necessary to describe the geometry of the simplicial manifold, at both the classical
and quantum level, and has to be of course invariant under the gauge group of spacetime $SL(2,\mathbb{C})$ or $Spin(4)$.
The natural choice is to obtain a $\mathbb{C}$-number for each 4-simplex, a function of the 10 representations labelling its
triangles, by fully contracting the tensors associated to its 
five tetrahedra pairwise summing over the parameters associated to the common triangles and respecting the symmetries of the
4-simplex, so:
\bes
A_\sigma\,:\,\otimes_i\,Inv\left( \mathcal{H}^{1_i}\otimes\mathcal{H}^{2_i}\otimes\mathcal{H}^{3_i}\otimes\mathcal{H}^{4_i}\right)\,\rightarrow\,\mathbb{C}.
\ees

The amplitude for a quantum 4-simplex is thus, in the Riemannian case:
\bes
A_\sigma\,=\,\mathcal{B}_{BC}\,=\,B^{J_1 J_2 J_3 J_4}_{k_1 k_2 k_3 k_4}\,B^{J_4 J_5 J_6 J_7}_{k_4 k_5 k_6 k_7}\,B^{J_7 J_3 J_8 J_9}_{k_7 k_3 k_8 k_9}\,B^{J_9 J_6 J_2 J_{10}}_{k_9 k_6 k_2 k_{10}}\,B^{J_{10} J_8 J_5 J_1}_{k_{10} k_8 k_5 k_1},    
\ees
with the analogous formula holding in the Lorentzian case. We thus obtain as amplitude what we may call a $10j$-symbol. Clearly 
it may be seen as the evaluation of a simple spin network with no open links and with the five vertices contracted following 
the contractions of the Barrett-Crane intertwiners. Of 
course, in the Lorentzian case the formula for the amplitude will involve several integrals (one for each intertwiner) over 
a non-compact domain and some regularization is needed, and we will discuss this issue after having obtained the full 
expression for the partition function of the Barrett-Crane model.

This amplitude is for fixed representations associated to the triangles, i.e. for fixed triangle areas; the full amplitude 
involves a sum over these representations with the above amplitude as a weight for each configuration.
In general, the partition function of the quantum theory describing the quantum geometry of a simplicial complex made of a 
certain number of 4-simplices, will be given by a product of these 4-simplex amplitudes, one for each 4-simplex in the 
triangulation, and possibly additional weights for the other elements of it, triangles and tetrahedra, with a sum over all
the representations assigned to triangles; moreover, also the restriction to a fixed triangulation $\Delta$ of spacetime has to be
 lifted, since in this non-topological case it represents a restriction of the dynamical degrees of freedom of the quantum 
spacetime, and a suitably defined sum over triangulations has to be implemented.
We can then envisage a model of the form:
\bes
Z(\mathcal{M})\,=\,\sum_\Delta\,Z(\mathcal{M}, \Delta)_{BC}\,=\,\sum_\Delta\,\sum_{J}\prod_{triangles}A_{tr}\,\prod_{tetrahedra}A_{tet}\,\prod_{4-simpl}\mathcal{B}_{BC},
\ees 
or, if one sees all the data as assigned to the 2-complex $\Delta*$ dual to the triangulation $\Delta$ (with faces $f$ dual to triangles, edges $e$ dual to tetrahedra, and vertices $v$ dual to 4-simplices):
\bes 
Z(\mathcal{M})\,=\,\sum_{\Delta*}\,Z(\mathcal{M}, \Delta*)_{BC}\,=\,\sum_{\Delta*}\,\sum_{J}\prod_{f}A_{f}\,\prod_{e}A_{e}\,\prod_{v}\mathcal{B}^v_{BC},
\ees
clearly with the general structure of a spin foam model.

The above analysis does not allow for a complete specification of the various amplitudes (although one
 may certainly guess what amplitudes are most naturally assigned to triangles and tetrahedra), and we 
will show now how they can be fully determined by a lattice gauge theory derivation of the type 
performed in the 3-dimensional case.

We close this section by noting that there exists \cite{nouiroche} also a quantum deformed version of the 
Lorentzian Barrett-Crane model based on the q-deformed Lorentz algebra $U(sl(2,\mathbb{C})_q$ \cite{podlesworon} and realizing the q-deformed 
Barrett-Crane intertwiner as an integral on the quantum hyperboloid $H^3_q$ \cite{buffroche}, in turn described as a pile of fuzzy 2-spheres.
This allows the construction of q-deformed simple spin networks, out of which the amplitudes for the spin foam model are defined, 
analogously to the undeformed case. Also, it is important to note that the quantum deformation implies that the simple 
representations of the algebra, including those labelled by a continuous parameter $\rho$, take values only in a finite 
interval in the real line; again, as in the three dimensional case we have discussed, the quantum deformations provides a 
natural cut-off on the allowed representations, and thus an elegant regularization of the spin foam model. We refer to the original 
paper for more details.
This model is not only extremely interesting from the purely technical point of view, for the non-triviality of the 
construction, but also potentially fundamental from the physical point of view, since it is believed to be a quantization 
of the Plebanski action in the presence of a cosmological constant; the situation would then be similar to the 3-dimensional 
case, with the Barrett-Crane model being the analogue of the Ponzano-Regge model and its quantum deformation being the 
analogue of the Turaev-Viro one. 

\section{Discretization and quantum constraints in the generalized BF-type action for gravity}
Before going on to give a derivation of the Barrett-Crane spin foam model, we want to clarify a further point. 
We have based the identification of the classical and quantum constraints on the bivectors reducing 
$BF$ theory to gravity, and consequently of the Hilbert spaces for triangles and tetrahedra, and 
finally of the  amplitudes for the 4-simplices, on the Plebanski formulation of a gravity theory. 
However, this is not the most general formulation of classical gravity as a constrained $BF$ theory, 
and  we want now to show what, if anything, changes when considering a more general action of the same
 kind. The results presented in this section were published in \cite{simplicity}.  
Recently, a new action for gravity as a constrained BF theory was proposed
\cite{CMPR} and it seemed that a discretization and
spin foam quantization of
it would lead to a model necessarily different from the Barrett-Crane
one. Also, a natural outcome would be a one-parameter ambiguity in the
corresponding spin foam model, related to the Immirzi parameter of
loop quantum gravity \cite{immirzi}\cite{tom&carlo}, and this
gives an additional reason to study the spin foam quantization of
the generalized action since it could help understand the link
between the current spin foam models and loop quantum gravity.
We show here that a careful discretization of the new
form of the constraints, an analysis of the field content of the theory,
at the classical level, and a spin foam quantization taking all this into
account, lead naturally to the Barrett-Crane model as a quantum theory
corresponding to that action. No one-parameter ambiguity arises in the
spin foam model and in the quantum geometry described by it. This suggests that the Barrett-Crane model is more
universal than
at first thought and that its continum limit may be described by several
different Lagrangians.
As we have seen, going from the field theory to the Barrett-Crane model
is achieved in 3 main steps. First, it is the {\it discretization} of
the two-form into bivectors associated to each face to the triangulation.
Then, we translate bivectors as elements of $so(4)^*$ or $so(3,1)^*$, using the function
\begin{equation}
\begin{array}{ccccc}
\theta &:& \Lambda^2 \R^4  & \rightarrow & so(4)^* \;\;\; or\;\;\;so(3,1)^* \\
 & & e \w f & \rightarrow & \theta(e\w f) \, (l) \rightarrow \eta(l e,f)
\end{array}
\end{equation}
where $\eta$ is the Riemannian or Lorentzian metric. Less formally, it is
the step we call {\it correspondence} between bivectors $B$ and elements
of the Lie algebra $J$. The last step is the {\it quantization} itself,
using
techniques from geometric quantization. This gives representation
labels to the faces of the triangulation and gives the Barrett-Crane model
(up to some normalisation factors).
In fact, there is an ambiguity at the level of the correspondence. We can
also choose to use the isomorphism $\theta \tr{o} *$ where $*$ is the
Hodge operator.
This leads to the so-called
flipped Poisson bracket, and it is indeed the right thing to do.
In the Riemannian case, this leads us to only ``real'' tetrahedra, whose
faces
are given by the bivectors and not the Hodge dual of the bivectors. This
amounts to selecting the sector $II$ of the theory which is the sector
we want \cite{bb}. In the Lorentzian case, such a check on the tetrahedra
has not been done yet, however the use of the flipped correspondence has a
nice consequence: it changes the sign of the area to being given by $-C_1$
instead of $C_1$ so that the discrete series of representations $(n,0)$
truly correspond to time-like faces and the continuous series $(0,\rho)$
to space-like faces, as implied by the algebraic
properties of these representations \cite{BC2}.
Using this ambiguity, we can generalize this correspondence. We have a family
of such isomorphisms given by $\theta \tr{o} (\alpha* +\beta)$. It is this
generalized correspondence we are going to use to deal with the
generalized BF-type action.
And, at the end, we will find again the same Barrett-Crane model

\subsection{Generalized BF-type action for gravity} \label{CMPR}
In \cite{CMPR} the following BF-type action was proposed for General
Relativity:
\bes
S=\int B^{IJ}\w F_{IJ} -\f{1}{2}\phi_{IJKL}B^{IJ}\w B^{KL}+\mu H
\label{action}
\ees
where $H=a_1\phi_{IJ}\,^{IJ}+a_2\phi_{IJKL}\epsilon^{IJKL}$, where $a_{1}$
and $a_{2}$ are arbitrary constants.
$B$ is a 2-form and $F$ is the curvature associated to the connection
$\omega$.
$\phi$ (spacetime scalar) and $\mu$ (spacetime 4-form) are Lagrange
multipliers, with $\phi$ having
the symmetries $\phi_{IJKL}=-\phi_{JIKL}=-\phi_{IJLK}=\phi_{KLIJ}$.
$\phi$ enforces the constraints on the $B$ field, while $\mu$ enforces
the condition $H(\phi)=0$ on $\phi$.
The $*$ operator acts on internal indices so that
$*B_{IJ}=1/2\,\epsilon_{IJKL}B^{KL}$ and $*^2=\epsilon$, with $\epsilon=1$
in the Riemannian case and $\epsilon=-1$ in the Lorentzian.
Before going on, we would like to point out that this is indeed the most
general action that can be built out of BF theory with a quadratic
constraint on the B field. This in turn means that the scalar constraint
$H=0$ is the most general one that can be constructed with a $\phi$ with
the given symmetry properties. This can be proven very easily. First of
all, we note that the two scalars $\phi_{IJ}\,^{IJ}$ and
$\phi_{IJKL}\epsilon^{IJKL}$ are linearly independent as it is immediate
to verify, so that what we have to prove is just that the space of scalars
made out of $\phi$ is 2-dimensional. There is an easy way to see this in
the Riemannian case: $\phi$ is a tensor in the 4-dimensional representation
of $so(4)$, which, using the splitting $so(4)\simeq su(2)\oplus su(2)$,
can be thought of as given by a sum of two 2-dimensional representations of
$su(2)$, that in turn can be decomposed into a sum of a 3-dimensional (and
symmetric) one and a 1-dimensional (and antisymmetric) one. Now we have
just to compute the tensor product of 4 such representations, paying
attention to keeping only the terms with the desired symmetry properties,
to see that we can have two and only two resulting singlets.   
The equations of motion for $\omega$ and $B$ are the same as those coming
from the Plebanski action, but the constraints on the field B now are:
\bes
B^{IJ}\w B^{KL} =\f{1}{6} (B^{MN}\w B_{MN}) \eta^{[I |K|} \eta^{J]L}
+\f{\epsilon}{12}(B^{MN}\w *B_{MN}) \epsilon^{IJKL}
\label{whole}
\ees

\bes
2a_2 B^{IJ}\w B_{IJ} -\epsilon a_1 B^{IJ}\w *B_{IJ}=0
\label{simple}
\ees

The solution of these constraints \cite{prieto}, for non-degenerate $B$
($B^{IJ}\w *B_{IJ}\neq 0$), is:
\bes
B^{IJ}=\alpha *(e^I \w e^J) + \beta\, e^I \w e^J
\label{B}
\ees
with:
\bes
\f{a_2}{a_1}=\f{\alpha^2+\epsilon\beta^2}{4\alpha\beta}
\label{a1a2}
\ees
Inserting this solution into (\ref{action}), we get:
\bes
S=\alpha \int  *(e^I \w e^J)\w F_{IJ}  +
\beta \int e^I \w e^J \w F_{IJ} \label{Holst}
\ees
so that there is a coupling between the geometric sector given by $*(e\w
e)$
(General Relativity)
and the non-geometric one given by $e\w e$.
Nevertheless, we note that the second
term vanishes on shell so that the equations of motion ignore the
non-geometric part and are still given by the Einstein equations.
In the usually studied case $a_1=0$, \cite{DP-F}\cite{FKP},
we are back to the Plebanski action, the sectors of
solutions being given by $\alpha=0$ and $\beta=0$ so that
we have either the General Relativity sector or the non-geometric sector
$e\w e$.
In the particular case $a_2=0$, in the Riemannian case, the only
solution to \Ref{a1a2} is $\alpha=\beta=0$
so that only degenerate tetrads are going to contribute.
On the other hand, in the Lorentzian case we have instead
$\alpha=\pm\beta$.

Looking at \ref{a1a2}, once we have chosen a pair $(\alpha,\beta)$,
we see that we have four posible sectors as with the Plebanski action
\cite{DP-F}. In the Riemannian case, we can exchange $\alpha$ and $\beta$.
Under this transformation, the B field gets changed into
its Hodge dual, so we can trace back this symmetry to the fact that we can
use both $B$ and $*B$ as field variables in our original action, without
any change in the physical content of the theory. We
can also change  $B\rightarrow -B$ without affecting the physics of our
model.
This gives us the following four sectors:
\bes
(\alpha,\beta) \quad (-\alpha,-\beta) \quad (\beta,\alpha) \quad
(-\beta,-\alpha)
\ees
In the Lorentzian case, the same $*$-symmetry brings us the following four
sectors:
\bes
(\alpha,\beta) \quad (\beta,-\alpha) \quad
(-\alpha,-\beta) \quad (-\beta,\alpha) 
\ees
The canonical analysis of the action \Ref{Holst} was performed in
\cite{holst}, leading to the presence of the Immirzi parameter of
loop quantum gravity given by $\gamma=\alpha/\beta$ and related to $a_1$
and $a_2$ by:

\bes
\f{a_2}{a_1}=\f{1}{4}\left(\gamma +\f{\epsilon}{\gamma}\right)
\ees
We can notice that we have two sectors
in our theory with different Immirzi parameters: $\gamma$
and $\epsilon/\gamma$, corresponding to a symmetry exchanging $\alpha$ and
$\epsilon\beta$.
The full symmetry group of the theory is then $Diff(M)\times
SO(4)\times Z_{2}\times Z_{2}$, with $SO(4)$ replaced by
$SO(3,1)$ in the Lorentzian case.  The $Z_{2}\times Z_{2}$
comes from the existence of the four sectors of solutions and
is responsible for their interferences at the quantum level.

We want to study the relationship between this new action and the usual
Plebanski action, and its spin foam quantization, in order to understand
if and in which cases the corresponding spin foam model at the quantum
level is still given by the Barrett-Crane one.
To this aim it is important to note that the constraints \Ref{whole} and
\ref{simple}
can be recast in an equivalent
form
leading to the same set of solutions,
for $a_2\neq 0$, $B$ non-degenerate, and
$\left(\f{a_1}{2a_2}\right)^2\neq\epsilon$
(this excludes the purely
selfdual and the purely anti-selfdual cases).
The situation is analogous to
that analyzed in \cite{DP-F} for the Plebanski action. The
new constraint is
\bes
\left(\epsilon_{IJMN}-\f{a_1}{a_2}\eta_{[I\mid M\mid}\eta_{J]N}
\right)
B^{MN}_{cd}B^{IJ}_{ab}=
e\,\epsilon_{abcd}
\left(1-
\epsilon\left(\f{a_1}{2a_2}\right)^2
\right)
\label{simpl2}
\ees
where
\bes
e=\f{1}{4!}\epsilon_{IJKL}B^{IJ}\w B^{KL}
\ees

This constraint can then be discretised to give the simplicity constraint
and the intersection constraint leading to the Barrett-Crane model, as we
will see in section \Ref{class}. But we can already
notice that in the case $(ab)=(cd)$, \Ref{simple2} gives an
equivalent to \Ref{simple}:

\bes
2a_2\f{1}{2}\epsilon_{IJKL}B^{IJ}_{ab}B^{KLab}
-\,a_1B^{IJ}_{ab}B^{ab}_{IJ}=0
\label{simple2}
\ees
In the following, we mainly use this second form of the constraints,
discussing the discretization and possible
spin foam quantization of the first one
in section \Ref{reis}.
Looking at the constraints on $B$ it is apparent that they are not anymore
just simplicity constraints as in the Plebanski case, so that a
direct
discretization of them for $B$ would not give the Barrett-Crane
constraints. We can see this by translating the constraints into
a condition on the Casimirs of $so(4)$ (or $so(3,1)$), using the
isomorphism between bivectors and Lie algebra.
We can naively replace
$B^{IJ}_{ab}$ with the canonical
generators $J^{IJ}$ of $so(4)$ (or $so(3,1)$), giving the correspondence:

\bes
B^{IJ}_{ab}B^{ab}_{IJ} &
\rightarrow & J^{IJ}J_{IJ}=2C_1\\
\f{1}{2}\epsilon_{IJKL}B^{IJ}_{ab}B^{KLab} &
\rightarrow & \f{1}{2}\epsilon_{IJKL}J^{IJ}J^{KL}
=2C_2
\ees
Using this, \Ref{simple2} gets transformed into:
\bes
2a_2 C_2 - \, a_1 C_1 =0
\label{mixed}
\ees
or equivalently:
\bes
2\alpha\beta C_1 = (\alpha^2+\beta^2)C_2
\ees
Then we would conclude as suggested in \cite{CMPR}
that we should use non-simple representations
in our spin foam model.
Actually the situation would be even worse than
this, since it happens that in general (for arbitrary values of $a_1$ and
$a_2$) no spin foam model can be constructed using only representations
satisfying \Ref{mixed} in the Riemannian case. 
More precisely,
using the splitting of the algebra $so(4)\simeq su(2)_+\oplus su(2)_-$,
the two Casimirs are:

\bes
C_1&=&j^+(j^++1)+j^-(j^-+1) \nonumber \\
C_2&=&j^+(j^++1)-j^-(j^-+1)
\ees

Apart from the case $a_1=0$ which gives us the Barrett-Crane simplicity
constraint $C_2=0$, the equations
\Ref{mixed} have an infinite number of solutions only when
$2a_2=\pm a_1$, in which cases we get representations of the form
$(j^+,0)$ (or $(0,j^-)$). This could be expected since the constraint
\Ref{simple} with this particular value of the parameters implies that the
$B$ field is selfdual or anti-selfdual when non-degenerate.
In the other cases, \Ref{mixed}
can be written as $j^+(j^++1)=\lambda j^-(j^-+1)$, with
$2a_2=(1+\lambda)/(1-\lambda)a_1$ and in general has no solutions.
For particular values of $\lambda$, it can have one and only one solution.
This would lead us to an ill-defined spin foam model using only one
representation $(j^+_0,j^-_0)$.
However, using the framework set up in \cite{MikeIntert},
it can be easily proven that
it is not possible to construct an intertwiner for
a spin network built out of a single representation
(a single representation is not stable under
change of tree expansion for the vertex) so that no spin foam model
can be created.

In the Lorentzian case, the situation is more complex. The representations
are labelled by pairs $(j\in\N/2,\rho\ge0)$ and the two Casimirs
are \cite{P-R3}\cite{Ruhl}:

\bes
C_1&=&j^2-\rho^2-1 \nonumber \\
C_2&=&\f{1}{2}j\rho
\ees
The equation \Ref{mixed} now reads:
\bes
\rho^2+\f{a_2}{a_1}j\rho-j^2+1=0
\ees

and admits the following solutions if $a_1\ne 0$:

\bes
\rho=\f{1}{2}\left[-\f{a_2}{a_1}j
\pm \sqrt{\left(\f{a_2}{a_1}\right)^2 j^2 +4(j^2-1)}\right]
\ees
So we always have some solutions to the mixed simplicity condition
\Ref{mixed}. However, instead of having a discrete series of
representations $(n,0)$ and a continuous series $(0,\rho)$, which are said
to correspond
to space-like and time-like degrees of freedom, as in the simple case
$a_1=0$, we end up with a (couple of) discrete series representations with
no direct interpretation. We think it is not possible to construct a
consistent vertex using them (it is also hard to imagine how
to construct a field theory over a group manifold formulation
of such a theory whereas, in the case of simple representations,
the construction was rather straightforward \cite{P-R,P-R2,P-R3}
adapting the Riemannian case to the Lorentzian case).
But this should be investigated.
As we will show, on the contrary, a careful analysis of the field content
of the theory, and of the correspondence between Lie algebra elements and
bivectors shows that not only a spin foam quantization is possible, but
that the resulting spin foam model should again be based on the
Barrett-Crane quantum constraints on the representations.

\subsection{Field content and relationship with the Plebanski action}
\label{class}
We have seen that the $B$ field, even if subject to constraints which are
more complicated than the Plebanski constraints, is forced by them to be
in 1-1 correspondence with the 2-form built out of the tetrad field,
$*(e\w e)$, which in turn is to be considered the truly physical field of
interest, since it gives the geometry of the manifold through the Einstein
equations. To put it in another way, we can argue that the physical
content of the theory, expressed by the Einstein equations, is independent
of the fields we use to derive it. In a discretized context, in
particular, we know that the geometry of the manifold is captured by
bivectors associated to 2-dimensional simplices, and constrained to be
simple. Consequently we would expect that in this context we should be
able to put the $B$ field in 1-1 correspondence with another 2-form field,
say $E$, then discretize to give a bivector for each triangle, in such a
way that the mixed constraints \Ref{simple2} would imply the simplicity of
this new bivector and the other Barrett-Crane constraints. If this
happens, then it would also mean that the action is equivalent to the
Plebanski action in terms of this new 2-form field, at least for what
concerns the constraints. This is exactly the case, as we are going to
prove.

The correspondence between $B$ and $E$ is actually suggested by the form
of the solution \Ref{B}. We take

\bes
B_{ab}^{IJ}\,=\,\left( \alpha\,I\,+\,\epsilon\,\beta\,*\right)E^{IJ}_{ab}
 \label{transf} 
\ees
This is an invertible transformation, so that it really gives a 1-1
correspondence, if and only if $\alpha^{2}-\epsilon\beta^{2}\neq 0$, which
we will assume to be the case in the following.

Formally, we can do such a change of variable directly on the action
and express the action itself in terms of the field $E$, making
apparent that the constraints are just the Plebanski constraints:

\bes
\left\{
\begin{array}{ccc}
B^{IJ}& =&\alpha E^{IJ}+\epsilon\beta *E^{IJ} \\
\tl{\phi}_{IJKL}&=&
(\alpha+\epsilon\beta\f{1}{2}\epsilon_{IJ}\,^{AB})\phi_{ABCD}
(\alpha+\epsilon\beta\f{1}{2}\epsilon_{KL}\,^{CD})
\end{array}
\right.
\label{change}
\ees

After this change, the action \Ref{action} becomes:

\bes
S\,=\,
\f{1}{|\alpha^2-\epsilon\beta^2|^3}\,
\int\,(\alpha E^{IJ}+\epsilon\beta*E^{IJ})\wedge F_{IJ}\,
-\,\f{1}{2}\tl{\phi}_{IJKL}E^{IJ}\wedge E^{KL}
\,+\,\mu\epsilon^{IJKL}\tl{\phi}_{IJKL}\;\;\;\;
\label{after}
\ees
so that we have the Plebanski constraints on the $E$ field and we can
derive
directly from this expression the Holst action \Ref{Holst}. In the
Riemannian
case, decomposing  into selfdual and anti-selfdual components can be
quite useful to understand the structure of the theory. This is done
in the appendix and shows that the previous change of variable is simply
a rescaling of the selfdual and anti-selfdual parts of the $B$
field.
Let us now look at the discretization of the constraints. We will follow
the
same procedure as in \cite{DP-F}. 
Using the 2-form $B$, a bivector can be associated to any 2-surface $S$ in
our manifold by integrating the 2-form over the surface:

\bes
B^{IJ}(S)\,=\,\int_{S}B^{IJ}\,=\,\int_{S}\left(
\alpha\,I\,+\,\epsilon\,\beta\,*\right)E^{IJ}\,
=\,\alpha\,E^{IJ}(S)\,+\,\epsilon\,\beta\,*E^{IJ}(S)
\label{dis}.
\ees
Note that this automatically implements the first of the Barrett-Crane
constraints for the bivectors $E(S)$ and $B(S)$ (a change of orientation
of the surface $S$ will change the sign of the bivectors).

We now take a triangulation of our manifold, such that $B$ is constant
inside each 4-simplex, i.e. $dB=0$, and we associate a bivector to each
triangle of the triangulation using the procedure above.
This is equivalent to assuming $E$ constant in the 4-simplex
since the map \Ref{transf} is invertible.
Then we can use
Stokes' theorem to prove that the sum of all the bivectors $E$ associated
to the 4 faces $t$ of a tetrahedron $T$ is zero:


\bes
0 = \int_{T}dE = \int_{\partial T}E
= \int_{t_1}E +\int_{t_2}E +\int_{t_3}E +\int_{t_4}E
 = E(t_1)+E(t_2)+E(t_3)+E(t_4)\;\;\;\;\;
\ees
meaning that the bivectors $E$ satisfy the fourth Barrett-Crane
constraints (closure constraint).
Let us consider the constraints in the form \Ref{simpl2}.
Using \Ref{transf}, we obtain a much simpler
constraint on the $E$ field:
\bes
\left(\alpha^{2} \,+\, \epsilon\,\beta^{2}\right) \,
\epsilon_{IJKL}\,E^{IJ}_{ab}\,E^{KL}_{cd}
\,=\,
\,e\,\epsilon_{abcd}
\label{discon}
\ees

Now,
$e=\f{1}{4!}\epsilon_{IJKL}B^{IJ}\w B^{KL}=
\f{1}{4!}(\alpha^2+\epsilon\beta^2)\epsilon_{IJKL}E^{IJ}\w E^{KL}$
is a sensible volume
element, since we assumed that the $B$ field is non-degenerate.
After imposing the equations of motion,
it appears that it is also the ``right'' geometric
volume element, i.e. the one constructed out of the tetrad field.
More precisely, equation \Ref{discon}
implies the simplicity of the field $E$. Thus there exist a tetrad field
such that $E^{IJ}=\pm e^I\w e^J$ or $E^{IJ}=\pm*(e^I\w e^J)$ and
this tetrad field is the one defining the metric after imposing the
Einstein equations.
The scalar $e$ is then proportional to
$\epsilon_{IJKL}e^{I}\wedge e^{J}\wedge e^{K}\wedge
e^{L}=det(e)$. Consequently, up to a factor, the 4-volume  
spanned by two faces $t$ and $t'$ of a 4-simplex is given by:

\bes
V(t,t')\,=\,\int_{x\in t\,;\,y\in t'}\,e\,\epsilon_{abcd}\,dx^{a}\w
dx^{b}\w dy^{c}\w dy^{d}
\ees
Then, integrating equation \Ref{discon} gives directly:
\bes
\epsilon_{IJKL}\,E^{IJ}(t)\,E^{KL}(t')
=\f{1}{\left( \alpha^{2}\,+\,\epsilon\,\beta^{2}\right)}\,V(t,t')
\label{cons}
\ees
Considering only one triangle $t$:
\bes
\epsilon_{IJKL}\,E^{IJ}(t)\,E^{KL}(t)=0
\ees
so that for any of the bivectors $E$ the selfdual part has the same
magnitude as the anti-selfdual part,
so that the $E(t)$ are simple bivectors.
This corresponds to the second of the Barrett-Crane constraints
(simplicity constraint).
For two triangles sharing an edge, we similarly have:
\bes
\epsilon_{IJKL}\,E^{IJ}(t)\,E^{KL}(t')=0
\ees
This can be rewritten as:
\bes
\epsilon_{IJKL}(E^{IJ}(t)+E^{IJ}(t'))(E^{KL}(t)+E^{KL}(t')) \nonumber \\
-\epsilon_{IJKL}\,E^{IJ}(t)\,E^{KL}(t)-
\epsilon_{IJKL}\,E^{IJ}(t')\,E^{KL}(t')=0
\ees
and this, together with the simplicity constraint,
implies that the sum of the
two bivectors associated to the two triangles is again a simple bivector.
This implements the third of the Barrett-Crane constraints
(intersection constraint).
Let us note that in the Riemannian case, we can use the decomposition
into selfdual and anti-selfdual components to write the previous
constraints (for $t=t'$ or $t$ and $t'$ sharing an edge) as:

\bes
\delta_{IJ}\left[
E^{(+)I}(t)E^{(+)J}(t')-E^{(-)I}(t)E^{(-)J}(t')\right]\,=\,0
\ees
thus showing that the simplicity for the bivectors $E(t)$ and $E(t)+E(t')$
is that their selfdual part and anti-self part have the same magnitude.
We note that in the Lorentzian case this decomposition implies
a complexification of the fields, so the physical interpretation is
somehow more problematic. However it presents no problems formally , and
corresponds to the splitting of the Lie algebra of $so(3,1)$, to which the
bivectors in Minkowski space are isomorphic, into $su(2)_{C}\oplus
su(2)_{C}$.

Thus, it is clear that the complicated constraints \Ref{whole} and
\Ref{simple2} for the field $B$ are just the Plebanski constraint for the
field $E$, associated to $B$ by means of the transformation \Ref{transf},
and, when discretized, are exactly the Barrett-Crane constraints. The
fields characterizing the 4-geometry of the triangulated manifold are then
the bivectors $E(t)$.
This result, by itself, does not imply necessarily that a spin foam
quantization of the generalized BF-type action gives the Barrett-Crane
model, but it means anyway that the geometry is still captured by a field
which, when discretized, gives a set of bivectors satisfying the
Barrett-Crane constraints. This in turn suggests strongly that the
Barrett-Crane constraints characterize the quantum geometry also in this
case, even if a first look at the constraints seemed to contradict this,
and consequently the simple representations are the right representations
of $so(4)$ and $so(3,1)$ that have to be used in constructing the spin
foam.   

\subsection{Spin foam quantization and constraints on the representations}
\label{quant}
We have already proven that the spin foam quantization cannot be performed
using a naive association between the $B$ field and the canonical
generators of the Lie algebra. On the other hand, we have seen that the
$B$ field can be put in correspondence with a field $E$ such that the
constraints on this are the Barrett-Crane constraints. This suggest that a
similar transformation between Lie algebra elements would make everything
work again, giving again the simplicity conditions for the
representations, as in the Barrett-Crane model.

In light of the natural way to
associate a bivector to a triangle in a gravitational context, using the
frame field, and also because the 2-forms $*(e^{I}\w e^{J})$ are a basis
for the space of 2-forms, we could argue that it is the bivector coming
from $*(e\w e)$ that has to be associated to the canonical generators of
the Lie algebra.    
That this is the right choice can be proven easily. In fact the
isomorphism between bivectors and Lie algebra elements is realized
choosing a basis for the bivectors such that they are represented by 4 by
4 antisymmetric matrices, and interpreting these matrices as being the
4-dimensional representation of Lie algebra elements. If this is done for
the basis 2-forms $*(e\w e)$, the resulting matrices give exactly the
canonical generators of $so(4)$ or $so(3,1)$, so that $*(e \w e)\,
\leftrightarrow\,J$.    

Then equation \Ref{B} suggests us that the field $B$ has to be associated
to elements $\tl{J}$ of the Lie algebra such that:
\bes
B \leftrightarrow \tl{J}^{IJ}\,=\,\alpha\, J^{IJ}\,+\,\epsilon\,\beta
*J^{IJ}
\label{modif}
\ees

This is simply a change of basis (but not a Lorentz rotation),
since the transformation is invertible,
provided that $\alpha^{2}-\epsilon\beta^{2}\neq 0$. In some sense, we can
say that working with the $B$ in the action is like working with a
non-canonical basis in the Lie algebra, the canonical basis being instead
associated to $*(e\w e)$.

Now we consider the constraint \Ref{simple2} on the $B$ field, and use the
correspondence above to translate it into a constraint on the
representations of the Lie algebra.
The Casimir corresponding to $\f{1}{2}\epsilon_{IJKL}B_{ab}^{IJ}B^{abKL}$
is
\bes
\tl{C}_2=\f{1}{2}\epsilon_{IJKL}\tl{J}^{IJ}\tl{J}^{KL}
=2\alpha \beta C_1 +(\alpha^2+\epsilon\beta^2)C_2
\ees
where $C_1$ and $C_2$ are the usual Casimirs associated to the canonical
generators $J^{IJ}$ (being $C_{1}=j^+(j^+ +1) + j^-(j^- +1)$ and
$C_{2}=j^+(j^+ +1) - j^-(j^- +1)$ in the Riemannian case, and $C_{1}=j^2
-\rho^2 -1$ and $C_{2}=\f{1}{2}j\rho$ in the Lorentzian case), while the
Casimir
associated to $B^{IJ}_{ab}B_{IJ}^{ab}$ is:
\bes
\tl{C}_1=\tl{J}^{IJ}\tl{J}_{IJ}
=(\alpha^2+\epsilon\beta^2)C_1 + 2\epsilon\alpha \beta C_2
\ees

Substituting these expressions into \Ref{simple2}, we get:
\bes
2\alpha\beta \tl{C}_1=(\alpha^2+\epsilon\beta^2)\tl{C}_2\,\,
\Rightarrow\,\,
(\alpha^2-\epsilon\beta^2)^2\,C_2\,=\,0
\ees

In the assumed case $\alpha^2\ne \epsilon\beta^2$,
we find the usual Barrett-Crane
simplicity condition $C_2=0$ with a restriction to the simple
representations of $so(4)$ ($j^+=j^-$)
or $so(3,1)$ ($n=0$ or $\rho=0$). We see that, at least for what
concerns the representations to be used in the spin foam model, the whole
modification of the inital action
is absorbed by a suitable redefinition of the correspondence between
the field $B$ and the generators of the Lie algebra.

Now we want to discuss briefly how general is the association we used
between the $B$ field and Lie algebra elements, i.e. how many other
choices would give still the Barrett-Crane simplicity constraint on the
representations starting from the constraint \Ref{simple2} on the $B$
field.
Suppose we associate to $B$ a generic element $\tl{J}$ of the Lie algebra,
related to the canonical basis by a generic invertible transformation
$\tl{J}=\Omega J$. Any such transformation can be split into
$\tl{J}^{IJ}=\Omega^{IJ}_{KL}J^{KL}=(\alpha I+\epsilon\beta
*)^{IJ}_{MN}U^{MN}_{KL}J^{KL}=(\alpha I+\epsilon\beta
*)^{IJ}_{MN}J'^{MN}$, for $\alpha^2\ne \epsilon\beta^2$, so shifting all
the ambiguity into $U$. Inserting this into the constraints we obtain the
condition $C'_{2}=0$, where $C'_2=J'*J'$. If we now require that this
transformation should still give the Barrett-Crane simplicity constraint
$C_2=0$, then this amounts to requiring $C'_2=\lambda C_2$ for a generic
$\lambda$. But if $C_2=0$ and the transformation preserves the second
Casimir modulo rescaling, then it should preserve, modulo rescaling, also
the first one. This means that the transformation $U$ preserves, modulo
rescaling, the two bilinear forms in the Lie algebra which the two
Casimirs are constructed with, i.e. the \lq\lq identity" and the
completely antisymmetric 4-tensor in the 6-dimensional space of
generators. Consider the Riemannian case. The set of transformations
preserving the first is given by $O(6)\times Z_{2}\simeq
SO(6)\times Z_{2}\times Z_{2}$, while the set of
transformations preserving the second is given by
$O(3,3)\times Z_{2}\simeq
SO(3,3)\times Z_{2}\times Z_{2}$, so that the $U$
preserving both are given by the intersection of the two groups, i.e. by
the transformations belonging to a common subgroup of them. Certainly a
common subgroup is given by $SO(3)\times
SO(3)\times Z_{2}\times Z_{2}\simeq
SO(4)\times Z_{2}\times Z_{2}$, and we can conjecture that
this is the largest one, since it covers all the symetries of the original
action, so that any solution of the theory, like \Ref{B} should be defined
up to such transformation, and we expect this to be true also in the
association between fields and Lie algebra elements. Of course we can in
addition rescale the $\tl{J}$ with an arbitrary real number. The argument
in the Lorentzian case goes similarly.

Coming back to the spin foam model corresponding to the new generalized
action, our results prove that it should still be based on the simple
representations of the Lie algebra, and that no ambiguity in the choice of
the labelling of the spin foam faces results from the more general form of
the constraints, since this can be naturally and unambiguously re-absorbed
in the correspondence between the $B$ field and the Lie algebra elements.
This suggests strongly that the resulting spin foam model corresponding to
the classical action here considered is still the Barrett-Crane model, as
for the Plebanski action, but it is not completely straightforward to
prove it due to the more complicated form of the action \Ref{after}.
Anyway a motivation for this is provided by the fact that our analysis
shows that the physics is still given by a set of bivectors $E$, in 1-1
correspondence with the field $B$ on which the generalized action is
based, and that this bivectors satisfy the Barrett-Crane constraints at
the classical level, with the translation of them at the quantum level
being straightforward. The action is in fact that of BF theory plus
constraints, which, as shown in this section, at the quantum level are
just the Barrett-Crane constraints on the representations used as
labelling in the spin foam, so that a spin foam quantization procedure of
the type performed in \cite{OW} seems viable.

\subsection{Alternative: the Reisenberger model} \label{reis}
We have seen that a natural discretization and spin foam quantization of
the constraints in this generalized BF-type action leads to the
Barrett-Crane model. In this section we want to discuss and explore a bit
the alternatives to our procedure, and the cases not covered in our
previous analysis. 

In order to prove the equivalence of the two form of the constraints
\Ref{simple} and \Ref{simple2}, we assumed that $a_2\neq 0$ and that
$\left(\f{2a_2}{a_1}\right)\neq\epsilon$, which, in terms of $\alpha$ and
$\beta$ is requiring that $\alpha^2\neq\epsilon\beta^2$
($\alpha\neq\pm\beta$ in the Riemannian case, and $\alpha\neq\pm i\beta$ in
the Lorentzian one). Later, the transformations we used both at the
classical level ($B\rightarrow E$) and at the quantum (better, Lie
algebra) level ($\tl{J}\rightarrow J$) were well-defined, i.e. invertible,
provided that $\alpha^2\neq\epsilon\beta^2$, again.
Actually, the only interesting condition is the last one, since we already
mentioned at the beginning (section \Ref{action}) that the case $a_2=0$
leads to  considering only degenerate $B$ fields and degenerate
tetrads (in the Riemannian case). 

The case $\left(\f{2a_2}{a_1}\right)\neq\epsilon$ or
$\alpha^2\neq\epsilon\beta^2$ corresponds, at the canonical level, to the
Barbero's choice of the connection variable (with Immirzi parameter
$\gamma=\pm 1$) in the Riemannian case, and to Ashtekar variables
($\gamma=\pm i$) in the Lorentzian case. It amounts to formulating the
theory using a selfdual (or anti-selfdual) 2-form field $B$, or
equivalently a selfdual (or anti-self dual) connection. In fact, a look at
the constraints \Ref{simple} shows clearly that, in this case, they
exactly imply the selfduality (or anti-selfduality) of the field $B$.
There is no rigorous way to relate the Barrett-Crane conditions and spin
foam model
to the classical action when this happens, at least using our procedure,
since the equivalence of \Ref{simple} and \Ref{simple2} cannot
be proved and the transformations we used are not invertible.
The constraints on the field $B$ just
state that $B$ is the (anti-)selfdual part of a field $E$ (and
in this case, the action \Ref{action} corresponds to the (anti-)selfdual
Plebanski action for $E$).
We note however that if we use still the constraints in the form
\Ref{simple2} in the Riemannian case, in spite of the fact that we are not
able to prove their equivalence with the original ones, and translate them
into a constraint on the representations of $so(4)$ using the naive
correspondence $B\rightarrow J$, where $J$ is the canonical basis of the
algebra, we get: $C_1=\pm C_2$ for $2a_2=\pm a_1$. The first case leads to
$j^-=0$ and the second to $j^+=0$, so we are reduced from $so(4)$ to
$su(2)_L$ or to $su(2)_R$, with a precise correspondence between the
(anti-)selfduality of the variables used and the (anti-)selfduality of the
representations labelling the spin foam.
There exist a spin foam model in these cases as well.
It is the Reisenberger
model for left-handed (or right-handed) Riemannian gravity
\cite{mike3}\cite{mike2}, whose relationship with the Barett-Crane model is
unfortunately not yet clear.
This model \cite{mike3}\cite{Re97b} can be associated to a different
discretization of the generalized action \Ref{action} using the original
form
\ref{simple} of the constraints, as it was
analyzed in \cite{DP-F} for the Plebanski action.
This is indeed the only other (known)
alternative to our procedure, at least in
the Riemannian case. 

As before we define the volume spanned by the
two triangles $S$ and $S'$:

\bes
V(S,S')=\int_{x\in S, y\in S'}e\,\epsilon_{abcd}\,
dx^a\w dx^b\w dy^c\w dy^d
\ees

To use \Ref{simple}, we decompose the 2-form B inside
the 4-simplex into a sum of 2-forms associated to the faces (triangles)
of the 4-simplex \cite{DP-F}\cite{mike3}\cite{F-K}:

\bes
B^{IJ}(x)=\sum_S B^{IJ}_S(x)
\ees
where $B^{IJ}_S(x)$ is such that
\bes
\int B^{IJ}_S\w J=B^{IJ}[S]\int_{S*}J
\ees
with $J$ is any 2-form and $S*$ the dual face of $S$ (more precisely
the wedge dual to $S$, i.e the part of the dual face to $S$
lying inside the considered 4-simplex).
Then, it is clear that:
\bean
\int_S B^{IJ}_{S'}=\delta_{S,S'}B^{IJ}[S] \\
\int B^{IJ}_S\w B^{KL}_{S'}=B^{IJ}[S]B^{KL}[S']\epsilon(S,S')
\eean
where $\epsilon(S,S')$ is the sign of the oriented volume $V(S,S')$.
More precisely $\epsilon(S,S')=\pm 1$ if $S$, $S'$ don't share any edge,
and $\epsilon(S,S')=0$ if they do. 

Using that, we can translate \Ref{whole} and \Ref{simple} into:
\bes
\tl{\Omega}^{IJKL}=\Omega^{IJKL}-
\f{1}{6}\eta^{[IK}\eta^{J]L}\Omega^{AB}_{AB}-
\f{1}{24}\epsilon^{IJKL}\epsilon_{ABCD}\Omega^{ABCD}=0
\ees

and 
\bes
4a_2\Omega^{AB}_{AB}=a_1\epsilon_{ABCD}\Omega^{ABCD}
\ees
where
\bes
\Omega^{IJKL}=\sum_{S,S'}B^{IJ}[S]B^{KL}[S']\epsilon(S,S')
\label{defomega}
\ees
These are the $so(4)$ analogs of the Reisenberger constraints.
Let us note that the constraints involve associations triangles
not sharing any edge whereas the Barrett-Crane procedure
was to precisely study triangles sharing an edge i.e being in
the same tetrahedron. This is one of the reasons why it is hard to link
these two models.
Then following \cite{mike2}\cite{F-K}, it is possible to calculate
the amplitude associated to the 4-simplex and the corresponding spin foam
model. 
For this purpose, it is useful to project these constraints on the selfdual
and anti-selfdual sectors as in \cite{DP-F}:

\bes
\tl{\Omega}^{ij}_{++}=
\Omega^{ij}_{++}-\delta^{ij}\f{1}{3}tr(\Omega_{++})=0
\ees

\bes
\tl{\Omega}^{ij}_{--}=
\Omega^{ij}_{--}-\delta^{ij}\f{1}{3}tr(\Omega_{--})=0
\ees

\bes
\Omega^{ij}_{+-}=0
\ees

\bes
\Omega_0=
(\alpha-\beta)^2tr(\Omega_{++})+
(\alpha+\beta)^2tr(\Omega_{--})=0
\ees

The two first constraints are the same as in the Reisenberger model
for $su(2)$ variables. The two last constraints link the two sectors
(+ and -) of the theory, mainly stating that there is no correlation
between them except for the constraint $\Omega=0$.
Indeed,  only that last constraint is modified by the introduction
the Immirzi parameter. Following the notations of \cite{mike2},
we replace the field $B^{IJ}$ by the generator $J^{IJ}$ of $so(4)$
(this is a formal quantization of the discretised BF action, see
\cite{mike2}
for more details) and we define the projectors $P_{1,2,3,4}$
(or some Gaussian-regularised projectors) on the
kernel of the operators corresponding to the four above constraints.
Then, the amplitude for the vertex $\nu$ is
a function of
the holonomies on the 1-dual skeleton of the three-dimensional
frontier $\pp\nu$ of the vertex. It is given by
projecting a universal state (or topological state since that without
the projectors, the amplitude gives the $so(4)$ Ooguri topological
model)
and then integrating it over
the holonomies around the ten wedges $\{h_l\}_{l=1\dots10}$:

\bes
a(g_{\pp\nu})=\int dh_l \prod_{s\,\tr{\tiny wedge}}
\sum_{j_s}tr^{j^1\otimes\dots\otimes j^{10}}\left[
P_1P_2P_3P_4\bigotimes_s(2j_s+1)U^{(j_s)}(g_{\pp s})
\right]
\ees
It seems that a vertex including the Immirzi parameter is
perfectly well-defined in this case. Let us analyse this
more closely.
As only one constraint involves the Immirzi parameter,
we will first limit ourselves to studying its action.
Replacing the field $B$ by the generators $J^{IJ}$, we get

\bes
\Omega_0&=&
\sum_{S,S'}\epsilon(S,S')\left[
(\alpha-\beta)^2
J^{+ i}[S]J^{+ i}[S']
+(\alpha+\beta)^2 \,(anti-selfdual)
\right]
\nonumber  \\
&=&
\sum_{S,S'}\epsilon(S,S')\f{1}{2}\left[
(\alpha-\beta)^2
\left((J^{+ i}[S]+J^{+ i}[S'])^2-J^{+ i}[S]^2-J^{+ i}[S']^2\right)+ \right. \nonumber \\
&+& \left. (\alpha+\beta)^2 \,( anti-selfdual )
\right]
\ees

So $\Omega_0$ can be expressed in term of the ($su(2)$) Casimir
of the representations associated to each wedge $S$ and
the Casimir of their tensor products.
The difference from the Barrett-Crane
model is that it is the tensor product of the representations
associated to two triangles which do {\it not} belong to the same
tetrahedron, so there is little hope of finding directly
an equivalent of the simplicity/intersection constraints.
Nevertheless, we can do a naive analysis.
Casimirs will always give numbers $j(j+1)$. But we need to choose a basis
of $j^1\otimes\dots\otimes j^{10}$. So calculating the action of
$\Omega_0$ might involve some change of basis and thus some Clebsch-Gordon
coefficient. However, those are still rational. 
So we conjecture that the amplitude of the vertices will be zero
(no state satisfying $\Omega_0=0$)
except if $\alpha=\beta$, $\alpha=0$ or $\beta=0$ as in our
first approach to quantizing the $B$-field constraints in
the Barrett-Crane framework.
If this conjecture is verified, we will have two possibilities:
either the discretization procedure
is correct and we are restricted to a few consistent cases, or we need to
modify the discretization or quantization procedures.

Resuming, the situation looks as follows. We have the most general BF-type
action for gravity, depending on two arbitrary parameters, both in the
Riemannian and Lorentzian signature. In both signatures, and for all the
values of the parameters except one (corresponding to the Ashtekar-Barbero
choice of canonical variables), the constraints which give gravity from BF
theory can be expressed in such a form that a spin foam quantization of
the theory leads to the Barrett-Crane spin foam model. In the Riemannian
case, for all the values of the parameters, a different discretization,
and quantization procedure, leads to the Reisenberger model, but it also
seems that the last spin foam model is non-trivial (i.e. non-zero vertex
amplitude) only for some particular values of the parameters $\alpha$ and
$\beta$. These two spin foam models may well turn out to be equivalent,
but they are a priori
different, and their relationship is not known at this stage. There is no
need to stress that an analysis of this relationship would be of paramount
importance.

\subsection{The role of the Immirzi parameter in spin foam models}
\label{area}
We would like to discuss briefly what our results suggest regarding the
role of the Immirzi parameter in the spin foam models, stressing that this
suggestion can at present neither be well supported nor disproven by
precise calculations.
As we said in section \Ref{CMPR}, the BF-type action \Ref{action}, after
the imposition of the constraints on the field $B$, reduces to a
generalized Hilbert-Palatini action for gravity, in a form studied within
the canonical approach in \cite{holst}. The canonical analysis performed
in that work showed that this action is the Lagrangian counterpart of the
Barbero's Hamiltonian formulation \cite{barbero} introducing the Immirzi
parameter \cite{immirzi} in the definition of the connection variable and
then in the area spectrum. This led to the suggestion \cite{CMPR} that
the spin foam model corresponding to the new action \Ref{action} would
present non-simple representations and an arbitrary (Immirzi) parameter as
well.  
 
On the contrary we have shown that the spin foam model corresponding to
the new action is given again by the Barrett-Crane model, with the
representations labelling the faces of the 2-complex being still the
simple representations of $so(4)$ or $so(3,1)$. The value of the area of
the triangles in this model is naturally given by the (square root of the)
first Casimir of the gauge group in the representation assigned to the
triangle, with no additional (Immirzi) parameter. From this point of view
it can be said that the prediction about the area spectrum of spin foam
models and canonical (loop) quantum gravity do not coincide. 

However both the construction of the area operator and its diagonalization
imply working with an Hilbert space of states, and not with their
histories as in the spin foam context, and the canonical structure of the
spin foam models, like the Barrett-Crane one, is not fully understood yet.
Consequently, the comparison with the
loop quantum gravity approach and results is not straightforward. In fact,
considering for example the Barrett-Crane model, it assigns an Hilbert
space to boundaries of spacetime, and these correspond, in turn, to
boundaries of the spin foam, i.e. spin networks, so that again a spin
network basis spans the Hilbert space of the theory, as in loop quantum
gravity. The crucial difference, however, is that the spin networks  used
in the Barrett-Crane model are constructed out of (simple) representations
of $so(4)$ or $so(3,1)$, i.e. the full local gauge group of the theory,
while in loop quantum gravity
(for a nice introduction see \cite{rovelli&gaul})
the connection variable used is an
$su(2)$-valued connection resulting from reducing the gauge group
from $so(4)$ or $so(3,1)$ to that subgroup, in the process of the
canonical $3+1$ decomposition, so that the spin network basis uses only
$su(2)$ representations. Consequently, a comparison of the results in spin
foam models could possibly be made more easily
with a covariant (with respect to the gauge
group used) version of loop quantum gravity, i.e. one in which the group
used is the full $so(4)$ or $so(3,1)$.

The only results in this sense we are aware of were presented in
\cite{alex}\cite{alex2}.
In these two papers, a Lorentz covariant version of loop
quantum gravity is sketched at the algebraic level. However, the
quantization was not yet achieved and 
the Hilbert space of the theory (``spin networks'') not
constructed because of problems arising from the non-commutativity of the
connection variable used.
Nevertheless, two results were derived from the formalism.
The first one is that the path integral of the theory
formulated in the covariant variables is independent 
of the Immirzi parameter \cite{alex}, which becomes
an unphysical parameter whose role is to regularize the theory.
The second result was the construction of an area operator acting
on the hypothetic ``spin network'' states of the theory \cite{alex2}:
\bes
{\cal A} \approx   l_P^2\sqrt{-C(so(3,1))+C(su(2))}
\label{spectrum}
\ees
where $C(so(3,1))$ is a quadratic Casimir of the Lie algebra $so(3,1)$
(the Casimir $C_1$ to a factor)
and $C(su(2))$ the Casimir of the spatial pull-back $so(3)$. So a
``spin network'' state would be labelled by both
a representation of $so(3,1)$
and a representation of $su(2)$. As we see, the area, whose spectrum
differ from both the the spin foam and the loop quantum gravity one, is
independent
of the Immirzi parameter. However, we do not know yet if such an area
spectrum has any physical meaning, since no Hilbert space has been
constructed for the theory.
Nevertheless, it suggests that a ``canonical'' interpretation
of a spin foam might not be as straightforward as it is believed. Instead of
taking as spatial slice an $SO(3,1)$ spin networks by cutting a spin foam, we might have to also project the $SO(3,1)$ 
structure onto
an $SU(2)$ one; the resulting $SU(2)$ spin network being our space and
the background $SO(3,1)$ structure describing its space-time embedding.

It is clear that this issue has not found at present any definite
solution, and it remains rather intricate. 
Thus all we can say is that our results (that do not regard directly the
issue of the area spectrum) and those in
\cite{alex2} suggest that the appearence of the Immirzi parameter in
loop quantum gravity is an indication of the presence of a quantum
anomaly, as discussed in \cite{tom&carlo}, but not of a fundamental one,
i.e. not one originating from the breaking up of a classical symmetry at
the quantum level, and indicating that some new physics takes place.
Instead what seems to happen is that the symmetry is broken by a
particular choice of quantization procedure, and that a fully covariant
quantization, like the spin foam quantization or the manifestly Lorentzian
canonical one, does not give rise to any one-parameter ambiguity in the
physical quantities to be measured, i.e. no Immirzi parameter. Of course,
much more work is needed to understand better this issue, in particular
the whole topic of the relationship between the canonical loop quantum
gravity approach and the covariant spin foam one is to be explored in
details, and to support or disprove this idea. A crucial step is the appearance of the same kind of
simple spin networks that appear in the Barrett-Crane model in the context of covariant canonical loop
quantum gravity, based on the full Lorentz group, obtained in \cite{alexetera}, about which we will say more later on.

\section{A lattice gauge theory derivation of the Barrett-Crane spin foam model} \label{sec:lgtBC}
We will now derive the Barrett-Crane spin foam model from a discretization of
 BF theory, imposing the constraints that reduce this theory to gravity 
(the Barrett-Crane constraints) at the quantum level, i.e. at the level of the representations of 
the gauge group used. We perform explicitely the calculations in the Riemannian case basically only for simplicity of 
notation, since the same procedure can be used in the Lorentzian case. We will write down explicitely the Lorentzian version of
the model in a later section. Of course the best thing to do would be to discretize and quantize directly the Plebanski 
action, using the discretized expression we have written down in section \ref{sec:pleb}. This is a bit more difficult,
 due to the non-linearity of the additional term in the B field (similar problems exist for the discretization of the 
BF theory with a cosmological constant, see \cite{OLough}), but progress along these lines has been done in \cite{alejandro},
 leading to the same result as in the simpler procedure, as we are going to show in the following. The results presented in 
this section were published in \cite{OW}.  

\subsection{Constraining of the BF theory and the state sum for a single 4-simplex} \label{sec:1BC}
Consider our 4-dimensional simplicial manifold, and the complex 
dual to it, having a vertex for each 4-simplex of the triangulation, an 
edge (dual link) for each tetrahedron connecting the two different 4-simplices that share it, and 
a (dual) face or plaquette for each triangle in the triangulation (see figures 4.3 and 4.4).

\begin{figure}
\begin{center} 
\includegraphics[width=9cm]{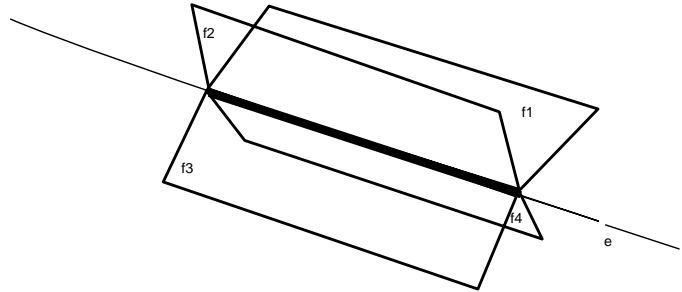}
\caption{A dual edge e with the four dual faces meeting on it}
\end{center}
\end{figure}

\begin{figure}
\begin{center}
\includegraphics[width=9cm]{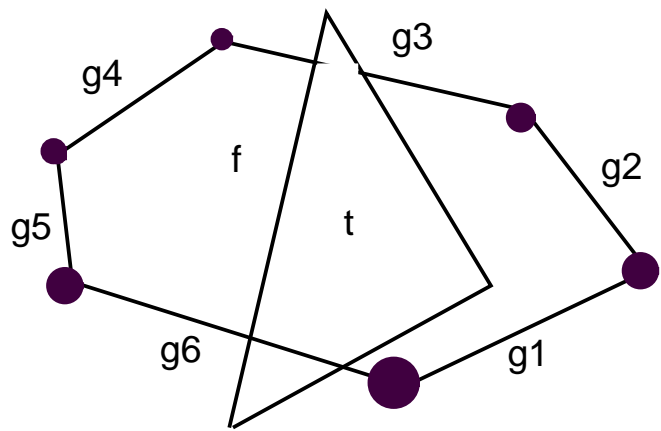}
\caption{The dual plaquette f for the triangle t}
\end{center}
\end{figure}

We introduce a dual link variable $g_{e}=e^{i\omega(e)}$ for 
each dual link $e$, through the holonomy of the $so(4)$-connection $\omega$
along the link. Consequently the product of dual link variables along the boundary
$\partial f$ of a dual plaquette $f$ leads to a curvature located at the centre of  
the dual plaquette, i.e. at the center of the triangle t, as we have already discussed in section \ref{sec:pleb}.
Dealing also with the $B$ field as we did there \cite{Baez2,Kawa,OW,FK}, we have \cite{OW} the following expression for the discretized
partition function of $Spin(4)$ BF theory:
\bes
Z_{BF}(Spin(4))\,=\,\int_{Spin(4)}dg\,\prod_{\sigma}\sum_{J_{\sigma}}\,\Delta_{J_{\sigma}}\,\chi_{J_{\sigma}}(\prod_{e}
g_{e}) \ees
where the first product is over the plaquettes in the dual complex (remember the 1-1 correspondence between triangles and plaquettes), the sum is over (the highest weight of) the 
representations of $Spin(4)$, $\Delta_J$ being their dimension, and $\chi_{J_{\sigma}}(\prod g)$ is the
character (in the representation $J_{\sigma}$) of the product of the
group elements assigned to the boundary edges of the dual plaquette
$\sigma$.

The partition function for the SO(4) BF theory is
consequently obtained by considering only the representation for which
the components of the vectors $J_{\sigma}$ are all integers. 

We see that the integral over the connection
field corresponds to the integral over group elements assigned to the
links of the dual lattice, while the integration over the $B$ field
is replaced by a sum over representations of the group. The meaning of
this can be understood by recalling from the previous sections that, roughly speaking, the $B$ field
turns into a product of tetrad fields after the imposition of the
constraints reducing BF theory to gravity, so giving the geometrical
information about our manifold, and that this information, in spin
foam models, has to be encoded into the algebraic language of
representation theory, and be given by the representations of the
gauge group labelling our spin foam. 

A generic 4-simplex has 5
tetrahedra and 10 triangles in it (see figure 4.5).
 Each dual link goes from a
4-simplex to a neighbouring one through the shared tetrahedron, so we
have 5 dual links coming out from a 4-simplex.

\begin{figure}
\begin{center}
\includegraphics[width=8cm]{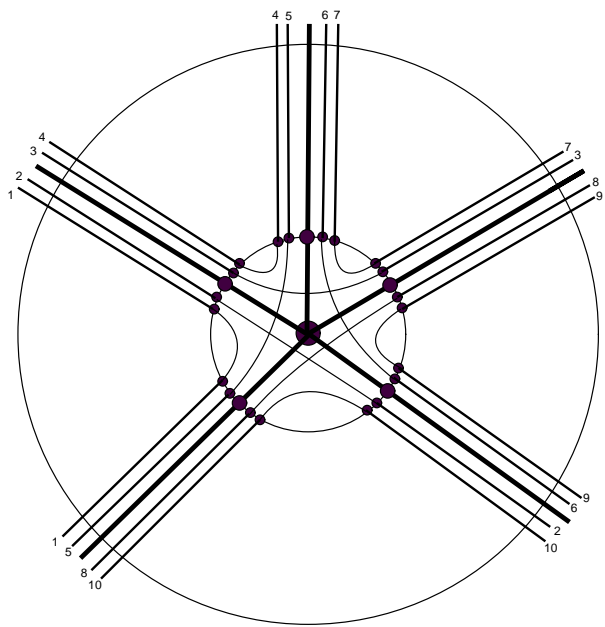}
\caption{Schematic representation of a 4-simplex; the thick lines represent the 5 tetrahedra and the thin lines the triangles}
\end{center}
\end{figure}
 
Now we refine the procedure above assigning two dual
link variables to each dual link, dividing it into two segments going
from the centre of each 4-simplex to the centre of the boundary
tetrahedron, i.e. we assign one group element $g$ to each of them (see
figure 4.6). Instead of the full plaquette, then, we are considering \lq\lq wedges" \cite{mike3}, i.e. the parts of the plaquettes
living inside a 4-simplex. In this way it is possible to deal with manifold with boundaries when the plaquette is truncated 
by the boundary, leaving only the wedges inside each 4-simplex.

\begin{figure}
\begin{center}
\includegraphics[width=8cm]{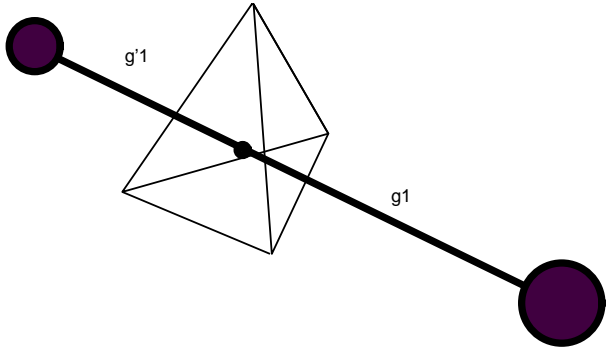} 
\caption{The dual link corresponding to the tetrahedron on which 2 4-simplices meet}
\end{center}
\end{figure}

In other words, a dual plaquette is given by a number, say, m of dual
links each divided into two segments, so there are 2m dual link
variables on the boundary of each plaquette. When a tetrahedron sharing the triangle to which the plaquette corresponds 
is on the boundary of the manifold, the plaquette results in being truncated by the boundary, and there will be edges 
exposed on it (not connecting 4-simplices). To each of these exposed edges we also assign a group variable.

The group variables assigned to the boundary links of each wedge give a curvature associated to it. Also the $B$ field, or 
better a representation of the gauge group is assigned to each wedge. So we have for the 
partition function of BF theory the same expression as above, but with a sum over wedges replacing the sum over plaquettes.

We now make use of the character decomposition formula which
decomposes the character of a given representation of a product of
group elements into a product of (Wigner) D-functions in that representation:
\bes
\chi_{J_{\sigma}}(\prod_{\tilde{l}\in \partial\tilde{P}}
g_{e}(\tilde{l}))\,=\,\sum_{\{k\}}\prod_{i}D_{k_{i}k_{i+1}}^{J_{\sigma}}\left(g_{e_{i}}\right),\;\;\;{\it
with}\;\;\;k_{1}=k_{2m+1},
\ees
where the product on the $i$ index goes around the boundary of the
dual plaquette surrounding the triangle labelled by $J_{\sigma}$, and
there is a D-function for each group element assigned to a dual link and to the edges exposed on the boundary .
We choose real representations of Spin(4) (this is always possible).
Consider now a single 4 simplex. Note that in this case all the tetrahedra are on the boundary of the manifold, which is
 given by the interior of the 4-simplex.
Writing down explicitly all the products of D-functions and labelling
 the indices appropriately, we can write down the partition function for the
Spin(4) BF theory on a manifold consisting of a single 4-simplex in the
following way:
\bes 
\lefteqn{Z_{BF}(Spin(4))=} \nonumber \\ &=&\sum_{\{J_{\sigma}\},\{k_{e}\}}\left(\prod_{\sigma}\Delta_{J_{\sigma}}\right)
\prod_{e}\int_{Spin(4)}dg_{e}\,D_{k_{e1}m_{e1}}^{J_{1}^{e}}D_{k_{e2}m_{e2}}^{J_{2}^{e}}D_{k_{e3}m_{e3}}^{J_{3}^{e}}
D_{k_{e4}m_{e4}}^{J_{4}^{e}}\;\left(\prod_{\tilde{e}}D_{il}^{J}\right)\;\;\;\;\;\;\;\;. 
\ees
The situation is now as follows: we have a contribution for each of the 5 edges of the dual complex, corresponding to 
the tetrahedra of the triangulation, each of them made of a product of the 4 D-functions for the 4 representations 
labelling the 4 faces incident on an edge, corresponding to the 4 triangles of the tetrahedron. There is an extra product 
over the faces with a weight given by the dimension of the representation labelling that face, and the indices of the 
Wigner D-functions refer one to the centre of the 4-simplex, one end of the dual edge, and the other to a tetrahedron 
on the boundary, the other end of the dual edge. There is also an additional product of D-functions, one for each group
 element assigned to an edge exposed on the boundary, and not integrated over because we are working with fixed connection
 on the boundary. The index of the $D$ function on the dual link referring to the tetrahedron on the boundary is contracted
 with one index of a D-function for an element
attached to (and only to) a link which is exposed on the boundary. The other index of each matrix for an exposed link 
(referring to a triangle) is contracted with the index coming from the D-function
referring to the same triangle (see figure).  

\begin{figure}
\begin{center}
\includegraphics[width=8cm]{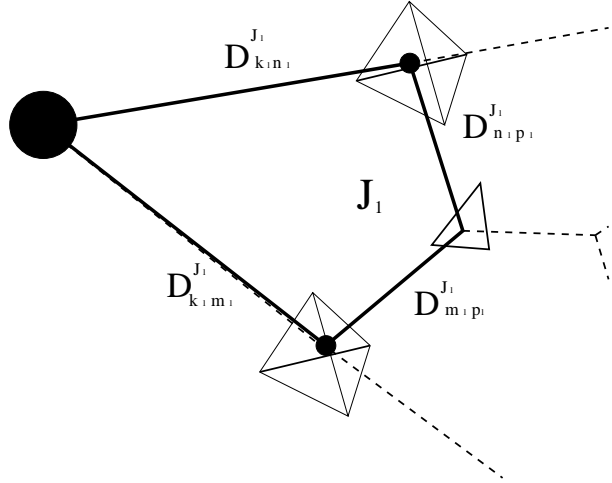}
\caption{The assignment of representation functions to the links of a wedge, with both interior links and links exposed on the boundary}
\end{center}
\end{figure}

It is crucial to note that the group elements attached to the links
exposed on the boundary for each wedge are {\it not} integrated over,
since we are working with fixed connection on the boundary, a boundary
condition which can be easily shown to not require any additional boundary
term in the classical action (see \cite{OLough} for the 3-dimensional
case).

Now we want to go from BF theory to gravity (Plebanski) theory by
imposing the Barrett-Crane constraints on the BF partition
function. These are quantum constraints on the representations of SO(4) which
are assigned to each triangle of the triangulation, so they can be
imposed at this \lq\lq quantum" level.
The constraints are essentially two: the simplicity constraint,
saying that the representations by which we label the triangles are to
be chosen from the simple representations of SO(4) (Spin(4)), and the closure
constraint, saying that the tensor assigned to each tetrahedron has to
be an invariant tensor of SO(4) (Spin(4)). As we have chosen real representations, it is easier to impose the 
first constraint, and the decomposition constraint will be imposed automatically in the following. We can 
implement the simplicity constraint
at this level by requiring that all the representation functions have
to be invariant under the subgroup SO(3) of SO(4), so realizing these representations in the space of harmonic 
functions over the coset ${SO(4)}/{SO(3)}\simeq S^{3}$, which is a 
complete characterization of the simple representations. We then implement the closure 
constraint
by requiring that the amplitude for a tetrahedron is invariant under a
general SO(4) transformation.
We note that these constraints have the effect of breaking the topological invariance of the theory.
Moreover, from now on we can replace the integrals over Spin(4) with integrals over SO(4), and the sum with a sum 
over the SO(4) representations only. In the Lorentzian case
everything works the same way, with $SL(2,C)$ and $SU(2)$
instead of $SO(4)$ and $SO(3)$ \cite{BC2,P-R2}, with the simple
representations given in this case by those labelled only by the
continuous parameter $\rho$.

Note that we are {\it not} imposing any projection in the boundary terms,
so that these are the same as those in pure BF theory. However, the
projection over simple representations in the edge amplitude imposes
automatically the simplicity also of the representations entering in the
D-functions for the exposed edges.

Consequently we write:
\bes 
\lefteqn{Z_{BC}=\sum_{J_{\sigma},\{k_{e}\}}\left(\prod_{\sigma}\Delta_{J_{\sigma}}\right)}
\nonumber \\ && \prod_{e}\int_{SO(4)}dg_{e}\int_{SO(3)}dh_{1}\int_{SO(3)}dh_{2}\int_{SO(3)}dh_{3}\int_{SO(3)}dh_{4}\int_{SO(4)}dg'_{e}
\nonumber \\ && D_{k_{e1}m_{e1}}^{J_{1}^{e}}(g_{e}h_{1}g'_{e})D_{k_{e2}m_{e2}}^{J_{2}^{e}}(g_{e}h_{2}g'_{e})D_{k_{e3}m_{e3}}^{J_{3}^{e}}(g_{e}h_{3}g'_{e})D_{k_{e4}m_{e4}}^{J_{4}^{e}}(g_{e}h_{4}g'_{e}) \left(\prod_{\tilde{e}}D\right)\;\;\;\;\;\;\;\;\nonumber \\ 
&=& \sum_{J_{\sigma},\{k_{e}\}}\left(\prod_{\sigma}dim_{J_{\sigma}}\right)\prod_{e}\,A_{e}\left(\prod_{\tilde{e}}D\right).  
\ees
Let us consider now the amplitude for each edge $e$ of the dual
complex (corresponding to a tetrahedron of the 4-simplex):
\bes
\lefteqn{A_{e}=\int_{SO(4)}dg_{e}\int_{SO(3)}dh_{1}\int_{SO(3)}dh_{2}\int_{SO(3)}dh_{3}\int_{SO(3)}dh_{4}\int_{SO(4)}dg'_{e}\,}
\nonumber \\ 
&& D_{k_{e1}m_{e1}}^{J_{1}^{e}}(g_{e}h_{1}g'_{e})D_{k_{e2}m_{e2}}^{J_{2}^{e}}(g_{e}h_{2}g'_{e})D_{k_{e3}m_{e3}}^{J_{3}^{e}}(g_{e}h_{3}g'_{e})D_{k_{e4}m_{e4}}^{J_{4}^{e}}(g_{e}h_{4}g'_{e})
\label{eq:partfunc}\ees
for a particular tetrahedron (edge) made out of the triangles
1,2,3,4. Performing explicitely all the integrals, the amplitude for a
single tetrahedron on the boundary turns out to be:
\bes 
A_{e}\,=\,\sum_{simple\;I,L}\frac{1}{\sqrt{\Delta_{J_{1}}\Delta_{J_{2}}\Delta_{J_{3}}\Delta_{J_{4}}}}C_{k_{1}k_{2}k_{3}k_{4}}^{J_{1}J_{2}J_{3}J_{4}I}C_{m_{1}m_{2}m_{3}m_{4}}^{J_{1}J_{2}J_{3}J_{4}L}
\nonumber
\\
=\frac{1}{\sqrt{\Delta_{J_{1}}\Delta_{J_{2}}\Delta_{J_{3}}\Delta_{J_{4}}}}B_{k_{1}k_{2}k_{3}k_{4}}^{J_{1}J_{2}J_{3}J_{4}}B_{m_{1}m_{2}m_{3}m_{4}}^{J_{1}J_{2}J_{3}J_{4}},
\ees
where the $B$'s are the (un-normalized) Barrett-Crane intertwiners,
$\Delta_{J}$ is the dimension of the representation $J$ of $SO(4)$, and all the representations for faces and edges in the
sum are now constrained to be simple. Considering the usual decomposition $Spin(4)\simeq SU(2)\times SU(2)$ a representation
 $J$ of $SO(4)$ corresponds, as we said, to a pair of $SU(2)$ representations $(j,k)$, so that
 its dimension is $(2j+1)(2k+1)$. This means that a simple representation $J$ of $SO(4)$ would have dimension $(2j+1)^{2}$, 
where $j$ is the corresponding $SU(2)$ representation. If we had performed only the integrals over $SO(3)$ we would have obtained 
the invariant vectors $w^J$'s contracting one index in each $D$ function inside the integrals over the group, and we would have had 
exactly the integral expression we have given above for the Barrett-Crane intertwiners. 

Explicitly, the (un-normalized) Barrett-Crane intertwiners are given by \cite{BC,DP-F-K-R, P-R} (this definition differs by 
a factor $\frac{1}{\sqrt{\Delta_1\Delta_2\Delta_3\Delta_4}}$ from the definition as an integral we have given in the first place, but this is not at all crucial here):
\bes
B_{k_{1}k_{2}k_{3}k_{4}}^{J_{1}J_{2}J_{3}J_{4}}\,=\,\sum_{simple\;I}C_{k_{1}k_{2}k_{3}k_{4}}^{J_{1}J_{2}J_{3}J_{4}I}\,=\,\sum_{simple\;I}\sqrt{\Delta_{I}}C_{k_{1}k_{2}k}^{J_{1}J_{2}I}C_{k_{3}k_{4}k}^{J_{3}J_{4}I}
\ees
where $I$ labels the representation of $SO(4)$ that can be thought as assigned to the tetrahedron whose triangles are 
instead labelled by the representations $J_{1},...,J_{4}$, $C_{k_{1}k_{2}k_{3}k_{4}}^{J_{1}J_{2}J_{3}J_{4}I}$ is an ordinary $SO(4)$ intertwiner between four representations, and finally $C_{k_{1}k_{2}k}^{J_{1}J_{2}I}$ are Wigner $3j$-symbols.

Note that the simplicity of the representations labelling the
tetrahedra, i.e. that appearing in the decomposition of the intertwiner into trivalent ones (the third of the Barrett-Crane constraints) comes 
automatically, without the need to impose it explicitly.
We note also that because of the restriction to the simple representations of the 
group, the result we end up with is independent of having started from the Spin(4) or the SO(4) BF partition 
function.

We see that each tetrahedron on the boundary of the 4-simplex contributes with two Barrett-Crane
intertwiners, one with indices referring to the centre of the 4-simplex
and the other indices referring to the centre of the tetrahedron
itself (see figure 4.8).

\begin{figure}
\begin{center}
\includegraphics[width=8cm]{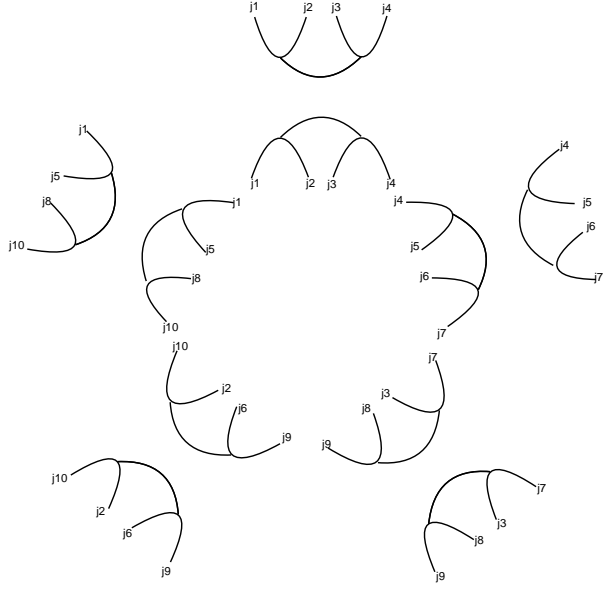}
\caption{Diagram of a 4-simplex, indicating the two Barrett-Crane intertwiners assigned to each tetrahedron}
\end{center}
\end{figure}

The partition function for this theory (taking into account all the
different tetrahedra) is then given by:
\bes 
\lefteqn{Z_{BC}=\sum_{\{J\},\{k\},\{n\},\{l\},\{i\},\{m\}}\Delta_{J_{1}}...\Delta_{J_{10}}\frac{1}{(\Delta_{J_{1}}...\Delta_{J_{10}})}}
\nonumber
\\
&&
B_{k_{1}k_{2}k_{3}k_{4}}^{J_{1}J_{2}J_{3}J_{4}}B_{l_{4}l_{5}l_{6}l_{7}}^{J_{4}J_{5}J_{6}J_{7}}B_{n_{7}n_{3}n_{8}n_{9}}^{J_{7}J_{3}J_{8}J_{9}}B_{h_{9}h_{6}h_{2}h_{10}}^{J_{9}J_{6}J_{2}J_{10}}B_{i_{10}i_{8}i_{5}i_{1}}^{J_{10}J_{8}J_{5}J_{1}} \nonumber
\\ && B_{m_{1}m_{2}m_{3}m_{4}}^{J_{1}J_{2}J_{3}J_{4}}B_{m_{4}m_{5}m_{6}m_{7}}^{J_{4}J_{5}J_{6}J_{7}}B_{m_{7}m_{3}m_{8}m_{9}}^{J_{7}J_{3}J_{8}J_{9}}B_{m_{9}m_{6}m_{2}m_{10}}^{J_{9}J_{6}J_{2}J_{10}}B_{m_{10}m_{8}m_{5}m_{1}}^{J_{10}J_{8}J_{5}J_{1}}\left(\prod_{\tilde{e}}D\right).\;\;\;\;\;
\ees

Now the product of the five Barrett-Crane intertwiners with indices
$m$ gives just the Barrett-Crane amplitude for the 4-simplex to which the
indices refer, given by a 15j-symbol constructed out of the 10 labels of the triangles and the 5 labels of the 
tetrahedra (see figure 4.9), 
\begin{figure}
\begin{center}
\includegraphics[width=8cm]{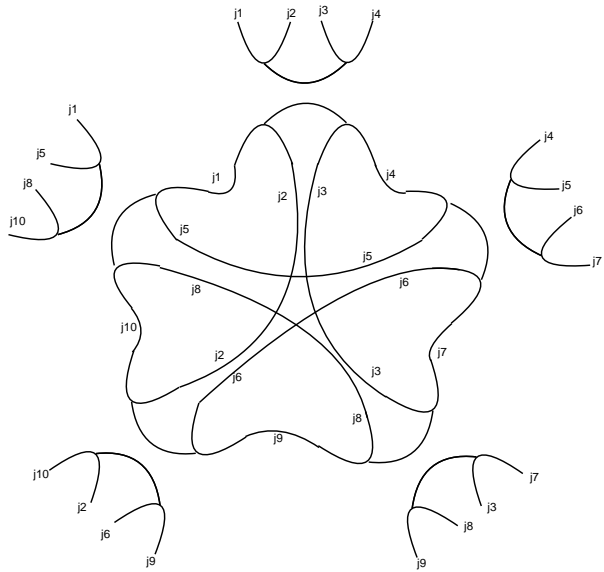}
\caption{Schematic representation of the Barrett-Crane amplitude for a 4-simplex}
\end{center}
\end{figure}
so that we can write down explicitly the state sum
for a manifold consisting of a single 4-simplex as:
\bes
Z_{BC}=\sum_{\{j_{f}\},\{k_{e'}\}}\prod_{f}\Delta_{j_{f}}\prod_{e'}\frac{B_{k_{e'1}k_{e'2}k_{e'3}k_{e'4}}^{j_{e'1}j_{e'2}j_{e'3}j_{e'4}}}{\sqrt{\Delta_{j_{e'1}}\Delta_{j_{e'2}}\Delta_{j_{e'3}}\Delta_{j_{e'4}}}}\prod_{v}\mathcal{B}_{BC}\;\left(\prod_{\tilde{e}}D\right), \label{eq:BC4s}
\ees
where it is understood that there is only one vertex, $\mathcal{B}_{BC}$ is the Barrett-Crane amplitude for a 
4-simplex, and the notation $e'i$ means that we are referring to the face $i$ (in some given ordering) of the 
tetrahedron $e'$, which is on the boundary of the 4-simplex, or equivalently to the i-th 2-simplex of the four which are
 incident to the dual edge (1-simplex) $e'$ of the spin foam (dual 2-complex), which is open, i.e. not ending on any 
other 4-simplex. Also the D-functions for the exposed edges are constrained to be in the simple representation.
The boundary terms are given by one Barrett-Crane
intertwiner for each tetrahedron on the boundary, and one D-function for
each group element on each of the exposed edges, contracted with the
intertwiner to form a group invariant, plus a \lq\lq
regularizing" factor in the denominator.

\subsection{Gluing 4-simplices and the state sum for a general manifold with boundary} \label{sec:BC1} 
Now consider the problem of gluing two 4-simplices together along a common tetrahedron, say, 1234. 
Having the state sum for a single 4-simplex, we consider two adjacent 4-simplices separately, so considering the
common tetrahedron in the interior twice (as being in the boundary of two different 4-simplices), and glue 
them together along it.  

The gluing is done simply by multiplying the two single partition functions, and
imposing that the values of the representations and of the projections (the $k_{e'i}$'s) of the
common tetrahedron are of course the same in the two partition
functions (this comes from the integration over the group elements assigned to the exposed edges that
 are being glued and become part of the interior, and thus have to be integrated out).
More precisely, the gluing of 4-simplices is now simply done by multiplying the partition
functions for the individual 4-simplices, and integrating over the group
variables that are not anymore on the boundary of the manifold, and
required to be equal in the two 4-simplices, again because we are working
with fixed connection on the boundary, so that the boundary data of the
two 4-simplices being glued have to agree.
These group variables appear only in two exposed edges each, and the
orthogonality between D-functions forces the representations corresponding
to the two wedges to be equal:

\bes
\int_{SO(4)}dg
D^{J}_{kl}(g)\,D^{J'}_{mn}(g)\,=\,\f{1}{\Delta_{J}}\delta_{km}\delta_{ln}\delta_{JJ'};
\label{eq:ort} \ees    
moreover, the factors $\f{1}{\Delta_{J}}$ compensate for having two wedges
corresponding to the same triangle, so that to each plaquette of the dual
complex, or triangle of the triangulation, corresponds still only a factor
$\Delta_{J}$ in the partition function. Finally, the equality of the
matrix indices in the previous relation forces the Barrett-Crane
intertwiners corresponding to the same shared tetrahedron to be fully
contracted.

Everything in the state sum is unaffected by the gluing, except for the
common tetrahedron, which now is in the interior of the manifold.
In this naive sense we could say that this way of gluing is local,
because it depends only on the parameters of the 
common tetrahedron, i.e. it should be determined only by the two
boundary terms which are associated with it when it is 
considered as part of the two different 4-simplices that are being glued.
What exactly happens for the amplitude of this interior tetrahedron
is:
\bes
\sum_{\{m\}}\frac{B_{m_{1}m_{2}m_{3}m_{4}}^{J_{1}J_{2}J_{3}J_{4}}}{\sqrt{\Delta_{J_{1}}\Delta_{J_{2}}\Delta_{J_{3}}\Delta_{J_{4}}}}\frac{B_{m_{1}m_{2}m_{3}m_{4}}^{J_{1}J_{2}J_{3}J_{4}}}{\sqrt{\Delta_{J_{1}}\Delta_{J_{2}}\Delta_{J_{3}}\Delta_{J_{4}}}}=\sum_{\{m\},I,L}\frac{C_{m_{1}m_{2}m_{3}m_{4}}^{J_{1}J_{2}J_{3}J_{4}I}C_{m_{1}m_{2}m_{3}m_{4}}^{J_{1}J_{2}J_{3}J_{4}L}}{\left(\Delta_{J_{1}}\Delta_{J_{2}}\Delta_{J_{3}}\Delta_{J_{4}}\right)^{2}}\nonumber
\\ = \sum_{I,L}\frac{\sqrt{\Delta_{I}\Delta_{L}}C_{m_{1}m_{2}m}^{J_{1}J_{2}I}C_{m_{3}m_{4}m}^{J_{1}J_{2}I}C_{m_{1}m_{2}n}^{J_{1}J_{2}L}C_{m_{3}m_{4}n}^{J_{3}J_{4}L}}{\left(\Delta_{J_{1}}\Delta_{J_{2}}\Delta_{J_{3}}\Delta_{J_{4}}\right)}
\nonumber \\ = \sum_{I,L}\frac{\Delta_{I}\delta_{IL}\delta_{mn}}{\left(\Delta_{J_{1}}\Delta_{J_{2}}\Delta_{J_{3}}\Delta_{J_{4}}\right)} = \sum_{I}\frac{\Delta_{I}}{\left(\Delta_{J_{1}}\Delta_{J_{2}}\Delta_{J_{3}}\Delta_{J_{4}}\right)}\,=\,\frac{\Delta_{1234}}{\Delta_{J_{1}}\Delta_{J_{2}}\Delta_{J_{3}}\Delta_{J_{4}}}\label{eq:edam}
\ees
where we have used the orthogonality between the intertwiners, and $I$ labels the interior edge (tetrahedron), while the 
quantity $\Delta_{1234}$ represent the number of possible intertwiners between the representations $J_1 - J_4$.

We see that the result of the gluing is the insertion of an
amplitude for the tetrahedra (dual edges) in the interior of the
triangulated manifold, and of course the disappearance of the boundary terms $B$ since the tetrahedron is not anymore 
part of the boundary of the new manifold (see figure 4.10). In other words, the gluing is not trivial, in the sense that the end result is
not just the product of pre-existing factors, but includes something
resulting from the gluing itself (the factor $\Delta_{1234}$).

\begin{figure}
\begin{center}
\includegraphics[width=8cm]{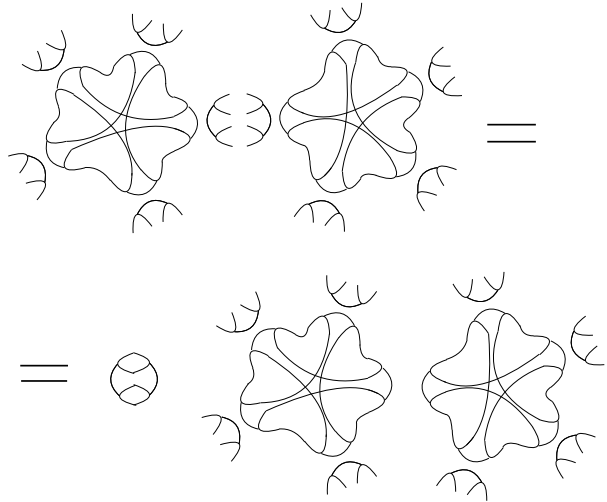}
\caption{The gluing of two 4-simplices along a common tetrahedron}
\end{center}
\end{figure}

We can now write down explicitly the state sum for a manifold with
boundary which is then constructed out of an arbitrary number of 4-simplices, and
has some tetrahedra on the boundary and some in the interior, with fixed connection on the boundary:
\bes
Z_{BC}=\sum_{j_{f},k_{e'}}\prod_{f}\Delta_{j_{f}}
\prod_{e'}\frac{B_{k_{e'1}k_{e'2}k_{e'3}k_{e'4}}^{j_{e'1}j_{e'2}j_{e'3}j_{e'4}}}
{\sqrt{\Delta_{j_{e'1}}\Delta_{j_{e'2}}\Delta_{j_{e'3}}\Delta_{j_{e'4}}}}\prod_{e}\frac{\Delta_{1234}}
{\left(\Delta_{j_{e1}}\Delta_{j_{e2}}\Delta_{j_{e3}}\Delta_{j_{e4}}\right)}\prod_{v}\mathcal{B}_{BC}
\;\left( \prod_{\tilde{e}}D\right),\,\,\,\,\,\,\,\label{eq:final}
\ees
where the $\{e'\}$ and the $\{e\}$ are the sets of boundary and interior edges of the spin foam, respectively, 
while the $\tilde{e}$ are the remaining exposed edges, where the boundary connection
data are located. The partition function is then a function of the
connection, i.e. of the group elements on the exposed edges.

It is important to note that the number of parameters which determine the gluing and that in the end characterize the 
tetrahedron in the interior of the manifold is five (4 labels for the faces and one for the tetrahedron itself), which is 
precisely the number of parameters necessary in order to determine  a first quantized geometry of a tetrahedron 
\cite{bb}, as we have seen. 

Moreover, the partition function with which we ended, apart from the boundary terms, is the one obtained in 
\cite{P-R} by studying a generalized matrix model that we will describe in the following, and shown to be
finite at all orders in the sum over the 
representations \cite{P-R,Per}. The same finiteness was proven for the Lorentzian counterpart of this model \cite{CPR}.
 
One can proceed analogously in the Lorentzian case, using the integral
representation of the Barrett-Crane intertwiners (the resulting expression
is of course more complicated, but with the same structure), and their
formula for the evaluation of relativistic (simple) spin networks. All
the passages above, in fact, amount to the evaluation of spin networks,
which were proven to evaluate to a finite number \cite{bb2}, so the procedure above
can be carried through similarly and sensibly.
   
We see that, starting from a ill-defined BF partition function, the
imposition of the constraints has made the resulting partition function
for gravity finite both in the Riemannian and Lorentzian cases
\cite{P-R,Per,CPR}.

Several comments are opportune at this point.

The gluing procedure used above is consistent with the formalism developed for general spin foams \cite{Baez,Baez2}, saying that when we glue two manifolds $\mathcal{M}$ and $\mathcal{M}'$ along a common 
boundary, the partition functions associated to them and to the composed manifold satisfy 
$Z(\mathcal{M})Z(\mathcal{M}')=Z(\mathcal{M}\mathcal{M}')$, as is easy to verify; see \cite{Baez,Baez2}
 for more details. 

Also, this result suggests that the best way of deriving a state sum  that implements the 
Barrett-Crane constraints from a generalized matrix model (we will discuss this topic in the next chapter) is like in \cite{P-R}, i.e. imposing the constraints on the 
representations only in the interaction term of the field over a group manifold, because this derivation leads to the 
correct edge amplitudes coming from the gluing, and these amplitudes
are not present in \cite{DP-F-K-R}.
 
Regarding the regularization issue, it seems that even starting from a discretized action in which the sum over the 
representations is not convergent, we end up with a state sum which
is finite at all orders, according to the results of
\cite{P-R,Per}. Anyway another way to regularize completely the state sum model, making it finite at all orders, is to use a quantum group at a root of unity so that the 
sum over the representations is automatically finite due to the finiteness of the number of representations of any such quantum group; in this case we have only to replace the elements of the state sum coming from the recoupling 
theory of SO(4) (intertwiners and 15j-symbols) with the corresponding objects for the quantum deformation of it. This procedure would produce a 4-dimensional analogue of the Turaev-Viro model.    

The structure of the state sum and the form of the boundary terms is a very close analogue of that discovered in 
\cite{CCM1,CCM2} for SU(2) topological field theories in any 
number of dimensions, the difference being the group used, of course, and the absence of any constraints on the 
representations so that the topological invariance is maintained.
   
Of course, in order to obtain a complete spin foam model for gravity we should implement a sum over triangulations
 or over dual 2-complexes, in some form, and both the implementation itself and the issue of its convergence are still open
 problems, not completely solved yet;  we note that a way to  implement naturally a sum over spin foams giving also a sum
 over topologies is given by the generalized matrix models or group field theories that we will discuss in the following.

Finally, the results reviewed of \cite{FKP,FK} allow for a generalization of this procedure for
deriving the Barrett-Crane spin foam model to any dimension \cite{OW}.

\subsection{Generalization to the case of arbitrary number of dimensions} \label{sec:lgtBChigh}
It is quite straightforward to generalize our procedure and results
to an arbitrary number of dimensions of spacetime. 
Much of what we need in order to do it is already at our disposal. It was shown in \cite{FKP} that it is possible to 
consider gravity as a constrained BF theory in any dimension (incidentally, the same was proposed for supergravity 
\cite{Eza,L-S,L-S2}), and also the concept of simple spin networks was generalized to any dimensions in 
\cite{FK}, with the representations of $SO(D)$ ($Spin(D)$) required to be invariant under a general transformation of 
$SO(D-1)$, so that they are realized as harmonic functions over the homogeneous space 
$SO(D)/SO(D-1)\simeq S^{D-1}$, and so that the spin network itself can be thought as a kind of Feynmann 
diagram for spacetime. 

Also the construction of a complete hierarchy of discrete topological field theories in every dimension of spacetime 
performed in \cite{CCM2}, with a structure similar to that one we propose for the Barrett-Crane model, represents 
an additional motivation for doing this.

We take as the discretized  partition function for BF theory in an arbitrary 
D-dimensional (Riemannian) (triangulated) spacetime for any compact group $G$ and in particular for SO(D) the expression:

\bes
Z_{BF}\,=\,\left( \prod_{e}\int_{G}dg_{e}\right)\prod_{f}\sum_{J_{f}}\,\Delta_{J_{f}}\,\chi_{J_{f}}
\left(\prod_{e'\in\partial f}g_{e'}\right)
\ees
where $e'$ is a dual link on the boundary of the dual plaquette $f$ (face of the spin foam) associated with a triangle 
$t$ in the triangulation, $e$ indicates the set of dual links, and the character is in the representation $J_{f}$ of the 
group $G$.

A way to justify this heuristically (for a more rigorous discretization leading to the same result, see \cite{FK})  is the 
following. 

We start from a discretized action like the one we used before
\bes S_{BF}\,=\,\sum_{t}\,B(t)\,F(t), \ees and with: $e^{iF(t)}=\prod_{e'\in\partial f}g_{e'}$. 
With this action the partition function for the theory becomes:
\bes
\lefteqn{Z_{BF}\,=\,\int_{G}\mathcal{D}A\int_{G}\mathcal{D}B(t)\,e^{i\sum_{t}B(t)F(t)}} \nonumber \\ 
&=&\,\int_{G}\mathcal{D}A\int_{G}\mathcal{D}B(t)\prod_{t}e^{i B(t)F(t)}\,=\,\int_{G}\mathcal{D}A\,\prod_{t}\,\delta\left( e^{iF(t)}\right) \nonumber \\
&=&\, \prod_{e}\int_{G}dg_{e}\prod_{f}\delta\left( \prod_{e'\in\partial f}g_{e'}\right) 
\ees
with the notation as above, and having replaced the product over the $(D-2)$-simplices with a product over the faces of the 
dual triangulation (plaquette), that is possible because they are in 1-1 correspondence.

Now we can use the decomposition of the delta function of a group
element into a sum of characters, obtaining:
\bes
Z_{BF}\,=\,\left( \prod_{e}\int_{G}dg_{e}\right)\prod_{f}\sum_{J_{f}}\,\Delta_{J_{f}}\,\chi_{J_{f}}\left(\prod_{e'\in\partial f}g_{e'}\right)
\ees
i.e. the partition function we were trying to derive.

From now on we can proceed as for SO(4) in 4 dimensions. Consider $G=SO(D)$, and the $J$'s as the highest 
weight labelling the representations of that group.

We can decompose the characters into a product of D-functions, and rearrange the sums and products in the 
partition function to obtain:
\bes
Z_{BF}\,=\,\sum_{\{J_{f}\}, \{k\},\{m\}}\prod_{f}\Delta_{f}\prod_{e}\,A_{e}\left(\prod_{\tilde{e}}D\right) \ees
where
\bes
A_{e}\,=\,\int_{SO(D)}dg_{e}D_{k_{e1}m_{e1}}^{J_{e1}}(g_{e})...D_{k_{eD}m_{eD}}^{J_{eD}}(g_{e}) \ees
where $ei$ labels the i-th of the D faces incident on the edge $e$.

Now we can apply our procedure and insert here the Barrett-Crane constraints:
\bes
\lefteqn{A_{e}\,=\,\int_{SO(D)}dg_{e}\int_{SO(D-1)}dh_{1}...\int_{SO(D-1)}dh_{D}\int_{SO(D)}dg'_{e}}
\nonumber \\ &&D_{k_{e1}m_{e1}}^{J_{e1}}(g_{e}h_{1}g'_{e})...D_{k_{eD}m_{eD}}^{J_{eD}}(g_{e}h_{D}g'_{e}). \ees

Performing the integrals, and carrying on the same steps as in 4 dimensions, leads to the
 analogue of the formula (~\ref{eq:final}) in higher dimensions (or
 alternatively to the analogue of case A in \cite{DP-F-K-R}):
\bes
\lefteqn{Z_{BC}^{D}\,=\,\sum_{\{j_{f}\},\{J_{e}\},\{J_{e'}\},\{k_{e'}\},\{K_{e'}\}}\,\prod_{f}
 \Delta_{j_{f}}\prod_{e'}\sqrt{\Delta_{J_{e'1}}...\Delta_{J_{e'(D-3)}}}}\nonumber\\ &&\frac{C^{j_{e'1}j_{e'2}J_{e'1}}_{k_{e'1}k_{e'2}K_{e'1}}C^{j_{e'3}J_{e'2}J_{e'1}}_{k_{e'3}K_{e'2}K_{e'1}}...C^{j_{e'(D-2)}J_{e'(D-4)}J_{e'(D-3)}}_{k_{e'(D-2)}K_{e'(D-4)}K_{e'(D-3)}}C^{j_{e'(D-1)}j_{e'D}J_{e'(D-3)}}_{k_{e'(D-1)}k_{e'D}K_{e'(D-3)}}}{\Delta_{j_{e'1}}...\Delta_{j_{e'D}}} \nonumber \\ && \prod_{e}\frac{\Delta_{J_{e1}}...\Delta_{J_{e(D-3)}}}{\left( \Delta_{j_{e1}}...\Delta_{j_{eD}}\right)^{2}}\,\prod_{v}\mathcal{B}_{BC}^{D}\;\left(\prod_{\tilde{e}}D\right).
\ees
There are $D$ faces (corresponding to $(D-2)$-dimensional simplices) incident on each edge (corresponding to 
(D-1)-dimensional simplices) $e$ ($(D-1)$-simplex in the interior) or $e'$ ($(D-1)$-simplex on the boundary).  
There are $D+1$ edges for each vertex (corresponding to a $D$-dimensional simplex), and consequently $D(D+1)/2$ 
faces for each $D$-simplex. Each edge is labelled by a set of $(D-3)$ $J$'s. $\mathcal{B}_{BC}^{D}$ is the higher 
dimensional analogue of the Barrett-Crane amplitude, i.e. (the $SO(D)$ analogue of) the $\frac{3}{2}(D+1)(D-2)J$-symbol constructed out of the $D(D+1)/2$ labels of the faces and the $(D-3)(D+1)$ labels of the edges.

Again, this result is a very close analogue of the state sum for a topological field theory in general dimension, obtained 
in \cite{CCM2}.

Of course, everywhere we are summing over only simple representations of $SO(D)$, i.e. representations of $SO(D)$ that
 are of class 1 with respect to the subgroup $SO(D-1)$ \cite{VK}.

\section{Derivation with different boundary conditions and possible variations of the procedure}
In this section, we extend the derivation described above to other
choices of boundary conditions, following an analogous study for
3-dimensional gravity \cite{OLough}, obtain the corresponding boundary
terms in the partition function for a single 4-simplex and then apply
again the gluing procedure to get the full partition function for the
triangulated manifold. Apart from giving the correct boundary terms in
this case, this serves as a consistency check for the previous derivation.
In fact it is of course to be expected that the amplitudes for the
elements in the interior of the manifold, the edge amplitudes in
particular, should not be affected by the choice of boundary conditions in
the 4-simplices (having boundaries) whose gluing produces them. The
result is that the derivation above and in \cite{OW} is indeed consistent, and
we get again the Perez-Rovelli version of the Barrett-Crane model. We then
examine a few alternatives to the procedure used above and in \cite{OW},
exploiting the freedom left by that derivation. In particular, we study
the effect of imposing the projection over the simple representations also
in the boundary terms, since this may (naively) recall the imposition of the
simplicity constraints in the kinetic term in the field theory over the
group manifold, leading to the DePietri-Freidel-Krasnov-Rovelli version of
the Barrett-Crane model \cite{DP-F-K-R}. Instead, this leads in the
present case to several drawbacks, as we discuss, and to a model which is
not the DePietri-Freidel-Krasnov-Rovelli version and it is not consistent,
in the sense specified above, with respect to different choices of
boundary conditions.  Moreover, we study and discuss the model that can be
obtained by not imposing the gauge invariance of the edge amplitude (as
required in \cite{OW}), since it was mentioned in \cite{hendryk},
explaining why we do not consider it a viable version of the Barrett-Crane
model, and finally the class of models that can be obtained by
imposing the two projections (simplicity and gauge invariance) more
that once. All the calculations in this section will be performed explicitely
for the Riemannian case, but are valid in (or can be easily extended to)
the Lorentzian case as well. The results presented in this section were published in \cite{boundary}.

\subsection{Fixing the boundary metric} \label{sec:bound}
Let us now study the case in which we choose to fix the B field on the
boundary (i.e. by the metric field), and let us analyse first the
classical action.

We note here that the partition function we will obtain in this section,
being a function of the representations $J$ (or $\rho$) of the group
$SO(4)$ (or $SL(2,C )$ assigned to the boundary, and representing
the B (metric) field, can be thought of as the Fourier transform of the one we ended up with in the previous section, being
instead a function of the group elements, representing the connection field.

The $so(4)$ Plebanski action is:
\bes
S\,=\,\int_{\mathcal{M}}B\wedge F\,-\,\frac{1}{2}\phi\,B\,\wedge\,B
\ees
so that its variation is simply given by:
\bes
\delta S\,=\,\int_{\mathcal{M}}\delta B\wedge \left(F\,-\,\phi\,B\right) +
\delta A \wedge \left(dB\,+\,A\wedge B\,+\,B\wedge A\right) \,-\,
\int_{\partial \mathcal{M}} B\wedge\delta A,
\ees
and we see that fixing the connection on the boundary does not require any
additional boundary term to give a well-defined variation, i.e. the field
equations resulting from it are not affected by the presence of a
boundary.

On the other hand, if we choose to fix the B field on the boundary, we
need to introduce a boundary term in the action:
\bes
S\,=\,\int_{\mathcal{M}}B\wedge
F\,-\,\frac{1}{2}\phi\,B\,\wedge\,B\,+\,\int_{\partial\mathcal{M}}
B\wedge A
\ees
so that the variation leads to:
\bes
\delta S\,=\,\int_{\mathcal{M}}\delta B\wedge \left(F\,-\,\phi\,B\right) +
\delta A \wedge \left(dB\,+\,A\wedge B\,+\,B\wedge A\right) \,+\,
\int_{\partial \mathcal{M}} \delta B\wedge A,
\ees
and to the usual equations of motion.  

Now we want to find what changes in the partition function for a single
4-simplex if we decide to fix the B field on the boundary, and then to
study how the gluing proceeds in this case. 

The additional term in the partition function resulting from the
additional term in the action is $\exp{\int_{\partial\mathcal{M}}B\wedge
A}$. We have to discretize it, expressing it in terms of representations
$J$ and group elements $g$ on the boundary, and then multiply it into the
existing amplitude. The connection terms on the boundary are then to be
integrated out, since they are not held fixed anymore, while the sums over
the representations have to be performed only on the bulk ones, i.e. only
on the representations labelling the triangles in the interior of the
manifold (none in the case of a single 4-simplex).
A natural discretization \cite{OLough} for the additional term is:
\bes
\exp{\int_{\partial\mathcal{M}}B\wedge
A}\,=\,\prod_{\tilde{e}}\chi^{J}(g_{\tilde{e}})\,=\,\prod_{\tilde{e}}D_{kk}^{J}(g_{\tilde{e}})
\label{eq:bt}
\ees  
where the representation $J$ is the one assigned to a wedge with edges
exposed on the boundary, and $g_{\tilde{e}}$ is actually the product
$g_{1}g_{2}$ of the group elements assigned to the two edges exposed on the
boundary, and the product runs over the exposed parts of the wedges. 

We multiply the partition function given in the previous section by this extra term, and
integrate over the group elements, simultaneously dropping the sum over
the representations, since all the wedges are on the boundary, and thus
all the representations are fixed. 

Using again the orthogonality of the D-functions, eq.~\ref{eq:ort}, the
result is the following:
\bes
Z_{BC}=\,\prod_{f}\Delta_{J_{f}}\prod_{e'}\frac{B_{k_{e'1}k_{e'2}k_{e'3}k_{e'4}}^{J_{e'1}J_{e'2}J_{e'3}J_{e'4}}}{\Delta_{J_{e'1}}\Delta_{J_{e'2}}\Delta_{J_{e'3}}\Delta_{J_{e'4}}}\prod_{v}{\bf
B}_{BC}
\label{eq:4s'}
\ees
where also the $k$ indices are fixed by the only constraint (coming
again from the integration over the group above) that the $k$s appearing
in different Barrett-Crane intertwiners but referring to the same triangle
must be equal. Of course we see that the partition function is now a
function of the representations $J$ on the boundary and of their
projections. The different power in the denominator of the boundary terms
is necessary to have consistency in the gluing procedure, as we will see.
Also, note that we did not impose any projection over the simple
representations in the boundary terms, i.e. in the D-functions coming from
the additional boundary term in the action, since we decided not to impose
it in the D-functions for the exposed edges. We will
analyse the alternatives to this choice in the next section.

Now we proceed with the gluing of 4-simplices. The different 4-simplices
being glued have to share the same boundary data for the common
tetrahedron, i.e. the representations $J$ and the projections $k$ in the
Barrett-Crane intertwiner referring to it have to agree. The gluing is
performed again by simply multiplying the partition functions for the two
4-simplices and summing over the $k$s, because they are now attached to a
tetrahedron in the interior of the manifold. In this way the Barrett-Crane
intertwiners corresponding to the same tetrahedron in the two 4-simplices
get contracted with each other, and they give again a factor
$\Delta_{1234}$ as before. The factors in the denominators of the
(ex-)boundary terms are multiplied to give a factor
$1/(\Delta_{J_{1}}\Delta_{J_{2}}\Delta_{J_{3}}\Delta_{J_{4}})^{2}$, but
since we have a factor of $\Delta_{J_{i}}$ for each wedge and for each
4-simplex, the factor in the denominator of the amplitude for the interior
tetrahedron is again
$1/(\Delta_{J_{1}}\Delta_{J_{2}}\Delta_{J_{3}}\Delta_{J_{4}})$. 

In the end the partition function for a generic manifold with boundary,
with the boundary condition being that the metric field is fixed on it,
is:
\bes
Z_{BC}=\sum_{\{J_{f}\}}\prod_{f}\Delta_{J_{f}}\prod_{e'}\frac{B_{k_{e'1}k_{e'2}k_{e'3}k_{e'4}}^{J_{e'1}J_{e'2}J_{e'3}J_{e'4}}}{\Delta_{J_{e'1}}\Delta_{J_{e'2}}\Delta_{J_{e'3}}\Delta_{J_{e'4}}}\prod_{e}\frac{\Delta_{1234}}{\Delta_{J_{e1}}\Delta_{J_{e2}}\Delta_{J_{e3}}\Delta_{J_{e4}}}\prod_{v}
{\bf B}_{BC}.
\ees

It is understood that the sum over the representations $J$s is only over
those labelling wedges (i.e. faces) in the interior of the manifold, the
others being fixed.

We recall that this can be understood as the probabiblity amplitude for
the boundary data, the representations of $SO(4)$ (or $SL(2,C)$)
in this case or the $SO(4)$ group elements as in the previous section, in
the Hartle-Hawking vacuum. If the boundary data are instead divided into
two different sets, then the partition function represents the transition
amplitude from the data in one set to those in the other.
The Lorentzian case, again, goes similarly, with the same result.

We see that, apart from the boundary terms, we ended up again with the
Perez-Rovelli version of the Barrett-Crane model. This was to be expected,
since the bulk partition function should not be affected by the boundary
conditions we have chosen for the single 4-simplices before performing the
gluing, but the fact that this is indeed the case represents a good
consistency check for the whole procedure we used to obtain the
Barrett-Crane model from a discretized BF theory. 

\subsection{Exploring alternatives}
Let us now go on to explore the alternatives to the procedure we have just
used, to see whether there are other consistent procedures giving
different results, i.e. different versions of the Barrett-Crane model.      
In particular we would like to see, for example, whether there is any variation of the
procedure used above resulting in the DePietri-Freidel-Krasnov-Rovelli
version of the Barrett-Crane model \cite{DP-F-K-R}, i.e. the other version
that can be derived from a field theory over a group manifold.
Again, the analogous calculations in the Lorentzian case go through
similarly. 

We have seen in the previous section that two choices were involved in
the derivation we performed: the way we imposed the constraints, with one
projection imposing simplicity of the representations and the other
imposing the invariance under the group of the edge amplitude, and the way
we treated the D-functions for the exposed edges, i.e. without imposing
any constraints on them. We will now consider alternatives to these
choices, starting from the last one. A few other alternatives to the first
were considered in \cite{OW}.

\subsubsection{Projections on the exposed edges} 
We then first leave the edge amplitude ~\ref{eq:edam} as it is, and look
for a way to insert an integral over the $SO(3)$ subgroup in the boundary
representation functions.
The idea of imposing the simplicity projections in the D-functions for the
exposed edges may (naively) resemble the imposition of them in the kinetic term in
the action for the field theory over a group, leading to the
DePietri-Freidel-Krasnov-Rovelli version of the Barrett-Crane model
\cite{DP-F-K-R}, since in both cases there are precisely 4 of them for
each tetrahedron, and they represent the boundary data to be transmitted
across the 4-simplices (in the connection representation). Anyway, this is
not the case, as we are going to show.

There are two possible ways of imposing the projections, corresponding to
the two possibilities of multiplying the arguments of the D-functions by
an $SO(3)$ element from the left or from the right, corresponding to
projecting over an $SO(3)$ invariant vector the indices of the D-functions
referring to the tetrahedra or those referring to the triangles (see
figure 4.7), then integrating over the subgroup as in ~\ref{eq:edam}, having
for each boundary term: 
\bes
\lefteqn{B_{k_{1}k_{2}k_{3}k_{4}}^{J_{1}J_{2}J_{3}J_{4}}D^{J_{1}}_{k_{1}m_{1}}(g_{1})D^{J_{2}}_{k_{2}m_{2}}(g_{2})D^{J_{3}}_{k_{3}m_{3}}(g_{3})D^{J_{4}}_{k_{4}m_{4}}(g_{4})\rightarrow}
\nonumber \\ &\rightarrow&
B_{k_{1}k_{2}k_{3}k_{4}}^{J_{1}J_{2}J_{3}J_{4}}D^{J_{1}}_{k_{1}l_{1}}(g_{1})D^{J_{2}}_{k_{2}l_{2}}(g_{2})D^{J_{3}}_{k_{3}l_{3}}(g_{3})D^{J_{4}}_{k_{4}l_{4}}(g_{4})w^{J_{1}}_{l_{1}}w^{J_{2}}_{l_{2}}w^{J_{3}}_{l_{3}}w^{J_{4}}_{l_{4}}w^{J_{1}}_{m_{1}}w^{J_{2}}_{m_{2}}w^{J_{3}}_{m_{3}}w^{J_{4}}_{m_{4}}
\nonumber \ees

or:

\bes
\lefteqn{B_{k_{1}k_{2}k_{3}k_{4}}^{J_{1}J_{2}J_{3}J_{4}}D^{J_{1}}_{k_{1}m_{1}}(g_{1})D^{J_{2}}_{k_{2}m_{2}}(g_{2})D^{J_{3}}_{k_{3}m_{3}}(g_{3})D^{J_{4}}_{k_{4}m_{4}}(g_{4})\rightarrow}
\nonumber \\ &\rightarrow&
B_{k_{1}k_{2}k_{3}k_{4}}^{J_{1}J_{2}J_{3}J_{4}}w^{J_{1}}_{k_{1}}w^{J_{2}}_{k_{2}}w^{J_{3}}_{k_{3}}w^{J_{4}}_{k_{4}}w^{J_{1}}_{l_{1}}w^{J_{2}}_{l_{2}}w^{J_{3}}_{l_{3}}w^{J_{4}}_{l_{4}}D^{J_{1}}_{l_{1}m_{1}}(g_{1})D^{J_{2}}_{l_{2}m_{2}}(g_{2})D^{J_{3}}_{l_{3}m_{3}}(g_{3})D^{J_{4}}_{l_{4}m_{4}}(g_{4})
\nonumber
\ees
where in the first case the second set of invariant vectors is contracted
with one coming from another boundary term, giving in the end no
contribution to the amplitude, while in the second case there is a
contraction between the Barrett-Crane intertwiners and these vectors,
giving a different power in the denominator, and the
disappearence of the intertwiners from the amplitude.

Let us discuss the first case. The effect of the projection is to break
the gauge invariance of the amplitude for a 4-simplex, and to decouple the
different tetrahedra on the boundary. In fact the amplitude for a
4-simplex is then:
 
\bes
Z=\sum_{\{J_{f}\},\{k_{e'}\}}\prod_{f}\Delta_{J_{f}}\prod_{e}
\frac{B_{k_{e1}k_{e2}k_{e3}k_{e4}}^{J_{e1}J_{e2}J_{e3}J_{e4}}}
{\left(\Delta_{J_{e1}}\Delta_{J_{e2}}\Delta_{J_{e3}}\Delta_{J_{e4}}\right)^{\frac{1}{2}}}D^{J_{e1}}_{k_{e1}l_{e1}}(g_{e1})
...D^{J_{e4}}_{k_{e4}l_{e4}}(g_{e4}) \nonumber \\ w^{J_{e1}}_{l_{e1}}w^{J_{e2}}_{l_{e2}}w^{J_{e3}}_{l_{e3}}w^{J_{e4}}_{l_{e4}}w^{J_{e1}}_{m_{e1}}w^{J_{e2}}_{m_{e2}}w^{J_{e3}}_{m_{e3}}w^{J_{e4}}_{m_{e4}}\prod_{v}{\bf
B}_{BC}
\label{eq:4s3}
\ees
which is not gauge invariant but only gauge covariant. 

This would be enough for discarding this variation of the procedure used
above as not viable. Nevertheless, this does not lead to any apparent
problem when we proceed with the gluing as we did previously. In fact, as
can be easily verified, the additional invariant vectors $w$ do not
contribute to the gluing, when the connection is held fixed at the
boundary, and the result is again the ordinary Perez-Rovelli version of
the Barrett-Crane model. The edge amplitude, i.e. the amplitude for the
tetrahedra in the interior of the manifold, is again given by
~\ref{eq:ampl}.
However, the inconsistency appears when we apply the \lq\lq consistency
check" used previously, i.e. when we study the gluing with different
boundary conditions. In fact, when the field $B$ is held fixed at the
boundary, we have to multiply again the partition function ~\ref{eq:4s3}
by the additional boundary terms ~\ref{eq:bt}, this time imposing the
simplicity projections here as well. 
The resulting 4-simplex amplitude is:     

\bes
Z=\,\prod_{f}\Delta_{J_{f}}\prod_{e'}\frac{B_{k_{e'1}k_{e'2}k_{e'3}k_{e'4}}^{J_{e'1}J_{e'2}J_{e'3}J_{e'4}}}{\left(\Delta_{J_{e'1}}\Delta_{J_{e'2}}\Delta_{J_{e'3}}\Delta_{J_{e'4}}\right)^{\frac{3}{2}}}\prod_{v}{\bf
B}_{BC}
\label{eq:4s3'}
\ees
and the gluing results in an edge amplitude for the interior tetrahedra: 
\bes
A_{e}=\frac{\Delta_{1234}}{\left(\Delta_{J_{1}}\Delta_{J_{2}}\Delta_{J_{3}}\Delta_{J_{4}}\right)^{2}}. 
\ees

This proves that this model is not consistent, since the result is
different for different boundary conditions, and shows also that, as we
said above, the \lq\lq consistency check" is not trivially satisfied by
every model. 
 
Considering now the second variation, we see that imposing the
simplicity constraint this way gives the same result as if we had imposed
it directly in the edge amplitude (~\ref{eq:edam}), having
$A_{e}(GR)=P_{h}P_{g}P_{h}A_{e}(BF)$. This, however, breaks the gauge
invariance of the edge amplitude, for which we were aiming when we imposed
the additional projection $P_{g}$. In turn this results in a breaking of
the gauge invariance of the 4-simplex amplitude. Because of this we do
not
explore any further this variation, but rather study directly the simpler
case in which we do not impose the projection $P_{g}$ at all in the edge
amplitude. Then we will give more reasons for imposing it. 

\subsubsection{Imposing the projections differently} \label{sec:otherpossi}
We then study the model obtained by dropping the projection $P_{g}$ in
(~\ref{eq:edam}), and not imposing any additional simplicity projection on
the D-functions for the exposed edges, since we have just seen that this
would lead to inconsistencies (more precisely, projecting the indices
referring to the triangles would lead to inconsistencies, while projecting
those referring to the tetrahedra would give exactly the same result as
not projecting at all, as can be verified).

The edge amplitude replacing (~\ref{eq:edam}) is then:

\bes
A_{e}=\frac{B_{k_{1}k_{2}k_{3}k_{4}}^{J_{1}J_{2}J_{3}J_{4}}}{\left(\Delta_{J_{1}}\Delta_{J_{2}}\Delta_{J_{3}}\Delta_{J_{4}}\right)^{\frac{1}{4}}}\,w^{J_{1}}_{m_{1}}w^{J_{2}}_{m_{2}}w^{J_{3}}_{m_{3}}w^{J_{4}}_{m_{4}}
\ees
and the partition function for a single 4-simplex is:

\bes
Z_{BC}=\sum_{\{J_{f}\},\{k_{e'}\}}\prod_{f}\Delta_{J_{f}}\prod_{e'}\frac{w^{J_{1}}_{m_{1}}w^{J_{2}}_{m_{2}}w^{J_{3}}_{m_{3}}w^{J_{4}}_{m_{4}}}{\left(\Delta_{J_{e'1}}\Delta_{J_{e'2}}\Delta_{J_{e'3}}\Delta_{J_{e'4}}\right)^{\frac{1}{4}}}\prod_{v}{\bf
B}_{BC}\;\left(\prod_{\tilde{e}}D\right), 
\ees
where the D-functions for the exposed edges are contracted not with the
Barrett-Crane intertwiners but with the $SO(3)$ invariant vectors
$w^{J}_{m}$. Consequently the partition function itself is not an
invariant under the group. However, let us go a bit further to see which
model results from the gluing.
Proceeding to the usual gluing, the resulting edge amplitude for interior
tetrahedra is simply:

\bes
\frac{1}{\sqrt{\Delta_{J_{e'1}}\Delta_{J_{e'2}}\Delta_{J_{e'3}}\Delta_{J_{e'4}}}} 
\ees
and the gluing itself looks rather trivial in the sense that in the end it
just gives a multiplication of pre-existing factors, with nothing new
arising from it. The gluing performed starting from the partition function
with the other boundary conditions gives the same result, again only if we
do not project the D-functions for the exposed edges. 

This is the \lq\lq factorized" edge amplitude considered in
\cite{hendryk}, and singled out by the requirement that the passage from
$SO(4)$ BF theory to gravity is given by a pure projection operator.
Indeed, we have just seen how this model is obtained using only the
simplicity projection, and dropping the $P_{g}$, which is responsible for
making the combined operator $P_{g}P_{h}$ not a projector
($P_{g}P_{h}P_{g}P_{h}\neq P_{g}P_{h}$). 

On the other hand, the additional projection $P_{g}$ introduces an
additional coupling of the representations for the four triangles
forming a tetrahedron. This coupling
can be understood algebraically directly from the way the $P_{g}$ operator
acts, since it involves all the four wedges incident on the same edge (see
equation (~\ref{eq:edam})), or recalling that the gauge invariance of the edge
amplitude (corresponding to the tetrahedra of the simplicial manifold)
admits a natural interpretation as the closure constraint for the
bivectors $B$ in terms of which we quantize both BF theory and gravity in
the Plebanski formulation. This is the constraint that the bivectors
assigned to the triangles of the tetrahedron, forced
to be simple bivectors because of the simplicity constraint $P_{h}$,
sum to zero. Thus we can argue more
geometrically for the necessity of the $P_{g}$ projection saying that
the model has to describe the geometric nature of the
triangles, but also the way they are \lq\lq coupled" to form \lq\lq
collective structures", like tetrahedra. Not imposing the $P_{g}$
projection results in a theory of not enough coupled triangles. For
this reason we do not consider this as a
viable version of the Barrett-Crane model.

But if the $P_{g}$ projector is necessary, then the procedure used above, giving the Perez-Rovelli version of the
Barrett-Crane model, can be seen as the minimal, and most natural, way of
constraining BF-theory to get a quantum gravity model. At the same time,
exactly because combining the projectors $P_{h}$ and $P_{g}$ does not give
a projector operator, \lq\lq non-minimal" models, sharing the same
symmetries as the Perez-Rovelli version, and implementing as well the
Barrett-Crane constraints, but possibly physically different from it, can
be easily constructed, imposing the two projectors more than once. It is
easy to verify that, both starting from the partition function for a
single 4-simplex with fixed boundary connection or with the B field fixed
instead, imposing the combined $P_{g}P_{h}$ operator $n$ times ($n\geq
1$), the usual
gluing procedure will result in an amplitude for the interior tetrahedra:

\bes
A_{e}\,=\,\frac{\Delta_{1234}^{2n-1}}{\left(\Delta_{J_{1}}\Delta_{J_{2}}\Delta_{J_{3}}\Delta_{J_{4}}\right)^{n}}.
\ees   
Of course, the same kind of model could be obtained from a field theory
over a group, with the usual technology. However, the physical
significance of this variation is unclear (apart from the stronger
convergence of the partition function, which is quite apparent).

To conclude, let us comment on the De Pietri-Freidel-Krasnov-Rovelli
version of the Barrett-Crane model. It seems that there is no natural (or
simple) variation of the procedure we used leading to this version of the
Barrett-Crane state sum, as we have seen. In other words, starting from
the partition function for BF theory, there appears to be no simple way to
impose the Barrett-Crane constraints at the quantum level, by means of
projector operators as we did, and to obtain a model with an amplitude for
the interior tetrahedra of the type:
\bes
A_{e}\,=\,\frac{1}{\Delta_{1234}} \label{eq:dpfkredam}
\ees     
as in \cite{DP-F-K-R}. Roughly, the reason can be understood as follows:
for each edge, the $P_{h}$ projection has the effect of giving a factor
involving the product of the dimensions of the representations in the
denominator, and of course of restricting the allowed representations to
the simple ones, while the $P_{g}$ projector is responsible for having a
Barrett-Crane intertwiner for the boundary tetrahedra, which in turn
produces the factor $\Delta_{1234}$ after the gluing. The imposition of
more of these projections in the non-minimal models can only change the
power with which these same elements appear in the final partition
function, as we said. So it seems that the imposition of these projectors
can not create a factor like $\Delta_{1234}$ in the denominator, which, if
wanted, has apparently to be inserted by hand from the beginning. The
un-naturality of this version of the Barrett-Crane model from this point
of view can probably be understood noting that in the original field
theory over group formulation \cite{DP-F-K-R} the imposition of the
operator $P_{h}$ in the kinetic term of the action, giving a kinetic
operator that is not a projector anymore, makes the coordinate space (or
\lq\lq connection" \cite{RR2}) formulation of the partition
function highly complicated, and this formulation is the closest
analogue of our lattice-gauge-theory-type of derivation. However, an intriguing logical
possibility that we think deserves further study is that the edge
amplitude (~\ref{eq:dpfkredam}) may be ``expanded in powers of
$\Delta_{1234}$'', so that it may arise from a (probably asymptotic) series in which the
n-th term results from imposing the $P_{g}P_{h}$ operator $n$ times with the
result shown above. More generally, many different models can be
constructed (consistently with different boundary conditions) in this way (combining models with different powers of $P_{g}P_{h}$), all based on the simple representations of
$SO(4)$ or $SL(2,C)$, having the same fundamental symmetries,
and the Perez-Rovelli version of the Barrett-Crane model as the
``lowest order'' term, with the other orders as ``corrections'' to
it, even if interesting models on their own right. This possibility will be investigated in the future.   

\section{Lattice gauge theory discretization and quantization of the Plebanski action}
We have just seen how the Barrett-Crane model can be obtained by writing down the partition function for quantum BF theory 
and then imposing suitable projections on the configurations of the group variables we are integrating. Although plausible, 
and justifiable using several arguments, this procedure is not easily related to the straightforward path integral 
quantization of the Plebanski action. In \cite{alejandro}, an argument has been given that relates the Barrett-Crane model 
to a discretization of the $SO(4)$ Plebanski action. We outline it here since it complements nicely the derivation we have presented.   

Consider the discretized Plebanski action as given in \ref{eq:Pleb}:
\bes
S(B,\omega)\,=\,\sum_t \tr B_t\,\Omega_t\,+\,\sum_{t,t'}\,\phi_{IJKL}\,B^{IJ}(t)\,B^{KL}(t').
\ees
A partition function based on this action for a single 4-simplex is:
\bes
Z_{Pl}\,=\,\prod_w \int dB_w\,\prod_e dg_e\,d\phi\,e^{i\,\sum_w \tr B_w\,\Omega_w(g,h)\,+\,\sum_{w,w'}\,\phi_{IJKL}\,B^{IJ}(w)\,B^{KL}(w')}\,= \\
=\,\prod_w \int dB_w\,\prod_e dg_e\,\delta(\epsilon_{IJKL}B^{IJ}(w)\,B^{KL}(w'))\,e^{i\,\sum_w \tr B_w\,\Omega_w(g,h)} ,
\ees
where we have used the wedges as defined previously and assigned group variables $g$ to the dual edges and group variables $h$
to the exposed ones, and the curvature is defined, we recall, by means of the holonomy or product of group elements around 
the boundary links of each wedge.
The crucial point is now to find a suitable re-writing of the bivectors $B$ that allows for a solution of the constraints.
The idea is then to substitute the bivector $B^{IJ}(w)$ associated to each wedge with the right invariant vector field 
$- i X^{IJ}(h_w)$ acting on the corresponding discrete holonomy (group element). This is justified by the action of this 
vector on the BF amplitude \cite{alejandro}. Therefore the constraints in the delta function are to be expressed as 
polynomials in the $X(h)$ field. The partition function then becomes:
\bes
Z_{Pl}(h_w)\,=\,\left(\prod_e \int dg_e\right) \delta(constr(X(h))\,\prod_w \delta( g_w h_w g'_w)\,= \nonumber \\
 =\,\delta(constr(X(h))\,\left(\prod_e\int dg_e\right)\,\prod_w \delta(g_w h_w g'_w) 
\ees
where we have used the fact that the $X$ fields act on the boundary connections $h$ only. The group elements $g_e$
 can be integrated over just as in the pure BF theory case, so that the effect of the constraints and of the delta function 
is just to restrict the set of configurations allowed in the BF partition function, to those satisfying the constraints.

The constraints in turn are defined by the differential equations:
\bes
\epsilon_{IJKL}\,X^{IJ}(h_w)\,X^{KL}(h_{w'})\,Z_{Pl}(h_w)\,=\,0.
\ees 
This set of equations can indeed be explicitely solved \cite{alejandro}, and the important thing is that the configurations 
satisfying these equations are precisely those satisfying the Barrett-Crane constraints; in other words, the equations 
so expressed in terms of the right invariant vector field $X$ are equivalent to the Barrett-Crane constraints, as it was to
be expected since they both come from a discretization of the constraints on the $B$ field in the Plebanski action, and lead
to the same conditions on the representations of $Spin(4)$ that represent the quantum counterpart of the same $B$ field. 

Therefore, also with this procedure coming directly from the Plebanski action we obtain for a single 4-simplex the same 
partition function, with field connection on the boundary, that we have obtained above \Ref{eq:BC4s}. Then, using the same gluing 
procedure, we would obtain the same complete partition function for the full simplicial manifold we have given, with the same 
edge amplitude resulting from the gluing of 4-simplices.

\section{The Lorentzian Barrett-Crane spin foam model: classical and quantum geometry}

We now want to discuss in more detail the Lorentzian Barrett-Crane model, with particular attention to the geometric meaning
of its amplitudes and of the variables appearing in it, and to its corresponding quantum states, given of course by  spin 
networks; we will also identify in the structure of the amplitudes a discrete symmetry that characterizes the model as a
definition of the physical inner product between these quantum gravity states or a-causal transition amplitudes, i.e. as a 
realization of the projector operator onto physical states. The presentation will be based on \cite{causal}. 

As we have said, the derivation we have performed in the Riemannian case can be similarly perfomed in the Lorentzian, with 
minor modifications. The $Spin(4)$ group is replaced of course by the group $Sl(2,\mathbb{C})$, and, if we are to construct 
a model with all the triangles and tetrahedra being spacelike, thus involving only simple representations labelled by a 
continuous parameter, the invariance we impose for each triangle (representation) is again under an $SU(2)$ subgroup of the 
Lorentz group. This leads to a realization of the representations in the space of functions on the upper hyperboloid in 
Minkowski space, and to the corresponding formula for the Barrett-Crane intertwiner as a multiple integral on this homogeneous 
space, that we have given in \Ref{eq:BCintLor}.

The main technical difference with respect to the Riemannian case is that, while we can follow the same steps we have followed 
in the Riemannian case from the very beginning up to formula \Ref{eq:partfunc}, we are not able in the Lorentzian case to 
perform the integrals over $SL(2,\mathbb{C})$, but only those over the subgroup $SU(2)$, that give the contraction with the 
invariant vectors $w^\rho$'s, and we are thus left with integrals over the hyperboloid $H^3$. However, these integral are 
exactly those entering in the definition of the Lorentzian Barrett-Crane intertwiner, and moreover, we shall see that this 
integral form of the spin foam model makes the geometric meaning of the model itself very clear and more transparent.
We have indeed, for each edge in the boundary of the wedges $1,2,3,4$ (tetrahedron bounded by the triangles $1,2,3,4$):
\bes
A_{e}&=&\int_{SL(2,\mathbb{C})}dg_{e}\int_{SU(2)}dh_{1}\int_{SU(2)}dh_{2}\int_{SU(2)}dh_{3}\int_{SU(2)}dh_{4}\int_{SL(2,\mathbb{C})}dg'_{e}\,
\nonumber \\ 
&&D_{j_1k_{1} j'_1m_{1}}^{n_1 \rho_{1}}(g_{e}h_{1}g'_{e})D_{j_2k_{2} j'_2m_{2}}^{n_2 \rho_{2}}(g_{e}h_{2}g'_{e})D_{j_3k_{3}
 j'_3m_{3}}^{n_3 \rho_{3}}(g_{e}h_{3}g'_{e})D_{j_4k_{4} j'_4m_{4}}^{n_4 \rho_{4}}(g_{e}h_{4}g'_{e})\,= \nonumber
\\ &=&\,\int_{H^3}dx_{e}\,\,D_{j_1k_{1} j''_1l_{1}}^{0 \rho_{1}}(g_{e})D_{j_2k_{2} j''_2l_{2}}^{0 \rho_{2}}(g_{e})
D_{j_3k_{3} j''_3l_{3}}^{0 \rho_{3}}(g_{e})D_{j_4k_{4} j''_4l_{4}}^{0 \rho_{4}}(g_{e})
w^{\rho_1}_{j''_1 l_1}...w^{\rho_4}_{j''_4 l_4} \nonumber \\ &&\int_{H^3}dx'_{e}\,w^{\rho_1}_{j'''_1 p_1}...w^{\rho_4}
_{j'''_4 p_4}
D_{j'''_1 p_{1} j'_1m_{1}}^{0 \rho_{1}}(g'_{e})D_{j'''_2p_{2} j'_2m_{2}}^{0 \rho_{2}}(g'_{e})D_{j'''_3p_{3} j'_3m_{3}}
^{0 \rho_{3}}(g'_{e})D_{j'''_4p_{4} j'_4m_{4}}^{0 \rho_{4}}(g'_{e})\,= \nonumber \\
&=&\,B^{\rho_1 \rho_2 \rho_3 \rho_4}_{j_1k_1 j_2k_2 j_3k_3 j_4k_4}\,B^{\rho_1 \rho_2 \rho_3 \rho_4}
_{j'_1m_1 j'_2m_2 j'_3m_3 j'_4m_4}
\ees

The Barrett-Crane intertwiners are then contracted just as in the Riemannian case to form simple spin networks giving the 
expression for the 4-simplex amplitudes and, after gluing, for the internal tetrahedra. The relevant formula is that for the
contraction of two representation functions in a simple representation to form the invariant kernel $K^\rho$ \cite{VK, Ruhl, gelfand},
 bi-invariant under $SU(2)$, one for each triangle in each 4-simplex: 

\bes
K^\rho(x_e \cdot x'_e)\,&=&\,D_{00 jk}^{\rho}(x_{e})\,D_{jk 00}^{\rho}(x'_{e})\,= \nonumber \\ 
&=& \langle \rho\; j=0 k=0 \mid D^{\rho}(x_e \cdot x'_e) \mid \rho\; j=0 k=0\rangle\,=\,D^{\rho}_{0000}(x_e x_e^{'-1}).\;\;\;\;\;\;\;\;\;
\ees

The resulting model,
based on continuous representations, is:

\bes
Z\,=\,(\prod_f
\int_{\rho_f}d\rho_f\,\rho_f^2)\,(\prod_{v,
e_v}\int_{H^+}dx_{e_v})\,\prod_e
\mathcal{A}_e(\rho_k)\,\prod_v\mathcal{A}_v (\rho_k, x_i)
\label{eq:Z}
\ees
with the amplitudes for edges (tetrahedra) (to be
considered as part of the measure, as we argue again below) and
vertices (4-simplices) being given by:

\bes
A_e(\rho_1,\rho_2,\rho_3,\rho_4)\,=\,\int_{H^+} dx_1 dx_2
K^{\rho_1}(x_1,x_2)K^{\rho_2}(x_1,x_2)K^{\rho_3}(x_1,x_2)K^{\rho_4}(x_1,x_2)\;\;\;\;
\\ \nonumber \\
A_v(\rho_k,
x_i)=K^{\rho_1}(x_1,x_2)K^{\rho_2}(x_2,x_3)K^{\rho_3}(x_3,x_4)K^{\rho_4}(x_4,x_5)K^{\rho_5}(x_1,x_5)
\nonumber \\ K^{\rho_6}(x_1,x_4)
K^{\rho_7}(x_1,x_3)K^{\rho_8}(x_3,x_5)K^{\rho_9}(x_2,x_4)
K^{\rho_{10}}(x_2,x_5)\;\;\;\; \label{eq:ampl}.
\ees
The variables of the model are thus the representations $\rho$ associated
to the triangles of the
triangulation (faces of the dual 2-complex), there are ten of them for
each 4-simplex as can be seen
from the expression of the vertex amplitude, and the vectors $x_i \in
H^+$, of which there is one for
 each of the five tetrahedra in each 4-simplex amplitude, so that two of
them correspond to each
interior tetrahedron along which two different 4-simplices are glued. We
will discuss the geometric
meaning of all these variables in the following section.
As written, the partition function is divergent, due to the infinite
volume of the Lorentz group (of the hyperboloid), and the immediate
way to avoid this trivial divergence is just to remove one of the
integration over $x$ for each edge and 4-simplex amplitude, or, in
other terms, to gauge fix one variable. It was shown that, after this
gauge fixing is performed, the resulting partition function (for fixed
triangulation) is finite \cite{bb2, CPR}.

The functions $K$ appearing in these amplitudes have the explicit
expression:

\bes
K^{\rho_k}(x_i,x_j)=\frac{2\sin(\eta_{ij}\rho_k/2)}{\rho_k\sinh\eta_{ij}}
\label{eq:Ker} \ees where $\eta_{ij}$  is the hyperbolic distance
between the points $x_i$ and $x_j$ on the hyperboloid $H^+$.

The amplitudes describe an interaction among the $\rho$'s that couples
different 4-simplices and
different tetrahedra in the triangulation whenever they share some
triangle, and an interaction among
the different tetrahedra in each 4-simplex.
In the presence of boundaries, formed by a certain number of
tetrahedra, the partition function above has to be supplemented by
boundary terms given by a function
$C^{\rho_1\rho_2\rho_3\rho_4}_{(j_1k_1)(j_2k_2)(j_3k_3)(j_4k_4)}(x)$
for each boundary tetrahedron \cite{boundary}, being defined as:

\bes
C^{\rho_1\rho_2\rho_3\rho_4}_{(j_1k_1)(j_2k_2)(j_3k_3)(j_4k_4)}(x)\,=\,D^{\rho_1}_{00j_1k_1}(x)\,D^{\rho_2}_{00j_2k_2}(x)D^{\rho_3}_{00j_3k_3}(x)\,D^{\rho_4}_{00j_4k_4}(x)
\label{xintertwiner}
\ees
where $x$ is a variable in $\sl/SU(2)$ assigned to each boundary
tetrahedron, $D^{\rho_i}_{00j_ik_i}(x)$ is a representation matrix for
it in the representation $\rho_i$ assigned to the $i-th$ triangle of
the tetrahedron, and the matrix elements refer to the canonical basis
of functions on $SU(2)$, where the $SU(2)$ subgroup of the Lorentz
group chosen is the one that leaves invariant the $x$ vector thought
of as a vector in $\mathbb{R}^{3,1}$, with basis elements labelled by a
representation $j_i$ of this $SU(2)$ and a label $k_i$ for a vector in
the corresponding representation space.

It is this term that gives rise to the amplitude for internal
tetrahedra after the gluing of two 4-simplices along it (it can be
seen as just a half of that amplitude with the integration over
the hyperboloid dropped) \cite{OW,boundary}.

Also, we stress that the partition function above should be understood
as just a term within some sum over triangulations or over 2-complexes to be defined, for
example, by a group field theory formalism. Only this sum would
restore the full dynamical content of the quantum gravity theory.

Let us now describe the classical and quantum geometry of the
Barrett-Crane model, i.e. the geometric meaning of the
variables appearing in it, the classical picture it furnishes, and the
quantum version of it. Conceptually, the spin foam
formulation of the quantum geometry of a manifold is more fundamental than
the classical approximation of it one uses, and
this classical description should emerge in an appropriate way only in a
semi-classical limit of the model; however, the
derivation of the model from a classical theory helps in understanding the
way it encodes geometric information both at the
classical and quantum level, even before this emergence is properly
understood.

\subsection{Geometric meaning of the variables of the model} \label{sec:geo} 
In this part, we describe the physical-geometrical content of the
mathematical variables appearing in the model. We summarize and
collect many known facts and show how the resulting
picture that of a discrete piecewise flat
space-time, whose geometry is described by a first order Regge
calculus action (as we then explain in the next section), in order to explain the physical intuition
on which the rest of our work is based.

The variables of the model are the irreducible representations $\rho$'s,
associated to each face of the 2-complex, and the
$x$ variables, associated to each edge, one for each of the two vertices
it links, as we discussed above.

Consider the $\rho$
variables. They result from the assignment of bivector operators to each
face of the 2-complex, in turn coming from the
assignment of bivectors to the triangles dual to them. Given these
bivectors, the classical expression for the area of the
triangles is, as we have already discussed: $A_t^2=B_t\cdot B_t=B_t^{IJ}B_{tIJ}$, and this, after the
geometric quantization procedure outlined above,
translates into:
$A_t^2=-C_1=-J^{IJ}(\rho_t)J_{IJ}(\rho_t)=\rho_t^2+1>0$\cite{BC, BC2}.
Thus the $\rho$'s determine the areas of the triangles
of the simplicial manifold dual to the 2-complex. The same result, here
obtained by geometric quantization of the bivector
field, can also be confirmed by a direct canonical analysis of the
resulting quantum theory, studying the spectrum of the
area operator acting on the simple spin network states that constitute the
boundary of the spin foam, and that represent the
quantum states of the theory. We describe all this in the following.
Moreover, from the sign of the (square of the) areas above,
it follows that all the triangles in the model are spacelike, i.e. their
corresponding bivectors are timelike, and consequently
all the tetrahedra of the manifold are spacelike as well, being
constituted by spacelike triangles only\cite{BC2}.
As a confirmation, one may note that we have chosen a particular
intertwiner -the simple or Barrett-Crane intertwiner- between
simple representations such that
it decomposes into only continuous simple representations, which 
translates in algebraic language to the fact that timelike bivectors add up
to timelike bivectors.
Clearly, also this sign
property of the areas can be confirmed by a canonical analysis.

Consider now the $x$ variables. They have the natural
interpretation of normals to the tetrahedra of the manifold, and
the tetrahedra being spacelike, they have values in $H^+$, i.e.
they are timelike vectors in $\mathbb{R}^{3,1}$\cite{barrettintegral,BW}. The reason for them being in
$\mathbb{R}^{3,1}$ is easily explained. The Barrett-Crane model corresponds
to a simplicial manifold, and more precisely to a piecewise flat
manifold, i.e patches (the 4-simplices) of flat space-time glued
together along their common tetrahedra\cite{Regge}.
To each flat patch or
4-simplex (a piece of $\mathbb{R}^{3,1}$) is attached a local reference
frame. In other words, we are using the equivalence principle,
replacing the usual space-time points by the 4-simplices: at each
4-simplex, there exists a reference frame in which the space-time
is locally flat. This explains why the normals to the tetrahedra
are vectors in $\mathbb{R}^{3,1}$ and also why there are two different
normals for each tetrahedron: they are the same vector seen in two
different reference frames. How does the curvature of spacetime
enter the game? Having a curved space-time means that these
reference frames are not identical: we need a non-trivial
connection to rotate from one to another (see also \cite{hendryk}). In this sense, the
non-flatness of spacetime resides in the tetrahedra, in the fact
that there are two normals $x,y\in H^+$ attached to each
tetrahedron, one for each 4-simplex to which the tetrahedron
belongs. The (discrete) connection is uniquely defined,
up to elements of the $SU(2)$ subgroup that leaves invariant
the normal vector on which the connection acts, by the
Lorentz transformation $g$ rotating from one of these normals to
the other ($g\cdot x=y$):
it is a pure boost connection mapping
two points in $H^+$ into one another\footnotemark. Thus, the association of two
variables $x$ and $y$ to each edge (tetrahedron) of the 2-complex
(triangulation) is the association of a connection variable $g$ to
the same edge (tetrahedron) (this is of course the connection variable
associated to the edges, and then constrained by the simplicity
constraints, that is used in all the various derivations of the Barrett-Crane model
\cite{RR, RR2, P-R, P-R2, alejandro, OW, boundary, hendryk}).
From the product of connection
variables around a closed loop in the dual complex, e.g. the
boundary of a face dual to a triangle, one obtains a measure of
the curvature associated to that triangle, just as in traditional
simplicial formulations of gravity (i.e. Regge calculus) or in
lattice gauge theory. This set of variables, connections on links
and areas on faces, is the discrete analogue of the set of
continuous variables $(B(x), A(x))$ of the Plebanski formulation
of gravity.

\footnotetext{The reconstruction of a (discrete) connection
or parallel transport from the two normals
associated to each tetrahedron was also discussed in \cite{hendryk}
in the context of the Riemannian Barrett-Crane model.
}

Of course, we have a local Lorentz invariance at each ``manifold point" i.e
each
4-simplex. Mathematically, it corresponds to the Lorentz invariance of
the amplitude associated to the 4-simplex. Physically, it says that
the five normals $(x_A,x_B,\dots,x_E)$ to the five tetrahedra of a given
4-simplex are given up to a global Lorentz transformation:
$(x_A,x_B,\dots,x_E)$ is equivalent to $(g\cdot x_A,g\cdot
x_B,\dots,g\cdot x_E)$. This
is also saying that the local reference frame
associated to each 4-simplex is given up to a Lorentz transformation, as
usual.
Now, given
two adjacent 4-simplices, one can rotate one of the two in order to
get matching normals on the common tetrahedron. If it is possible to
rotate the 4-simplices to get matching normals everywhere, then the
(discrete) connection would be trivial everywhere (i.e. the identity
transformation) and we would get the
equivalent of a classically flat space-time. However, unlike in 3
dimensions where we have only one normal attached to each triangle, in
4 dimensions, we do not want only flat space-times and the ``all
matching normals" configuration is only one particular configuration
among the admissible ones in the Barrett-Crane model.

The presence of such Lorentz invariance shows that the true physical
variables of the model are indeed pairs of $x$ vectors and not
the vectors themselves. One may use this invariance to fix one of the
vectors in each 4-simplex and to express then the
others with respect to this fixed one; in other words the geometric
variables of the model, for each vertex (4-simplex) are
the hyperbolic distances between two vectors corresponding to two
tetrahedra in the 4-simplex, measured in the hyperboloid $H^+$,
i.e. the variables $\eta$ appearing in the formula \Ref{eq:Ker}. These in
turn have the interpretation, in a simplicial context,
of being the dihedral angles between two tetrahedra sharing a triangle, up
to a sign depending on whether we are dealing with
external or internal angles, in a Lorentzian context (see \cite{BarFoxon}).
These angles may also be seen as the counterpart
of a connection variable inside each 4-simplex.

\medskip
To sum up the classical geometry underlying the Barrett-Crane model,
we have patches (4-simplices) of flat spaces glued together into a curved space.
This curvature is introduced through the change of frame associated
to each patch, which can be identified with the change of time normal
on the common boundary (tetrahedron) of two patches.
Now a last point is the {\it size} of the flat space patches, or in other words
the (space-time) volume of a fixed 4-simplex in term of the 10 $\rho$
representations defining it. This volume can be obtained as the wedge
product of the bivectors associated to two (opposite) triangles $-$ not
sharing a common edge $-$ of the 4-simplex, which reads:
\be
{\cal V}^{(4)}=
\f{1}{30}\sum_{t,t'}\f{1}{4!}
\epsilon_{IJKL}sgn(t,t')B_t^{IJ}B_{t'}^{KL}
\ee
where $(t,t')$ are couples of triangles of the 4-simplex and $sgn(t,t')$
register their relative orientations.
After quantization, the $B$ field are replaced by the generators $J$ and
this formula becomes:
\be
{\cal V}^{(4)}=
\f{1}{30}\sum_{t,t'}\f{1}{4!}
\epsilon_{IJKL}sign(t,t')J_t^{IJ}J_{t'}^{KL}.
\ee
There is another formula to define the volume, which can be seen as more suitable
in our framework in which we use explicitely the (time) normals to the tetrahedra:
\be
({\cal V}^{(4)})^3=\f{1}{4!}\epsilon^{abcd}
N_a\wedge N_b\wedge N_c\wedge N_d
\ee
where is the oriented normal with norm $|N_i|=v^{(3)}_i$ the
3-volume of the corresponding tetrahedron (more on the 4-volume
operator in the Barrett-Crane model can be found in \cite{danhend}).

\subsection{Simplicial classical theory underlying the model}
\label{area-angle}
Thus the classical counterpart of the quantum geometry of the
Barrett-Crane model, or in other words the classical description
of the geometry of spacetime that one can reconstruct at first from the
data encoded in the spin foam, is a simplicial geometry
described by two sets of variables, both associated to the triangles in the
manifold, being the areas of the triangles
themselves and the dihedral angles between the two normals to the two
tetrahedra sharing each triangle. A classical simplicial
action that makes use of such variables exists and it is given by the
traditional Regge calculus action for gravity. However,
in the traditional second-order formulation of Regge calculus both the
areas and the dihedral angles are thought of as
functions of the edge lengths, which are the truly fundamental variables
of the theory. In the present case, on the other hand, no variable
corresponding to the edge length is present in the model and both the
dihedral angles and the areas of triangles have to be seen as
fundamental variables. Therefore the underlying classical theory for
the Barrett-Crane model is a first order formulation of Regge Calculus
based on angles and areas.

A formulation of first order Regge calculus was proposed by Barrett
\cite{balone} in the Riemannian case based on the action: \be
S(l,\theta)\,=\,\sum_t\,A_t(l)\,\epsilon_t\,=\,\sum_t\,A_t(l)\,(
2\pi\,-\,\sum_{\sigma(t)}\,\theta_t(\sigma)) \ee where the areas
$A_t$ of the triangles $t$ are supposed to be functions of the
edge lengths $l$, $\epsilon_t$ is the deficit angle associated to
the triangle $t$ (the simplicial measure of the curvature) and
$\theta_t(\sigma)$ is the dihedral angle associated to the
triangle $t$ in the 4-simplex $\sigma(t)$ containing it. The
dihedral angles $\theta$, being independent variables, are
required to determine, for each 4-simplex, a unique simplicial
metric, a priori different from the one obtained by means of the
edge lengths (actually, unless they satisfy this constraint, the
ten numbers $\theta$'s can not be considered dihedral angles of
any 4-simplex, so we admit a slight language abuse here).
Analytically, this is expressed by the so-called Schl{\"a}fli
identity: \be \sum_t\,A_t\,d\theta_t\,=\,0. \ee The variations of
this action are to be performed constraining the angles to satisfy
such a requirement, and result in a proportionality between the
areas of the triangles computed from the edge lengths and those
computed from the dihedral angles: \be
A_t(l)\,\propto\,A_t(\theta). \ee

The meaning of this constraint is then to assure the agreement of
the geometry determined by the edge lengths and of that determined
by the dihedral angles, and can be considered as the discrete
analogue of the ``compatibility condition'' between the $B$ field
and the connection, basically the metricity condition for the
connection, in the continuum Plebanski formulation of gravity.
Note, however, that this agreement is required to exist at the
level of the areas only. This constraint may also be implemented
using a Lagrange multiplier and thus leaving the variation of the
action unconstrained (see \cite{balone}). In this case, the full
action assumes the form: \be S\,=\,\sum_t\,A_t(l)\,\epsilon_t\,+\,
\sum_\sigma\,\lambda_\sigma\,det\Gamma_{ij}(\theta), \ee where
$\lambda_\sigma$ is a Lagrange multiplier enforcing the mentioned
constraint for each 4-simplex, and the constraint itself is
expressed as the vanishing of the determinant of the matrix
$\Gamma_{ij}=-\cos{\theta_{ij}}=-\cos{(x_i\cdot x_j)}$, where the
$x$'s variables are the normals to the tetrahedra in the 4-simplex
introduced above.

The main difference with the situation in the Barrett-Crane model
is that, in this last case, the areas are not to be considered as
functions of the edge lengths, but independent variables. The
relationship with the dihedral angles, however, remains the same,
and this same first order action is the one to be considered as
somehow ``hidden'' in the spin foam model. We will discuss more
the issue of the simplicial geometry and of the classical action
hidden in the Barrett-Crane model amplitudes, and of its
variation, when dealing with the Lorentzian case.

We
point out that other formulations of first order Regge Calculus exist,
with different (but related) choices for the fundamental variables of
the theory \cite{CaAd89}\cite{Katsy}\cite{Katsy2}. Also, the idea of using the areas
as fundamental variables was put forward at first in \cite{carloPR} and
then studied in
\cite{Make1}\cite{Make2}. The possibility of describing simplicial
geometry only in terms of areas of triangles, inverting the
relation between edge lengths and areas and thus expressing all
geometric quantities (including the dihedral angles) in terms of the latter, was analysed in 
\cite{BRW}\cite{MW}\cite{ReggeWill}.

\subsection{Quantum geometry: quantum states on the boundaries and quantum amplitudes}

From this classical geometry of the Barrett-Crane model, it is easier
to understand the quantum geometry defined in terms of (simple) spin
networks geometry states. Similarly to the Ponzano-Regge case, the
boundary states are Lorentz invariant functionals of both a boundary
connection and the normals on the boundary \cite{boundary2} and a
basis of the resulting Hilbert space is given by the simple spin networks.

More precisely, let us consider a (coloured) spin foam with boundary.
The boundary will be made of 4-valent vertices glued with each other
into an oriented graph. Such structure is dual to a 3d triangulated
manifold. Each edge of the graph is labelled by a (face)
representation $\rho$ and each vertex corresponds to a (simple)
intertwiner between the four incident representations. On each edge $e$,
we can put a group element $g_e$ which will correspond to the boundary
connection \cite{boundary} and we can decompose the simple intertwiner
such that a "normal" $x_v\in{\cal H}_+$ lives at each vertex $v$
\cite{boundary2}.

\begin{figure}
\begin{center}
\includegraphics[width=7cm]{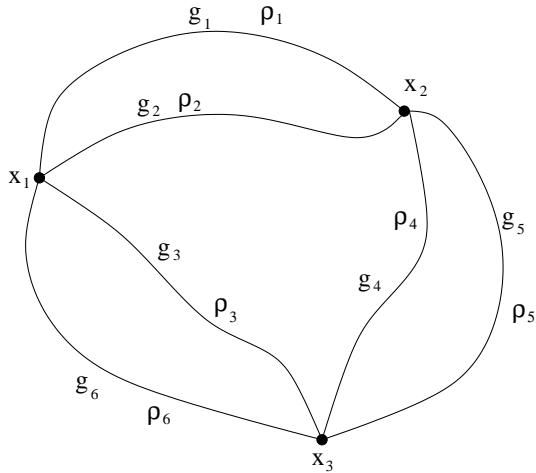}
\caption{A (closed) Lorentz spin network}
\end{center}
\end{figure}

That way, the boundary state is defined by a Lorentz invariant functional

\bes
\phi(g_e,x_v)=\phi(k_{s(e)}^{-1}g_ek_{t(e)},k_v.x_v)
\textrm{ for all } k_v\in \sl
\label{gaugeinv}
\ees
where $s(e)$ and $t(e)$ are respectively the source vertex and the
target vertex of the oriented edge $e$. This imposes an $SU(2)$
invariance at each vertex $v$ once we have fixed the normal
$x_v$. One should go further and impose an $SU(2)$ invariance for each
edge incident to the vertex in order to impose the simplicity
constraints \cite{boundary2}.
One endows this space of functionals with the
$\sl$ Haar measure:
\bes
\mu(\phi)=\int_{SL(2,\mathbb{C})^E}\textrm{d}g_e\,
\phi^*(g_e,x_v)\,\phi(g_e,x_v).
\ees
This measure is independent of the choice for the $x_v$ due to the gauge
invariance \Ref{gaugeinv}.
Then an orthonormal basis of the Hilbert space of $L^2$
functions is given by the simple spin networks~:

\bes
s^{\{\rho_e\}}(g_e,x_v)=
\prod_{e}K_{\rho_e}(x_{s(e)},g_e.x_{t(e)})=
\prod_{e}\langle\rho_ex_{s(e)}(j=0)|g_e|\rho_ex_{t(e)}(j=0)\rangle,\;\;\;
\label{simplespinnet}
\ees
where $\mid \rho x(j=0)\rangle$ is the vector of the $\rho$ representation
invariant under $SU(2)_x$ (the $SU(2)$ subgroup leaving the vector $x$
invariant). The general notation is $\mid \rho xjm\rangle$ for the vector
$|m\rangle$ in the $SU(2)$ representation space $V^j$ in the
decomposition $R^\rho=\oplus_j V^j_{(x)}$ of the $SL(2,\mathbb{C})$
representation $\rho$ into representations of $SU(2)_x$.

We should point out that this same Hilbert space for kinematical states
comes out of the canonical analysis of the (generalised)
Hilbert-Palatini action in a explicit covariant framework.
Indeed requiring no anomaly of the diffeomorphism invariance of the
theory, the Dirac brackets, taking into account the second
class constraints (simplicity constraints), of the connection
$\SA_i^X$ ($i$ is the space index and $X$ is the internal Lorentz index)
and the triad $P^i_X$, read \cite{alex2}:

\bes
\left\{
\begin{array}{ccc}
\{I\SA,I\SA\}_D &=&0 \\
\{P,P\}_D &=& 0 \\
\{ \SA_i^X,P_Y^j\}_D&=&\delta_i^j I_Y^X.
\label{diracbracket}
\end{array}
\right.
\ees
where $I$ projects the Lorentz index $X$ on its boost part, this
latter being defined relatively to the time normal $x$ (which is built
from the triad field).
One can try to {\it loop quantize} this theory: one would like to
consider spin networks of the connection $\SA$ (cylindrical
functions), but in fact, one needs so-called projected spin networks
which depend on both the connection $\SA$ and the time normal field
\cite{boundary2}.
Then it turns out that
quantizing these commutations relations at a finite number of points
(the vertices $v$ of the graph)
lead to the imposition of the simplicity constraints at the vertices and
the same space of simple spin networks \ref{simplespinnet} as shown in
\cite{alexetera}.
The graph, representing the quantum geometry state, then has edges
labelled with $SL(2,\Cb)$ simple representations $\rho\ge0$, which
corresponds to an area carried by this edge given by:
\be
area\sim\sqrt{\rho^2+1}.
\label{areaspectrum}
\ee
The restriction to a finite number of points is natural from the spin foam
viewpoint since the space-like slices are made of tetrahedra glued together and
that these same tetrahedra are considered as the points of this 3d-slice.
This explicit relation between the spin foam setting and the canonical theory
is likely to provide us with information on the dynamics of gravity
in both theories.

We have described up to now closed boundaries. What happens if we deal
with an {\it open boundary}? Then one needs to introduce open spin
networks. Let us consider a graph with open ends.
At all vertices will still live an $H^+$ element (time normal).
The edges in the interior will be
defined as previously: on each edge live a $\rho$ representation and an
$\sl$ group element. On the exterior edges, we still have an
$\sl$ representation and a group element, but we introduce some
new label, at the (open) end of the edge, which is a vector in the
(edge) representation. Within its $\sl$ representation, this
vector $v_e$ can be defined in an orthonormal basis
by its $SU(2)$ (sub)representation $j$ and its label
$m$ witin this representation. Then the overall functional is defined
as
\bes
s^{\{\rho_e,j_{ext},m_{ext}\}}(g_e,x_v)=
\prod_{e\in int}K_{\rho_e}(x_{s(e)},g_e.x_{t(e)})
\prod_{e\in ext}\langle\rho_ex_{int(e)}(j=0)|
g_e|\rho_e v_e=(j_e,m_e)\rangle,\;\;\;\;
\ees
where $int(e)$ is the (interior) vertex of the exterior edge $e$.
The scalar product is defined by the integration over all $g_e$
variables. And we get an orthonormal basis labelled by
$\{\rho_e,j_{ext},m_{ext}\}$.

\medskip
We are then ready to explain the geometry behind the amplitude of
the Barrett-Crane model. There are two different ingredients~: the
4-simplex amplitude and the eye diagram corresponding to the
tetrahedron amplitude. The 4-simplex amplitude defines the
dynamics of the space-time; it is to be understood as an operator
going from the Hilbert space of quantum states of a (past)
hypersurface to the Hilbert space of quantum states of a (future)
hypersurface obtained from the first one by a Pachner move
corresponding to the 4-simplex. The eye diagram is a statistical
weight which can be understood as the number of quantum states of
a tetrahedron defined by the areas of its 4 triangles. Indeed, as we have discussed in section \Ref{sec:qg4d}, a
(classical) tetrahedron is defined by 6 numbers (the 6 lengths of
its edges as an example) and is not fully determined by the 4
areas of its faces: we miss the values of the area of the 2
interior parallelograms \cite{bb}. This indeterminacy persists at
the quantum level. In fact, at the quantum level we can specify,
in addition to the 4 representations corresponding to the 4
triangle areas, an additional variable for each tetrahedron as a
basis for the space of intertwiners between the four
representations in a decomposition of the 4-valent vertex.
However, we are forbidden to specify a further variable
corresponding to the other parallelogram area, by a kind of
indeterminacy relation that forces us to leave it completely
undetermined \cite{bb}. More precisely, let us note $B_1,\dots B_4$
the bivectors of the four triangles of the tetrahedron. Then, one
can construct the areas $A_i=|B_i|$ of the four triangles, the
intermediate areas $A_{ij}=|B_i+B_j|$ of the internal
parallelograms and the volume squared of the tetrahedron
$U=B_1\cdot (B_2\times B_3)$. When quantizing, the bivectors
become $\sl$ generators in some representation $\rho_i$ and
$A_i^2$ become the Casimir of these representations. As for
$A_{ij}^2$, it lives in $\rho_i\otimes\rho_j$ and is diagonal on
each (sub)representation $\rho_{ij}$ in this tensor product. Such
an intermediate representation defines an intertwiner between all
four $\rho_1,\dots,\rho_4$. When computing the commutator of two
such internal areas one gets, as we already mentioned in section \Ref{sec:qg4d}: \bes [A_{12}^2,A_{13}^2]=4U. \ees 
This
way, $A_{12}^2$ and $A_{13}^2$ appear as conjugate variables: if
one fixes the value of $A_{12}$ then $A_{13}$ is undetermined.
This can be understood a purely algebraic level: if one defines
the intertwiner between $\rho_1,\dots,\rho_4$ through $\rho_{12}$,
then this intertwiner decomposes on a whole range of possible
$\rho_{13}$, without fixing it. Thus, the tetrahedron is never
fully specified: the quantum tetrahedron is a fuzzy tetrahedron,
with a fluctuating volume $U$ ($U$ depends on both $\rho_{12}$ and
$\rho_{13}$). Then the eye diagram basically measures the number
of possible states for a tetrahedron (each corresponding to a
possible choice of the unspecified parameter) that do not
correspond to different configuations in the partition function.
This explains why we consider it as part of the measure in the
partition function. This statistical interpretation is
well-defined in the Riemannian case for which the dimension of the
space of states is finite, and for which all the geometric quantization procedure leading to the identification of the space 
of states for a quantum tetrahedron was carried out. In the Lorentzian case, the parameters
are continuous and the dimensions become infinite, so that it is
better to think in terms of the connection: we replace the data of
the 2 interior areas by the Lorentz transformation between the
frames of the two 4-simplices sharing the tetrahedron, which
defines the connection internal to the tetrahedron and is
equivalent to the choice of two $H^+$ elements (time-normals). The
{\it eye diagram} can be obtained by integrating over the possible
connections ($\sl$ group elements relating the two normals
attached to the tetrahedron) living at the tetrahedron. This
\lq\lq eye diagram" weight allows all possible connections, thus
it randomizes the connection and correspondingly the space-time
curvature. More precisely, it appears as the {\it amplitude associated
to a connection} living at a tetrahedron defined by its four
triangle representations (areas). We are aware that \cite{baezriem}
proposed another (natural) interpretation for another tetrahedron
weight in the spin foam amplitude which does not fit with our
point of view. Here, we interpret the eye diagram as a localised
quantum fluctuation corresponding to what we could call a quantum
refractive wave. Refractive waves, as introduced in \cite{refrac},
are a new class of gravitational waves. They allow a discontinuity
in the metric while areas are still
well-defined. This area matching condition is exactly what we have
in the Barrett-Crane model: two touching 4-simplices will agree on
the four face areas of their common tetrahedron while disagreeing
on its complete geometry (different time normals, different
volumes). Moreover it is this indeterminacy which is responsible
for the non-topological character of the Barrett-Crane model.
Indeed, in the topological BF model the
weight associated to the eye diagram, obtained by full contraction of two $SO(3,1)$ or $SO(4)$ intertwiners without any 
restriction on the representations summed over, is simply (and always) one.

\medskip
One main difference with the Riemannian Barrett-Crane model is that
the degeneracy of the quantum Lorentzian 4-simplices is
unlikely, at least for what concerns the group representations used, so that we would not have to consider any
contribution to the amplitude from such 4-simplices. This is due
to the Plancherel measure. A degenerate 4-simplex would have some
of its triangles of zero area. However, $\rho=0$ has Plancherel
measure $\rho^2=0$ and equivalently has a zero probability. One
can be more precise and look at one particular triangle area
probability graph. As we simply would like to get an idea of
what is happening, to simplify the calculations, one can suppose
the space to be isotropic around the triangle. Then the amplitude
for the area $\rho$ is

\bes {\cal A}(\rho)=\rho^2\left(
\f{\sin(\rho\theta)}{\rho\sinh(\theta)} \right)^N \qquad
\theta=\f{1}{N}(2\pi-\theta_0) \ees where $N$ is the number of
4-simplices sharing this same triangle and $\theta_0$ is the
deficit angle corresponding to the curvature around the given
triangle, $\theta_0=0$ corresponding an approximatively flat
space-time. First, the amplitude is 0 for $\rho=0$. Then, we see
that we have a discrete series of maxima which hints toward the
possibility of generating dynamically a discrete area spectrum.
Moreover, this spectrum scales with the curvature. Nevertheless,
there is a big difference of the amplitude of each extremum, which
increases with $N$ i.e when we refine the triangulation.
Therefore, for important $N$, the most probable area $\rho$ is of
order 1 (likely between 1 and 6) and the other extrema are
negligible. This is a reason why it is likely that the relevant
regime will be for plenty of small 4-simplices at small $\rho$ and
not in the asymptotical limit $\rho\arr\infty$.

\medskip
\subsection{The Barrett-Crane model as a realization of the
projector operator}
Now we turn to the interpretation of the amplitudes defined by the
Barrett-Crane model. What kind of transition amplitudes do they
represent? We have seen that a path integral realization can be given,
in the relativistic particle case, both for the projector operator
over physical states, thus realizing in a covariant manner the
physical inner product among canonical states, and for the Feynman
propagator, or causal transition amplitude; moreover, we have seen
that a similar situation is present at least formally for quantum
gravity. Now we have a model that furnishes a rigorous construction
for a sum-over-histories formulation of quantum gravity. The issue is
then to realize whether we have in our hands a projector operator or
a causal evolution operator.

We will argue that the Barrett-Crane model is a realization of the
projection operator for quantum gravity, as argued also for example in \cite{ansdorf} and
in \cite{carloprojector}.
In \cite{ansdorf}, in fact, a discretization of the projection operator
was proposed, leading to an expression very close to that provided by
spin foam models.

In the 4-dimensional case, for the topology $\Sigma\times [0,1]$, with
$\Sigma$ compact, using a regular lattice with spatial sections $V$
and with time spacing
$\epsilon$ and spatial spacing $l$, this would read:

\bes
P &:=&\int\mathcal{D}N\,e^{i\,\int_{0}^{1} dt\int d^3
x\,N(x,t)\,\mathcal{H}(x)}\,
\rightarrow\,
P_{\epsilon,l}\,=\,\int\mathcal{D}N\,
e^{i\,\sum_{t=0}^{1/\epsilon}\sum_{x\in
V}\,\epsilon\,l^3\,N(x)\,\mathcal{H}(x)} \nonumber \\ &=&
\,\prod_{t}\prod_{x}\,\int_{-\infty}^{+\infty}dT\,e^{i\,\epsilon\,l^{3}\,T\,
\mathcal{H}}\,=\,\prod_{t}\prod_{x}\,\int_{0}^{+\infty}dT\,
\left(
e^{i\,\epsilon\,l^3\,T\,\mathcal{H}}\,+\,e^{-i\,\epsilon\,l^3\,T\,\mathcal{H}}
\right) \nonumber \\ &=&\,\prod_{t}\prod_{x}\,U_{\epsilon,l}(t,x).
\ees

The expression above is then interpreted as a sum over triangulations
and the operators $U_{\epsilon,l}$, being (a sum over) local evolution operators,
implement the action of the projection operator on a given spin
network and are identified with the quantum operators corresponding to
given Pachner moves in the spin foam case \cite{ansdorf}. The integral over $T$ is
then the analogue of the integral over the algebraic variables in the
spin foam model.

More generally, the idea is, we stress it again, to regard the
partition function for a spin foam model as giving the physical
inner product between (simple) spin network states as the action
of a projector over kinematical spin network states: \be
\langle \Psi_2 \mid \Psi_1 \rangle_{phys}\,=\,\langle \Psi_2 \mid
\mathcal{P}\mid \Psi_1
\rangle_{kin}\,=\,\sum_\Delta\,\lambda(\Delta)\,Z_\Delta, \ee
where the full partition function for the given spin foam model
has been split into a sum over triangulations (or their dual
2-complexes) $\Delta$ of partition functions associated with each
given one.

Still at a formal level, one may thus give a discretized
expression for the projection operator as a sum over triangulations
of projectors associated with each triangulation. In general one
may expect then this projector to have the form: \bes
Z(\Psi_1,\Psi_2)&=&_{kin}\langle \Psi_2\mid P_{\mathcal{H}=0}\mid
\Psi_1\rangle_{kin}=\int_{-\infty}^{+\infty}dT e^{i\,T\,\int
dx\mathcal{H}(x)}
=\sum_\Delta\lambda(\Delta)\sum_\sigma\int_{-\infty}^{+\infty}dT_\sigma
e^{i\,T_\sigma\,\mathcal{H}_\sigma} \nonumber\\
&=&\sum_\Delta\lambda(\Delta)\sum_\sigma\sum_{t_\sigma}\int_{-\infty}^{+\infty}dT_{t_\sigma}
e^{i\,T_{t_\sigma}\,\mathcal{H}_{t_\sigma}}\nonumber \\ &=&\,
\sum_\Delta\lambda(\Delta)\sum_\sigma\sum_{t_\sigma}\int_{0}^{+\infty}dT_{t_\sigma}\left(
e^{i\,T_{t_\sigma}\,\mathcal{H}_{t_\sigma}}\,+\,
e^{-\,i\,T_{t_\sigma}\,\mathcal{H}_{t_\sigma}}\right)\,\,\,\,\,\,\,\,\,
\label{eq:projdiscr},
\ees
depending on whether the discretization procedure associates the
relevant integral (i.e. the relevant integration variable $T$ or some
analogue of it) to the
4-simplices $\sigma$ or more specifically to each triangle $t_\sigma$ within
each 4-simplex $\sigma$.

To obtain a more conclusive argument for considering a given spin foam
model as realizing the projection over physical states one has however
to go beyond this and try to identify the distinguishing features of
the projector operator in the particular spin foam model under
consideration.
What are then these distinguishing features?
We have already discussed how the difference between a path
integral realization of the projector operator and a path integral
formula for the causal transition amplitude is marked uniquely by
the presence in the former of a $Z_2$ symmetry relating positive
and negative lapses (or proper times), both in the relativistic
particle case and in the formal path integral quantization of
gravity in the metric formalism.
We have also seen this symmetry present in the Ponzano-Regge
model, which is the 3-dimensional model corresponding to the
4-dimensional Barrett-Crane one.
Moreover, we have also noted that a similar and related $Z_2$
symmetry has to be present in the Barrett-Crane model as well as a
result of the quantization procedure that originated its
construction.
The task is then to locate clearly where, in the peculiar structure of
the amplitudes of the model, this symmetry is implemented and how.
Let us come back then to the expression for the Barrett-Crane
partition function \ref{eq:Z}.
We see that, considering the amplitude
for the faces and those for the internal edges as part of the measure,
as we suggest is the most reasonable thing to do, the model has
exactly the form \ref{eq:projdiscr}, since the amplitude for the vertices
of the 2-complex, encoding the dynamics of the theory and thus related
to the Hamiltonian constraint, is given by a product of terms each
associated to the triangles in the 4-simplex dual to the given vertex.

Even more remarkably, the $K$ functions associated with each
triangle are given by a sum of two exponentials with opposite
signs (with weighting factors in the denominators), resembling the
structure of the formal discretization given above.
Can we interpret the $K$ functions, because of their structure, as encoding the
same $Z_2$ symmetry in the proper time (or in the lapse) that
characterizes the projection operator?
The answer is yes.

Consider in fact that, as we noted above, the symmetry
$T\rightarrow -T$ in the proper time is the same as the symmetry
$N\rightarrow -N$ in the lapse function, in turn originating from
the symmetry $B\rightarrow -B$ in the bivector field, that we
expect to be present in the model at the quantum level. A change
in the sign of $B$ in our discretized context corresponds to a
change in the orientation of the triangles on the simplicial
manifold, and to a consequent change in the orientation of the
tetrahedra. Therefore we expect the sum over both signs of proper
time, assuring the implementation of the Hamiltonian constraint
but also erasing causality requirements from the model, to be
realized in our simplicial model as a sum over both orientations
of the triangles and of the tetrahedra in the manifold. This is
indeed what happens.

In fact, let us analyse more closely the $Z_2$ symmetry of the $K$
functions.
We use the {\it unique} decomposition of $\sl$
representation functions of the 1st kind (our $K$ functions) into representation
functions of the 2nd kind (we denote them $K^{\pm}$)\cite{Ruhl} and we write the $K$ function as
\bes
K^{\rho_{ij}}(x_i,x_j)&=&\frac{2\sin(\eta_{ij}\rho_{ij}/2)}{\rho_{ij}\sinh\eta_{ij}}\,
= \nonumber \\ &=&\,\frac{e^{i\,\eta_{ij}\,\rho_{ij}/2}}{i\rho_{ij}\sinh\eta_{ij}}\,-
\,\frac{e^{-\,i\,\eta_{ij}\,\rho_{ij}/2}}{i\rho_{ij}\sinh\eta_{ij}}\,=
\nonumber \\
&=&\,K^{\rho_{ij}}_{+}(x_i,x_j)\,+\,K^{\rho_{ij}}_{-}(x_i,x_j)\,= \nonumber \\
&=&\,K^{\rho_{ij}}_{+}(x_i,x_j)\,+\,K^{\rho_{ij}}_{+}(-x_i,-x_j)
\nonumber \\
&=&\,K^{\rho_{ij}}_{+}(x_i,x_j)\,+\,K^{\rho_{ij}}_{+}(x_j,x_i)\,= \nonumber \\ &=&\,
K^{\rho_{ij}}_{+}(\eta_{ij})\,+\,K^{\rho_{ij}}_{+}(-\eta_{ij}),
\label{eq:Ksplit}
\ees
which makes the following alternative expressions
of the same $Z_2$ symmetry manifest: 
\bes
K^{\rho_{ij}}(x_i,x_j)\,&=&\,K^{\rho_{ij}}(\eta_{ij})\,=
\,K^{\rho_{ij}}(-\eta_{ij})\,=\,K^{\rho_{ij}}(-x_i,-x_j)\,= \nonumber \\
&=&\,K^{\rho_{ij}}(x_j,x_i)\,=\,K^{-\rho_{ij}}(x_i,x_j). \ees

 We see that the symmetry
characterizing the projection operator is indeed implemented as a
symmetry under the exchange of the arguments of the $K$ functions
associated to each triangle in each 4-simplex (so that the
resulting model does not register any ordering among the
tetrahedra in each 4-simplex), or as a symmetry under the change
of sign (orientation) of the two tetrahedra sharing the given
triangle (so that the resulting model does not register the
orientation of the triangle itself), and these in turn are
equivalent to a change in sign of the hyperbolic distance between
the two points on the upper hyperboloid $H^+$ identified by the
vectors normal to the tetrahedra themselves. Also it is a symmetry between a representation labelled by $\rho$ and its 
dual labelled by $-\rho$.
It is also possible
to consider negative hyperbolic distance on the upper hyperboloid
$H^+$ as positive distances on the lower hyperboloid $H^-=\{
x\in\Rb^{3,1} | x\cdot x = 1, x^0 < 0\}$ in Minkowski space, so
that the symmetry we are considering is in the way the model uses
both the upper and lower hyperboloids in Minkowski space. Note
also the analogy between \ref{eq:Ksplit}, for the kernels $K$ and
\Ref{eq:had} for the Hadamard Green function for the relativistic
particle (maybe it is possible to make this analogy more precise
by studying the quantization of relativistic particles on the
upper hyperboloid $H^+$, but we do not analyze this possibility in
this thesis).

It is interesting to note also that the reality and positivity of
the partition fuction \cite{BaezChristensen}, a puzzling feature
indeed if the model was to represent a sum-over-geometries
realization of a causal transition amplitude, i.e. the matrix
elements in a spin network basis of some (unitary?) evolution operator, is
perfectly understandable if the model represents a path integral
realization of the projector operator (it is always possible to
find a basis such that the matrix elements of the projector
operator are real and positive). This symmetry can be traced back
exactly to the reality of the $K$ functions and this in turn can
be clearly seen as due to the symmetry worked out above.

Let us stress finally, and clarify, an important conceptual point. What we have identified is a symmetry under change of orientation 
which is present in the quantum amplitudes and makes the spin foam model not sensitive to the orientation of the various 
geometric elements of the simplicial manifold. The same situation is present in 3-dimensions for the Ponzano-Regge and the 
Turaev-Viro models, and in 4-dimensions for the Riemannian Barrett-Crane model. Of course, in the Riemannian case, no 
causal interpretation is possible, for the absence of a preferred time direction and of the notion of light cone and 
causality itself, 
therefore in this case we can just talk of dependence or independence from the orientation and that is all.
In the Lorentzian case, this characterization of the symmetry is still available, of course, and we can talk of whether the 
model discriminates between a given orienation and its opposite for triangles, tetrahedra or 4-simplices; however, we argue
 that at the fundamental quantum level, in absence of any time variable, there is no other notion of causality than the 
notion of {\bf ordering} and orientation, i.e. that it is the relative orientation of the various geometric elements, and then 
the ordering of the boundary states resuting from this, that generates a causal structure for spacetime. This explains why
we talk of causality when dealing with orientation dependence in this Lorentzian context. 
 
\section{Asymptotics of the $10j$-symbol and connection with the Regge action}
As we have already discussed, the dynamics of quantum states is encoded, in spin foam models, in the vertex amplitudes, i.e.
in the amplitudes associated with 4-simplices of the triangulation and with vertices of the dual 2-complex. Although the 
links with Plebanski actions we have shown and the way these amplitudes are constructed give support to the hope that they 
really encode all the dynamical properties of the gravitational field, we would like a further connection with gravity to
 arise in a suitable semi-classical limit of these amplitudes. In particular, we would like a similar situation to the 
3-dimensional case to be verified also in this 4-dimensional one, with the large spin asymptotics of the $10j$-symbol giving 
the exponential of the Regge action for simplicial gravity as a good approximation to the quantum amplitude, showing then
a clear link with the path integral approach to quantum gravity.

Let us consider the Riemannian case first (see \cite{llasym,baezasym,bs}). The vertex amplitude is 
given by a $10j$-symbol defined by the contraction of five
Barrett-Crane intertwiners, one for each tetrahedron in the 4-simplex, and given by the integral formula \Ref{eq:BCintRie}.
This integral formula gives us also an integral formula for the evaluation of the (simple spin network corresponding to the) 
vertex amplitude (analogous to the integral expression for the vertex amplitude in the Lorentzian Barrett-Crane model):

\bes
A_v(j)\,&=&\,(-1)^{\sum_{t}2j_t}\,\int_{(S^3)^5}\prod_t\,K^{j_t}(\phi_t)\,dx_1...dx_5\,\nonumber \\ 
&=&\,(-1)^{\sum_t 2j_t}\int_{(S^3)^5}\prod_t\frac{\sin(2j_t+1)\phi_t}{\sin\phi_t}dx_1...dx_5
\ees
where, just as in the Lorentzian case, the $x$ variables are interpreted as normals to the tetrahedra in the 4-simplex, 
so that $\phi_t=\arccos x \cdot y$ is the dihedral angle associated to the triangle $t$ common to the tetrahedra with normals
$x$ and $y$, the kernels $K$ result from the contraction of two representation matrices, the quantities $a_t = 2j_t + 1$ are 
the areas of the triangles, and we have neglected a factor $\frac{1}{\Delta_{j_1}..\Delta_{j_{10}}}$ in the denominator, 
since it does not affect the results to be described.

It is known \cite{BaezChristensen} that the integral above is positive, so we take a rescaling with parameter $N$ of the ten areas 
and consider its absolute value only in order to study its asymptotics for large $N$. As in the 3-dimensional case, there is
a region in the integration domain where the integrand is divergent, for $\phi_t =0,\pi$, so we split the integral into two 
contributions, one corresponding to the \lq\lq safe" sector of the integration domain, which we call it $I_>$, i.e. for 
$\phi_t \in [\epsilon, \pi - \epsilon]$, with $\epsilon << 1$ and one close to 
these critical points, i.e. for $\phi_t \in [0,\epsilon]\cup[\pi-\epsilon,\pi]$, corresponding to degenerate geometries for 
the 4-simplex, since they correspond to the 4-simplex being \lq\lq squeezed" to a 3-dimensional hypersurface, which we call 
$I_<$.

Consider $I_>$ first. Writing the kernels as a sum of exponentials we obtain an expression of the form:
\bes
I_>\,=\,\int_{(S^3)^5}\sum_{\sigma_t} f(x_1,..,x_5) \prod_t e^{i\,\sigma_t\, N\, a_t\,\phi_t}\,=\,\int_{(S^3)^5}\sum_{\sigma_t} f(x_1,..,x_5) \, e^{i\,\sum_t\sigma_t \,N \,a_t\,\phi_t}
\ees
where $\sigma = \pm 1$ and $f(x_1,..,x_5)=\frac{1}{\prod_t \sin\phi_t}$, and the quantity in the exponential can be 
intepreted as an action $S(\sigma,x,a)$ whose stationary points dominate the integral. Using the fact that the dihedral angles are
 constrained to satisfy the Schlafli identity we mentioned above, the stationary phase analysis of the integral, involving 
the variations of the action, in the limit $N\rightarrow\infty$ gives as a result \cite{llasym,bs,baezasym}:
\bes
I_>\,\sim\,\sum_i\,\frac{1}{N^\frac{9}{2}}\frac{f(x_i)}{\sqrt{H(x_i)}}\,\cos\left(\sum_t\, N\,a_t\,\phi_t(x_i)\,+\,\frac{\pi}{4}s(H(x_i))\right),
\ees
where the sum is over the statonary points $x_i$ of the action $S$,
$H(x_i)$ is the determinant of the matrix of second derivatives of $S$ 
computed at the points $x_i$, $s(H)$ is the signature of this matrix, and, most important, the resulting action, 
appearing in the argument of the cosine, once we have fixed the signs $\sigma$ to the values given by the stationary points 
(these result to be all positive or all negative), is just the contribution of a single 4-simplex to the Regge calculus 
action for gravity.

The result then shows that, in this asymptotic limit, the non-degenerate configuration in the spin foam model give a 
contribution of the order $O(N^{-\frac{9}{2}})$ to the amplitude and with a term proportional to the exponential of the 
Regge calculus action for simplicial gravity, just as the Ponzano-Regge model in 3-dimensions. This result was obtained first 
in \cite{BW} and then confirmed recently in \cite{llasym,baezasym,bs}.

One needs however to worry about the degenerate configurations as well, to see whether they actually dominate the amplitude
 or not. In the 3-dimensional case we have seen that their contribution is of the same order of magnitude as the
non-degenerate ones. 
In this case, however, the result is different. In fact we obtain, for the integral $I_<$ \cite{llasym,bs,baezasym}:
\bes
I_<\,\sim\,\frac{1}{N^2}\int_{(\mathbb{R}^3)^5}du_1..du_5\prod_t \frac{\sin(a_t \mid u_{t1}\,-\,u_{t2}\mid)}{\mid u_{t1} - u_{t2}\mid}
\ees
so we see that its contribution is of the order $O(N^{-2})$ so that it tends to dominate the amplitudes in the asymptotic 
limit.

This result may be obtained also defining \lq\lq degenerate spin networks" corresponding to these degenerate configurations 
and based on the symmetry group of the hypersurface to which the 4-simplex is confined when degenerate, i.e. the Riemannian 
group $\mathbb{E}^3$, since an appropriate formalism for their evaluation can be defined in full \cite{baezasym,bs}.

This result is, in itself, quite disappointing, since it may imply, if the asymptotic approximation to the amplitudes 
so computed is indeed what describes the semi-classical limit of the theory, that the Barrett-Crane spin foam model does not
reproduce classical simplicial gravity in the semiclassical limit but it is dominated only by degenerate geometries.

However, much more work is needed to understand the real implications of this result, for several reasons: it may well be that
classical gravity is not to be obtained by this method or in the asymptotic limit; recall in fact that the amplitudes have to
be used within a sum over triangulations and it may well be that the semi-classical limit in this wider context has to be 
approched and studied by very different methods than this asymptotic one; moreover, we have argued that the Barrett-Crane 
model corresponds in fact to a first order formulation of simplicial gravity, with two sets of variables, areas and dihedral 
angles to be treated on equal footing, while the procedures we described held fixed the areas and considered the contribution 
of different dihedral angle configurations to the amplitude; maybe a more symmetric procedure is needed, based on something 
like a WKB approximation for two sets of coupled variables. 

The Lorentzian case has been studied in details in \cite{bs}, to which we refer, and the results are analogous to those 
described above for the Riemannian. The degenerate and non-degenerate configurations give the same type of contribution to 
the asymptotic limit, and, most important, with the same order of magnitude. The implication of this result have then still 
to be understood also in this Lorentzian context.

\chapter{The group field theory formulation of spin foam models in 4d}
In this chapter we want to describe in detail how the Barrett-Crane model we have derived and discussed in the previous 
chapter can be obtained using the technology of field theories over group manifolds, in analogy with the group field theory
derivation of the Ponzano-Regge model we have already discussed. We deal with both the Riemannian and Lorentzian cases, and 
try to explain more at length the origin of the simplicial complexes as Feynman graphs of the group field theory; we discuss
the subtleties of this correspondence and whether these simplicial complexes are or are not simplicial manifolds; finally, 
we discuss the physical interpretation of the group field theories from a quantum gravity point of view.\footnotemark
\footnotetext{For a comparison of this approach to quantization with the canonical one and the 
spin foam procedure in the simple case of 2d BF theory, see \cite{LivPerRov}.}  

Matrix models were invented more than 15 years
 ago to give a precise formulation of 2d quantum gravity (or \lq\lq zero-dimensional string theory") 
\cite{D, ADF, KKM, BKKM, DS}. The group field theories may be seen as a generalization of these matrix models, where tensors
replace matrices as fundamental variables, and as such were constructed in 3 dimensions by Boulatov \cite{Boul}, whose model
 gives the Ponzano-Regge-Turaev-Viro (spin foam) formulation of 3d quantum gravity, as we have seen, and in 4 dimensions 
by Ooguri \cite{Oog}, giving the Crane-Yetter spin foam formulation of 4d BF theory 

\cite{CY, CKY}, a quantum topological field theory. It could be expected that imposing 
in a proper way the Barrett-Crane 
constraints in the Ooguri model would give in the end the Barrett-Crane state sum instead of the Crane-Yetter one, so giving
 quantum gravity in 4 dimensions instead of a topological theory. This is indeed the case, as we will see. The first 
derivation of the Barrett-Crane model from a field theory over a group manifold was given in \cite{DP-F-K-R}, and an 
alternative one was proposed in \cite{P-R}. A general formalism for deriving a whole class of spin foam models from a 
group field theory was developed in \cite{RR, RR2}, proving the power and versatility of the formalism itself, and 
further study on this topic is in \cite{DP1, DP2}. The model of \cite{P-R} reproduces exactly the 
Barrett-Crane state sum in the form given when we derived it from a lattice gauge theory perspective,
 while the model in \cite{DP-F-K-R} provides an 
alternative choice of amplitudes for the edges of the spin foam, and other variations leading to different choices are 
possible as we will show.

Before turning to a more detailed description of a field theory giving the Barrett-Crane state sum, we would like to point 
out
what are the most interesting aspects of this approach. First of all
the purely algebraic and combinatorial nature of the model, and its
background independence, are made manifest, since the field theory
action is defined in terms of integrals over a group of scalar
functions of group elements only, while its mode expansions involve
only the representation theory of the group used, and its Feynman
graphs (describing \lq\lq interactions") require only notions of
combinatorial topology to be constructed. Then these Feynman graphs
are interpreted as encoding the geometry of a 4-dimensional (Riemannian) spacetime, but no geometric notion enters in their
 definition. The basic tool involved in the definition of the models is the harmonic analysis of functions on the group, and 
the resulting representation theory, so no reference is made to any spacetime geometry or topology even, and this is why they allow for a fully background independent formulation of a theory of quantum spacetime. Moreover, this 
approach permits a natural implementation of a suitable sum over spin foams, i.e. the key 
ingredient missing in the other derivations and necessary for a
 complete definition of a spin foam model for 4-dimensional quantum gravity, since in the 4-dimensional case, as we stressed
 already, any choice of a fixed triangulation represents a truncation of the dynamical degrees of freedom of the 
gravitational field.    
 
\section{General formalism} 
The fundamental idea of this approach is that one can represent a tetrahedron in 4-dimensions, to which a state of the 
quantum theory can be associated, by a function of four group variables (where the group $G$ is of course the Lorentz group 
$sl$ in the Lorentzian case and $Spin(4)$ in the Riemannian) to be thought of as associated to its four triangular faces. 
The action is then chosen in such a way to mimic the combinatorial structure of a simplicial manifold, i.e. with an 
interaction term given by a product of five fields since a 4-simplex has five tetrahedra and a 4-simplex can be interpreted 
as an elementary interaction of tetrahedra. The kinetic term is of course quadratic in the fields, as is usual in field 
theories. These considerations are already almost enough to write down the models for the group field theory actions
 giving rise to the topological models corresponding to $BF$ theory in different dimensions. When it comes to gravity the
 situation is more complicated since there are many different ways of imposing the constraints that reduce BF theory to 
gravity, as conditions on the fields in the action; this ambiguity is what originates the different models, all 
understandable as different versions of the Barrett-Crane model, that we are going to discuss.

The field is a (scalar) function of four group elements: $\phi(g_1, g_2,g_3,g_4)$, that can be simply written as 
$\phi(g_i)$. We may require different symmetry properties on this field with respect to permutations of its four arguments, 
i.e. we may require: $\phi(g_1,g_2,g_3,g_4)=\phi(g_{\pi(1)},g_{\pi(2)},g_{\pi(3)},g_{\pi(4)})$, where $\pi$ is a generic 
permutation of $(1234)$ or an even permutation only, etc. Different choices give rise to different models, as we will see.
We also define group projectors $P_g$ and $P_h$ by means of which we impose additional conditions on the fields: we impose 
invariance under the group $G$ (gauge invariance) applying the projector $P_g$:
\bes
P_g\,\phi(g_1, g_2, g_3, g_4)\,=\,\int_G dg\,\phi(g_1 g, g_2 g, g_3 g, g_4 g),  
\ees
and invariance under an appropriately chosen subgroup $H$ applying the projector $P_h$:
\bes
P_h\,\phi(g_1,g_2,g_3,g_4)\,=\,(\prod_{i=1..4}\int_H dh_i)\,\phi(g_1 h_1, g_2 h_2, g_3 h_3, g_4 h_4).
\ees

One then defines an action for the field, which in 4 dimensions is a $\phi^5$ action, i.e. of the form $S=\int \phi^2 + 
\lambda \phi^5$, where the integral is over the group elements that can be thought of as representing the triangles 
of a simplicial complex; more precisely:
\bes
S[\phi]=\frac{1}{2}\int dg_{i}d\tilde{g}_{i}\phi(g_{i})\mathcal{K}(g_{i},\tilde{g}_{i})\phi(\tilde{g}_{i})-
\frac{\lambda}{5}\int dg_{ij}\mathcal{V}(g_{ij})\phi(g_{1j})\phi(g_{2j})\phi(g_{3j})\phi(g_{4j})\phi(g_{5j}),
\ees
with $\phi(g_{1j})=\phi(g_{12},g_{13},g_{14},g_{15})$ and so on. $\mathcal{K}$ and $\mathcal{V}$ are the kinetic operator 
(whose inverse, in the space of gauge invariant fields, is the propagator of the theory) and the vertex (or potential) 
operator respectively. 
Alternatively, one can expand the field in modes, i.e. apply the rules of the harmonic analysis of functions on groups, and 
re write the action in terms of these modes, so that a conjugate expression of the kinetic and potential operators has to be 
used. The modes of the field will be functions of four group representations, instead of the four group elements.

The partition function of the theory is given, as in the usual quantum field theory literature, by an integral over the 
field values of the exponential of the action, and can be re-expressed in terms of its Feynman graphs (as an expansions in 
powers of $\lambda$) as:
\bes
Z\,=\,\int\,\mathcal{D}\phi\,e^{-\,S[\phi]}\,=\,\sum_{\Gamma}\frac{\lambda^{v[\Gamma]}}{5!^v\, v!\,sym[\Gamma]}\,Z[\Gamma], 
\label{eq:Z2}
\ees
where $v[\Gamma]$ and $sym[\Gamma]$ are the number of vertices and the order of symmetries  (number of automorphisms) of the
 Feynman graph $\Gamma$. This last number $sym[\Gamma]$ can be shown \cite{DP2} to be: 
$sym[\Gamma,v] = 2^{5\,v/2}\cdot(5\,v/2)!$. 

So we are interested in finding the form of the amplitude for a generic Feynman graph, i.e. the Feynman rules of the theory.
We will see that this amplitude for each Feynman graph is exactly the partition function for fixed triangulation that we 
have derived from a lattice gauge theoretic approach above, for a triangulation to be put in correspondence with a given 
interaction graph of the group field theory. In order to understand how this correspondence is to be found, let us analyse 
more closely the construction of the Feynman graphs.

\begin{figure}
\begin{center}
\includegraphics[width=8cm]{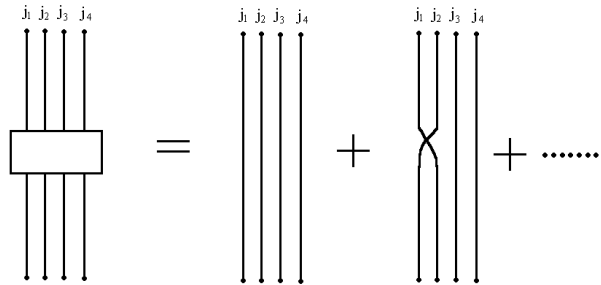}
\caption{The propagator of the theory; each of the four strands carries a (simple) representation of the group and 
the box stands for a symmetrization of the four arguments, i.e. for a sum over given permutations of the ordering of the
arguments}
\end{center}
\end{figure}

\begin{figure}
\begin{center}
\includegraphics[width=7cm]{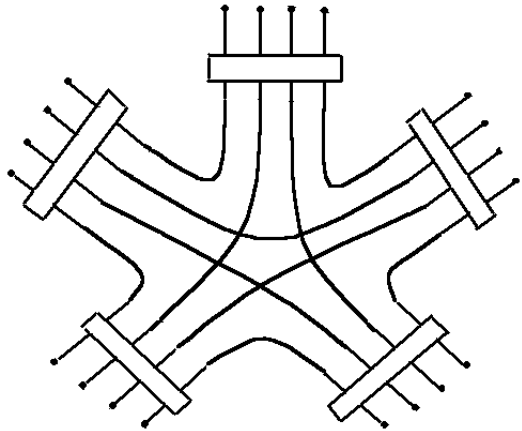}
\caption{The vertex of the theory; it has the combinatorial structure of a 4-simplex and has a (simple)
 representation of the group at each open end}
\end{center}
\end{figure}

The propagator (see figure 5.1) can be represented by four straight parallel lines, representing the four arguments of the field 
or of its modes, and has a group variable or a group representation) at the two ends of each line,
while the vertex operator (~\ref{eq:V}) (see figure 5.2 ) has the combinatorial structure (given by the way it pairs the 
arguments of the fields) of a 4-simplex, with 5
vertices and 4 lines coming out of each of them (five tetrahedra-propagators with four triangles-lines each), again with a 
group variable or group representation at the two ends of each line. Everywhere at
 the open ends of propagators and vertices are the
four group variables that are the arguments of the field. 
All the possible Feynman graphs are obtained (see figure 5.3) connecting a number of
vertices with the propagators,
constructing in this way what is called a \lq\lq fat graph" for each possible permutation of the lines in the propagator.

\begin{figure}
\begin{center}
\includegraphics[width=8cm]{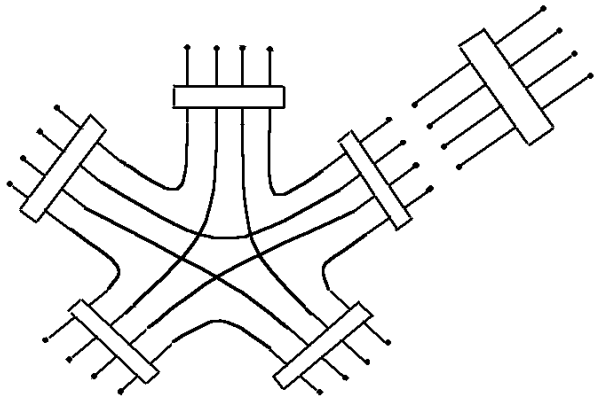}
\caption{The construction of a Feynman graph of the theory, connecting a propagator with a vertex}
\end{center}
\end{figure}
 
Each of the strands of a propagator is connected to a strand in one of
the five \lq\lq open sites" of the vertex, in such a way that the
orientations in vertices and propagators match. Moreover, because of
the symmetry of the field, each propagator really corresponds to many
different terms given by the different ways in which its four carried
indices can be permutated. Which are these possible ways depends of course by the symmetries under permutations we have 
imposed on the field, so that all possible permutations appear in the propagator, or only all the even ones, etc.

In this way, the sum over Feynman graphs $\Gamma$ can be re-written as a sum over fat graphs $\gamma$, with an amplitude for
 each fat graph, as: 
\bes
Z(\lambda)\,=\,1\,+\,\sum_\gamma\,w(\gamma)\cdot \lambda^{v(\gamma)}\,Z(\gamma),
\ees
where $w(\gamma) = \frac{l(\gamma)}{v(\gamma)!\cdot(5!^{v(\gamma)}\cdot (4!)^{e(\gamma)-v(\gamma)}}$, $w(\gamma)$ is the 
number of inequivalent ways of labelling the vertices of $\gamma$ with $v(\gamma)$ symbols and $e(\gamma)$ is the number of 
links in $\gamma$.
 
Now, each strand of the fat graph can go through several propagators
and several vertices, and at the end closes on itself, and
consequently forms a cycle. This cycle is labelled (in ``momentum space") by the representation assigned to the strand that 
encoles it. The abstract object formed by these
cycles, the edges and the vertices is a 2-complex, moreover, since
each face (cycle) is labelled by a representation of $G$,
this is a labelled 2-complex, i.e. a spin foam. So we see that there is
a 1-1 correspondence between the Feynman graphs of our field theory
and spin foams. 
This means that the sum in (~\ref{eq:Z2}) can be interpreted as a sum over spin foams (labelled 2-complexes), the amplitude
 $Z(\Gamma)$ as an amplitude for each spin foam, and the partition function itself defines a spin foam model. In turn, each 
2-complex can be thought of as dual to a triangulation, i.e. a simplicial complex. In general it is not true that a 
simplicial complex is determined by the dual 2-skeleton, but it is true in this case since the complex is obtained by gluing
of codimension-1 faces of simplices.
Knowing this, we can interpret the vertices as referring to 4-simplices, and the propagators to the tetrahedra they share, 
an interpretation that we had anticipated but that we now see as justified.
At this stage attention must be paid to the properties of the simplicial complexes that are in correspondence with the 
Feynman graphs. When working with a fixed triangulation, one could always assume the properties wanted for it, in particular
it was possible to assume that the complex would satisfy the conditions necessary to make it a proper manifold. Now, the 
theory  itself generates the simplicial complexes we work with as a sum over Feynman graphs as we have just said and we have
 to check whether these conditions are fulfilled or not.

Before discussing these manifold conditions, let us say more about which kind of complexes arise in the expansion.
First of all, if we define the orientation of a vertex of the 2-complex as an ordering of its adjacent edges up to even 
permutations, and an orientation of the edges as the one induced by the vertex, we see that the orientation of two vertices 
linked by an edge is consistent only if the edge is induced opposite orientation by the two vertices, and we can define a 
2-complex orientable if it admits a consistent orientation of all its vertices. Equivalently, a 2-complex is orientable if
all its loops are even. In particular, a 2-complex dual to a traingulation of an orientable manifold is orientable. Given 
this definition, it is easy to check \cite{DP-F-K-R} that if we require the field to be invariant under even permutations 
only of its arguments, the propagators above contain odd permutations only, while the pairing of strands determines always 
an odd permutation. If we then consider a loop of edges in a Feynman graph, we cross an equal number of vertices and edges, 
and therefore the strands undergo an even number of odd permutations. This means that for each loop we have always an even 
permutation and the 2-complex is orientable. Therefore the field theory based on a field invariant under even permutations 
only generates a sum over orientable 2-complexes only. this is not true for other choices pf symmetries for the field.
Second, it is also easy to check that the pairing of lines in vertices and propagators, taken into account the different 
possibilities involved in each propagator, originates not only all the triangulations of a given topology obtained with a 
given number of top-dimensional simplices, but also all the triangulations of all the possible topologies, indeed 
corresponding to all the possible pairing of lines, i.e. to all the possible gluing of triangles.
Therefore, the group field theory generates a sum over topologies as well as a sum over different triangulations of the same
topology, so gives rise to a fully background independent theory of spacetime.

Let us now turn to the manifold conditions \cite{DP1, DP2}. The condition  for a space $M$ to be a manifold is that each 
point
in it has to have a closed neighbourhood homeomorphic to a disk $D^4$ (in 4-dimensions). There is also another way to 
formulate the same condition: consider the space $M^\partial$ (closed with boundary) constructed by gluing the polyhedra
obtained by removing the open star of the original vertices, then the condition is that $M$ is a manifold iff the boundary of 
$M^\partial$ is a disjoint union of 3-spheres. 

There is nothing to check for the point in the interior of the simplices or on tetrahedra. As a consequence, in dimensions 2 
and 3 the condition is always fulfilled. In dimension four, we have to check this condition on the barycenters of triangles 
and edges. A complete characterization of these condition in terms of conditions on the fat graphs can be given 
\cite{DP1,DP2}
and it can be shown that in the expansion generated by the group field theory not all the fat graphs correspond to 
simplicial manifold, but some of them correspond to simplicial manifolds with conical singularities. Whether the field 
theory has then to be suitably modified so to incorporate only simplicial manifold, or whether the inclusion of conifolds 
has to be considered as a feature and not a problem of this approach, is at present not understood.
On this see also \cite{craneconi,craneconi2}.

\section{Riemannian group field theories}   
We present now different versions of the Riemannian Barrett-Crane model, existing in the literature, and as obtained by group 
field theory, and then give the most general structure of this kind of model.
These different versions arise because of the non-trivial interplay between the two projectors $P_g$ and $P_h$ we defined 
above. 

The starting point is the field $\phi(g_1,g_2,g_3,g_4)$. One can then impose on it the projector $P_h$ making it a field 
over four copies of the homogeneous space $G/H$ and then impose gauge invariance by means of the projector $P_g$, so that
the basic object of the theory is the field: $P_g P_h \phi(g_1,g_2,g_3,g_4)$. 
We recall that the projector $P_h$ imposes invariance under a given subgroup $H$ of $G$, and restricts the representations 
involved in these models to be the simple (class 1) representations with respect to this subgroup, while the projector $P_g$
imposes gauge invariance (invariance under the full group $G$).
Alternatively, one can decide to work with 
gauge invariant fields from the beginning, having as basic object the field $P_g \phi(g_1,g_2,g_3,g_4)$. On this field one 
can  then impose further projectors. Another choice then is involved: we can decide to impose more projectors and in 
different combinations in the kinetic and in the potential terms in the action. We will see that the models existing in the 
literature all work with gauge invariant fields as basic objects, but differ in the way they impose further projectors in 
the kinetic and potential terms. These difference are more manifest in the conjugate representation of the fields in terms
of representations of the group, i.e. using their mode expansion.

\subsection{The Perez-Rovelli version of the Riemannian Barrett-Crane model}
A first possibility is to work with gauge invariant fields $P_g\phi$, and then impose the combination of projectors $P_gP_h$
in the interaction term of the action only.

The action is then: 
\bes
\lefteqn{S[\phi]\,=\,\frac{1}{2}\,\int\,dg_{1}...dg_{4}\,\left[P_{g}\phi(g_{1}, g_{2}, g_{3}, g_{4})\right]^{2}\,+} 
\nonumber \\ &+&\,\frac{\lambda}{5!}\,\int \,dg_{1}...dg_{10}\,\left[P_{g}P_{h}P_g\phi(g_{1}, g_{2}, g_{3}, g_{4})\right]\,
\left[P_{g}P_{h}P_g\phi(g_{4}, g_{5}, g_{6}, g_{7})\right]\, \nonumber \\ &+&\left[P_{g}P_{h}P_g\phi(g_{7}, g_{3}, g_{8}, 
g_{9})
\right]\,\left[P_{g}P_{h}P_g\phi(g_{9}, g_{6}, g_{2}, g_{10})\right]\,\left[P_{g}P_{h}\phi(g_{10}, g_{8}, g_{5}, g_{1})
\right],
\;\;\;\;\;\;\;\;\;\;.    \label{eq:act}
\ees
The removal of the combination of projectors $P_g P_h$ would lead to a model directly analogous to the Boulatov model giving 
the Ponzano-Regge spin foam mdoel, called Ooguri model, and corresponding to 4-dimensional BF theory.
We work here with the field $\phi$, but it is clear that one could redefine $\phi'=P_g\phi$ without affecting the result, 
but changing only the expression for kinetic and potential operators.
Using the notation anticipated above, the kinetic term in \lq\lq coordinate space" is given by:
\bes
\mathcal{K}(g_{i},\tilde{g}_{i})\,=\,\sum_{\sigma}\int d\gamma \,\prod_{i}\delta\left( g_{i}\,\gamma\,\tilde{g}_{\sigma(i)}
^{-1}\right), \label{eq:K}
\ees 
which corresponds to a projector in to the space of gauge invariant fields, and is such that it is equal to its inverse, so 
that the propagator is $\mathcal{K}$ itself. In the formula above the product is over the four arguments of the field and 
the sum is over the possible even permutations of them. If we had used the redefinition $\phi\rightarrow\phi'$ the kinetic 
operator would have been only a product of deltas without any integral over $G=SO(4)$.

The vertex operator is:
\bes
\mathcal{V}(g_{ij})\,=\,\frac{1}{5!}\,\int d\beta_{i}d\tilde{\beta}_{i}dh_{ij}\,\prod_{i<j}\,\delta\left( g_{ji}^{-1}
\tilde{\beta}_{i}h_{ij}\beta_{i}^{-1}\beta_{j}h_{ji}\tilde{\beta}_{j}^{-1}g_{ij}\right) \label{eq:V}
\ees
where $\beta$ and $\tilde{\beta}$ are $SO(4)$ integration variables, and $h_{ij}\in SO(3)$. Again the product is over the 
arguments of the fields entering  in the interaction term.

The amplitude for a Feynman graph is then given by a product of interaction terms $\mathcal{V}(g_{ij})$, one for each vertex 
of the graph, connected by propagators $\mathcal{P}(g_i,\tilde{g}_i)=\mathcal{K}$, one for each edge of the graph, with an 
integral over the original group elements common to both kinetic and potential terms.
Doing this for a generic Feynman graph, using the 2-complex corresponding to it, and recalling the correspondence between
 this and a simplicial complex, it is easy to check that this model gives an amplitude identical to the one defined in 
section \Ref{sec:lgtBC} for the Barrett-Crane model, i.e. a delta function of a product of group elements assigned to the 
boundary links of each face of the 2-complex, with the same projections inserted, meant to impose the constraints that 
reduce BF theory to gravity. This fact is even more evident in the momentum formulation of the model

Because of the Peter-Weyl theorem we can
expand any $\mathcal{L}^{2}(G)$ function $\phi(g)$ over the group
$G=SO(4)$ in terms of matrices $D^{\Lambda}_{\alpha\beta}(g)$ of the
irreducible representations $\Lambda$ of $SO(4)$ (repeated indices are
summed over), so we can expand our basic field $P_g\phi(g_{1},g_{2},g_{3},g_{4})$ as:
\bes
P_g\phi(g_{1},g_{2},g_{3},g_{4})=\sum_{J_{1},J_{2},J_{3},J_{4}\Lambda}\phi^{J_{1}J_{2}J_{3}J_{4}}_{\alpha_{1}\beta_{1}
\alpha_{2}
\beta_{2}\alpha_{3}\beta_{3}\alpha_{4}\beta_{4}}D^{J_{1}}_{ \beta_{1}\gamma_{1} }(g_{1})...D^{J_{4}}_{ \beta_{4}\gamma_{4}}
(g_{4})\,C^{J_1J_2J_3J_4\Lambda}_{\gamma_1\gamma_2\gamma_3\gamma_4}\,
C^{J_1J_2J_3J_4\Lambda}_{\alpha_1\alpha_2\alpha_3\alpha_4}.
\ees  
and then redefine the field components as:
\bes
\Phi^{J_{1}J_{2}J_{3}J_{4}\Lambda}_{\beta_{1}\beta_{2}\beta_{3}\beta_{4}}\,\equiv\,\phi^{J_{1}J_{2}J_{3}J_{4}}
_{\alpha_{1}\beta_{1}\alpha_{2}\beta_{2}\alpha_{3}\beta_{3}\alpha_{4}\beta_{4}}\,
C^{J_{1}J_{2}J_{3}J_{4}\Lambda}_{\alpha_{1}\alpha_{2}\alpha_{3}\alpha_{4}},
\ees
where the $C$'s are $SO(4)$ intertwiners.

Inserting this expansion into the action, we have for the kinetic term in the action: 
\bes
\mathcal{K}\,=\,\sum_{J_{1},J_{2},J_{3},J_{4},\Lambda}\,\Phi^{J_{1}J_{2}J_{3}J_{4}\Lambda}_{\alpha_{1}\alpha_{2}\alpha_{3}
\alpha_{4}}\Phi^{J_{1}J_{2}J_{3}J_{4}\Lambda}_{\beta_{1}\beta_{2}\beta_{3}\beta_{4}}\,\left( \Delta_{J_{1}}\Delta_{J_{2}}
\Delta_{J_{3}}\Delta_{J_{4}}\right)^{-1}\,\delta_{\alpha_{1}\beta_{1}}...\delta_{\alpha_{1}\beta_{1}},
\ees
and for the potential term:
\bes
\mathcal{V}\,=\,\frac{1}{5}\sum_{J}\sum_{\Lambda}\Phi^{J_{1}J_{2}J_{3}J_{4}\Lambda_{1}}_{\alpha_{1}\alpha_{2}\alpha_{3}
\alpha_{4}}\Phi^{J_{4}J_{5}J_{6}J_{7}\Lambda_{2}}_{\alpha_{4}\alpha_{5}\alpha_{6}\alpha_{7}}\Phi^{J_{7}J_{3}J_{8}J_{9}
\Lambda_{3}}_{\alpha_{7}\alpha_{3}\alpha_{8}\alpha_{9}}\Phi^{J_{9}J_{6}J_{2}J_{10}\Lambda_{4}}_{\alpha_{9}\alpha_{6}
\alpha_{2}\alpha_{10}}\Phi^{J_{10}J_{8}J_{5}J_{1}\Lambda_{5}}_{\alpha_{10}\alpha_{8}\alpha_{5}\alpha_{1}}\,
\nonumber \\ \left( \Delta_{J_{1}}...\Delta_{J_{10}}\right)^{-2}\,\mathcal{B}_{J_{1}...J_{10}}^{BC},
\ees
where $\mathcal{B}^{BC}$ is the Barrett-Crane amplitude given by a product of five 
fully contracted Barrett-Crane intertwiners. From these expressions we can read off the formulae for the propagator
\bes
\mathcal{P}_{\alpha_{1}\beta_{1}...\alpha_{4}\beta_{4}}\,=\,\delta_{\alpha_{1}\beta_{1}}...\delta_{\alpha_{4}\beta_{4}} 
\left(\Delta_{J_{1}}\Delta_{J_{2}}\Delta_{J_{3}}\Delta_{J_{4}}\right)
\ees
and for the vertex amplitude given by the Barrett-Crane amplitude, as we said, with the coefficient appearing in the fomula
 above. Using these formulae, and taking into account the extra sum over the $\Lambda$s, labeling the intertwiners of the 
4 representations $J_{1},J_{2},J_{3},J_{4}$, we have for each Feynman graph $J$ the amplitude:
\bes
Z(J)\,=\,\sum_{J}\,\prod_{f}\Delta_{J_{f}}\,\prod_{e}\frac{\Delta_{1234}}{\left( \Delta_{J_{e_{1}}}\Delta_{J_{e_{2}}}
\Delta_{J_{e_{3}}}\Delta_{J_{e_{4}}}\right)}\,\prod_{v}\mathcal{B}^{BC}_{v}
\ees
where $\Delta_{1234}$ is the number of possible intertwiner between the representations $J_1,J_2,J_3,J_4$,
the sum is over the simple (because of the projection operators $P_{h}$ imposed above) representations of $SO(4)$ and
 the products are over the faces, edges and vertices of the Feynman graph, or equivalently over the faces, edges and 
vertices of the corresponding spin foam.
Moreover, we see that we have obtained precisely the Barrett-Crane state sum model as derived in section \Ref{sec:lgtBC}, with
a precise matching of the face, edge and vertex amplitudes; this is the Perez-Rovelli version of the Barrett-Crane model, 
as obtained in \cite{P-R}.  

We close by noting that the same kind of action, but with the group used being $SO(D)$, the field being a function of D 
group variables, and the projector $P_h$ imposing 
invariance under an $SO(D-1)$ subgroup, would result in the higher-dimensional generalization of the Riemannian 
Barrett-Crane model we have sketched in section \Ref{sec:lgtBChigh}.

\subsection{The DePietri-Freidel-Krasnov-Rovelli version of the Barrett-Crane model}
Instead of imposing the combination of projectors $P_g$ and $P_h$ only in the interaction term, we can impose it also in the 
kinetic term, using the combination $P_hP_g$, defining the action to be:
\bes
\lefteqn{S[\phi]\,=\,\frac{1}{2}\,\int\,dg_{1}...dg_{4}\,\left[P_h P_g\phi(g_{1}, g_{2}, g_{3}, g_{4})\right]^{2}\,+} 
\nonumber \\ &+&\,\frac{\lambda}{5!}\,\int \,dg_{1}...dg_{10}\,\left[P_{h}P_g\phi(g_{1}, g_{2}, g_{3}, g_{4})\right]\,
\left[P_{h}P_g\phi(g_{4}, g_{5}, g_{6}, g_{7})\right]\, \nonumber \\ &\times&\left[P_{h}P_g\phi(g_{7}, g_{3}, g_{8}, 
g_{9})
\right]\,\left[P_{h}P_g\phi(g_{9}, g_{6}, g_{2}, g_{10})\right]\,\left[P_{h}P_g\phi(g_{10}, g_{8}, g_{5}, g_{1})
\right],
\;\;\;\;\;\;\;\;\;\;.    \label{eq:act2}
\ees

In this case one can see that the kinetic operator of the theory is given, in coordinate space, by:
\bes
\mathcal{K}(g_{i},\tilde{g}_{i})\,=\,\sum_{\sigma}\int dh_i d\gamma d\tilde{\gamma} \,
\prod_{i}\delta\left( g_{i}\,\gamma\,h_i\,\tilde{\gamma}\,\tilde{g}_{\sigma(i)}^{-1}\right), \label{eq:K2}
\ees 
while the vertex is:
\bes
\mathcal{V}(g_{ij})\,=\,\frac{1}{5!}\,\int d\beta_{i}d\tilde{\beta}_{i}dh_{ij}\,\prod_{i<j}\,\delta\left( g_{ji}^{-1}
\tilde{\beta}_{i}h_{ij}\beta_{i}^{-1}\beta_{j}h_{ji}\tilde{\beta}_{j}^{-1}g_{ij}\right) \label{eq:V2}.
\ees

Now the kinetic operator is not a projector anymore, and therefore the definition of its inverse in coordinate space, giving
 the propagator of the theory, is much more complicated. The analysis is much simpler in momentum space.

The field, including the projectors, can be expanded in modes, again using the Peter-Weyl theorem, as follows:
\bes
P_h P_g\phi(g_{1},g_{2},g_{3},g_{4})=\sum_{J_{1},J_{2},J_{3},J_{4},\Lambda}\phi^{J_{1}J_{2}J_{3}J_{4}}
_{\alpha_{1}\beta_{1}\alpha_{2}
\beta_{2}\alpha_{3}\beta_{3}\alpha_{4}\beta_{4}}D^{J_{1}}_{ \beta_{1}\gamma_{1} }(g_{1})...D^{J_{4}}_{ \beta_{4}\gamma_{4}}
(g_{4})\,C^{J_1J_2J_3J_4\Lambda}_{\gamma_1\gamma_2\gamma_3\gamma_4}\,\nonumber \\
C^{J_1J_2J_3J_4\Lambda}_{\delta_1\delta_2\delta_3\delta_4}\,w^{J_1}_{\delta_1}w^{J_1}_{\alpha_1}...w^{J_4}_{\delta_4}w^{J_4}
_{\alpha_4}\,= \nonumber \\
=\,\sum_{J_{1},J_{2},J_{3},J_{4},\Lambda}\Phi^{J_{1}J_{2}J_{3}J_{4}}_{\beta_{1}\beta_{2}\beta_{3}\beta_{4}}D^{J_{1}}
_{\beta_{1}\gamma_{1} }(g_{1})...
D^{J_{4}}_{ \beta_{4}\gamma_{4}}(g_{4})\,C^{J_1J_2J_3J_4\Lambda}_{\gamma_1\gamma_2\gamma_3\gamma_4}\,\left( \Delta_1\,
\Delta_2\,\Delta_3\,\Delta_4\right)^{-\frac{1}{4}}.\,\,\,\,\,\,\,\,
\ees  
with the evident redefinition of the field. 

Using this expansion, the kinetic term in the action assumes the form:
\bes
\mathcal{K}\,=\,\sum_{J_{1},J_{2},J_{3},J_{4},\Lambda}\,\Phi^{J_{1}J_{2}J_{3}J_{4}}_{\alpha_{1}\alpha_{2}\alpha_{3}
\alpha_{4}}\Phi^{J_{1}J_{2}J_{3}J_{4}}_{\beta_{1}\beta_{2}\beta_{3}\beta_{4}}\,\left( \Delta_{J_{1}}\Delta_{J_{2}}
\Delta_{J_{3}}\Delta_{J_{4}}\right)^{-\frac{3}{2}} \,\Delta_{1234}\,\delta_{\alpha_{1}\beta_{1}}...\delta_{\alpha_{4}
\beta_{4}},
\ees
and the potential term is:
\bes
\mathcal{V}\,=\,\frac{1}{5!}\sum_{J}\Phi^{J_{1}J_{2}J_{3}J_{4}}_{\alpha_{1}\alpha_{2}\alpha_{3}
\alpha_{4}}\Phi^{J_{4}J_{5}J_{6}J_{7}}_{\alpha_{4}\alpha_{5}\alpha_{6}\alpha_{7}}\Phi^{J_{7}J_{3}J_{8}J_{9}
}_{\alpha_{7}\alpha_{3}\alpha_{8}\alpha_{9}}\Phi^{J_{9}J_{6}J_{2}J_{10}}_{\alpha_{9}\alpha_{6}
\alpha_{2}\alpha_{10}}\Phi^{J_{10}J_{8}J_{5}J_{1}}_{\alpha_{10}\alpha_{8}\alpha_{5}\alpha_{1}}\,
\left( \Delta_{J_{1}}...\Delta_{J_{10}}\right)^{-\frac{3}{2}}\,\mathcal{B}_{J_{1}...J_{10}}^{BC}.\,\,\,\,\,\,
\ees

The propagator can then be read off immediately from the expression of the kinetic term:
\bes
\mathcal{P}\,=\,\left( \Delta_{J_{1}}\Delta_{J_{2}}
\Delta_{J_{3}}\Delta_{J_{4}}\right)^{3/2}\,\Delta_{1234}^{-1}\,\delta_{\alpha_{1}\beta_{1}}...\delta_{\alpha_{4}\beta_{4}},
\ees

Building up the amplitudes for each Feynman graph as usual, the result is:
\bes
Z(F)\,=\,\sum_{J}\,\prod_{f}\Delta_{J_{f}}\,\prod_{e}\frac{1}{\Delta_{1234}}\,\prod_{v}\mathcal{B}^{BC}_{v}.
\ees

The function $\Delta_{1234}$ can be thought of as the norm square of the Barrett-Crane intertwiners between the 
representations $J_1, J_2,J_3,J_4$, since $B^{J_1J_2J_3J_4}_{\gamma_1\gamma_2\gamma_3\gamma_4}B^{J_1J_2J_3J_4}
_{\gamma_1\gamma_2\gamma_3\gamma_4}=\Delta_{1234}$; therefore this model, as it is easy to check, can be interpreted as 
having no edge amplitude at all, but simply using normalized Barrett-Crane intertwiners wherever they are involved in the
definition of the vertex amplitudes. 

We note, however, that no lattice gauge theory derivation of this model is known, due to the difficulties mentioned in 
section \Ref{sec:otherpossi}.

\subsection{Convergence properties and spin dominance}
Let us discuss now some properties of the models we have just derived. 
Not much is understood about the 
properties of the sum over 2-complexes; it is almost certainly a  divergent sum, but because it is interpreted as a 
perturbative expansion of a field theory partition function, this is not surprising nor would it be too upsetting, if it 
could be shown 
that the sum itself is just an asymptotic series of some known and computable (and possibly meaningful) function. i.e. if it
were possible to give it a non-perturbative definition and meaning. Unfortunately, no result of this kind is available yet,
unlike the 3-dimensional case \cite{llsum}. 

Anyway, convergence of the partition function or of the generic transition amplitudes, although useful, since it 
simplifies the technical matters, may well not be a necessity at all, since when computing expectation values of 
observables (which is were 
physics is) one usually works with expressions like: $\langle O \rangle=\frac{Z(O)}{Z}=\frac{\sum_F O A(F)}{\sum_F A(F)}$, 
and the fact that both numerator and denominator diverge may not preclude the meaningfulness of $\langle O\rangle$. Also, 
if one wants a canonical interpretation of the theory, one is interested in defining the scalar product 
between states $\langle \Psi \mid \Psi'\rangle$ where a normalization by means of a quotient with respect to the modulus 
of the vacuum state (partition function in the denominator) is not necessarily to be used, and for computing this kind 
of quantities, the well-posedness of the sum over histories definition of the states and of the scalar product may be 
necessary. The situation, thus, is not so clear.  

Still from a field theoretic point of view, another attractive possibility would 
be that each term of the expansion, i.e. each amplitude for a given Feynman graph, is either finite or regularizable (made 
finite). This would imply nothing about the behaviour of the full partition function, but will be important if the single 
terms in the perturbation expansion of the group field theory, i.e. the amplitudes for fixed triangulations, can be given a
 physical interpretation on their own. For example, if we interpret them as it is natural as path integral expressions for 
gravity on a fixed triangulation, i.e. path integrals for discrete gravity, their finiteness would be quite remarkable.

Luckily, it is possible to study these issues both analytically and numerically, thanks to the tools developed in 
\cite{baezriem,BaezChristensen}, i.e. an efficient algorithm for computing simple Riemannian spin networks (and thus 
spin foam amplitudes).
 
We start from the DePietri-Freidel-Krasnov-Rovelli version of the Barrett-Crane model. 
It can be shown numerically that the partition function for fixed triangulation in this version of the model diverges very 
rapidly even for the simplest triangulation, i.e. those involving very few vertices. In particular for the simple case of
a 6-vertices triangulation of a 4-sphere, after having imposed a spin cut-off $J$, it can be extimated a dvergence of the 
order $O(J^{23})$, so spectacularly rapid indeed. This divergence is mainly due to the rapid grow of the face amplitudes
with increasing spins, while the $10j$-symbols decay in $J$ as $O(J^{-2})$ and the $4j$-symbol or $\Delta_{1234}$ or \lq\lq 
eye diagram" grows as $O(J)$, 
so there may be a slower grow only for triangulations which involve many vertices and edges compared to faces; since for 
non-degenerate triangulations the number of tetrahedra is always $5/2$ times the number of 4-simplices, it may be possible 
only for triangulations were a very high number of 4-simplices share any given triangle.
Because of these convergence properties, the amplitudes are dominated by configurations with extremely high values of the
representations assigned to the triangles, which would naively (but possibly) correspond to very large areas of these.

In the Perez-Rovelli version, the situation is basically the opposite. 
Analytically it can be shown \cite{Per} that, for any triangulation of a pseudomanifold (manifold with possible conical 
singularities) such that each triangle lies in at least three 4-simplices, the sum over representations giving the quantum 
amplitude for the given triangulation is indeed finite, i.e. convergent. This implies, at least for this (large) class of 
triangulations or Feynman graphs, perturbative finiteness to all orders from the point of view of the group field theory or
from the lattice gauge theory point of view (that we have used to derive this model in section \Ref{sec:lgtBC}) a well 
defined partition function for any choice of discretization of the theory. Remarkable and quite striking.
More precisely, the following upper bound for the amplitude $Z(J)$ of
an arbitrary 2-complex $J$ was found \cite{Per}:
\bes
\mid Z(\Delta)\mid \leq \prod_{f\in J}\sum_{j_{f}}\left(\Delta_{j_{f}}\right)^{-1}\,=\,(\zeta(2)-1)^{F_{\Delta}}\approx(0.6)
^{F_{\Delta}},
\ees
where the product is over the faces of the 2-complex, the sum is over
their associated representations, $\zeta$ is the Riemann zeta
function, and $F_{\Delta}$ is the total number of faces in the 2-complex
$\Delta$.

Numerical tools can be applied here as well. In this way the covergence can be confirmed \lq\lq empirically". Moreover, the
numerical simulations show a very rapid convergence, and this due to the presence of the $\Delta_J$ factors in the 
denominator of the edge amplitudes, that tends to suppress high spins. In turn we have seen in section \Ref{sec:lgtBC} that 
this form of the edge amplitudes can be geometrically interpreted as coming from the gluing of 4-simplices along common 
tetrahedra, because it is this gluing, performed in the most simple way. This also implies a clear dominance of zero or very 
low spins, such that any configuration with more than a very few faces labelled by even spin $J=1/2$ will be suppressed.
The amplitude for any spin foam with all the faces labelled by zero spins evaluates to $1$, while even the very 
close spin foam with all faces labelled by zero except four neighbouring ones labelled by $J=1/2$ have amplitude of the 
order $2^{-20}$. The most generic but relevant configuration of the model would then be given by spin foams with most of 
the faces labelled by zero spins and a few small islands of faces labelled by slightly higher spins. 
The interpretation of this result is not trivial, since there is no complete agreement on the form of the area spectrum 
for the triangles, either given by $J(J+1)$ or by $2J+1$, so that the dominant zero spins configurations may correspond to 
either zero areas for all the triangles in the triangulations or triangulations where all the triangles are of Planck size.
However, it implies that the asymptotic analysis of the $10j$-symbol is not so relevant for this model.

Using numerical tools a different version of the Barrett-Crane model was proposed in \cite{baezriem}, designed so to be at 
the very limit of convergence, i.e. convergent but very slowly, and thus not dominated by low spins; the model is basically 
just as the DePietri-Freidel-Krasnov-Rovelli version, but with trivial amplitudes for the faces of the spin foams ($A_f=1$); 
however, no proper derivation is known for this model.

\subsection{Generalized group field theory action} 
The two versions of the Barrett-Crane mode we have just presented, as derived from a group field theory, differ in the way 
the projectors imposing the Barrett-Crane contraints are used, i.e. in the particular combination of them which is chosen, 
and in whether they are imposed only in the 
potential term in the action or also in the kinetic one. Now we want to discuss what happens when one does not make any 
explicit choice regarding this, and see what is the resulting model.

The analysis is best performed in the momentum representation, i.e. using the harmonic decomposition of the field.

We have first to see what is the expansion for the field depending on how we impose the projectors on it.
We can work with gauge invariant fields from the beginning, i.e. imposing first and both in the kinetic and potential terms
the projection over gauge invariant fields, having $P_g\phi(g_1,g_2,g_3,g_4)$. 
On this gauge invariant field we can impose the projector $P_h$, obtaining an object like $P_h P_g \phi$. This object has 
the mode expansion:
\bes
&& P_h\,P_g\,\phi(g_1,g_2,g_3,g_4)\,= \nonumber \\ &=&\,\sum_{J_{1},J_{2},J_{3},J_{4}\Lambda}\phi^{J_{1}J_{2}J_{3}J_{4}}_{\alpha_{1}\beta_{1}
\alpha_{2}
\beta_{2}\alpha_{3}\beta_{3}\alpha_{4}\beta_{4}}D^{J_{1}}_{ \beta_{1}\gamma_{1} }(g_{1})...D^{J_{4}}_{ \beta_{4}\gamma_{4}}
(g_{4})\,C^{J_1J_2J_3J_4\Lambda}_{\gamma_1\gamma_2\gamma_3\gamma_4}\,
C^{J_1J_2J_3J_4\Lambda}_{\delta_1\delta_2\delta_3\delta_4}\,w^{J_1}_{\delta_1}\,w^{J_1}_{\alpha_1}...w^{J_4}_{\delta_4}
w^{J_4}_{\alpha_4}\,\nonumber  \\ &=&\,\sum_{J_{1},J_{2},J_{3},J_{4}\Lambda}\Phi^{J_{1}J_{2}J_{3}J_{4}}_{\beta_{1}
\beta_{2}\beta_{3}\beta_{4}}D^{J_{1}}_{ \beta_{1}\gamma_{1} }(g_{1})...D^{J_{4}}_{ \beta_{4}\gamma_{4}}
(g_{4})\,C^{J_1J_2J_3J_4\Lambda}_{\gamma_1\gamma_2\gamma_3\gamma_4}\,\left(\Delta_1..\Delta_4\right)^{-1/4}.
\ees  
with the obvious redefinition of the field.

As we have anticipated, this field used everywhere in the action, without any additional projection imposed, gives rise to 
the DePietri-Freidel-Krasnov-Rovelli version we derived above.

The opposite choice, i.e. to impose first the projector over the subgroup $H$ and then over the full group $G$, having 
as field $P_gP_h\phi$, gives a different mode
expansion: 
\bes
&&P_g\,P_h\,\phi(g_1,g_2,g_3,g_4)\,= \nonumber \\ &=& \sum_{J_{1},J_{2},J_{3},J_{4}\Lambda}\phi^{J_{1}J_{2}J_{3}J_{4}}_{\alpha_{1}\beta_{1}
\alpha_{2}
\beta_{2}\alpha_{3}\beta_{3}\alpha_{4}\beta_{4}}D^{J_{1}}_{ \beta_{1}\gamma_{1} }(g_{1})...D^{J_{4}}_{ \beta_{4}\gamma_{4}}
(g_{4})\,w^{J_1}_{\gamma_1}\,w^{J_1}_{\delta_1}...w^{J_4}_{\gamma_4}
w^{J_4}_{\delta_4}\,C^{J_1J_2J_3J_4\Lambda}_{\delta_1\delta_2\delta_3\delta_4}\,
C^{J_1J_2J_3J_4\Lambda}_{\alpha_1\alpha_2\alpha_3\alpha_4}\, \nonumber  \\ &=& \sum_{J_{1},J_{2},J_{3},J_{4}\Lambda}
\Phi^{J_{1}J_{2}J_{3}J_{4}\Lambda}_{\beta_{1}\beta_{2}\beta_{3}\beta_{4}}D^{J_{1}}_{ \beta_{1}\gamma_{1} }(g_{1})...
D^{J_{4}}_{ \beta_{4}\gamma_{4}}(g_{4})\,w^{J_1}_{\gamma_1}...w^{J_4}_{\gamma_4}\,\left(\Delta_1..
\Delta_4\right)^{-1/4}.
\ees
and it can be easily checked that this choice of field gives a model that still uses only simple representations of $SO(4)$ 
and gauge invariant objects, so it has the same symmetries of the Barrett-Crane model, but contains only trivial amplitudes
for the vertices of the Feynman graphs ($A_v = 1$). Therefore we do not analyse this possibility further.

Another possibility is to start from the gauge invariant field $P_g\phi$ and impose on it the full combination of projectors
$P_gP_h$, working with the field $P_g P_h P_g \phi$; this act does not alter the symmetries of the field or of the action, 
but leads to a different mode expansion of course, and to different quantum amplitudes. This combination of projectors is
the one present in the Perez-Rovelli model, in the interaction term.  

There doesn't seem to be any other possibility than working with the field $P_h P_g \phi$ or with the field $P_g P_h P_g 
\phi$; then we can still choose to impose such a combination only in the kinematic or potential terms or in both.

However, there is more we can do, in order to obtain different models, all sharing the same symmetries, i.e. all of them 
obtained simply by imposition of the two projectors $P_g$ and $P_h$.
The crucial observation is that the combination $P_g P_h$ or $P_h P_g$ of the two projectors is {\it not} a projector 
itself. This means
that imposing it more than once is not at all equivalent to imposing it once only. We can thus consider models where the 
combination $P_g P_h$ is imposed, say, $n$ times obtaining the field $(P_g P_h)^n P_g\phi$, and different amplitudes, if we 
impose it a different number of times in the kinetic and in the potential terms. In fact, it is clear that it is not the 
number $n$ itself that makes a difference but the \lq\lq asymmetry" between the treatment of the kinetic and of the 
potential terms, since any given number of projectors can be re-absorbed in the definition of the field without affecting 
the result.

Suppose then that we impose the combination $P_g P_h$ $n$ times in the kinetic term and $m$ times in the interaction term, 
starting form the already gauge invariant field $\phi$. Of course it would be possible to impose it differently in all the 
fields appearing in the action, but this will spoil the symmetry of the theory treating differently the fields and it seems
not sensible.
The action would then be:
\bes
\lefteqn{S[\phi]\,=\,\frac{1}{2}\,\int\,dg_{1}...dg_{4}\,\left[(P_{g} P_h)^n P_g\phi(g_{1}, g_{2}, g_{3}, g_{4})\right]^{2}
\,+} \nonumber 
\\ &+&\,\frac{\lambda}{5!}\,\int \,dg_{1}...dg_{10}\,\left[(P_{g}P_{h})^m P_g\phi(g_{1}, g_{2}, g_{3}, g_{4})
\right]\,\left[(P_{g}P_{h})^m P_g\phi(g_{4}, g_{5}, g_{6}, g_{7})\right]\, \nonumber \\ 
&&\left[(P_{g}P_{h})^m P_g\phi(g_{7}, g_{3}, g_{8}, g_{9})
\right]\,\left[(P_{g}P_{h})^m P_g\phi(g_{9}, g_{6}, g_{2}, g_{10})\right]\,
\left[(P_{g}P_{h})^m P_g\phi(g_{10}, g_{8}, g_{5}, g_{1})\right],
\;\;\nonumber.    \label{eq:act3}
\ees 

The mode expansion for the field is, as it is easy to verify:
\bes
(P_g P_h)^n P_g\phi(g_{1},g_{2},g_{3},g_{4})\,=\,\Phi^{J_{1}J_{2}J_{3}J_{4}\Lambda}_{\beta_{1}\beta_{2}\beta_{3}\beta_{4}}
D^{J_{1}}_{ \beta_{1}\gamma_{1} }(g_{1})...D^{J_{4}}_{ \beta_{4}\gamma_{4}}(g_{4})\,C^{J_1J_2J_3J_4\Lambda}
_{\gamma_1\gamma_2\gamma_3\gamma_4}\,\Delta_{1234}^{n-1}\left( \Delta_1\,\Delta_2\,\Delta_3\,\Delta_4\right)^{-\frac{n}{2}}
\,\nonumber.
\ees 
for $n\neq 0$; for $n =0$ it holds the expansion for $P_g\phi$ we have given when dealing with the Perez-Rovelli model.

Therefore we get, for the kinetic term:
\bes
\mathcal{K}\,=\,\sum_{J_{1},J_{2},J_{3},J_{4},\Lambda}\,\Phi^{J_{1}J_{2}J_{3}J_{4}\Lambda}_{\alpha_{1}\alpha_{2}\alpha_{3}
\alpha_{4}}\Phi^{J_{1}J_{2}J_{3}J_{4}\Lambda}_{\beta_{1}\beta_{2}\beta_{3}\beta_{4}}\,\left( \Delta_{J_{1}}\Delta_{J_{2}}
\Delta_{J_{3}}\Delta_{J_{4}}\right)^{-n-1}\,\Delta_{1234}^{2n-1}\,\delta_{\alpha_{1}\beta_{1}}...
\delta_{\alpha_{4}\beta_{4}},\,\,\,\,\,
\ees
or 
\bes
\mathcal{K}\,=\,\sum_{J_{1},J_{2},J_{3},J_{4},\Lambda}\,\Phi^{J_{1}J_{2}J_{3}J_{4}\Lambda}_{\alpha_{1}\alpha_{2}\alpha_{3}
\alpha_{4}}\Phi^{J_{1}J_{2}J_{3}J_{4}\Lambda}_{\beta_{1}\beta_{2}\beta_{3}\beta_{4}}\,\left( \Delta_{J_{1}}\Delta_{J_{2}}
\Delta_{J_{3}}\Delta_{J_{4}}\right)^{-1}\,\delta_{\alpha_{1}\beta_{1}}...
\delta_{\alpha_{4}\beta_{4}},
\ees
for $n=0$, and for the interaction term:
\bes
\mathcal{V}\,=\,\frac{1}{5!}\sum_{J}\sum_{\Lambda}\Phi^{J_{1}J_{2}J_{3}J_{4}\Lambda_{1}}_{\alpha_{1}\alpha_{2}\alpha_{3}
\alpha_{4}}\Phi^{J_{4}J_{5}J_{6}J_{7}\Lambda_{2}}_{\alpha_{4}\alpha_{5}\alpha_{6}\alpha_{7}}\Phi^{J_{7}J_{3}J_{8}J_{9}
\Lambda_{3}}_{\alpha_{7}\alpha_{3}\alpha_{8}\alpha_{9}}\Phi^{J_{9}J_{6}J_{2}J_{10}\Lambda_{4}}_{\alpha_{9}\alpha_{6}
\alpha_{2}\alpha_{10}}\Phi^{J_{10}J_{8}J_{5}J_{1}\Lambda_{5}}_{\alpha_{10}\alpha_{8}\alpha_{5}\alpha_{1}}\,\nonumber \\
\left( \Delta_{J_{1}}...\Delta_{J_{10}}\right)^{-m-1}\,\prod_e \Delta_{e1234}^{m-1}\,\,\mathcal{B}_{J_{1}...J_{10}}^{BC}.
\ees
or 
\bes
\mathcal{V}\,=\,\frac{1}{5!}\sum_{J}\sum_{\Lambda}\Phi^{J_{1}J_{2}J_{3}J_{4}\Lambda_{1}}_{\alpha_{1}\alpha_{2}\alpha_{3}
\alpha_{4}}\Phi^{J_{4}J_{5}J_{6}J_{7}\Lambda_{2}}_{\alpha_{4}\alpha_{5}\alpha_{6}\alpha_{7}}\Phi^{J_{7}J_{3}J_{8}J_{9}
\Lambda_{3}}_{\alpha_{7}\alpha_{3}\alpha_{8}\alpha_{9}}\Phi^{J_{9}J_{6}J_{2}J_{10}\Lambda_{4}}_{\alpha_{9}\alpha_{6}
\alpha_{2}\alpha_{10}}\Phi^{J_{10}J_{8}J_{5}J_{1}\Lambda_{5}}_{\alpha_{10}\alpha_{8}\alpha_{5}\alpha_{1}}\,\nonumber \\
\left( \Delta_{J_{1}}...\Delta_{J_{10}}\right)^{-1}\,\mathcal{C}_{J_{1}...J_{10}\Lambda_1..\Lambda_5}.
\ees
for $m = 0$, where $\mathcal{C}$ is a $15j$-symbol built up using the 10 representations $J$ of the faces and the five 
representations $\Lambda$ for the edges. 

The most general model resulting from a group field theory, using the combination of projectors $P_g P_h$ acting on the 
field $P_g \phi$ is a version of the Barrett-Crane model

\bes
Z(\Delta)\,=\,\sum_{J}\,\prod_{f}\Delta_{J_{f}}\,\prod_{e}A_e(m,n)\,\prod_{v}\mathcal{B}^{BC}_{v}.
\ees
with edge amplitudes
\bes
A_e(m,n)\,=\,\frac{\Delta_{e1234}^{2(m - n)}}{\left(\Delta_1\Delta_2\Delta_3\Delta_4\right)^{m - n}}\;\;\;\;\;\;\;m,n\neq 0
\\
A_e(0,n)\,=\,\frac{\Delta_{e1234}^{1 - 2n}}{\left(\Delta_1\Delta_2\Delta_3\Delta_4\right)^{- n}}\;\;\;\;\;\;\;\;m=0,n\neq 0
\\
A_e(m,0)\,=\,\frac{\Delta_{e1234}^{2m - 1}}{\left(\Delta_1\Delta_2\Delta_3\Delta_4\right)^{m}}\;\;\;\;\;\;\;\;m\neq 0,n = 0
\ees

We see that the case $n=0, m=1$ corresponds to the Perez-Rovelli version of the Barrett-Crane model, as it should be; we 
note also that the case $n= 0, m \neq 0$ corresponds to the generalized versions of the Barrett-Crane model that can also be
obtained from a lattice gauge theory derivation as we have discussed in section \Ref{sec:otherpossi}.
Note also that the version of the Barrett-Crane model with trivial edge amplitudes $A_e$ belongs to this class and it is 
obtained by setting $m=n\neq 0$.
As anticipated and as logical, the result only depends on the difference $m-n$, which is then the true parameter labelling 
each representative of this class of models.

Of course the DePietri-Freidel-Krasnov-Rovelli version is not in this class of models, since it involves the combination
$P_h P_g$ acting directly on the field $\phi$. Therefore it belongs to a different class of models obtained by imposing 
$P_h P_g$ $n$ times in the kinetic term and $m$ times  in the interaction.

Note that $P_g (P_h P_g)^n \phi = (P_g P_h)^n P_g \phi$, so that any additional projector $P_g$ to the combination $P_h P_g$
will lead us back to the other class of models.
This also implies that the previously described class is obtained by requiring gauge invariance to the field 
$(P_h P_g)^n \phi$ which is the basic object of the other class (to which the DePietri-Freidel-Krasnov-Rovelli version 
belongs); the fact that this requirement is not trivially satisfied, so that imposing it we leave the class of models, hints 
to the non-gauge invariance of the field used in this second class of models. Of course, this non-gauge invariance of the
field may well be compensated by the form of the action, leading to gauge invariant quantum amplitudes for the Feynman 
graphs, as indeed happens in this case.

The relevant mode expansion (of course now it is $m, n \neq 0$ necessarily) is:
\bes
(P_h P_g)^n \phi(g_{1},g_{2},g_{3},g_{4})\,=\,\sum_{J_{1},J_{2},J_{3},J_{4},\Lambda}\Phi^{J_{1}J_{2}J_{3}J_{4}}
_{\beta_{1}\beta_{2}\beta_{3}\beta_{4}}D^{J_{1}}_{\beta_{1}\gamma_{1} }(g_{1})...
D^{J_{4}}_{ \beta_{4}\gamma_{4}}(g_{4})\,C^{J_1J_2J_3J_4\Lambda}_{\gamma_1\gamma_2\gamma_3\gamma_4}\,\nonumber \\ \,\left( \Delta_1\,
\Delta_2\,\Delta_3\,\Delta_4\right)^{-\frac{1}{4}}\,\Delta_{1234}^{n - 1}.
\ees  
so that the kinetic and potential terms are:
\bes
\mathcal{K}\,=\,\sum_{J_{1},J_{2},J_{3},J_{4}}\,\Phi^{J_{1}J_{2}J_{3}J_{4}}_{\alpha_{1}\alpha_{2}\alpha_{3}
\alpha_{4}}\Phi^{J_{1}J_{2}J_{3}J_{4}}_{\beta_{1}\beta_{2}\beta_{3}\beta_{4}}\,\left( \Delta_{J_{1}}\Delta_{J_{2}}
\Delta_{J_{3}}\Delta_{J_{4}}\right)^{-\frac{2n + 1}{2}} \,\Delta_{1234}^{2n - 1}\,\delta_{\alpha_{1}\beta_{1}}...
\delta_{\alpha_{1}
\beta_{1}},
\ees
and
\bes
\mathcal{V}\,=\,\frac{1}{5!}\sum_{J}\Phi^{J_{1}J_{2}J_{3}J_{4}}_{\alpha_{1}\alpha_{2}\alpha_{3}
\alpha_{4}}\Phi^{J_{4}J_{5}J_{6}J_{7}}_{\alpha_{4}\alpha_{5}\alpha_{6}\alpha_{7}}\Phi^{J_{7}J_{3}J_{8}J_{9}
}_{\alpha_{7}\alpha_{3}\alpha_{8}\alpha_{9}}\Phi^{J_{9}J_{6}J_{2}J_{10}}_{\alpha_{9}\alpha_{6}
\alpha_{2}\alpha_{10}}\Phi^{J_{10}J_{8}J_{5}J_{1}}_{\alpha_{10}\alpha_{8}\alpha_{5}\alpha_{1}}\,
\left( \Delta_{J_{1}}...\Delta_{J_{10}}\right)^{-\frac{2 m + 1}{2}}\,\nonumber \\ \prod_e \Delta_{e1234}^{m -1}\,
\mathcal{B}_{J_{1}...J_{10}}^{BC}.
\ees

Therefore also in this case we get of course a Barrett-Crane model:
\bes
Z(\Delta)\,=\,\sum_{J}\,\prod_{f}\Delta_{J_{f}}\,\prod_{e}A_e(m,n)\,\prod_{v}\mathcal{B}^{BC}_{v}.
\ees
with edge amplitudes
\bes
A_e(m,n)\,=\,\frac{\Delta_{e1234}^{2(m - n) - 1}}{\left(\Delta_1\Delta_2\Delta_3\Delta_4\right)^{m - n}}
\;\;\;\;\;\;\;m,n\neq 0.
\ees
again depending only by the parameter $m-n$, with the DePietri-Freidel-Krasnov-Rovelli version corresponding to the case
$m=n$ or $m-n = 0$.

Interestingly, the Perez-Rovelli version of the Barrett-Crane model belongs to this class as well, for $m - n = 1$. 
Indeed, all this second class of models is in one-to-one correspondence with the subclass $n =0$ of the first class,
except, strikingly, for the DePietri-Freidel-Krasnov-Rovelli model. 
This correspondence and all the relations between the two classes deserve to be understood better.

We conclude by noting that there exists even another class of models obtained by using the combination $(P_g P_h)^n\phi$ in
the kinetic term only (we have seen that using it in the interaction term leads to trivial vertex amplitudes), while using
the combination $(P_g P_h)^m P_g\phi$ in the interaction term; this choice leads to an edge amplitude of the form:
\bes
A_e(m,n)\,=\,\frac{\Delta_{e1234}^{2(m - n) + 1}}{\left(\Delta_1\Delta_2\Delta_3\Delta_4\right)^{m - n + 1/2}}
\;\;\;\;\;\;\;m,n\neq 0
\\
A_e(m,n)\,=\,\frac{\Delta_{e1234}^{- n + 2}}{\left(\Delta_1\Delta_2\Delta_3\Delta_4\right)^{- n + 1/2}}
\;\;\;\;\;\;\;n\neq 0.
\ees
No other combination of projectors seems allowed.

The overall result of this analysis is that we have a very large class of models, all with the same amplitude for the 
vertices of the Feynman graphs (spin foams), but differing in their edge amplitudes, i.e. in their propagators.
They all share the same symmetries, involving the same set of representations of $SO(4)$ and of course all 
implementing the Barrett-Crane constraints at the quantum level, and depend on two new discrete parameters $m$ and
$n$, that, when non zero, enter always combined as $m - n = M$. These models generalize both the Perez-Rovelli version and 
the DePietri-Freidel-Krasnov-Rovelli version of the Barrett-Crane model, while it seems that there is no easy field
 theoretic derivation possible for the version \lq\lq at the brick of convergence" proposed by Baez et al. 

This opens several new possibilites for further research: it allows first of all to deal with the group field theory 
formulation of the Barrett-Crane model in greater generality; furthermore, it may play a crucial role in a possible 
application of renormalization group techniques to spin foam models, in the context of group field theory formulations of 
them, and one can envisage a renormalization group flow for the group field theory action $S(\lambda,M)$, with a running of
both the parameter $\lambda$ (coupling constant) corresponding to a renormalization of the vertices of the Feynman graphs 
and of the parameter $M$ corresponding to a renormalization of the propagators, and thus playing an analogue role of a 
discretized mass for the quanta of the field over the group (or homogeneous space); also, one can think of imposing a
 general function of the combination of projectors $P_g P_h$, maybe coming from considerations about the a classical 
gravity action (Plebanski action), and then expand this general function in powers of $P_g P_h$, thus obtaining a linear
combination of the elements in the class of models we have introduced. All these possibilities are currently being 
investigated.

\section{Lorentzian group field theories}
We now consider the Lorentzian Barrett-Crane model, and its realization via a group field theory.
The actual form of the Barrett-Crane model in the Lorentzian case
differ depending on the way we choose to realize our simple
representations, with a consequently different geometrical
interpretation of the resulting model, as we have already discussed. In fact \cite{VK} the simple
representations of the Lorentz group can be realized in the space of
square integrable functions on hyperboloids in Minkowski
space. These are: $Q_{1}$, or Lobachevskian space, characterized by
$x^{\mu}x_{\mu}=1$ and $x^{0}>0$, the positive null cone $Q_{0}$, with
$x^{\mu}x_{\mu}=0$ and $x^{0}>0$, and the imaginary Lobachevskian
space $Q_{-1}$, with $x^{\mu}x_{\mu}=-1$. 

The relevant Barrett-Crane models are those using the realizations on
$Q_{1}$, which involves only spacelike representations of the type
$(0,\rho)$, and $Q_{-1}$, which instead involves both timelike $(j,0)$
and spacelike $(0,\rho)$ representations. The fist case was developed
in \cite{BC2,P-R2}, while the second one appeared in \cite{P-R3}. In both cases one can give
an explicit formula for the evaluation of spin networks, and for the
amplitudes appearing in the spin foam partition function, as we will
see in the following.  

The procedure is the same as the one explained in the previous section; we consider a field 
$\phi(g_{i})=\phi(g_{1},g_{2},g_{3},g_{4})$
over 4 copies of $SL(2,\mathbb{C})$, assumed to be completely
symmetric in its four arguments, and we define the two following
projectors:
\be
P_{g}\phi(g_{i})\,=\,\int_{SL(2,\mathbb{C})}dg\,\phi(g_{1}g,g_{2}g,g_{3}g,g_{4}g)
\ee
imposing the Barrett-Crane closure constraint, and
\be
P_{h^{\pm}}\phi(g_{i})\,=\,\int_{U^{\pm}}..\int_{U^{\pm}}dh_{1}..dh_{4}\,\phi(g_{1}h_{1},g_{2}h_{2},g_{3}h_{3},g_{4}h_{4})
\ee
imposing the simplicity constraint, where $U^{\pm}$ are two subgroups
of $SL(2,\mathbb{C})$; more precisely, $U^{+}=SU(2)$ and
$U^{-}=SU(1,1)\times\mathbb{Z}_{2}$. 

Choosing the first sign in the
projector appearing in the action leads to the Barrett-Crane model
based on $Q_{1}$ and containing only spacelike simple representations
$(0,\rho)$, since $SL(2,\mathbb{C})/SU(2)\simeq Q_{1}$, while the
choice of the second sign in the projector leads to the Barrett-Crane model
on $Q_{-1}$, with both series of simple representations $(0,\rho)$ and
$(j,0)$, since the coset space
$SL(2,\mathbb{C})/SU(1,1)\times\mathbb{Z}_{2}$ is isomorphic to the
hyperboloid $Q_{-1}$ with the opposite points identified.

Therefore we have an additional freedom in defining the model, other than the combination of projectors we use in the 
kinetic and potential terms; this last freedom is of course still present and can be exploited, as we have done in the 
previous section, to construct the Lorentzian analogue of the class of models we have obtained in the Lorentizan case; 
however, exactly because not much changes in this Lorentzian context with respect to the Riemannian case, we do not discuss
this generalized Lorentzian models in the following.

We consider only the Lorentzian analogues of the Perez-Rovelli version of the Barrett-Crane model, the one based on $Q_1$ 
and the one based on $Q_{-1}$.

The action for the theory is again: 
\bes
S^{\pm}[\phi]\,=\,\int
dg_{i}\,[P_{g}\phi(g_{i})]^{2}\,+\,\frac{\lambda}{5!}[P_{g}P_{h^{\pm}}P_g\phi(g_{i})]^{5},
\ees
where $\lambda$ is an arbitrary real constant, and the fifth power stands for
\bes
[\phi(g_{i})]^{5}\equiv\,\phi(g_{1},g_{2},g_{3},g_{4})\phi(g_{4},g_{5},g_{6},g_{7})\phi(g_{7},g_{3},g_{8},g_{9})
\phi(g_{9},g_{6},g_{2},g_{10})\phi(g_{10},g_{8},g_{5},g_{1})
\ees

Carrying out the same steps as in the previous section, with
particular care to the technical problems coming from the fact that
$SL(2,\mathbb{C})$ is not compact, we can compute the expressions for
the propagator and vertex of the theory, both in ``coordinate'' and in
``momentum'' space, and the amplitude for each Feynman diagram, which
in turn is, as before, in 1-1 correspondence with a combinatorial
2-complex $J$.

The general structure of this amplitude is of course the same spin
foam structure we encountered before, with amplitudes for faces,
edges, and vertices of the 2-complex, and an integral over the
continuous representations instead of a sum, but of course the particular
form of this amplitude and the representations involved are different
for the two cases based on $Q_{1}$ or $Q_{-1}$.
 
In the first case ($Q_{1}$) the amplitude is given by \cite{P-R2,
P-R3}:
\bes
A^{+}(J)\,=\,\int_{\rho_{f}}d\rho_{f}\prod_{f}\rho_{f}^{2}\prod_{e}A_{e}^{+}(\rho_{1},\rho_{2},\rho_{3},\rho_{4})
\prod_{v}A^{+}_{v}(\rho_{1},...,\rho_{10}) \label{eq:ampllor}
\ees
with
\bes
A_{e}^{+}(\rho_{1},\rho_{2},\rho_{3},\rho_{4})\,=\,\int_{Q_{1}} dx_{1}dx_{2}
K^{+}_{\rho_{1}}(x_{1},x_{2})K^{+}_{\rho_{2}}(x_{1},x_{2})K^{+}_{\rho_{3}}(x_{1},x_{2})K^{+}_{\rho_{4}}(x_{1},x_{2})
\nonumber \label{eq:edg}
\ees
and 
\bean 
\lefteqn{A^{+}_{v}(\rho_{1},...,\rho_{10})\,=\,\int_{Q_{1}}dx_{1}...dx_{5}K^{+}_{\rho_{1}}(x_{1},x_{5})K^{+}_{\rho_{2}}
(x_{1},x_{4})K^{+}_{\rho_{3}}(x_{1},x_{3})K^{+}_{\rho_{4}}(x_{1},x_{2})}
\\ && K^{+}_{\rho_{5}}(x_{2},x_{5})K^{+}_{\rho_{6}}(x_{2},x_{4})K^{+}_{\rho_{7}}(x_{2},x_{3})K^{+}_{\rho_{8}}(x_{3},x_{5})
K^{+}_{\rho_{9}}(x_{3},x_{4})K^{+}_{\rho_{10}}(x_{4},x_{5}),
\eean
where, for $\eta_{ij}$ being the hyperbolic distance on $Q_1$ of the points $x_i$ and $x_j$,
\bes
K_\rho^+(x_i,x_j)\,=\,K^+_\rho(\eta_{ij})\,=\,\frac{2\sin(\eta\rho/2)}{\rho\sin\eta}
\ees
is the $K$ function we had already introduced.
 
Note that all these expression correspond to the evaluation of a
relativistic spin network, i.e. the contraction of a certain number of Lorentzian Barrett-Crane intertwiners, and
require the dropping of one of the integrals to be regularized, because of the non-compactness of the Lorentz group. Of
course, only representations of the type $(0,\rho)$ appear. The results of \cite{bb2} show that this simple regularization 
is enough for the simple spin networks appearing in the Barrett-Crane model, and for a much larger class of them, to have a
finite evaluation, in spite of the non-compactness of the domains of the integrals involved in their definition, which is
wha makes this result highly non-trivial.

For the model based on $Q_{-1}$ we have instead:
\bes
 A^{-}(J)\,=\,\sum_{j_{f}}\int_{\rho_{f}}d\rho_{f}\prod_{f}(\rho_{f}^{2}+j_{f}^{2})\prod_{e}A_{e}^{-}(\rho_{1},...,\rho_{4};
j_{1},...,j_{4})\prod_{v}A^{-}_{v}(\rho_{1},...,\rho_{10};j_{1},...,j_{10}),\,\,\,\,\,\, \label{eq:ampllor2}
\ees
where it is understood that each face of the 2-complex is labelled either
by a representation $(0,\rho)$ or by a representation $(j,0)$, so that
either $j_{f}=0$ or $\rho_{f}=0$ for each face $f$. 
In this case, we have:
\bes
A_{e}^{-}(\rho_{1},...,\rho_{4};j_{1},...,j_{4})\,=\,\int_{Q_{-1}} dx_{1}dx_{2}
K^{-}_{j_{1}\rho_{1}}(x_{1},x_{2})K^{-}_{j_{2}\rho_{2}}(x_{1},x_{2})K^{-}_{j_{3}\rho_{3}}(x_{1},x_{2})K^{-}_{j_{4}\rho_{4}}
(x_{1},x_{2})
\nonumber \label{edgeamplor}
\ees
and 
\bean 
\lefteqn{A^{-}_{v}(\rho_{1},...,\rho_{10};j_{1},...,j_{10})\,=\,\int_{Q_{-1}}dx_{1}...dx_{5}K^{-}_{j_{1}\rho_{1}}
(x_{1},x_{5})K^{-}_{j_{2}\rho_{2}}(x_{1},x_{4})K^{-}_{j_{3}\rho_{3}}(x_{1},x_{3})K^{-}_{j_{3}\rho_{4}}(x_{1},x_{2})}
\\ && K^{-}_{j_{5}\rho_{5}}(x_{2},x_{5})K^{-}_{j_{6}\rho_{6}}(x_{2},x_{4})K^{-}_{j_{7}\rho_{7}}(x_{2},x_{3})
K^{-}_{j_{8}\rho_{8}}(x_{3},x_{5})K^{-}_{j_{9}\rho_{9}}(x_{3},x_{4})K^{-}_{j_{10}\rho_{10}}(x_{4},x_{5}).
\eean
Again it is understood that one of the integrations has to be dropped,
and an explicit expression for the functions
$K^{-}_{j_{f}\rho_{f}}(x_{1},x_{2})$ can be given \cite{gelfand,Ruhl,P-R3}, of course, differing in the two cases for $j=0$ 
or $\rho=0$.

The properties of the Riemannian model and its physical interpretation, discussed above, hold also 
for the Lorentzian one, including the finiteness.
In fact it was proven \cite{CPR}, using the results of \cite{bb2}, that, for any non-degenerate and finite (finite number 
of simplices) triangulation, the amplitude (~\ref{eq:ampllor}) is finite, in the sense that the integral over the simple 
continuous representations of the Lorentz group converges absolutely. This convergence is made possible by the simplicity 
constraint on the representations and by the presence of the edge amplitude (~\ref{eq:edg}), as in the Riemannian case. The 
analysis of the amplitude \ref{eq:ampllor2} was not yet carried out. 

Since, as we have seen, the amplitude for a given 2-complex is interpretable as a term in a Feynman expansion of the field
theory, its finiteness means that the field theory we are considering is finite order by order in perturbation theory, which
 is truly remarkable for a theory of Lorentzian quantum gravity. This result is even more remarkable if we think that the 
sum over the representations of the gauge group is a precise implementation of the sum over geometries proposal for quantum 
gravity we mentioned in the introduction, and that the path integral implementing this in the original approach was badly 
divergent in the Lorentzian case.  Of course, finiteness only does not mean necessarily correctness and in particular the 
choice of the edge amplitude ~\ref{eq:edg} has still to be justified by geometrical or physical considerations, apart from 
those coming from a lattice gauge theory derivation. 

Regarding questions of spin dominance, the analysis of the Lorentzian models by numeical techniques is of course much more
complicated than for the Riemannian ones; however, a few things can still be said; just as in the Riemannian case, the 
convergence of the model based on spacelike 
representations only is likely to imply dominance of the lower representations in the model, but now the $\rho = 0$ 
representation is not allowed, due to the Plancherel measure, as we have discussed when we have introduced the Lorentzian 
Barrett-Crane model; this avoids certainly a possible complication, which is present instead in the Riemannian case, but 
makes again the relevance of the asymptotic results on the Lorentzian $10j$-symbols not so clear.

\section{Quantum field theoretic observables: quantum gravity transition amplitudes}
A very difficult (conceptual and technical) problem in quantum gravity
(but the situation in the classical theory is not much better) is the
definition and computation of the observables of the theory (see \cite{carloObs}), which are required
to be fully gauge (diffeomorphism) invariant. As mentioned when dealing with 3-dimensional spin foam mdoels, the group field
 theory permits us to define and compute a natural set of observables for
the theory, namely the n-point functions representing transition
amplitudes between eigenstates of geometry, i.e. spin networks (better
s-knots. Clearly these are purely quantum observables and have no analogue in the classical theory. 

As in chapter 3, and repeating to same extent what we have said there for completeness, we start the discussion 
of these observables, following
\cite{P-R-Obs}, first in the context of the canonical theory, and then within the group field theory framework.

The kinematical state space of the canonical theory is given by a Hilbert space of s-knot states, solutions of the
gauge and diffeomorphism constraints, $\mid s\rangle$, including the
vacuum s-knot $\mid 0\rangle$. On this space a projection operator $P:\mathcal{H}_{diff}\rightarrow\mathcal{H}_{phys}$
from this space to the physical state space of the solutions of the
Hamiltonian constraint, $\mid s\rangle_{phys}=P\mid s\rangle$, is defined. The
operator $P$ is assumed to be real, meaning that $\langle s_{1}\cup
s_{3}\mid P\mid s_{2}\rangle=\langle s_{1}\mid P\mid s_{2}\cup
s_{3}\rangle$, and this implies an invariance of the corresponding amplitude under the exchange of its two arguments; in a
  Lorentzian context this would represent physically the invariance under
exchange of past and future boundaries, so would characterize an a-causal transition amplitude. 
The $\cup$ stands for the disjoint union of two s-knots, which is another s-knot. The quantities
\bes
W(s,s')\equiv \,\,_{phys}\langle s\mid s'\rangle_{phys}\,=\,\langle s\mid P\mid s'\rangle
\ees
are fully gauge invariant (invariant under the action of all the
constraints) objects and represent transition amplitudes between
physical states. However, the explicit and rigorous definition of the
operator $P$ is very difficult to achieve, and we are about to see how
a covariant approach, based on the group field theory formalism, allows
to construct explicitely the transition amplitudes we want without the
use of any explicit projector operator. 
 
We can introduce in $\mathcal{H}_{phys}$ the operator
\bes
\phi_{s}\mid s'\rangle_{phys}\,=\,\mid s\cup s'\rangle_{phys}
\ees
with the properties of being self-adjoint (because of the reality of
$P$) and of satisfying $[\phi_{s},\phi_{s'}]=0$, so that we can define
\bes
W(s)\,=\,\,_{phys}\langle 0\mid \phi_{s}\mid 0\rangle_{phys}
\ees
and
\bes
W(s,s')\,=\,\,_{phys}\langle 0\mid \phi_{s}\phi_{s'}\mid
0\rangle_{phys}\,=\,W(s\cup s').
\ees
In this way we have a field-theoretic definition of the $W$'s as
n-point functions for the field $\phi$. 

Before going on to the realization of these functions in the context of
the field theory over a group, we point out in which sense they encode
the dynamics of the theory, an important property in our background
independent context.

Consider the linear space of (linear combination of) spin networks
(with complex coefficients), say, $\mathcal{A}$, with elements $A=\sum c_{s} s$. Defining on
$\mathcal{A}$ an algebra 
product (with values in the algebra itself) as $s\cdot s'=s\cup s'$, a star operation giving, for each
s-knot $s$, an s-knot $s^{*}$ with the same underlying graph and the
edges labelled by dual representations, and the norm as $\mid\mid
A\mid\mid= sup_{s}\mid c_{s}\mid$, then $\mathcal{A}$ acquires the
stucture of a $C^{*}$-algebra (assuming that the product is continuous in a suitably chosen topology, and other important 
technicalities, which are actually not proven yet). Moreover, $W(s)$ is a linear functional
on this algebra, which turns out to be positive definite. 

This permits the application of the GNS construction \cite{Haag} to reconstruct,
from $\mathcal{A}$ and $W(s)$, a Hilbert space $\mathcal{H}$,
corresponding to the physical state space of our theory, including a
vacuum state $\mid 0\rangle$, and a representation $\phi$ of
$\mathcal{A}$ in $\mathcal{H}$ with $W(s)=\langle 0\mid\phi\mid
0\rangle$.

Clearly $\mathcal{H}$ is just the Hilbert space $\mathcal{H}_{phys}$ of the
canonical theory, so this means that there is the possibility of
defining the physical state space, annihilated by all the canonical
constraints, without making use of the projection operator $P$. The
full (dynamical) content of the theory is given by the $W$ functions,
and these, as we are going to see now, can be computed in a fully
covariant fashion.

Using the field theory described above, its n-point functions are
given, as usual, by:

\be
W(g_{1}^{i_{1}},...,g_{n}^{i_{1}})=\int\mathcal{D}\phi\,\,\phi(g^{i_{1}}_{1})...\phi(g^{i_{n}}_{n})\,e^{-S[\phi]},
\ee
where we have used a shortened notation for the four arguments of the
fields $\phi$ (each of the indices $i$ runs over the four arguments of the field).

Expanding the fields $\phi$ in \lq\lq momentum space'', we have \cite{P-R-Obs} the following explicit (up to a
rescaling depending on the representations $J_{i}$)
expression in terms of the ``field components''
$\Phi_{J_{1}J_{2}J_{3}J_{4}\Lambda}^{\alpha_{1}\alpha_{2}\alpha_{3}\alpha_{4}}$:
\be
W_{J^{1}_{1}J^{2}_{2}J^{3}_{3}J^{4}_{4}\Lambda^{1}}^{\alpha^{1}_{1}\alpha^{2}_{2}\alpha^{3}_{3}\alpha^{4}_{4}}.....
_{J^{n}_{1}J^{n}_{2}J^{n}_{3}J^{n}_{4}\Lambda^{n}}^{\alpha^{n}_{1}\alpha^{n}_{2}\alpha^{n}_{3}\alpha^{n}_{4}}\,=
\,\int\mathcal{D}\phi\,\,\phi_{J^{1}_{1}J^{2}_{2}J^{3}_{3}J^{4}_{4}\Lambda^{1}}^{\alpha^{1}_{1}\alpha^{2}_{2}\alpha^{3}_{3}
\alpha^{4}_{4}}...\phi_{J^{n}_{1}J^{n}_{2}J^{n}_{3}J^{n}_{4}\Lambda^{n}}^{\alpha^{n}_{1}\alpha^{n}_{2}\alpha^{n}_{3}
\alpha^{n}_{4}}\,e^{-S[\phi]}.
\ee
However, the W functions have to be invariant under the gauge group
$G$ to which the $g$'s belong, and this requires all the indices $\alpha$ to be 
suitably paired (with the same representations for the paired indices)
and summed over. Each independent choice of indices and of their pairing defines an
independent W function. If we associate a 4-valent vertex to each
$\phi_{J^{1}_{1}J^{2}_{2}J^{3}_{3}J^{4}_{4}\Lambda^{1}}^{\alpha^{1}_{1}\alpha^{2}_{2}\alpha^{3}_{3}\alpha^{4}_{4}}$
with label $\Lambda^{1}$ at the vertex and $J^{i}_{i}$ at the i-th edge,
and connect all the vertices as in the chosen pairing, we see that we
obtain a 4-valent spin network, so that independent  n-point functions
$W$ are labelled by spin networks with n vertices.

To put it differently, to each spin network $s$ we can associate a
gauge invariant product of field operators $\phi_{s}$

\bes
\phi_{s}\,=\,\sum_{\alpha}\,\prod_{n}\,\phi_{J^{1}_{1}J^{2}_{2}J^{3}_{3}J^{4}_{4}\Lambda^{1}}^{\alpha^{1}_{1}
\alpha^{2}_{2}\alpha^{3}_{3}\alpha^{4}_{4}}.
\ees
This provides us with a functional on the space of spin networks
\bes
W(s)\,=\,\int\mathcal{D}\phi\,\phi_{s}\,e^{-S[\phi]}
\ees
that we can use, if positive definite, to reconstruct the full Hilbert space
of the theory, using only the field theory over the group, via the GNS
construction. This is
very important in light also of the difficulties in implementing the
Hamiltonian constraint in the canonical loop quantum gravity approach.

The transition functions between spin networks can be easily computed using a perturbative
expansion in Feynman diagrams. As we have seen above, this turns out
to be given by a sum over spin foams $\sigma$ interpolating between the n spin
networks representing their boundaries, for example:

\bes
W(s,s')\,=\,W(s\cup s')\,=\,\sum_{\sigma/\partial\sigma=s\cup
s'}A(\sigma).
\ees  

Given the interpretation of spin foams as quantum 4-geometries,
discussed in detail in the last sections, it is clear that this
represents an implementation of the idea of constructing a
quantum gravity theory as a sum over geometries, as sketched in the
introduction.

We refer again to \cite{Mikov} for a different way of using the field theoretical techniques, giving rise to a Fock space of spin networks on which creation and annihilation operators constructed from
 the field act. 
In \cite{Mikov}, the perturbative expression for the transition amplitude between states in which one sums over spin foams 
only for a given number of vertices, and does not make this number going to infinity, is considered as physical meaningful 
on its own. It 
would represent the transition amplitude between spin networks for a given number of time steps, each of these corresponding
 to a 4-simplex of planckian size in the triangulation. In this approach, then, time would be discrete as well, the time 
variable would be given by the number of 4-simplices, and the transition amplitudes of the theory would be finite. 
This idea is clearly related to approaches like unimodular gravity when one treats the 4-volume fixed when varying the 
action for gravity, and in a quantum context would correspond to constraining the transition amplitudes to be given by a sum
 over all those histories of the gravitational field such that the 4-volume is bounded from above. More work
 is needed however to understand to which extent this idea is viable. 

\section{A quantum field theory of simplicial geometry?}
We have seen that the formalism of field theories over group manifolds provides us with the most complete definition of a 
theory of a quantum spacetime: it gives a prescription for a sum over topologies and for a sum over all the possible 
triangulations of each given topology, and furnishes the quantum amplitudes that have to be assigned to each 4-dimensional 
configurations, being the spin foam amplitudes we have already derived from and connected to a classical action for gravity.
The result is a fully background independent theory, which is discrete, based on simplicial decompositions of spacetime, and
purely algebraic since all the ingredients come from the representation theory of the Lorentz group, and no reference is 
made to spacetime and to its geometry. moreover, it gives a picture of  spacetime (and spin foams) as emerging as a possible
Feynman graph for the interaction of quanta of geometry, which is extremely intriguing and attractive.
It seems there are enough reasons to take it seriously as the right and best definition of the theory we are looking for.

However, we have used the group field theory only in a rather partial way, in deriving the various versions of the 
Barrett-Crane model, since we have just used the expression for the kinetic and potential terms in the action to write down 
the amplitudes for the Feynman graphs of the theory, assuming and not proving, by the way, that the perturbative expansion
can be defined. There is more to a quantum field theory than its Feynman graphs. We lack for example a precise canonical
 formulation of the theory: a definition of its one-particle states, a physical interpretation of them and of their quantum 
numbers, a study of bound states and a notion of (the analogue of) particles and their antiparticles, a full construction of
a Fock space of states, including a rigorous definition of creation and annihilation operators (for some work on this see 
\cite{Mikov}), and of field-theoretic quantum observables other than the transition amplitudes; we lack a full 
characterization of these transition amplitudes and of their symmetries, with a full construction of causal and a-causal 
ones; we would need a better definition of its perturbative expansion; the list could go on.

We can just guess, at present, what the structure of this group quantum field theory would be.
Consider the field $\phi(g_1,g_2,g_3,g_4)$. It is a scalar function of four group elements, each to be thought of as 
corresponding to a triangle in a possible triangulation, or alternative as a bivector, considering that 
$\wedge^2\mathbb{R}^{3,1}$ is isomorphic to the Lie algebra of the Lorentz group. In 3 dimensions each of the three 
arguments of the field would correspond to a vector, and in two dimensions to a scalar, geometrically associated to lines 
and points, in a discrete setting. The reason why the basic objects of the theory are associated to co-dimension two 
surfaces, a fact that lies at the very basis of the spin foam approach, where the basic variables are representations 
labelling 2-cells, has to do with the curvature being a always a 2-form and being given always as parallel transport along 
closed path, and thus with very basic geometric considerations. It can be probably justfied further by other kinds of 
arguments. 

Consider then the function of one group variable only: $\phi(g)$; this would be the basic field in a second quantization of 
the bivectors or of the triangles along these lines; its \lq\lq Fourier decomposition", i.e. the result of the harmonic 
analysis on the group, $\phi(g)\,=\,\sum_J \phi^J D^J(g)$, would be the starting point for a definition of a Fock space for 
the quantized field, and this needs a definition of creation and annihilation operators starting from the field modes 
$\phi^J$, just as for ordinary quantum fields in Minkowski space the creation and annihilation operators are defined from 
the Fourier modes resulting from the Fourier expansion of the field. Of course, one may just postulate such a definition 
\cite{Mikov} and go ahead, and this may lead to interesting results as well, but we feel that a proper and complete 
derivation and justification of the definition used would greatly enhance the understanding of the subject. 
Recall for example that for scalar fields in Minkowski space, the commutation relations defining the creation and 
annihilation operators are obtained (or, better, justified) by the canonical commutation relations between the field and 
their conjugate momenta, in turn coming from the classical Poisson brackets; what the analogue of these relations is in this 
purely group theoretic context is not clear.
These operators would generate a Fock space of states of the field, each characterized by a given number of quanta each 
carrying a given representation $J$, that would play the role of the momentum (or of the energy) of the quantum particle, 
again in full analogy with ordinary quantum fields in Minkowski space.

Having done this, one could also consider the case for anti-particles, i.e. anti-bivectors or anti-triangles, defined, as
 different objects with respect to the particles, only in the
case of a complex field over the group, and having correpondent creation and annihilation operators, again coming from the 
harmonic analysis of the field; after all, what we have in the group field theory approach to spin foam models, or even 
just from the point of view of the simplicial geometry, is a relativistic (because we preserve Lorentz invariance) theory of
interacting triangles or bivectors, and we know that the presence of interaction that makes the theory of a single 
relativistic particle inconsistent, requiring the passage to a field theory and the existence of antiparticles.
The natural interpretation of these antiparticles would be that they correspond to \lq\lq particles" with opposite spacetime
orientations, again just as in the usual QFT; having a given oriented triangle or a given bivector, the opposite orientation
would be the corresponding anti-triangle or anti-bivector. Further analogic justification for such an interpretation could
of course be given (e.g. role of PCT symmetry in QFT).

The field could then actually be written also as: $\phi_1(g_1)\times\phi_2(g_2)\times\phi_3(g_3)\times\phi_4(g_4)$, and so 
it would describe four independent bivectors or, geometrically, triangles with no \lq\lq interaction" among them, if it was 
not for the constraint of gauge invariance, i.e. invariance under the (right) action of the Lorentz group, imposed by means
 of the projector $P_g$. Note that indeed this projector is imposed first and directly on the field $\phi$ (at least in the 
interaction term in the action) in all the models we considered. This constraint has the effect of tying together the four 
bivectors and the associated triangles, making the field a genuine function of the four of them. This is evident in the mode 
expansion of the field: without the gauge invariance, the field is expanded (we use a more sketchy notation) as:
 $\phi(g_1,g_2,g_3,g_4)=\phi^{1234}D^1(g_1)...D^4(g_4)\simeq\phi^1D^1(g_1)...\phi^4 D^4(g_4)=\phi^1(g_1)...\phi^4(g_4)$,
while the gauge invariant field $P_g\phi$ is expanded, with a suitable redefinition of the field components, as:
 $\phi(g_1,g_2,g_3,g_4)=\Phi^{1234\Lambda}D^1(g_1)...D^4(g_4)C^{1234\Lambda}$ and can {\it not} be splitted into the product 
of four independent fields. Therefore we we can interpret the invariance under the gauge group as the requirements that the 
quantum particles of our field, i.e. the bivectors or, geometrically, the triangles, come always combined 
together in the form of \lq\lq bound states" made out of four of them, these behaving in turn as the true quanta of the
 field. These would be bound states of particles and anti-particles, i.e. of bivectors and triangles with different 
orientations, and would correspond, in the simplicial geometry interpretations, to tetrahedra. Also for these tetrahedra
the notion of corresponding anti-particles, being tetrahedra with different \lq\lq orientation in spacetime" should be 
defined in much the same way as for the triangles. From this point of view one would expect it possible to define an action
of a spacetime inversion operator, equivalently of a charge conjugation operator, exchanging particles with anti-particles
on the field configuration and on its quantum amplitudes, but this may imply enlarging the group we are working with from 
the connected part of the full Lorentz group we are working with to include discrete transformations; also, because as we
 have seen the symmetries under permutations of the arguments of the field affect the orientation properties of the 
2-complexes and of the triangulations corresponding to the Feynman graphs, we could probably expect a non-trivial interplay
between the discrete symmetries of the full Lorentz group and these symmetries under permutations.

The role of the simplicity constraint in this setting is to impose the \lq\lq geometricity" of the triangles appearing in
the theory , i.e. the fact that the bivectors associated to them can be obtained as wedge product of edge vectors (the 
simplicial analogue of a tetrad field), and thus there exist a metric field that can be reconstructed from these simplicial 
data. It is imposed as a further constraint on the group elements associated to the triangles already forming a tetrahedron,
i.e. already required to constitute a bound state, as a restriction on their \lq\lq momenta" (representations of the group),
by adding a further projector $P_h$ to the gauge invariant field $P_g\phi$, in all the models we have described, and not by
working with simple bivectors from the start, i.e. from a field defined on four copies of the hyperboloid in Minkowski 
space, and then constrained to be gauge invariant, which would result in a field $P_gP_h\phi$. The quantum field theoretic 
interpretation of this fact is to be understood.

The group field theory action represents then the theory of interacting tetrahedra, each being the bound state of four 
triangles; these tetrahedra interact with the elementary interaction given by the fifth order interaction interpretable 
geometrically as a 4-simplex made of five tetrahedra; all this was anticipated when describing the group field theory 
formalism, and of course is at the root of the fact that the Feynman graphs can be interpreted as triangulations of 
spacetime pseudomanifolds, but a full quantum picture in terms of particles, antiparticles, bound states, and so on, would
 certainly make it more precise and would help in understanding much more of the whole formalism.

Among the issues that may be better understood if this picture is developed we cite: the physical meaning of the coupling 
constant appearing in the field theory action, the nature of the transition amplitudes of the theory, the role that 
causality plays in it and in spin foam models in general. 

What we would have, if all these ideas can be made more precise, is a formulation of a quantum theory of spacetime as a 
purely combinatorial and algebraic, thus background independent, quantum field theory of simplicial geometry. An attractive
and intriguing possibility indeed.

\chapter{An orientation dependent model: implementing causality} \label{sec:causal} 
When discussing the interpretation of the Barrett-Crane spin foam model as a realization of the projector operator from 
kinematical states to physical states, giving a definition of the inner product of the canonical theory, we have tied this
interpretation to the fact that the transition amplitudes defined by it do not register the orientation of the 2-complexes 
on which they are based, and thus of the spacetime triangulations dual to them. From an algebraic point of view, this is due 
to a symmetry under the exchange of each representation with its dual, which is present in the vertex amplitudes of the 
model, and, in general, in all the simple spin networks as defined by means of the Barrett-Crane intertwiners.
In a Lorentzian context the orientation of the elements of the triangulation, and thus the order among them, admits the 
interpretation in terms of causal structure of spacetime, and thus one can say that the transition amplitudes defined by the
Barrett-Crane model do not contain any notion of (quantum) causality.
 
We turn now to the problem of implementing causality in the
Barrett-Crane model, i.e. constructing a causal transition
amplitude starting from the a-causal one. Having identified the
$Z_2$ symmetry that erase causality from the model, the problem is
now to break this symmetry in a consistent and meaningful way. 
We have seen that this symmetry is implemented by means of a sum, for each face of the 2-complex (triangle in the 
triangulation), of both representation functions of the 2nd kind to give a representation function of the 1st kind. The
restriction implementing causality can be naturally guessed then as the restriction to just one of these terms in the sum, 
which one one chooses depending on the orientation of the 2-complex (triangulation).
 
Our strategy is to find this consistent restriction by requiring that
the resulting amplitude has stationary points corresponding to
good simplicial Lorentzian geometries. We analyse first the case
of a single 4-simplex and give at the end a causal amplitude for
the whole simplicial manifold. After we have obtained the causal
amplitude, we show how the resulting causal model fits in the
general framework of quantum causal sets (or histories).
The results presented in this chapter were published in \cite{causal}, and a briefer account of them is
 given in \cite{causal2}.

\section{A causal amplitude for the Barrett-Crane model}

\subsection{Lorentzian simplicial geometry}
First, we will carry out a stationary point analysis in the
Lorentzian setting as done in the Riemannian case in \cite{BW},
more precisely we study the action associated to a single
4-simplex and study its (non-degenerate) stationary point. They
will turn out to be oriented Lorentzian 4-simplices \cite{causal,bs}.

For this purpose, we first give some details about simplicial
Lorentzian geometry, introduce notations and derive the
Schl{\"a}fli identity encoding the constraint between the angle
variables in a first order Regge formalism based on the area-angle
variables. We denote by
$x_i\in{\cal H}^+$ the five un-oriented normals to the tetrahedra
of the 4-simplex and $n_i=\alpha_ix_i$ the oriented (outward)
normal with $\alpha_i=\pm 1$ (we can think also of $+x_i$ as
living in $\mathcal{H}^+$ and of $-x_i$ as living in $\cal{H}^-$).
We call {\it boost parameter}, associated to the triangle shared
by the $i$ and $j$ tetrahedra, the quantity
\be
\eta_{ij}=\cosh^{-1}(x_i\cdot x_j)\ge 0 \label{angledef}
\ee
and then the {\it dihedral Lorentz angle} is defined as
\be
\theta_{ij}=\alpha_i\alpha_j\eta_{ij}=
\alpha_i\alpha_j\cosh^{-1}(\alpha_i\alpha_jn_i.n_j).
\ee
When $\alpha_i\alpha_j=1$, $\theta_{ij}=\eta_{ij}$ is called an
interior angle and, in the opposite case, it is called an exterior
angle. Then, the (spherical) kernel in the $\rho$ representation
reads
\be
K_\rho(x_i,x_j)=\f{\sin\rho\eta_{ij}}{\rho\sinh\eta_{ij}}
=\sum_{\epsilon=\pm1} \f{\epsilon}{2i\rho\sinh\eta_{ij}}
e^{i\epsilon\rho\eta_{ij}}. \ee

The Schl{\"a}fli identity is the (differential of the)
constraint relating the 10 angles stating that they come from the
5 normals to the tetrahedra: the 5 normals $n_i$ defined a unique
(geometric) 4-simplex (up to overall scale) \cite{BC} and the 10
angles defined from them through \Ref{angledef} are therefore not
independent. They satisfy
\be
\sum_{i\ne j} 
A_{ij}\textrm{d}\theta_{ij}
=\sum_{i\ne j} \alpha_i\alpha_j
A_{ij}\textrm{d}\eta_{ij}
=0
\label{schlaflieq} \ee
where $A_{ij}$ are the areas of the 10 triangles of the defined
4-simplex. Following \cite{balone}, the proof is straightforward
from the definition of the angles. Let us call
$$
\gamma_{ij}=n_in_j=\alpha_i\alpha_j\cosh\eta_{ij}
=\alpha_i\alpha_j\cosh(\alpha_i\alpha_j\theta_{ij}).
$$
Then the closure of the 4-simplex reads \be
\sum_{i=1}^{i=5}|v^{(3)}_i|n_i=0 \ee where $v^{(3)}_i$ is the
3-volume of the tetrahedron $i$. This implies \be \sum_i
|v^{(3)}_i|\gamma_{ij}=0,\;\;\;\forall j, \ee that is the
existence of the null vector $(|v^{(3)}_i|)_{i=1\dots5}$ for the
matrix $\gamma_{ij}$. Differentiating this relation with respect
to the metric information (this will be the area variables in our
case) and contracting with the null vector, we obtain \be
\sum_{i\ne j} |v^{(3)}_i||v^{(3)}_j|\textrm{d}\gamma_{ij}=
\sum_{i\ne j} |v^{(3)}_i||v^{(3)}_j|\alpha_i\alpha_j
\sinh(\eta_{ij})\textrm{d}\eta_{ij} =0. \ee
Finally, one can easily show that \be
\sinh(\eta_{ij})|v^{(3)}_i||v^{(3)}_j|=
\f{4}{3}|{\cal V}^{(4)}|A_{ij} \ee where ${\cal V}^{(4)}$ is the
4-volume of the 4-simplex. Then one can conclude with the
Schl{\"a}fli identity \Ref{schlaflieq}.

The first order action for Lorentzian Regge calculus, with the
constraint on the dihedral angles enforced by means of a Lagrange
multiplier $\mu$ \cite{balone}, then reads:

\be
S_{R}\,=\,\sum_t\,A_t\,\epsilon_t\,+\,\sum_\sigma\,\mu_\sigma\,det(\gamma^\sigma_{ij})\,
=\,\sum_t\,A_t\,\sum_{\sigma/t\in\sigma}\,\theta_{ij}(\sigma)\,+\,\sum_\sigma\,\mu_\sigma\,det(\gamma^\sigma_{ij})
\ee
where the areas and the angles are considered as independent variables.

\subsection{Stationary point analysis and consistency conditions on
the orientation}
Let us now turn to the Barrett-Crane amplitude and study its stationary points.
For a given {\bf fixed} triangulation $\Delta$, with triangles $t$, tetrahedra $T$
and 4-simplices $s$, the amplitude reads
\be
A(\Delta)=\sum_{\epsilon_t=\pm1}\int\prod_t\rho_t^2\textrm{d}\rho_t
\prod_T A_{eye}(\{\rho_t,t\in T\})
\prod_s\int_{({\cal H}^+)^4} \prod_{T\in s}\textrm{d}x^{(s)}_T
\left(\prod_{t\in s}
\f{\epsilon_t}{2i\rho_t\sinh\eta_t}\right)
e^{i\sum_{t\in s}\epsilon_t\rho_t\eta_t}
\ee
where $\rho_t$ are the representation labelling the triangles (faces of the
triangulations, dual to the faces of the 2-complex) and $A_{eye}(T)$ are the
amplitude of the eye diagram associated to the tetrahedron $T$
(obtained by gluing together two 4-intertwiners labelled by the 4 representations
living on the faces of $T$).

The action for a single (decoupled) 4-simplex is then \be
S=\sum_{t\in s}\epsilon_t\rho_t\eta_t, \ee with the angles
$\theta_t$ given by \Ref{angledef} and thus constrained by the
Schl{\"a}fli identity. Consequently, we have \be
\textrm{d}S=\sum_{t=(ij)\in s}\epsilon_t\rho_t\textrm{d}\eta_t
=\mu\times \sum_{i\ne j} \alpha_i\alpha_j
A_{ij}\textrm{d}\eta_{ij} \ee where $\mu\in\mathbf{R}$ is the
Lagrange multiplyer enforcing the constraint. Therefore, we obtain
the following equations defining the stationary points \be \left\{
\begin{array}{ccc}
\epsilon_{ij}\alpha_i\alpha_j&=&sign(\mu) \\
\rho_{ij}&=&|\mu|A_{ij}
\end{array}
\right.
\label{localorient}
\ee
This means that the area of the triangles are given (up to scale) by the
$\sl$ representation labels $\rho_{ij}$ and that we have a consistency
relation between the orientation of the tetrahedra $\alpha_i$, the orientation
of the triangles $\epsilon_{ij}$ and the global orientation
of the 4-simplex $sign(\mu)$.
This means that only some particular choices of values for $\epsilon$
corresponds to stationary points being represented by well-defined
Lorentzian simplicial geometries

These stationary oriented 4-simplices are supposed to be the main contribution
to the full path integral in the asymptotical limit $\rho_{ij}\rightarrow\infty$
up to degenerate 4-simplices as dealt with in
\cite{baezasym,bs,llasym}. Indeed, such degenerate 4-simplices
dominate the (standard) Riemannian Barrett-Crane model in the asymptotical limit.
Ways to sidestep this problems have been proposed through
modifications of the model
in \cite{baezasym,llasym}. However, it will maybe turn
out that the asymptotical
limit is not the physically relevant sector and that small
$\rho$'s would dominate the
path integral, when we take into account the coupling between
4-simplices (through the representations $\rho$'s); this seems also quite likely to happen, because of the results about the
convergence of the partition function for any fixed non-degenerate triangulation.
Nevertheless, we will not discuss this issue in the present work and
we will focus on the orientation issue and the resulting causal structure of the
discrete spin foam manifold.

\medskip

Through the stationary point analysis, we have analysed the
orientability of a single quantum 4-simplex, and obtained consistency
relations between the global oriention of the simplex, the
orientations of its 5 tetrahedra and the orientations of its 10
triangles (linking the tetrahedra).
Now, can we extend this orientation to the whole spin foam~?
This means having consistent orientations of all the
tetrahedra: if a tetrahedron is past-oriented for one 4-simplex,
then it ought to be future-oriented for the other.
Therefore a consistent orientation is a choice of $\mu_v$ (for
4-simplices) and $\alpha_{T,v}$ (for each tetrahedron $T$ attached to a
4-simplex $v$) such that:

\be
\forall T,\,\mu_{p(T)}\alpha_{T,p(T)}=-\mu_{f(T)}\alpha_{T,f(T)}
\label{globalorient}
\ee
where $v=p(T)$ and $v=f(T)$ are the two 4-simplices sharing $T$.

In fact, this imposes constraints on the signs around each loop of
4-simplices which are equivalent to requiring an orientable
2-complex. We have already discussed how to generate only orientable 2-complexes from a group field theory: the field 
over $\sl^4$ (or $Spin(4)$ in
the Riemannian case), which represents the quantum tetrahedron:
\bes
\phi(g_1,g_2,g_3,g_4)=\phi(g_{\sigma(1)},g_{\sigma(2)},
g_{\sigma(3)},g_{\sigma(4)})
\ees
has to be invariant under even permutations $\epsilon(\sigma)=1$ only. 
The consistency conditions on the sign factors appearing in the action associated to a 4-simplex amplitude, that we have 
just found, can then be thought as the counterpart, at the level of the amplitudes, of this requirement on the field.

\subsection{A causal transition amplitude}
Now that we know what are the consistency conditions on the
$\epsilon$ that correspond to well-defined geometric
configurations, we can fix these variables and thus break the
$Z_2$ symmetry they encode (average over all possible orientations
of the triangles) and that erases causality from the model. This
means choosing a consistent orientation
$\{\mu_v,\alpha_{T,v},\epsilon_t\}$ for all simplices satisfying
\Ref{localorient} and \Ref{globalorient}. In other words we fix
the $\epsilon_t$ in the amplitude, and thus the orientation, to
the values corresponding to the stationary points in the a-causal
amplitude (of course we do not impose the other conditions coming
from the stationary point analysis, i.e. we do ot impose the
equations of motion). Then, this leads to a {\it causal
amplitude}, constructed by picking from the general Barrett-Crane
model only the terms corresponding to the chosen orientations
$\epsilon_t$ of the triangles:

\bes
A_{causal}(\Delta)&=&\prod_s\int_{({\cal H}_+)^4}
\prod_{T\in s}\textrm{d}x^{(s)}_T
\prod_{t\in s}
\f{\epsilon_t}{2i\rho_t\sinh\eta_t}\int\prod_t\rho_t^2\textrm{d}\rho_t
\prod_T A^T_{eye}(\{\rho_t\}_{t\in T})
\prod_s\,e^{i\sum_{t\in s}\epsilon_t\rho_t\eta_t} \nonumber\\
&=&\prod_s\int_{({\cal H}_+)^4} \prod_{T\in s}\textrm{d}x^{(s)}_T
\prod_{t\in s}
\f{\epsilon_t}{2i\rho_t\sinh\eta_t}\int\prod_t\rho_t^2\textrm{d}\rho_t
\prod_T A^T_{eye}(\{\rho_t\}_{t\in T})\,e^{i\,\sum_t\,\rho_t\,\sum_{s|t\in
s}\theta_t(s)}\nonumber\\
&=&\prod_s\int_{({\cal H}_+)^4} \prod_{T\in s}\textrm{d}x^{(s)}_T
\prod_{t\in s}
\f{\epsilon_t}{2i\rho_t\sinh\eta_t}\int\prod_t\rho_t^2\textrm{d}\rho_t
\prod_T A^T_{eye}(\{\rho_t\}_{t\in T})\,e^{i\,S_{R}}.
\label{orientampli}
\ees
Let us note that the above amplitude is purely formal as it stands,
since the integrals appearing in it, both that in the representation label
$\rho$ and  those over the hyperboloids $\mathcal{H}_{+}$, do not
converge. Therefore a suitable regularization of the above expression
must be found in order for it to make sense.   

For a (simplicial) manifold $\Delta$ with boundaries, we need to take in
account the boundary term, which are exactly the same as for the usual
Barrett-Crane model.
This corresponds to a particular local causal structure defined by the
relative orientation of the 4-simplices: a 4-simplex is in the
immediate past of another if they share a tetrahedron whose
normal is future oriented for the first and past oriented for the
other.

Of course the above amplitude has to be understood within a sum over
oriented 2-complexes or triangulations:

\be
Z_{causal}\,=\,\sum_\Delta\,\lambda(\Delta)\,A_{causal}(\Delta).
\ee

An interesting question is: do we have
any closed time-like curves? Taking all $\mu$ positive, these
time-like curves correspond to a loop (closed sequence of 4-simplices
linked by common tetrahedra) along which all triangles have a
$\epsilon_t=-1$ orientation. A priori such configurations are allowed, and correspond to the fact that oriented complexes 
are slightly weaker structures than posets \cite{sorkinspeci}, and one can put forward some arguments suggesting that 
oriented complexes not corresponding to posets do not appear in a continuum approximation of any theory based on them and 
describing gravity \cite{sorkinspeci}. Therefore they may result in being automatically negligible when looking at the 
dominant contribution of the full spin foam transition amplitude, including a sum over oriented 2-complexes.
Nevertheless, we can take advantage of the fact that in the present
context we are working with a fixed triangulation, and just consider only
oriented triangulations without such pathological configurations
allowed.
This way, we get a proper (global) causal
structure on the spin foam. The Barrett-Crane model can then be
considered not simply as a sum over (combinatorial) manifolds $\Delta$
but more precisely as a sum over manifolds
provided with a (consistent) causal structure
$(\Delta,\{\mu_v,\alpha_T,\epsilon_t\})$.

\medskip
Let us now discuss the features of the causal amplitude, and compare
the resulting spin foam model with other approaches to Lorentzian quantum gravity.
We stress again that what we have is a spin foam model that takes into account the orientation of the 2-complexes, and that
the interpretation in terms of causality is indeed just an interpretation that one can use only in a Lorentzian context if
associates the notion of (quantum) causality with a notion of ordering.

Let us first re-write the causal partition function in a
simplified (and physically more transparent) form. Basically, it is just:

\bes
Z_{causal}\,=\,\sum_\Delta\,\lambda(\Delta)\,
\int\mathcal{D}\theta_t(\Delta)\,
\int\mathcal{D}A_t(\Delta)\,e^{i\,S_{R}^\Delta(A_t,\theta_t)}
\ees

What are its features?
\begin{itemize}
\item It is clearly a realization of the sum-over-geometries approach
to quantum gravity, in a simplicial setting; its ingredients are: a
sum over causally well-behaved triangulations (or equivalently
2-complexes), where the causal relations are encoded in the
orientation of each triangulation (2-complex); a sum over metric
data for each triangulation, being given by the areas of triangles and
dihedral angles, treated as independent variables, and with a {\it
precise assignment of a measure} (albeit a rather complicated one) for
them; an amplitude for each geometric configuration given by the
exponential of ($i$ times) the first order Lorentzian Regge action. It
can thus be seen as a path integral for Regge calculus, with an
additional sum over triangulations. We note that such an hybrid form of quantum Regge calculus, combining elements of the 
traditonal formulation with those characterizing the dynamical triangulations approach, has been recently studied as a 
viable approach on its own to quantum gravity in \cite{DRC1}. 

\item The geometric variables have a natural and unique algebraic characterization and
origin in the representation theory of the Lorentz group, and the only
combinatorial data used from the underlying triangulation is from the
``first 2 levels'' of it, i.e. from its dual 2-complex only; in other
words, the model is a spin foam model\cite{Baez}\cite{Oriti}.

\item It realises the general definition given in \cite{gupta} for a causal spin foam model (it is, to the best of our
knowledge, its first non-trivial example), except for the use of
the full Lorentz group instead of its $SU(2)$ subgroup; its
boundary states are thus covariant (simple) spin networks, and it
can be regarded as giving a path integral definition of the
evolution (so encoding the action of the Hamiltonian constraint in covariant loop quantum gravity.

\item Each ``first layer'' of the dual 2-complex being a causal set,
the model identifies it a the fundamental discrete structure on which
quantum gravity has to be based, as in the causal set approach
\cite{sorkincausal}; the main difference is that it contains additional metric
data intended to determine a consistent length scale (volume element),
which instead, in the traditional causal set approach is meant to be
obtained by ``counting only'', i.e. just in terms of the combinatorial
structure of the causal set (number of vertices, etc.); this is a
particular case included in the general form of the partition function
above; in fact, if we fix all the geometric data to some arbitrary
value (and thus neglect the integrals in the formula), then
any calculation of geometric quantities, e.g. the 4-volume of each
4-simplex, reduces to a counting problem; this is indeed what happens also in the canonical loop quantum gravity theory, 
regarding the spectrum of area and 3-volume operators; the causal model defines then a quantum amplitude for each causal set,
and we will analyse in details
the causal set reformulation of the model in the following section.

\item Also, if we fix all the geometric data to be those obtained from
a fixed edge length for any edge in the triangulation (e.g. the Planck
length), then what we obtain is the conventional and well-studied sum
over triangulations in the dynamical triangulations approach to
quantum gravity, for causally well-behaved Lorentzian triangulations; the
additional integrals, if instead left to be performed, can be
intepreted as providing a sum over proper times that is usually not
implemented in that approach; this may leave hopes for the expected
emergence of a smooth classical limit from the present simplicial
model, in light of the important results obtained recently
\cite{amblol1}\cite{amblol2}\cite{amblol3} on Lorentzian dynamical triangulations.
The difference, however, is that in the present context we are presumably summing over all triangulations of a given 
topology but also over all possible topologies, as dictated by the group field theory formulation of spin foam models, while
in Lorentzian dynamical triangulations the topology of the simplicial manifolds is fixed to be the cylindrical one 
$S^3\times \mathbb{R}$.

\end{itemize}

Let us finally comment on the analogy with the relativistic particle
case. We have seen that different quantum amplitudes may be
constructed for a relativistic particle: the Wightman function,
solution of all the constraints (defining a projector operator), distinguishing the order of its
arguments, and taking into account all the trajectories of interest
for a single paticle (not anti-particle); the Hadamard function,
solution of all the constraints, not registering the order of its
arguments, and including all the trajectories of both particle and
anti-particle solutions; the Feynman propagator, not a solution of the
constraint, registering the order of its arguments, again
including both particle and anti-particle solutions, and which reduces
to the appropriate Wightman function depending on the time-ordering of
its arguments. It is clear that for a single-particle theory and if we
do not consider anti-particles, then the Wightman function is all we
have, and all we need to define a physical transition amplitude
between states.  The Feynman propagator is of not much use in a theory of a
single particle, but needed in a theory of many particles. If we want
instead something which is a
solution
to the constraint, but not reflect any ordering, then the only answer is the Hadamard
function. It can be interpreted as a particle plus an
antiparticle amplitude, but the point is really just the ordering of the
arguments in the amplitude. Now, the question we have to ask in our
spin foam case is: is it a single particle or a multiparticle theory? does the amplitude we have in the
Barrett-Crane model reflect the ordering of the arguments or not? if we
impose it to reflect such an ordering, what kind of amplitude do we get?

Now, the BC amplitude is real and does not reflect any ordering, so it
cannot correspond to a Wightman function, and it must instead correspond
to the Hadamard function (for what the particle analogy is useful). The
one may say that in order to impose causality (ordering-dependence), I have
to reduce it to the Wightman function. In any case this will give us
something different from the Barrett-Crane amplitude as it is. How to
impose the restriction, then? The only possible
restriction imposing an ordering dependence seem to be the one we performed. This is basically
because in the spin foam we have only combinatorial data and
representations, and the only ordering we can impose is an ordering at the
level of the 2-complex. If we do it, and ask for an amplitude that
reflects it, we get the above causal amplitude.

Now, is there a way to interpret what we do in terms of the particle
analogy? Work on this issue is in progress, and the idea would be that what we have is a
multiparticle theory, where we have a particle on the
hyperboloid corresponding to each simple representation (there is indeed
an interpretation of the kernel $K$ as a Green function for a particle on
the hyperboloid); then the two parts of the kernel would
correspond to the two Wightman functions and the kernel itself to the Hadamard
function; so the two different orientations for the same
particle-face-representation correspond to something like a particle and
an antiparticle, or to the two Wightman functions $G^+$ and $G^-$. In this context the right amplitude
would be exactly that selecting a particle for one orientation and an
antiparticle for the opposite orientation, i.e. simply one or the
other Wightman function. This is what we do, in fact, and this is what the
Feynman propagator does in the particle case. 

Of course we should bear in mind that we should not rely too much on perturbative QFT, nor on the
particle analogy, but nevertheless they can give some insight;
moreover, we know from the group field theory that in some non-trivial
sense what we are dealing with is indeed a form of perturbative field
theory, and that the spin foam
for fixed 2-complex is exactly a perturbative
Feynman graph, although very peculiar, and the particle interpretation suggested above is indeed natural from the 
point of view of the group field theory, as we have discussed in the previous chapter.

\section{Barrett-Crane model as a quantum causal set model}
In the new formulation presented above, the Barrett-Crane model fits
into the general scheme of quantum causal sets (or quantum causal
histories) models, as developed in \cite{marsmo}\cite{fot}\cite{fot2}\cite{fotlogic}. It
represents, in fact, the first explicit and highly non-trivial example
of such a class of models. As
previously said,
all the metric information of a Lorentzian space-time can
be encoded in its causal set skeleton apart from its conformal factor,
and therefore causal sets are a natural context for a discrete and
finitary
space-time model \cite{sorkinspeci}\cite{sorkincausal}. Reformulating the
Barrett-Crane model in these terms helps in understanding the
underlying causal structure in the model, and gives insight on the
properties required for quantum causal
sets (especially the evolution operators).

Let us first recall the basic elements of this scheme. We have anticipated most of this when we have described the quantum 
causal histories approach to quantum gravity among the other approaches related to spin foam models in the introduction, but
we repeat it here for the sake of completeness and to make it easier to follow the reformulation of the causal Barrett-Crane 
model along these lines.

Consider a discrete set of events $\{p, q, r, s, ...\}$, endowed
with an ordering relation $\leq$, where the equal sign implies
that the two events coincide. The ordering relation is assumed to
be reflexive ($\forall q, q\leq q$), antisymmetric ($q\leq s,
s\leq q \Rightarrow q=s$) and transitive ($ q\leq r, r\leq s\,
\Rightarrow q\leq s$). These properties characterize the set as a
partially ordered set, or poset, or, interpreting the ordering
relation as a causal relation between two events, as a causal set
(see figure 6.1). In this last case, the antisymmetry of the ordering
relation, together with transitivity, implies the absence of
closed timelike loops. We also report a few definitions that will
be useful later:
\begin{itemize}
\item the causal past of an event $p$ is the set $P(p)=\{ r | r\leq p\}$;
\item the causal future of an event $p$ is the set $F(p)=\{ r | p\leq r\}$;
\item an acausal set is a set of events unrelated to each other;
\item an acausal set $\alpha$ is a complete past for the event $p$ if
$\forall r\in P(p), \exists s\in\alpha \,|\, r\leq s\,\,or\,\,s\leq r$;
\item an acausal set $\beta$ is a complete future for the event $p$ if
$\forall r\in F(p), \exists s\in\beta \, |\, r\leq s\,\,or\,\,s\leq r$;
\item the causal past of an acausal set $\alpha$ is
$P(\alpha)=\cup_i\,P(p_i)$ for all $p_i\in\alpha$, while its causal
future is $F(\alpha)=\cup_i\,F(p_i)$ for all $p_i\in\alpha$;
\item an acausal set $\alpha$ is a complete past (future) of the
acausal set $\beta$ if $\forall p\in P(\beta) (F(\beta))\,\,\exists
q\in\alpha \,\,|\,\,p\leq q\,\,or\,\,q\leq p$;
\item two acausal sets $\alpha$ and $\beta$ are a complete pair when
$\alpha$ is a complete past of $\beta$ and $\beta$ is a complete future
for $\alpha$.
\end{itemize}

From a given causal set, one may construct another causal set
given by the set of a-causal sets within it endowed with the
ordering relation $\rightarrow$, so that $\alpha\rightarrow\beta$
if $\alpha$ and $\beta$ form a complete pair, also required to be
reflexive, antisymmetric and transitive. This poset of a-causal
sets is actually the basis of the quantum histories model.

The quantization of the causal set is as follows. It can be seen as a functor
between the causal set and the categories of Hilbert spaces.

\begin{figure}
\begin{center}
\includegraphics[width=9cm]{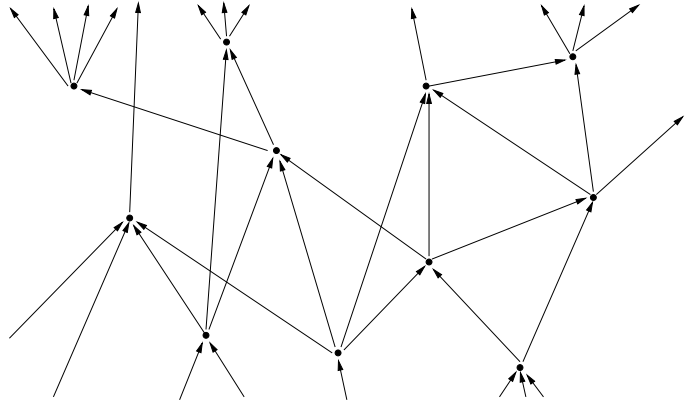}
\caption{A causal set}
\end{center}
\end{figure}

We attach an Hilbert space $\mathcal{H}$ to each node-event and
tensor together the Hilbert spaces of the events which are
spacelike separated, i.e. causally unrelated to each other; in
particular this gives for a given a-causal set
$\alpha=\{p_i,...,p_i,...\}$ the Hilbert space
$\mathcal{H}_\alpha=\otimes_i \mathcal{H}(p_i)$.

Then given two a-causal sets $\alpha$ and $\beta$ such that
$\alpha\rightarrow\beta$, we assign an evolution operator  between
their Hilbert spaces: \be
E_{\alpha\beta}\,:\,\mathcal{H}_{\alpha}\,\rightarrow\,\mathcal{H}_{\beta}.
\ee

In the original Markopoulou' scheme, the Hilbert spaces considered
are always of the same (finite) dimension, and the evolution
operator is supposed to be unitary, and fully reflecting the
properties of the underlying causal set, i.e. being reflexive:
$E_{\alpha\alpha}=Id_\alpha$, antisymmetric:
$E_{\alpha\beta}E_{\beta\alpha}=Id_\alpha
\Leftrightarrow\,E_{\alpha\beta}=E_{\beta\alpha}=Id_\alpha$, and
transitive: $E_{\alpha\beta}E_{\beta\gamma}=E_{\alpha\gamma}$.

The evolution operators mapping complete pairs that are causally
unrelated to each other, and can be thus tensored together, may also
be tensored together, so that there are cases when given evolution
operators may be decomposed into elementary components. More on the
dynamics defined by this type of models can be found in \cite{fot}\cite{fot2}.

Another possibility is to assign Hilbert spaces to the causal
relations (arrows) and evolution operators to the nodes-events.
This matches the intuition that an event in a causal set (in
spacetime) is an elementary change. This second possibility gives
rise to an evolution that respects local causality, while the
first one does not. This is also the possibility realised in the
causal Barrett-Crane model, as we are going to discuss. Also, this
hints at a fully relational reformulation of quantum mechanics
\cite{carlo96}\cite{Smolin}, since the Hilbert space between two
events, $a$ and $b$, admits the interpretation of describing the
possible states of ``$a$ seen by $b$'' (and reciprocally); this is
closely related also to the ``many-views'' category-theoretic
formulations of sum-over-histories quantum mechanics \cite{isham}.

Let us consider now how this scheme is implemented in the causal
Barrett-Crane model defined above.

The starting point is an oriented graph, i.e. a set of vertices linked
by arrows, restricted to be 5-valent (each vertex is source or target
of five arrows) and to not include closed cycles of
arrows. Interpreting the arrows as representing causal relations, this
is a causal set as defined above, the orientation of the links
reflects the ordering relation among the vertices, and it does not contains closed
timelike curves.

Because of the restriction on the valence, it can be decomposed into building blocks, in the sense that for each
vertex, only one out of four possibilities may be realised, depending
on how the arrows are related to the vertex itself (see figure 6.2).

\begin{figure}
\begin{center}
\includegraphics[width=9cm]{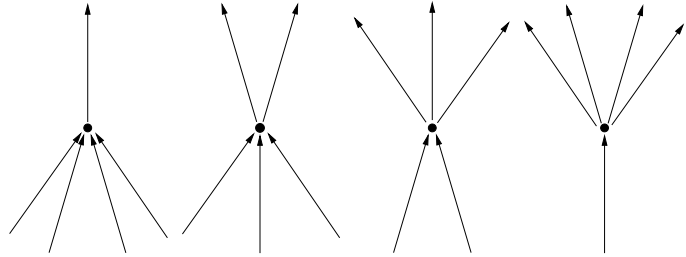}
\caption{Building blocks for the BC causal set}
\end{center}
\end{figure}

This causal set is just the first {\it stratum} of the 2-complex on
which the Barrett-Crane Lorentzian spin foam model is based, assuming it
to be oriented. As such, there is a dual simplicial interpretation of
the elements in it, since now the vertices are dual to 4-simplices and
the oriented links are dual to oriented tetrahedra, the tetrahedra in
a given 4-simplex can be considered as belonging to different
spacelike hypersurfaces so that the four
moves (in figure 6.2, above) represent the four possible Pachner moves ($4-1$, $3-2$, and
their reciprocal) giving the evolution of a 3-dimensional simplicial
manifold in time.

The crucial point, we note, is the identification of the direction of
the arrow in the causal set with the orientation of the tetrahedron it
refers to.

The quantization is then the assignment of Hilbert spaces, the
Hilbert spaces of intertwiners among four given continuous
(principal irreducible unitary) representations of the Lorentz
group, representing the possible states of the tetrahedra in the
manifold, to the {\it arrows} of the causal set (oriented edges of
the 2-complex)and of the causal Barrett-Crane amplitude defined
above to the nodes of the causal set, as the appropriate evolution
operator between the Hilbert spaces. We have seen in fact how it
registers the ordering of its arguments and how it can thus be
interpreted as the right transition amplitude between spin network
states.

More precisely it proceeds as follows.

We pass from the causal set defined above to the so-called {\it
edge-poset}, i.e. the causal set obtained associating a node to
each arrow of the previous one and a new causal relation (arrow)
linking each pair of nodes corresponding to not spacelike
separated arrows in the previous graph. The building blocks of
this new causal set, obtained directly form those of the previous
poset, are then as in figure 6.3.

\begin{figure}
\begin{center}
\includegraphics[width=9cm]{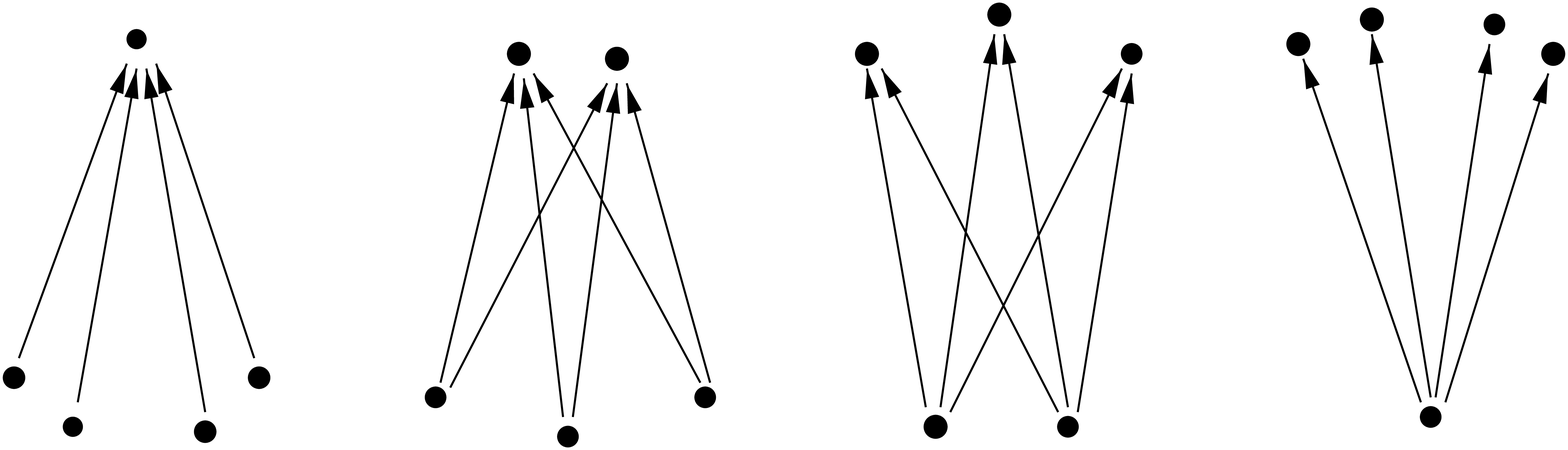}
\caption{Building blocks for the BC edge-poset}
\end{center}
\end{figure}
To each node in this new causal set it is associated the Hilbert
space of intertwiners between four simple representations of the
Lorentz group. More precisely, an ``intertwiner'' here is the open (simple)
spin network with a single vertex and with a fixed normal $x$,
so that it is labelled by four vectors living in the four representations
and the normal $x\in{\cal H_+}$ sitting at the vertex.
The Hilbert space of such ``interwiners'' is then
$L^2({\cal H}_+)_{\rho_1,\rho_2,\rho_3,\rho_4}$, where
we have specified the four intertwined representations. 
As we said above, Hilbert spaces that are
a-causal to each other can be tensored together. In particular, in
each of the building blocks the source nodes form a minimal
a-causal set and the target nodes as well; moreover, these form a
complete pair, so they are in turn linked by a causal relation in
the poset of a-causal sets defined on the edge-poset; to each of
these causal relations among complete pairs, i.e. to each of the
building blocks of the edge-poset, we associate the causal
Barrett-Crane amplitude defined earlier \Ref{orientampli}.

Taking two touching (but a-causally related) tetrahedra or
correspondingly two interwtiners, we can tensor them and glue them together.
Doing so, we get an open (simple) spin network with two vertices describing
the state of the system formed by these two tetrahedra. More precisely, let's note
$x_1$ ($x_2$) the normal of the first (second) intertwiner,
$\alpha_1,\dots,\alpha_4$ ($\beta_i$) its four representations and
$(j_i,m_i)$ ($(k_i,n_i)$) the vectors in the $\alpha_i$ ($\beta_i$)
representations living at the ends of the one-vertex spin network.
Let's suppose that the two intertwiners are glued along the edge
$\rho=\alpha_1=\beta_1$, then we need to identify the vectors
$(j_1,m_1)=(k_1,n_1)$ and sum over them. This yields the right
open spin networks with two vertices. At the level of the normals,
the resulting tensored Hilbert space
is $L^2(x_1\in{\cal H}_+)\times L^2(x_2\in{\cal H}_+)$.
In particular, the state
corresponding to any of the a-causal sets present in the building
blocks is given by an open simple spin network, with
representations of $SO(3,1)$, but also representations of a given
$SU(2)$ subgroup and a label (``angular momentum projection'') for
states in this subgroup on their open ends, and the causal
Barrett-Crane amplitude for a single vertex (4-simplex), with
appropriate boundary terms, interpolates between these states.

Of course, just as in the general case, states (and operators
between them) referring to a-causal sets that are not causally
related to each other can be tensored together again, and
composite states constructed in this way evolve according to
composite evolution operators built up from the fundamental ones.

Let us now discuss what are the properties of the evolution operator
(Barrett-Crane causal amplitude) in this quantum causal set context.

First of all we recall that in the original Markopoulou' scheme
\cite{fot} the single poset evolution operator $E_{\alpha\beta}$
is required to be reflexive, antisymmetric, transitive and unitary
(we have defined above what these properties mean in formulae).

However, we have stressed from the beginning that the
Barrett-Crane partition function for a given 2-complex (triangulation)
has to be understood as just one term in a sum over 2-complexes of all
the partition functions associated with them (with some given
weight). In this causal set picture this means that we have to
construct a sum over causal sets interpolating between given boundary
states, as outlined also in \cite{fot} and each of them will be then
weighted by the causal Barrett-Crane amplitude, made out of the
building block evolution operators as explained.

Therefore, given two a-causal sets $\alpha$ and $\beta$, the full
evolution operator between them will be given by an operator
$\mathcal{E}_{\alpha\beta}$ defined as: \be
\mathcal{E}_{\alpha\beta}\,=\,\sum_c\,\lambda_c\,E_{\alpha\beta}^c
\ee where we have labelled the previously defined evolution
operator for a single causal set with $c$ to stress its dependence
on the underlying graph, and we also included a possible
additional weight $\lambda_c$ for each causal set in the sum.
Recall also the formal expression for a path integral for quantum
gravity we outlined in the introduction, where we split the sum
over geometries into a sum over causal structures (causal sets)
and a sum over metric degrees of freedom defining a length scale
in a consistent way.

It is clear then that this is the real physical evolution operator
between states, and this is the operator whose physical properties
have to be understood\footnotemark \footnotetext{Let us note that
there is another possible way of getting rid of the dependence of
the theory on any fixed underlying causal set, which is to define
the complete model by a refining procedure of the initial causal
set, that increases progressively (possibly to infinity) the
number of vertices in it and thus defines the model as the limit
of the fixed poset one under this refinement. We tend to prefer
the solution provided by the sum over 2-complexes (causal sets)
only because, while there exist (acausal) models that furnish this
last sum, i.e. those based on the formalism of group field theory,
there is (to the best of our knowledge) no implementation yet for
this refining procedure. Also the first solution seems to us more
in agreement with the general idea of summing over geometries to
obtain the amplitudes for a quantum gravity theory}.

It is sensible to require the partial operators $E_{\alpha\beta}$ to
be reflexive, of course, since imposing this leads to an analogue
property for the full evolution operator (up to a factor dependng on
$\lambda$).

Also, we require both the fixed-poset evolution operator $E^c$ and the
full evolution operator $\mathcal{E}$ to be antisymmetric, to reflect
the property of the underlying causal set, in the first case, and as a
way to implement the physical requirement that imposing that the
evolution passes through a given state changes significantly the
evolution of a system, even if it then returns in the initial state.

For the other properties the situation is more delicate.

Doubts on the meaningfulness to require transitivity for the
single poset evolution operator were put forward already in
\cite{fot}. The reason for this is that this properties implies a
sort of directed triangulation invariance of the resulting model,
in that the evolution results in being at least partly independent
on the details of the structure of the causal set itself. In fact,
transitivity implies that, as far as the evolution operators are
concerned, a sequence of two causal relations linking the a-causal
sets $\alpha$ and $\beta$ and then $\beta$ and $\gamma$ is
perfectly equivalent to a single arrow linking directly $\alpha$
and $\gamma$. The causal set is in turn dual to a triangulation of
the manifold and this is why transitivity implies a sort of
triangulation invariance.

Therefore, while it is certainly meaningful to require transitivity
even for the single-poset evolution operators $E^c_{\alpha\beta}$ if
we are dealing with 3-dimensional gravity, which is a topological
field theory, it does not seem appropriate to impose it also in the
4-dimensional case, where any choice of a fixed causal set is a
restriction of physical degrees of freedom of the gravitational field.

On the other hand, the situation for the full evolution operator is
different. Even in the non-topological 4-dimensional case, we do
require that the evolution from a given acausal set $\alpha$ to
another $\gamma$ is independent of the possible intermediate states in
the transition, {\it if we sum over these possible intermediate
states}, i.e. we require:

\be
\sum_\beta\mathcal{E}_{\alpha\beta}\mathcal{E}_{\beta\gamma}\,=\,\mathcal{E}_{\alpha\gamma},
\ee which is nothing else than the usual composition of amplitudes
in ordinary quantum mechanics. We do not require the same property
to hold if we drop the sum over intermediate states (an
intermediate state corresponding to an intermediate  measurement
of some physical properties).

The property of unitarity of the evolution is of extreme
importance. We believe that the evolution defined by the causal
spin foam model has to be indeed unitary to be physically
acceptable.

In fact, the initial and final a-causal sets that represent the
arguments of the evolution operator, i.e. of the causal amplitudes
of the model, are always assumed to form a complete pair, and this
completeness should hold for any causal set summed over in the
definition of the full evolution operator mapping one a-causal set
into the other. Therefore, if the final a-causal set is a complete
future of the first, and the initial a-causal set is a complete
past for the final one (this is, we recall, the definition of a
complete pair), we expect all the information from the initial
state to flow to the final state. The requirement of unitarity is
thus nothing more than the requirement of the conservation of
information in the evolution.

We stress again, however, that the physical evolution operator is
$\mathcal{E}$, and not $E^c$, i.e. it involves a sum over causal sets.
Any term in this sum, on the contrary, represents a truncation of the
full dynamical degrees of freedom in the theory, and a restriction on
the flow of information.
Therefore, it is this operator that, we think, has to be required to be unitary.

But if this is the case, then the single-poset operator $E$ has to be
{\it not} unitary, as it is easy to verify (a sum of unitary operators
cannot be a unitary operator itself).

Therefore we require:
\bes
\sum_{\beta}\mathcal{E}_{\alpha\beta}\,\mathcal{E}^{\dagger}_{\alpha\beta}\,=\,
\sum_\beta\mathcal{E}_{\alpha\beta}\bar{\mathcal{E}}_{\beta\alpha}\,=\,Id_\alpha
\ees

which implies:
\bes
\sum_\beta E^c_{\alpha\beta}\,E^{c\dagger}_{\alpha\beta}\,=\,
\sum_\beta E^c_{\alpha\beta}\bar{E}^c_{\beta\alpha}\,\neq\,Id_\alpha.
\ees

\begin{figure}
\begin{center}
\includegraphics[width=7cm]{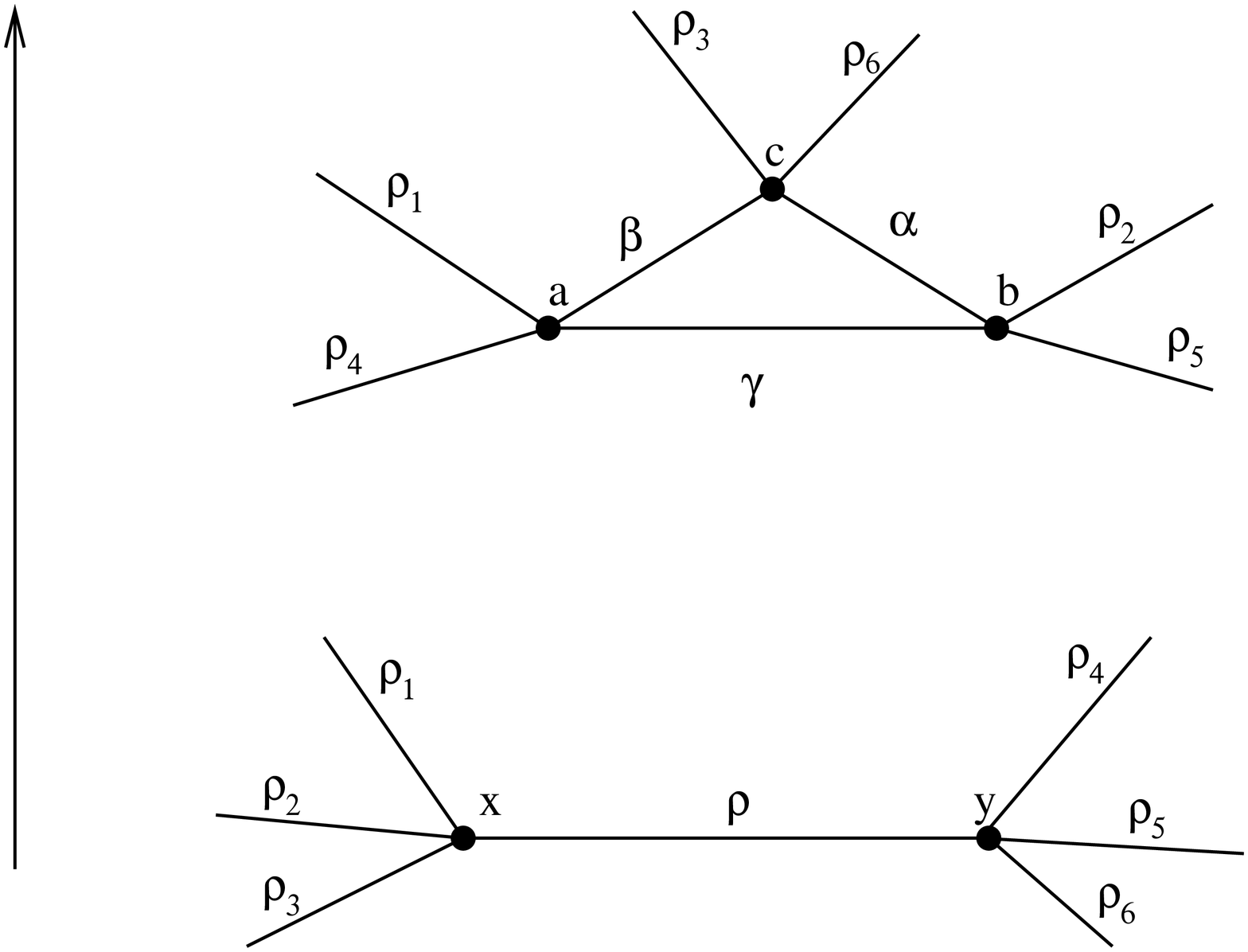}
\caption{4-simplex: the $2\rightarrow 3$ move}
\end{center}
\end{figure}

More precisely, let's consider a $2\rightarrow 3$
move as on figure 6.4 and its amplitude read directly from the oriented
spin foam amplitude \Ref{orientampli} (choosing $\mu=+1$ for the
global orientation of the 4-simplex):
\bes
{\cal A}(2\rightarrow 3)
&=&E^c_{\rho\rightarrow(\alpha,\beta,\gamma)} \nonumber \\
&=&
\int \textrm{d}x\,\textrm{d}y\,\textrm{d}a\,
\textrm{d}b\,\textrm{d}c \quad
\prod_{v,w\in\{x,y,a,b,c\}}\sinh^{-1}(\theta(v,w)) \nonumber \\
&&e^{+i\rho\theta(x,y)}
e^{+i\alpha\theta(b,c)}
e^{+i\beta\theta(c,a)}
e^{+i\gamma\theta(a,b)} \times \nonumber \\
&&e^{-i\rho_1\theta(x,a)}
e^{-i\rho_2\theta(x,b)}
e^{-i\rho_3\theta(x,c)}
e^{-i\rho_4\theta(y,a)}
e^{-i\rho_5\theta(y,b)}
e^{-i\rho_6\theta(y,c)}.
\ees
In fact, when one ``acts'' with this (oriented) 4-simplex on the region of a
(boundary) simple spin network made of two glued tetrahedra, one has to
take into account the shift between the two normals to the tetrahedra on the boundary
state and their values on the 4-simplex, so that the final evolution operator consists
of, first, Lorentz transformations to go to the initial normals to the normals attached
to the 4-simplex and then the transformation from a 2 vertex open spin network to a
3 vertex open spin network generated by the 4-simplex itself. This way, we recover the full
oriented spin foam amplitude with the eye diagram weights.
Changing the time orientation, we simply change $\mu$ from $1$ to $-1$, which
changes the above operator to its complex conjugate.

Let us now finally check that the causal amplitudes defined above
have all the properties we want (or not want) them to have.

We do not have at our disposal a well-motivated definition of the
complete evolution operator, coming from well understood physical
requirements or obtained within some known mathematical formalism,
e.g. a group field theory model. Therefore we must limit ourselves to
check the wanted (and not-wanted) properties of the fixed-poset
evolution operators. We will indeed check these properties in the
simplest case of the evolution operator associated to a single
building block of the causal set, i.e. a single choice of minimal
source and target acausal sets.

We recall that the properties we want such an operator to satisfy are:
reflexivity, antisymmetry, absence of transitivity and absence of
unitarity.
We satisfy reflexivity trivially just by defining the evolution
operator $E_{\alpha\alpha}$ to be the identity operator.

As for antisymmetry, we clearly have $E_{\alpha\beta}E_{\beta\alpha}\ne \textrm{Id}$.

Transitivity is NOT satisfied. Indeed, composing two  4-simplex
operators is not equivalent to a single 4-simplex operator. This is
straightforwardly due to the non-isomorphism of the initial and final (boundary)
states of a 4-simplex move.

Finally, most relevant, unitarity is NOT satisfied either.
Indeed starting from two open spin networks, made of 2 vertices or equivalently
2 glued tetrahedra, then the resulting states after a $2\rightarrow 3$ move should have
a identical scalar to the one of the initial states. Considering that the scalar product
of the initial spin networks is $\delta(\rho-\rho')$ and that the scalar product
of two 3 vertex open spin network is
$\delta(\alpha-\alpha')\delta(\beta-\beta')\delta(\gamma-\gamma')$, it is
straightforward to check that this is not the case.

The unitarity is reserved to the
evolution operator resulting from a sum over intermediate triangulations. Nevertheless,
unitarity is equivalent to conservation of the information. Intuitively, this is
violated by the fact that the Hilbert space of 2 vertex open spin networks is NOT
isomorphic to the one of 3 vertex open spin networks: they do not carry the same
information (not same number of internal representations). Is there still a way in which
unitarity is verified at this microscopic level. The answer resides in the
edge poset picture. Indeed, looking at Fig. 7, each arrow represents the same 4-simplex, but
each carry a different amplitude/operator which is a single exponential. This operator
takes the normal of the past tetrahedron of the arrow to the normal of the future
tetrahedron, and thus is an automorphism of $L^2({\cal H}_+)$.
This way, we can define an unitary evolution attached to the 4-simplex
but it
means associating not one but many unitary operators to it.

This concludes the formulation of the causal model based on the
causally restricted Barrett-Crane amplitudes as a quantum causal set model. We note that it seems also possible to refine
this reformulation in terms of algebra of operators associated to arrows in the fundamental causal set and completely 
positive maps associated to its nodes, along the lines of the most recent and complete definition of quantum causal 
histories formulated in \cite{HMS}, although the details of this refined formulation have not been worked out yet.

\chapter{The coupling of matter and gauge fields to quantum gravity in spin foam models}
The spin foam approach was originally developed for pure gravity, in
absence of any gauge or matter field, i.e. as a description of pure
quantum geometry. It is, of course, essential to understand how to
couple matter and gauge fields to these models of pure
gravity. From another point of view, one needs to understand whether and how matter and gauge fields may arise from pure
gravity configurations in some limit, if one takes a geometric approach to the nature of such fields.
Ultimately, one wishes to describe the Standard Model matter
and interactions in the metric background provided by quantum gravity
if a suitable classical limit of the gravity sector is
taken. Moreover, the coupling of spin foam gravity to matter might be
essential in order to understand various subtle and yet unsolved
questions in the area of quantum gravity, and it may even be
ultimately required in order to understand the classical limit and
therefore to decide which one of several conceivable spin foam models
of gravity, all having the same local symmetries, is the correct
choice. Finally, such a coupling may lead to new ways of approaching
some fundamental and yet unresolved issues of standard particle
physics, for example, the hierarchy problem, the cosmological constant
problem or a deeper understanding of renormalization. All these issues have been in fact on different occasions and 
different reasons argued to be possibly solved only if and when gravity is included in the picture.

In this chapter we discuss several ideas that have been put forward on the coupling of matter and gauge fields to gravity in 
a spin foam setting; some of them are rather speculative and far reaching, others are more concrete and less ambitious, but
in any case any evaluation of them at this stage may be only tentative since the whole subject of the coupling of matter to 
spin foam gravity is at an exploratory stage. We present first some heuristic argument by L. Smolin \cite{leestrings} 
relating spin networks and spin foams to strings, then the ideas about \lq\lq hypergravity" by L. Crane \cite{cranehyper} 
and the combination of gravity and matter degrees of freedom into a quantum topological theory, and an alternative approach,
 again by Crane \cite{craneconi}, on the interpretation as matter of the conical singularities appearing in the 
configurations of the group field theory approach to spin foam models; we review the coupling of matter and gauge field as 
proposed by A. Mikovic \cite{mikovicmatter} at the level of the group field theory action, with a corresponding modification
 of the resulting spin foam amplitudes, to describe also non-gravitational degrees of freedom in the same algebraic 
language; finally, and more extensively, we show the construction of a model for pure Yang-Mills theory (for any gauge 
group) coupled to quantum gravity in the spin foam formulation using ideas and techniques from lattice gauge theory. 

\section{Strings as evolving spin networks}
String theory is a promising candidate for a complete description of matter and gauge fields, going beyond the ideas and 
results of the Standard Model of fundamental interactions; moreover, it appears to be able to incorporate the 
gravitational interaction, in a consistent way, in a more unified description together with the other fields, at least at 
the perturbative level in which one describes these matter and gauge fields, and a weak gravitational field as well (i.e.
gravitational waves or at the quantum level gravitons), as moving on a classical background spacetime; many results are 
available, in this approach, also at the non-perturbative level, although the background dependence of these results is hard
to sidestep. Because of these achievements, one may take the attitude that a consistent theory of quantum gravity, even if 
formulated in a purely algebraic language as spin foams models are, must admit an approximate description in terms of a 
smooth classical spacetime (of course this part is not an assumption, but a necessary condition for a quantum gravity theory
to be considered complete) with interacting strings (and branes) moving on it as perturbations. Needless to say, this 
description is very far from being obtained, but a scheme of how it can be obtained is presented in \cite{leestrings}.

Consider the transition amplitude between two spin network states as given by spin foam models:
\bes
\langle \Gamma, j \mid \Gamma', j'\rangle\,=\,\sum_{\sigma\mid \Gamma, \Gamma'}\sum_{J_f\mid j, j'} \prod_f A_f(J_f)\,
\prod_v\,A_v(J_f,j,j'),
\ees
with the spin network states labelled by a graph $\Gamma$ colored by representation of a given group $j$, and sum over all
 the 2-complexes having the graphs $\Gamma$ and $\Gamma'$ as boundaries, and a sum over all the representations $J_f$ of the 
group associated to the faces $f$ of the 2-complexes except the boundary ones, $j$ and $j'$, held fixed as boundary data, 
with a quantum amplitude $A_f$ for each of them as a measure, and a vertex amplitude $A_v(J_f,j,j')$ depending on all the 
representations encoding the dynamics of these algebraic degrees of freedom. This is the general structure of all the spin 
foam models, causal and a-causal, in both the Lorentzian and Riemannian cases, for any gauge group. Here we may consider for 
concreteness the case of the gauge group being $SU(2)$.

In the dual picture, one can think of the states as discretized 3-surfaces, with representations labelling the triangles, 
evolving by a sequence of Pachner moves obtained by acting with a sequence of 4-simplices on the initial triangulation, 
to eah of these 4-simplices representing interactions and evolution we associate an amplitude $A_v$.

If we are interested in the study of how perturbations of a given classical configuration look like in this setting, we have
to give a rule for perturbing the states $\Psi$ of the theory first. Any perturbation changing the triangulation of the 
initial slice may be interpreted as evolution, so we consider a different type of perturbation, changing the values of the 
representations labelling the edges of the spin network graph and leaving the structure of the graph (triangulation) 
unaffected. The most fundamental consistent change of the labellings is obtained by taking a (Wilson) loop $\gamma$ 
labelled by the elementary representation 1 around and inserting it in the spin network graph in correspondence to a closed 
loop of edges and nodes,
where the insertion corresponds to taking the tensor product of the representations labelling any edge touched by the loop 
with the representation 1 of the loop, with the corresponding changes in the intertwiners at the nodes.
The new perturbed spin network can be decomposed in a sum of more elementary ones, using the decomposition of the tensor 
product of representations, as:
\bes
\mid \gamma * \Psi \rangle\,=\,\sum_{\delta = \pm 1} \prod_n C(n,j,\delta) \, \mid \Gamma j+\delta\rangle
\ees
where the coefficients $C(n,j,\delta)$, depending on the perturbations $\delta$ of the various spins $j$ and associated to 
the nodes of the graph $\Gamma$, can be explicitely computed using recoupling theory of $SU(2)$.
Because of the triangle inequalities and the properties of the vertex amplitudes, a perturbation of the initial state leads 
to a perturbation of the histories in the spin foam model, i.e. to a perturbations of the spin foams themselves; moreover,
because the perturbation applied is given by a loop, the perturbed histories contain a 2-surface $S$ in such a way that
whenever one takes a slice of the 2-complex $\sigma$ containing the 2-surface $S$, one has associated to it a spin network 
with an identificable loop of perturbation. This 2-surface has to be thought of as an embedded surface in the spacetime 
manifold one reconstructs from the spin foam configurations. Of course, if one is working within a causal spin foam model, 
one also requires 
the perturbation to be causal. The perturbations $\delta$ assumes the values $\pm 1$, as we have said, and thus what we have 
is a (complicated) spin system living on the 2-dimensional surface $S$, each configuration of which is described by a 
quantum amplitude calculated as an induced amplitude from the ones defining the spin foam model, defined by a partition 
function of the type:
\bes
W(S)\,=\,\sum_\delta\,\prod_{v \in S}\,\frac{A_v(J,\delta)}{A_v(J)}\,=\,\sum_\delta\,e^{i\,\sum_v S^{eff}_v(\delta,J)}\,=
\,\sum_\delta\,e^{i\,S_{eff}},
\ees
so that an effective action may be defined as well for the perturbations as:
\bes
S_{eff}\,=\,-i\,\log\left( \prod_{v\in S}\frac{A_v(\delta,J)}{A_v(J)}\right),
\ees
depending of course also on the \lq\lq background" representations $J$.

Under several reasonable assumptions, one can deduce some properties of this action. In fact, let us assume (with a degree 
of optimism) that there exists a smooth 4-dimensional spacetime $M$ with Minkowskian metric $\eta$ that is the classical 
limit of the fixed spin foam background configuration $(\sigma,J)$ that we are considering, in the sense that the causal 
structure, metric quantitites, and topology that one can reconstruct from $(\sigma, J)$ coincide, given a suitable embedding
 map, with those of $(M,g)$; let us assume also the causal structures of $\sigma$ and the representations $J$ are such that 
their distribution in $M$ allows a form of Poincare invariance of the distribution itself, and that there exists a suitable 
timelike cylindrical surface $s\in M$ corresponding to $S$ with large area and small extrinsic curvature.
Under these (strong) assumptions $S_{eff}=s_{eff}N_4(S\cap M)$, i.e. the effective action is given by a contribution for 
each
of the 4-simplices that $S$ crosses in (the discretization of) $M$, times the number of these 4-simplices.
In turn, by Poincare invariance and additivity, this number is proportional (with factor $K$) to the area $A(s,\eta)$ of $s$
 computed using the metric $\eta$, so that:
\bes
S_{eff}\,\sim\,\frac{K\,s_{eff}}{l_{Pl}^2}\,A(s,\eta).
\ees

Therefore we would have as perturbations of our quantum gravity configurations closed loops propagating along timelike 
surfaces embedded in spacetime, with an action for the perturbations proportional, with computable (quantum) factors, to the
Nambu action for bosonic strings.
Of course, the whole argument is very heuristic, and even if it turns out to be correct the effective action may describe 
a non-critical string theory as in QCD (likely to be the case, since there is no consistent string theory in 4d based on the 
pure Nambu action), but the possibilities and ideas put forward are extremely intriguing.   

\section{Topological hypergravity}
We have seen that the Barrett-Crane spin foam model is obtained as a restriction of the Crane-Yetter spin foam model 
describing topological quantum BF theory in 4d, by considering only those configurations in the latter given by the 
simple representations of the Lorentz group, with the motivations for this coming also from the classical formulation of 
General Relativity as a constrained BF theory. In a particular version of the model (the DePietri-Freidel-Krasnov-Rovelli 
version) in fact the lifting of this restriction and the consequent sum over all the (unitary) representations of the group
would give exactly the Crane-Yetter spin foam model.

It has been proposed \cite{cranehyper} that the fundamental dynamics of our universe is described by the Crane-Yetter state 
sum, and thus by 4-dimensional BF theory, with the gravity sector given by the non-topological quasi-phase encoded in the 
simple representations and matter fields arising from the non-simple representations left out in the Barrett-Crane model BC.
More precisely, there would exist a \lq\lq geometric quasi-vacuum" being a subspace $S$ of the total labelling space $T$ of 
the topological spin foam model CY such that the labels in $S$ admit a geometric interpretation in terms of discrete geometry
 of the underlying manifold, and such also that the time evolution in CY for initial data in $S$ is governed up to small 
corrections by the dynamics defined by BC. The conjecture is then that this quasi-vacuum is defined by the spin foam model BC.
Matter would then be represented by the labels in CY that are not in the definition of the state given by BC, with their 
low energy limit giving the known description in terms of fields. In regimes of high curvature or energy, these additional 
degrees of freedom would first appear, then become more and more relevant, but at the end they would tend to cancel out one 
another reproducing the absence of local degrees of freedom of the full topological field theory.
In principle, this conjecture could be tested by means of explicit calculations within the non-topological state sum BC, 
checking the stability of suitable initial conditions.

Let us try to be a bit more precise. Consider the Lorentzian Barrett-Crane model based on the representations $(0,\rho)$ of
$Sl(2,\mathbb{C})$ or of the quantum Lorentz algebra. It can be embedded (by passing to a quantum deformation at a root of 
unity) into a topological state sum where a series of copies of the labels for gravity appear, one set $(k,\rho)$ for each 
half-integer $k$. There would thus be a hierarchy of partners to gravity, alternatively bosonic and fermionic, for the 
various $k$. Adjacent partner sets would be naturally connected by an operation of tensoring with the representation 
$(0,1/2)$ or $(1/2,0)$, that would map a representation $(k,\rho)$ to a representation $(k+1/2,\rho)\oplus(k-1/2,\rho)$, 
and these are two fermionic maps, reminescent of those entering in supersymmetric field theories.
 
An additional motivation for studying this kind of possibility for the coupling of matter to spin foams is the connection
 with non-commutative geometries. In the Connes approach to non-commutative geometry \cite{CC} the Standard Model of 
fundamental interactions emerges naturally if the symmetry algebra of the theory is 
$W = \mathbb{C} \oplus H \oplus M^3(\mathbb{C})$. Now, this algebra is just the semisimple part of the quantum group 
$U_q sl(2,\mathbb{C})$ that we are using here to define the hypergravity spin foam model, if $q$ is a third root of unity.

Also, one can argue from both mathematical and physical reasons that the Feynmanology of the fundamental interaction in the
Standard Model would appear naturally having at its basis the morphisms in the tensor representation category we are using.  

Finally, one can re-phrase in this context the argument connecting spin foams and string theory we sketched in the previous
section: if the loops of representations we used there occupy very small circles of very small 2-simplices in the 
triangulation, we can approximate them as circles in the bundle of bivectors over the spacetime 4-manifold; this bundle is a
10-dimensional space, with base the 4-dimensional spacetime and fiber the 6-dimensional space of bivectors, and thus the 
same heuristic argument reproduced above, together with the presence now of a tower of fermionic and bosonic terms, 
connected by a natural functor, in each hypergravity multiplet, may suggest that something like a superstring theory would 
be obtained in the classical limit. But at present this is not much more than a suggestive possibility.

\section{Matter coupling in the group field theory approach}
A further approach to the coupling of matter fields to gravity in a spin foam setting uses the formalism of group field 
theory \cite{mikovicmatter}. The very basic idea is that, while the gravitational sector is described by the unitary 
representations of the Lorentz group, the matter sector is to be described by $SO(3)$ representations contained in a given 
finite-dimensional representation of it. 

This idea has a canonical analogue in the loop quantum gravity approach. In fact, consider an open spin networks 
$(\Gamma, \rho, j_i, k_i)$ defined as a partial contraction of Barrett-Crane intertwiners, with the open edges not
ending in any node labelled by an irreducible representation (and state) of an $SO(3)$ (or $SU(2)$) subgroup of the Lorentz
group, in addition to the irrep of the full group. The resulting expression is then a tensor from the incoming 
representation spaces to the outgoing ones, and in order to form a scalar (group invariant) out of it one needs to contract
 this tensor with vectors and co-vectors from these external representation spaces:
\bes
S(\Gamma, \rho, v^\rho_i)\,=\,B^{\rho_1..\rho_4}_{j_1k_1..j_4k_4}...B^{\rho_1..\rho_i}_{j_1k_1..j_ik_i}...v^{\rho_ij_i}
_{k_i}.
\ees 

Recall now that the action for the group field theory (when expanded in modes) has the structure:
\bes
S\,=\,\frac{1}{2}\sum_\rho\,\phi_{j_ik_i}^{\rho_i}\,\mathcal{K}_{j_ik_i j'_ik'_i}^{\rho_i\rho'_i}\,\phi_{j'_ik'_i}^{\rho'_i}
\,+\,\frac{1}{5!}\sum_\rho\,\phi_{j^1_ik^1_i}^{\rho^1_i}...\,\phi_{j^5_ik^5_i}^{\rho^5_i}\,
\mathcal{V}_{j^1_ik^1_i... j^5_ik^5_i}^{\rho_i^1...\rho^5_i}
\ees
with propagator and vertex given by the evaluation of the appropriate (simple) spin networks, call them $A_e$ and $A_v$.
 
We can now introduce a group action that describes fermions, with the relevant propagator and vertex again given by simple
spin networks, this time open and contracted on the external edges by fermions, i.e. by vectors in a spin $j$ 
representation of the Lorentz group:
\bes
S_F\,=\,\sum_\rho\,\psi^\rho_{jk}\,\mathcal{K}^{\rho\rho'}_{jkj'k'}\,\psi^{\rho'}_{j'k'}\;\;\;+\;\;\;(h.c.)\,\,\,+\,\,\,
\sum_\rho\,\phi_{j^1_ik^1_i}^{\rho^1_i}...\,\phi_{j^5_ik^5_i}^{\rho^5_i}\,\psi^\rho_{jk}\,\psi^{\rho'}_{j'k'}\,
\mathcal{V}_{j^1_ik^1_i... j^5_ik^5_ijkj'k'}^{\rho_i^1...\rho^5_i\rho\rho'}.\ees
In this way the fermions would propagate on complexes generated by the gravitational sector, and thus with five valent 
vertices.

The graphs of this theory would have the same basic structure of the ones for gravity alone, but with possible additions of 
lines describing fermions, in such a way that the propagator of the full theory will now be given by an \lq\lq eye diagram" 
with two external fermion lines, if fermions are present, and the vertex would be seven or five valent depending on 
whether they are present or not. 
The corresponding amplitudes will be given by open spin networks if fermions are considered, or the usual closed ones if 
they are not, and will be constructed by contraction of group intertwiners with the additional representations of the 
fermionic lines taken into account, so intertwining four simple representations for the pure gravity case and five 
representations for each vertex of the spin network with an external fermion line.

Of course, explicit formulae for these intertwiners and for the evaluation of the corresponding fermionic spin networks can
be given \cite{mikovicmatter}, in both the Riemannian and Lorentzian cases.
In the Riemannian case, the fermions are represented by $(1/2,0)$ or $(0,1/2)$ representations, and, if one uses the 
integral formula for the evaluation of the spin networks, with an $SU(2)$ character 
$K^{j,j}(\theta)=\frac{\sin((2j+1)\theta)}{\sin\theta}$ associated to each gravity link, each fermionic link is associated 
to an analogous (matrix element of a) representation function  $K^{1/2}_{ss'}(\theta)= D^{1/2}_{ss'}(\theta)$, being a 
so-called matrix spherical function, a generalization of the scalar propagator $K$, with an explicit expression available.
In the Lorentzian case, the situation is analogous, with the fermions corresponding again to representations $1/2$ of the 
$SU(2)$ subgroup of the Lorentz group, and the spinor analogue of the gravity $K$ functions used to characterize the 
fermionic links in the spin networks giving the spin foam amplitudes.

This construction can be generalized to include other types of matter including gauge bosons and their interactions with 
fermions. However, there are difficulties in implementing the internal gauge symmetry of the gauge fields and of their 
interactions with matter, their being massless (thus invarint under an $SO(2)$ and not $SO(3)$ subgroup of the Lorentz 
group, and to constrain them to satisfy the properties we would like them to have if we want them to reprduce some features
of the Standard Model.
In spite of these difficulties, this approach is very promising, especially because it fits naturally in the framework of
group field theories, and shares with them the purely algebraic and representation theoretic language.

Before closing this section, we would like to mention another idea about how matter may arise in a spin foam formulation of
quantum gravity, in the context of group field theory.
We have seen that the 2-complexes generated as Feynman graphs of the group field theory for 
4-dimensional gravity are always in 1-1 correspondence with simplicial complexes, but that these simplicial complexes are 
not always simplicial manifolds, since some of them correspond to conifolds, i.e. manifolds with conical singularities at
a finite number of points, where the given point has a neighbourood homeomorphic to a cone over some other lower dimensional
manifold. The proposal in \cite{craneconi} is to interpret this web of singularities in these particular Feynman graphs of 
the field theory as matter Feynman graphs, i.e. to interpret the low energy part of the geometry around these cones as 
particles, and the 3-manifold with boundary connecting them as interaction vertices. Matter would no longer be a separate
concept that exists in addition to space-time geometry, but it would rather appear as the structure of singularities in a
generalized geometry. This proposal, again, is intriguing, but must be analysed and developed further, so not much can be 
said on whether it is viable or not. However, some work on this idea of conical matter was done in \cite{craneconi2}, where 
it is argued that it furnishes several hints to the solution of many cosmological puzzles in the Standard Big Bang Scenario.

The approach based on the group field theory formalism, shared in part also by the hypergravity proposal, has the advantage
 that the degrees of freedom that appear in addition to the gravity ones are particular well-specified
representations of the frame group which immediately suggests their
interpretation as particles of a given spin. However, one has then to
explain why, say, spin one particles would appear as gauge bosons, and
whether these particles have, at least in some limit, the dynamics
given by ordinary Yang-Mills theory. One of the problems here is that
the concept of a gauge boson as a particle is ultimately a
perturbative concept and that we should be able to explain how the
Hilbert space of our (non-perturbative) model can be approximated by a
(perturbative) Fock space. Similar problems arise for spin-$1/2$
representations whose quantum states, at least in a regime in which
the gravity sector yields flat Minkowski space and in which the
Standard Model sector is perturbative, should admit a Fock space
representation and exhibit Fermi-Dirac statistics.

One might hope that there exists enough experience with lattice gauge
theory (see, for example \cite{Ro92,MoMu94}) in order to clarify these
issues, since we have seen that a particular approach to spin foam models uses ideas and techniques from lattice gauge 
theory. Unfortunately, one usually relies heavily on fixed hypercubic
lattices which represent space-time and which contain information
about a flat background metric. The construction of the weak field or
na{\"\i}ve continuum limit which makes contact with the perturbative
continuum formulation, relies heavily on the special properties of the
lattice. The variables of the path integral in lattice gauge theory
are the parallel transports $U_\ell=\mathrm{P}\exp(i\int_\ell A_\mu
dx^\mu)$ along the links (edges) $\ell$ of the lattice. In calculating
the weak-field limit of lattice gauge theory \cite{Ro92,MoMu94}, the
four components of the vector potential $A_\mu$ correspond to the four
orthogonal edges attached to each lattice point on the hypercubic
lattice. Even though the parallel transport is independent of the
background metric, the transition to the (perturbative) Fock space
picture does depend on it. For fermions, the situation is even less
transparent, and one faces problems similar to the notoriously
difficult question of how to put fermions on the lattice. Whereas in
the usual Fock space picture in continuous space-time, the spin
statistics relation appears as a consistency condition without any
transparent geometric justification, a unified approach to gravity
plus matter should provide a construction from which this relation
arises naturally, at least in a suitable perturbative limit. These are
deep and as yet unresolved questions.

\section{A spin foam model for Yang-Mills theory coupled to quantum gravity}
In the view of these conceptual and practical difficulties, we present
an alternative and essentially complementary construction for the
coupling of `matter' to spin foam gravity. The model we are going to present was proposed in \cite{danhend}. We concentrate on gauge
fields rather than fermions or scalars, i.e. on pure Yang-Mills
theory. For pure gauge fields, we can circumvent some of the above-mentioned
conceptual problems if we focus on the effective behaviour of gauge
theory. We rely on the weak field limit of lattice gauge theory and
make sure that the gauge theory sector approaches the right continuum
limit an effective sense when the lattice is very fine compared with
the gauge theory scale. The model we are going to describe is therefore phenomenologically
realistic if the gravity scale is much smaller than the gauge theory
scale.

We concentrate for simplicity on the Riemannian gravity case, but the same construction works for the Lorentzian metric as 
well,
although in this case the Yang-Mills sector of the model is less under control, because lattice Yang-Mills theory is best 
(only?) understood for Riemannian lattices.

The model realizes the coupling of pure lattice Yang-Mills
theory to the Barrett-Crane model of quantum gravity in the following
way. In order to find the relevant geometric data for Yang-Mills
theory, we analyze the continuum classical action for the gauge fields
and discretize it on a generic triangulation. The relevant geometric
data are then taken, configuration by configuration, from the
Barrett-Crane model.

As an illustration, consider a situation in which quantum gravity has
a classical limit given by a smooth manifold with Riemannian metric,
and assume that we study lattice Yang-Mills theory on this classical
manifold, using triangulations that are a priori unrelated to
the gravity model. Then we require that the continuum limit of this
lattice gauge theory agrees with continuum Yang-Mills theory on the
manifold that represents the classical limit of gravity.

In order to take the continuum limit including a non-perturbative
renormalization of the theory, one sends the bare coupling of
Yang-Mills theory to zero and at the same time refines the lattice in
a particular way~\cite{Ro92,MoMu94}. However, we do not actually pass
to the limit, but rather stop when the lattice gets as fine as the
triangulation on which the Barrett--Crane model is defined. We assume
that we have chosen a very fine triangulation for the Barrett--Crane
model and that the Barrett--Crane model assigns geometric data to it
that are consistent with the classical limit. This means that the path
integral of the Barrett-Crane model has to be dominated by configurations
whose geometry is well approximated by the Riemannian metric of the
smooth manifold in which the triangulation is embedded and which
represents the classical limit.

Therefore the geometries that the dominant configurations of the
Barrett--Crane model assign to the triangulation, should correspond to
the geometry of the triangulation of Yang-Mills theory if we approach
the continuum limit of Yang-Mills theory by refining the lattice for
Yang-Mills theory more and more. Our strategy is now to \emph{define}
Yang--Mills theory on the same very fine triangulation as the
Barrett-Crane model and, configuration by configuration, to use the
geometric data from the Barrett-Crane model in the discretized
Yang-Mills action.

To be specific, for pure $SU(3)$ Yang--Mills theory interpreted as
the gauge fields of QCD, the typical scale is $10^{-13}$cm. If the
fundamental triangulation is assigned geometric data at the order of
the Planck scale, the embedding into the classical limit manifold
provides the edges of the triangulation with metric curve lengths of
the order of $10^{-33}$cm. From the point of view of QCD, this is
essentially a continuum limit. In our case, however, the lattice is
not merely a tool in order to define continuum Yang--Mills theory
non-perturbatively, but we rather have a model with a very fine
triangulation that is physically fundamental. This model can be
approximated at large distances by a smooth manifold with metric and
Yang--Mills fields on it.

\subsection{Discretized pure Yang--Mills theory}
Let us now consider the classical continuum Yang--Mills action for
pure gauge fields on a Riemannian four-manifold $M$. The gauge group
is denoted by $G$ and its Lie algebra by $g$. 
\begin{equation}
\label{eq_ymaction}
  S=\frac{1}{4g_0^2}\int_M\tr(F_{\mu\nu}F^{\mu\nu})\sqrt{\det g}\,d^4x
   =\frac{1}{4g_0^2}\int_M\tr(F\wedge\ast F).
\end{equation}
Here we write $F=F_{\mu\nu}\,dx^\mu\wedge dx^\nu$,
$\mu,\nu=0,\ldots,3$, for the field strength two-form using any
coordinate basis $\{dx^\mu\}$. The action makes use of metric data as
it involves the Hodge star operation.

We reformulate this action in order to arrive at a path integral
quantum theory which can be coupled to the Barrett--Crane model. This
is done in two steps.
Firstly, we consider the preliminary step towards the Barrett--Crane
model in which classical
variables $B(t)=B^{IJ}(t)T_{IJ}\in so(4)$ are attached to the
triangles which are interpreted as the bi-vectors
$B^{IJ}(t)v_I\wedge v_J\in\Lambda^2(\mathbb{R}^4)$ that span the
triangles in $\mathbb{R}^4$.
Therefore we discretize the Yang--Mills action~\eqref{eq_ymaction} on
a generic combinatorial triangulation. We mention that for generic
triangulations with respect to a flat background metric, there exists
a formalism in the context of gauge theory on random
lattices~\cite{ChFr82b}. Here we need a formulation which does not
refer to any background metric.

We pass to locally orthonormal coordinates, given by the co-tetrad
one-forms ${\{e^I\}}_I$, $I=0,\ldots,3$, i.e. $dx^\mu=c^\mu_I e^I$,
and obtain
\begin{eqnarray}
\label{eq_ymstep2}
  S&=&\frac{1}{8g_0^2}\int_M\tr(F_{IJ}F_{KL})
    \epsilon^{KL}{}_{MN}e^I\wedge e^J\wedge e^M\wedge e^N\nonumber\\
   &=&\frac{1}{4g_0^2}\int_M\sum_{I,J,M,N}\tr(F_{IJ}^2)
    \epsilon_{IJMN}\ast(e^I\wedge e^J)\wedge\ast(e^M\wedge e^N),
\end{eqnarray}
where $F_{IJ}=F_{\mu\nu}c^\mu_Ic^\nu_J\,e^I\wedge e^J$. In the last
step, we have made use of the symmetries of the wedge product and of
the $\ast$-operation, and there are no summations other than those
explicitly indicated, in particular there is no second sum over $I,J$.

Discretization of~\eqref{eq_ymstep2} turns integration over $M$ into
a sum over all four-simplices,
\begin{equation}
  S=\sum_{\sigma}S_\sigma.
\end{equation}
Two-forms with values in $g$ and $so(4)$ correspond to a colouring
of all triangles $t$ with values $F(t)\in g$ and $B(t)\in so(4)$,
respectively.

Equation~\eqref{eq_ymstep2} resembles the preliminary step in the
construction of the four-volume operator in~\cite{Re97b}. The total
volume of $M$ is given by
\begin{equation}
  V=\int_M\sqrt{\det g}\,dx^1\wedge dx^2\wedge dx^3\wedge dx^4
   =\frac{1}{4!}\int_M
      \epsilon_{IJMN}\ast(e^I\wedge e^J)\wedge\ast(e^M\wedge e^N).
\end{equation}
Discretization results in
\begin{equation}
\label{eq_volume}
  V=\sum_{\sigma}\frac{1}{30}\sum_{t,t^\prime}\frac{1}{4!}
    \epsilon_{IJMN}\,s(t,t^\prime)T^{IJ}(t)T^{MN}(t^\prime),
\end{equation}
where the sums are over all four-simplices $\sigma$ and over all pairs
of triangles $(t,t^\prime)$ in $\sigma$ that do not share a common
edge. The wedge product of co-tetrad fields $\ast(e^I\wedge e^K)$ was
replaced by a basis vector $T^{IJ}$ of the ${\so(4)}^\ast$ that is
associated to the given triangle $t$, and the wedge product of two of
them is implemented by considering pairs $(t,t^\prime)$ of
complementary triangles with a sign factor $s(t,t^\prime)$
depending on their combinatorial orientations. Let $(12345)$ denote
the oriented combinatorial four-simplex $\sigma$ and $(PQRST)$ be a
permutation $\pi$ of $(12345)$ so that $t=(PQR)$ and $t^\prime=(PST)$
(two triangles $t,t^\prime$ in $\sigma$ that do not share a common
edge have one and only one vertex in common). Then the sign factor is
defined by $s(t,t^\prime)=s(\pi)$~\cite{Re97b}.

The boundary of a given four-simplex $\sigma$ is a particular
three-manifold and can be assigned a Hilbert space~\cite{Baez2} which
is essentially a direct sum over all colourings of the triangles $t$
of $\sigma$ with simple representations $V_{j_t}\otimes V_{j_t}$,
$j_t=0,\frac{1}{2},1,\ldots$ of $SO(4)$, of the tensor product of the Hilbert spaces associated to its 5 tetrahedra for 
given colorings. From~\eqref{eq_volume}, one
obtains a four-volume operator,
\begin{equation}
\label{eq_volume2}
  \hat V_\sigma = \frac{1}{30}\sum_{t,t^\prime}\frac{1}{4!}
    \epsilon_{IJMN}\hat T^{IJ}(t)\hat T^{KL}(t^\prime),
\end{equation}
where the $\so(4)$-generators $\hat T^{IJ}$ act on the representation
$V_{j_t}\otimes V_{j_t}$ associated to the triangle~$t$. $\hat
V_\sigma$ is an operator on the vector space
\begin{equation}
  \sym{H}_\sigma=\bigotimes_t V_{j_t}\otimes V_{j_t},
\end{equation}
of one balanced representation $V_{j_t}\otimes V_{j_t}$ for each
triangle $t$ in $\sigma$. The space $\sym{H}_\sigma$ is an
intermediate step in the implementation of the
constraints~\cite{bb} where only the simplicity condition has
been taken into account.

Observe that the sum over all pairs of triangles $(t,t^\prime)$
provides us with a particular symmetrization which can be thought of
as an averaging over the angles\footnote{An alternative expression for
the four-volume from the context of a first order formulation of Regge
calculus~\cite{CaAd89}, that we have anticipated in section \Ref{sec:geo} is given by,
\begin{equation}
  (V_\sigma)^{3} =
  \frac{1}{4!}\epsilon^{abcd}b_{a}\wedge b_{b}\wedge
  b_{c}\wedge b_{d}, \label{4sop2}.
\end{equation}
where the indices $a,b,c,d$ run over four out of the five tetrahedra
of the four-simplex $\sigma$ (the result is independent of the
tetrahedron which is left out), and the $b_a$ are vectors normal to
the hyperplanes spanned by the tetrahedra whose lengths are
proportional to the three-volumes of the tetrahedra. This formulation
favours the angles between the $b_a$ over the quantized areas and fits
into the dual or connection formulation of the Barrett--Crane model.}
that would be involved in an exact calculation of the volume of a
four-simplex.

\subsection{The coupled model}
We are interested in a discretization of the Yang-Mills
action~\eqref{eq_ymstep2} which can be used in a path integral
quantization, i.e. we wish to obtain a number (the value of the
action) for each combined configuration of gauge theory and the
Barrett-Crane model. In the case of the four-volume,
equation~\eqref{eq_volume2} provides us with an operator for each
four-simplex $\sigma$. The analogous operator obtained
from~\eqref{eq_ymstep2} reads,
\begin{equation}
\label{eq_ymoperator}
  \hat S_\sigma=\frac{1}{4g_0^2}\,\frac{1}{30}\sum_{t,t^\prime}\tr({F(t)}^2)
    \epsilon_{IJMN}\,s(t,t^\prime)\hat T^{IJ}(t)\hat T^{MN}(t^\prime).
\end{equation}
Since in~\eqref{eq_ymstep2} only the field strength components
$F_{IJ}$, but not $F_{MN}$ appear, we need $F(t)$ only for one of the
two triangles.

How to extract one number for each configuration from it? $\hat
S_\sigma$ is not merely a multiple of the identity operator so that it
does not just provide a number for each assignment of balanced
representations to the triangles. One possibility, natural from a lattice gauge theory point of view, but less from the spin 
foam one, is to generalize
the sum over configurations of the path integral so that it not only
comprises a sum over irreducible representations attached to the
triangles, but also a sum over a basis for each given
representation, and to take the trace of $\hat S_\sigma$
over the vector space $\sym{H}_\sigma$. The value of the action to use
in the path integral is therefore,
\begin{equation}
\label{eq_step1}
  S_\sigma=\frac{1}{\dim\sym{H}_\sigma}\tr_{\sym{H}_\sigma}(\hat S_\sigma).
\end{equation}
In a lattice path integral, the weight for the
Yang--Mills sector is therefore the product,
\begin{equation}
\label{eq_step2}
  \exp\bigl(i\sum_\sigma S_\sigma\bigr)=\prod_{\sigma}\exp(i\,S_\sigma),
\end{equation}
over all four-simplices. An alternative prescription
to~\eqref{eq_step1} and~\eqref{eq_step2} is given by,
\begin{equation}
\label{eq_step3}
  \prod_\sigma\frac{1}{\dim \sym{H}_\sigma}\tr_{\sym{H}_\sigma}\exp(i\hat S_\sigma).
\end{equation}
Whereas~\eqref{eq_step2} provides the average of the eigenvalues of
the operator $\hat S_\sigma$ in the exponent, the trace in
\eqref{eq_step3} can be understood as a sum over different
configurations each contributing an amplitude $\exp(i S_\sigma)$
with a different eigenvalue of the four-volume. We stick
to~\eqref{eq_step2} as this expression is closest to the classical
action.

We note that the operator $\hat S_\sigma$ of~\eqref{eq_ymoperator} is
Hermitean, diagonalizable and $\so(4)$-invariant. This can be seen
for each of its summands if one applies the splitting
$\so(4)\cong\su(2)\oplus\su(2)$ of $\so(4)$ into a self-dual and an
anti-self dual part. Then
\begin{equation}
  \epsilon_{IJMN}\hat T^{IJ}\otimes\hat T^{MN}=4\sum_{k=1}^3
    (\hat J^+_k\otimes\hat J^+_k+\hat J^-_k\otimes\hat J^-_k).
\end{equation}
Here $J^\pm_k$, $k=1,2,3$, denote the generators of the
(anti)self-dual $\su(2)$. Invariance under $\su(2)\oplus\su(2)$
follows from the fact that for a tensor product $V_j\otimes V_\ell$ of
irreducible $\su(2)$-representations,
\begin{equation}
  \sum_k\hat J_k\otimes\hat
  J_k=\frac{1}{2}\bigl(j(j+1)+\ell(\ell+1)-C^{(2)}_{V_j\otimes V_\ell}\bigr),
\end{equation}
where $C^{(2)}_{V_j\otimes V_\ell}$ denotes the quadratic Casimir
operator of $\su(2)$ on $V_j\otimes V_\ell$. This argument holds
independently for the self-dual and anti self-dual tensor factors.

There is a further possible choice for an extraction of the
four-simplex volume from the Barrett--Crane configurations. We could
insert the operator into the $10j$-symbol itself, i.e. in the
Barrett-Crane vertex amplitude. This means contracting 20 indices of the five Barrett-Crane intertwiners with the indices of 
the Yang-Mills operator, pairing those referring to the same triangle, and having two indices for each triangle, each
 contracted with a different intertwiner:  
\begin{equation}
\label{eq_volbbb}
  \prod_\sigma B_{i_1i_2i_3i_4}B_{j_4i_5i_6i_7}B_{j_7j_3i_8i_9}
    B_{j_9j_6j_2i_{10}}B_{j_{10}j_8j_5j_1}
    {[e^{i\hat S}]}_{i_1j_1i_2j_2\ldots i_{10}j_{10}}.
\end{equation}
This way of coupling Yang--Mills theory to the Barrett--Crane model is
more natural from the spin foam point of view since it just amounts to
a spin network evaluation, with an operator insertion. In fact it was shown in \cite{Mi02b} to be
derivable directly from a path integral for BF theory plus constraints (Plebanski constraints) plus a function of the B 
field, in this case the Yang-Mills action, using the formalism
of~\cite{F-K}. The result does not depend on the way we paired the indices, of course, and it results in an interaction
vertex amplitude which is both gauge and Lorentz ($SO(4)$) invariant. Therefore we consider it a much better option for the 
definition of the model, although the other possibility is of course to be kept in mind.

So far, we have prescribed how the Yang--Mills path integral obtains
its geometric information from the Barrett--Crane model which is
required to formulate the discretization of the
action~\eqref{eq_ymstep2}. The curvature term $\tr({F(t)}^2)$ of the
Yang-Mills connection in~\eqref{eq_ymoperator} can be treated as usual
in LGT with Wilson action.

Associate elements of the gauge group $g_e\in G$ to the edges $e$ of
the triangulation which represent the parallel transports of the gauge
connection. Calculate the holonomies $g(t)$ around each triangle $t$
for some given orientation. Then the curvature term arises at second
order in the expansion of the holonomy~\cite{Ro92,MoMu94},
\begin{equation}
  \Re\tr g(t) \sim \Re\tr\biggl(I + ia_tF(t) -
  \frac{a_t^2}{2}{F(t)}^2 + \cdots\biggr) = d-\frac{a_t^2}{2}\tr{F(t)}^2
  + \cdots,
\end{equation}
where $a_t$ denotes the area of the triangle $t$. Here the $\tr$ is
evaluated in a representation of dimension $d$ of $G$.

The area $a_t$ of a triangle $t$ is easily obtained from the data of
the Barrett--Crane model by $a^2_t=j_t(j_t+1)$, ignoring all
prefactors (or by the alternative choice
$a^2_t={(j_t+\frac{1}{2})}^2$).

For each four-simplex, we therefore obtain the Yang--Mills amplitude,
\begin{equation}
\sym{A}^{(YM)}_\sigma = \exp\biggl(\beta\sum_{t,t^\prime}
  \frac{\Re\tr g(t)-d}{j_t(j_t+1)}\epsilon_{IJMN}
  \frac{1}{\dim\sym{H}_\sigma}\tr_{\sym{H}_\sigma}\bigl(
    \hat T^{IJ}(t)\hat T^{MN}(t^\prime)\bigr)\biggr),
\end{equation}
or a coupled gravity-YangMills amplitude $A_\sigma(j,B_{BC},A_{YM})$ (as described in \ref{eq_volbbb}) depending on whether we
 have chosen the first or the second way to extract a number out of the Yang-Mills operator,
where $\beta$ is a coupling constant which absorbs all prefactors and
the bare gauge coupling constant. The fundamental area scale,
$\ell_P^2$, cancels because we have divided a four-volume by a square
of areas. Observe that the geometric coupling in the exponent, a
volume divided by a square of an area, is essentially the same as in
random lattice gauge theory~\cite{ChFr82b}.

Note two special cases. Firstly, for a flat gauge connection we have
$g(t)= I$ so that the Boltzmann weight is trivial,
$\sym{A}^{(YM)}_\sigma=1$. In this case, we recover the Barrett--Crane
model without any additional fields. Secondly, if a given
configuration of the Barrett--Crane model corresponds to a flat metric
and the triangulation is chosen to be regular, for example obtained by
subdividing a hypercubic lattice, then the four-volume is essentially the
area squared of a typical triangle,
\begin{equation}
  \sum_{t,t^\prime}\epsilon_{IJMN}\frac{1}{\dim\sym{H}_\sigma}\tr_{\sym{H}_\sigma}\bigl(
    \hat T^{IJ}(t)\hat T^{MN}(t^\prime)\bigr)
  \sim j_t(j_t+1)\cdot\mbox{const}.
\end{equation}
In this case, the Yang--Mills amplitude reduces to the standard
Boltzmann weight of lattice gauge theory,
\begin{equation}
  \sym{A}^{(YM)}_\sigma = \exp\biggl(\beta^\prime\sum_t(\Re\tr g(t)-d)\biggr).
\end{equation}

The model of Yang--Mills theory coupled to the Barrett--Crane model is
finally given by the partition function
\begin{equation}
\label{eq_coupledpartition}
  Z=\Bigl(\prod_e\int_G\,dg_e\Bigr)
    \Bigl(\prod_t\sum_{j_t=0,\frac{1}{2},1,\ldots}\Bigl)\,
    \Bigl(\prod_t\sym{A}^{(2)}_t\Bigr)\,
    \Bigl(\prod_\tau\sym{A}^{(3)}_\tau\Bigr)\,
    \Bigl(\prod_\sigma(\sym{A}^{(4)}_\sigma\,\sym{A}^{(YM)}_\sigma)\Bigr).
\end{equation}
In addition to the Barrett--Crane model of pure gravity, we now have
the path integral of lattice gauge theory, one integration over $G$
for each edge $e$, and the amplitude $\sym{A}^{(YM)}_\sigma$ of
Yang--Mills theory with one factor for each four-simplex in the
integrand; alternatively, in the second scheme we described, we have for each vertex (4-simplex) a composite amplitude 
for gravity and Yang-Mills, obtained as in \Ref{eq_volbbb}.

The observables of the gauge theory sector of the coupled model are,
as usual, expectation values of spin network functions under the path
integral~\eqref{eq_coupledpartition}.

\subsection{The coupled model as a spin foam model}
While the model~\eqref{eq_coupledpartition} is a hybrid involving a
lattice gauge theory together with a spin foam model of gravity, we
can make use of the strong-weak duality transformation of lattice
gauge theory~\cite{Diakonov, OePf} in order to obtain a single
spin foam model with two types of `fields'. What we have to do is basically to use Peter-Weyl theorem to express functions 
on the group in terms of irreducible representations of it, and perform the integrals over the gauge 
connection, $G$ group elements, using the formulae from the representation theory of compact groups; this is analogous to
the lattice gauge theory derivation of the Barrett-Crane model we have described above. 

Therefore we split the gauge theory amplitudes so that
\begin{equation}
\label{eq_trianglesplit}
  \prod_\sigma \sym{A}_\sigma^{(YM)}=\prod_t\sym{A}_t^{(YM)},
\end{equation}
where the second product is over all triangles and
\begin{equation}
\label{eq_ymweight}
  \sym{A}_t^{(YM)}:=\exp\Bigl(\beta n_t\frac{\Re\tr g(t)-d}{j_t(j_t+1)}
    \sum_{t^\prime}\epsilon_{IJMN}
    \frac{1}{\dim\sym{H}_\sigma}\tr_{\sym{H}_\sigma}(\hat
    T^{IJ}(t)\hat T^{MN}(t^\prime))\Bigr).
\end{equation}
Here $n_t$ denotes the number of four-simplices that contain the
triangle $t$, and the sum is over all triangles $t^\prime$ that do not
share an edge with $t$. 

We can apply the duality transformation (perform the integrals) to the gauge theory sector of
the coupled model~\eqref{eq_coupledpartition} and obtain
\begin{eqnarray}
\label{eq_coupledspinfoam}
  Z&=&\Bigl(\prod_t\sum_{\rho_t}\Bigr)
    \Bigl(\prod_e\sum_{I_e}\Bigr)
    \Bigl(\prod_t\sum_{j_t=0,\frac{1}{2},1,\ldots}\Bigr)
    \Bigl(\prod_t(\sym{A}^{(2)}_t{\hat{\sym{A}}}^{(YM)}_t)\Bigr)\nonumber\\
   &&\times \Bigl(\prod_\tau\sym{A}^{(3)}_\tau\Bigr)
    \Bigl(\prod_\sigma\sym{A}^{(4)}_\sigma\Bigr)
    \Bigl(\prod_v\sym{A}^{(YM)}_v(\{\rho_t,I_e\})\Bigr).
\end{eqnarray}
Here ${\hat{\sym{A}}}^{(YM)}_t$ are the character expansion
coefficients of $\sym{A}^{(YM)}_t$ as functions of $g(t)$. For
example, for $G=U(1)$, we have
\begin{equation}
  {\hat{\sym{A}}}_t^{(YM)}=I_{k_t}(\gamma)e^{-\gamma d},\qquad
  \gamma=\frac{\beta\,n_t}{j_t(j_t+1)}
    \sum_{t^\prime}\epsilon_{IJMN}
    \frac{1}{\dim\sym{H}_\sigma}\tr_{\sym{H}_\sigma}(\hat
    T^{IJ}(t)\hat T^{MN}(t^\prime)),
\end{equation}
where $I_k$ denote modified Bessel functions and the irreducible
representations are characterized by integers $k_t\in\Z$ for each
triangle $t$. Similarly for $G=SU(2)$,
\begin{equation}
  {\hat{\sym{A}}}_t^{(YM)}=2(2\ell_t+1)I_{2\ell_t+1}(\gamma)e^{-\gamma
  d}/\gamma,
\end{equation}
where $\ell_t=0,\frac{1}{2},1,\ldots$ characterize the irreducible
representations of $G$. Note that these coefficients depend via
$\gamma$ on the assignment of balanced representations $\{j_t\}$ to
the triangles. The path integral now consists of a sum over all
colourings of the triangles $t$ with irreducible representations of
the gauge group $G$ and over all colourings of the edges $e$ with
compatible intertwiners of $G$ as well as of the sum over all
colourings of the triangles with balanced representations of
$SO(4)$. Under the path integral, there are in addition amplitudes
$\sym{A}_v^{(YM)}$ for each vertex which can be calculated from the
representations and intertwiners at the triangles and edges attached
to $v$. The $\sym{A}_v^{(YM)}$ are very similar to the four-simplex
amplitudes, just using the gauge group intertwiners
attached to the edges incident in $v$. For more details,
see~\cite{OePf} where the $\sym{A}_v^{(YM)}$ are called
$C(v)$. The observables of lattice gauge theory can be evaluated as
indicated in~\cite{OePf}.

Observe that in~\eqref{eq_coupledspinfoam}, simplices at several
levels are coloured, namely triangles with irreducible representations
of the gauge group $G$ and with simple representations of $SO(4)$,
edges with compatible intertwiners of $G$ and tetrahedra with
Barrett--Crane intertwiners (hidden in the
$\sym{A}_\sigma^{(4)}$). The model~\eqref{eq_coupledspinfoam}
therefore does not admit a formulation using merely two-complexes. The
technology of the field theory on a group formulation would have to be
significantly extended, namely at least to generate three-complexes,
before it can be applied to the
model~\eqref{eq_coupledspinfoam}. Observe furthermore that we now have
amplitudes at all levels from vertices $v$ to four-simplices $\sigma$.

\subsection{Features of the model}
We now discuss briefly the main features of the coupled model we
propose. Firstly, it shares the main characteristics of spin foam
models for pure gravity: it is formulated without reference to any
background metric, using only the combinatorial structure of a given
triangulation of a four-manifold as well as algebraic data from the
representation theory of the frame group of gravity, here $SO(4)$,
and of the gauge group $G$ of Yang--Mills theory. The partition
function~\eqref{eq_coupledspinfoam} is well defined on any finite
triangulation and formulated in non-perturbative terms.

The general discretization procedure we have used in order to write
down lattice gauge theory in the geometry specified by the spin foam
model of gravity, is also applicable to other spin foam models of
geometry and, moreover, to theories other than pure gauge theory as
long as they can be reliably studied in a discrete setting.

The structure of the coupled state sum we propose, reflects the fact
that the action of classical gravity coupled to classical Yang--Mills
theory is the action of pure gravity plus the action of Yang--Mills
theory in curved space-time. Indeed, the amplitudes for the gravity
sector are unaffected by the coupling whereas those of the gauge
theory sector acquire a dependence on the representations labelling
the gravity configurations, i.e. they depend on the four-geometries
that represent the histories of the gravitational
field. Interestingly, the data we need in order to specify this
coupling, are only areas of triangles and volumes of four-simplices.

Since the labellings used in the coupled model~\eqref{eq_coupledspinfoam}
make use of more than two levels of the triangulation, there is no
easy way to a ``GUT-type'' unification of gravity and Yang--Mills
theory by just studying a bigger symmetry group which contains both
the frame group of gravity and the gauge group of Yang--Mills
theory. The problem is here that gauge theory in its connection
formulation lives on the edges and triangles of the given
triangulation while the $SO(4)$ $BF$-theory from which the
Barrett--Crane model is constructed, naturally lives on the
two-complex dual to the triangulation. Gravity and Yang--Mills theory
therefore retain separate path integrals and are coupled only by the
amplitudes. Nevertheless, it seems possible to re-arrange all the Yang-Mills data to fit only on the dual 2-complex, 
although no explicit calculations has been done yet on this problem.

Finally, the point of view of effective theories we have chosen in the
construction of the coupled model might mean that our strategy is
\emph{only} valid at an effective level, but not the final answer
microscopically. The model might, however, still form an important
intermediate step in the construction of the classical limit and be
relevant also to other microscopic approaches of coupling matter to
gravity if these models are studied at large distances.

We remark that the model does depend on the chosen triangulation
because already the Barrett--Crane model does. A practical solution
might be that the long range or low energy effective behaviour turns
out not to depend on the details of the triangulation. More strongly,
one can pursue approaches such as a refinement and renormalization
procedure or a sum over triangulations in order to make the microscopic
model independent of the triangulation.

\subsection{Interplay of quantum gravity and gauge fields from a spin foam perspective}
Several aspects of quantum gravity are obviously affected by the
presence of matter in the model, changing the answer to several
questions from the context of pure gravity. For example, it was
studied which is the dominant contribution to the path integral of the
Barrett--Crane model. Numerical calculations~\cite{BaezChristensen} using the
Perez--Rovelli version~\cite{P-R} of the Barrett--Crane model show
a dominance of $j_t=0$ configurations which correspond to degenerate
geometries if the $\sqrt{j_t(j_t+1)}$ are interpreted as the areas of
the triangles. One might think that this degeneracy can be avoided by
just using the alternative interpretation, taking $j_t+\frac{1}{2}$ to
be the areas, so that most triangles have areas of Planck
size. However, independent of this interpretation, also the
formulation of the Barrett--Crane model in the connection
picture~\cite{hendryk} indicates problems with geometrically degenerate
configurations. This situation may well change if matter is included
in the model, and it will also affect the construction of a classical
limit. Also the divergence of the partition function of the version of
\cite{DP-F-K-R} and the classical limit will be affected
by the presence of matter. The point is basically that now the sum over gravity representations is influenced by the 
presence of the geometric coupling of Yang-mills (4-volume over triangle area squared), and this may well change many of the 
known results.

Just as many questions in quantum gravity are affected by the presence
of matter and gauge fields, many issues in gauge theory have to be
rethought or rephrased when the coupling with gravity is considered.
Here we briefly discuss some of the questions we face if we compare
the gauge theory sector of the coupled
model~\eqref{eq_coupledspinfoam} with a realistic theory and interpret
it as the pure gauge fields of QCD.

In the standard formulation of lattice Yang--Mills theory, the
(hypercubic) lattice is considered as a purely technical tool in order
to define the continuum theory in a non-perturbative way. Starting
with some lattice with a spatial cut-off given by the lattice spacing
$a$, one wishes to construct a continuum limit in which the lattice is
refined while the relevant physical quantities are kept fixed. These
physical quantities are, for example, the masses of particles
$m_j=1/\xi_ja$ which are given by the Euclidean correlation
lengths $\xi_j$ which we specify in terms of multiples of the lattice
constant. In pure QCD, quantities of this type are the glue balls.

One tunes the bare parameters of the theory towards a critical point,
i.e. to a value where the relevant correlation lengths $\xi_j$
diverge.  This allows a refinement of the lattice, $a\rightarrow 0$,
while the observable masses $m_j$ are kept constant. Taking this limit
removes the cut-off and non-perturbatively renormalizes the
theory. 

In a model in which lattice Yang--Mills theory is coupled to gravity,
we are no longer interested in actually taking this continuum
limit. The triangulation is now rather a fundamental structure with a
typical length scale of the order of the Planck length, for example
obtained by dynamically assigning areas to the triangles as in the
Barrett--Crane model. Instead of the continuum limit, we now have to
consider a continuum approximation in which the long distance
behaviour of Yang--Mills theory (long distances compared with the
Planck length) is approximated by a continuum theory, very similar to
common situations in condensed matter physics in which there are
underlying crystal lattices.

Coming back to our example in which we interpret the gauge theory
sector as QCD, we first have to explain why the ratio $m_{\rm
Planck}/m_{\rm QCD}\sim 10^{20}$ is so big, where $m_{\rm QCD}$ is a
typical mass generated by QCD, or why the typical correlation length
of QCD, $\xi_{\rm QCD}\sim 10^{20}$, is so large in Planck units
(see~\cite{Wi02} for some not so common thoughts on this issue).

One solution would be to employ a fine-tuning mechanism in the
combined full quantum model. There could be a parameter (maybe not yet
discovered in the formulations of the Barrett--Crane model) which has
to be fine-tuned to make the combined model almost critical and to
achieve exactly the right correlation length $\xi_{\rm QCD}$. The
coupled model~\eqref{eq_coupledspinfoam} also contains the bare
parameter $\beta$ which enters the ${\hat{\sym{A}}}_t^{(YM)}$ and
which originates from the inverse temperature of lattice Yang--Mills
theory. This $\beta$ is another candidate for such a fine-tuning
procedure.

However, there might be a way of avoiding any fine tuning. Looking at
the structure of the Yang--Mills amplitude~\eqref{eq_ymweight}, one
could drop $\beta$ from that expression and rather consider an
effective quantity
\begin{equation}
  \beta_{\rm eff}=\frac{n_t\left<V\right>}{\left<a^2\right>},
\end{equation}
of Yang--Mills theory which originates from the geometric data of the
gravity sector, say, via suitable mean values for four-volume $V$ and
area square $a^2$. From perturbation theory at one loop, the typical
correlation length of QCD in lattice units scales with the bare
inverse temperature $\beta$ as
\begin{equation}
  \xi_{\rm QCD}=\xi_0\exp\bigl(\frac{8\pi^2}{11}\beta\bigr),
\end{equation}
where the prefactor $\xi_0$ depends on the details of the action and
of the lattice. A rough estimate shows that one can reach $\xi_{\rm
QCD}\sim 10^{20}$ already with $\beta\sim 10^1$. It is therefore
tempting to drop the last coupling constant from our toy model of QCD
and to make use of the gravity sector in order to provide an effective
coupling constant for QCD. As suggested in~\cite{Wi02}, one should
reverse the argument and ask what is the effective QCD coupling
constant at the Planck scale. In the coupled spin foam model this
corresponds to extracting $\beta_{\rm eff}$ from the small-$j$ regime
of the gravity sector. This might be an elegant way of generating a
large length scale and an almost critical behaviour without
fine-tuning.

The crucial question is whether the long distance behaviour of the
coupled model is stable enough even though the effective coupling
constant $\beta_{\rm eff}$ of the gauge theory sector is affected by
quantum fluctuations of the geometry. From random lattice gauge theory
on a triangulation with fixed geometry, i.e. without quantum
fluctuations, we expect that the large distance behaviour is described
by an almost critical lattice gauge theory and thus by universality
arguments largely independent of the microscopic details. If this
situation persists as the geometry becomes dynamical, then the gauge
theory sector would still automatically be almost critical. In
particular, a correlation function over $10^{20}$ triangles would have to be
independent of the microscopic quantum fluctuations of the
geometry. This question therefore forms a test of whether the coupled
model can solve the hierarchy problem, i.e. in our language whether it
can generate an exponentially large scale that is stable under
microscopic fluctuations of the quantum geometry. The same mechanism
would then also predict a dependency of the observed coupling
`constant' $\alpha_s$ on the geometry of space-time, i.e. potentially
explain varying constants in particle physics.

\chapter{Conclusions and perspectives}
\section{Conclusions...}
We conclude by giving a brief summary of the results obtained so far in the spin foam approach to 
quantum gravity, from our personal point of view, of course, and of what has been achieved. Then we discuss some of the many
points which are still missing, what is still poorly understood or not understood at all, what needs to be improved or 
included in the approach as a whole or in the particular model (the Barrett-Crane model) we have been focusing on. In the 
end, we try to offer a few perspectives on future work, on what we think are possible lines of further research in this 
area. Needless to say, all this will be tentative and broad, and again very much from a personal point of view.

The first point we would like to stress is that the very basic fact that we can speak of a spin foam approach is already
a result of not negligible importance, considering that we are basically forced to consider this type of models whether we 
start from a canonical approach, a sum-over-histories one, from simplicial quantum gravity, from discrete causal approaches
or from a lattice gauge theory point of view, or from purely mathematical considerations in terms of category theory. 
As it is quite clear, much work remains to be done and many things to be understood, but spin foam models seem to represent
a point of convergence of many apparently unrelated approaches, with a striking consequent convergence of results and ideas
that can boost further developments in a rapid way. 

In the 3-dimensional case we have a reasonably well extablished spin foam model for Riemannian quantum gravity (the 
Ponzano-Regge model) and its generalization to include a cosmological constant by means of a quantum deformation of the 
local symmetry group  (Turaev-Viro model). We have several derivations of both models and their construction is 
mathematically rigorous, with a clear connection to classical gravity in both the continuum and discrete case. We have some 
understanding of the observables of these models and of how to extract physical informations from them 
\cite{barrettobs, BNR}. The model, based on a fixed 3-manifold representing spacetime, can be generalized also to obtain a 
sum over topologies in a natural way using the formalism of group field theory, a sum that in turn is surprisingly 
possible to control and to make sense of. Also, the connection of these models with canonical loop quantum gravity is 
clear and the two approaches seem to complete one another in a very nice way. The Lorentzian version of these models is also 
known, and, although less developed and analysed (because it was developed more recently), seems to be on a similarly solid
basis. The basic reason for the quantity of the results obtained, and for their solidness, is of course the absence of local 
degrees of freedom of General Relativity in 3 dimensions, that simplifies considerably the technical aspects of the 
construction of a quantum theory of it.

In the 4-dimensional case, this simplification is not available, the classical theory is highly non trivial, and the 
technical difficulties one has to overcome in the quantization procedure are numerous, let alone the conceptual ones.
As a consequence, the results we have are less numerous and less well-extablished, but important nevertheless.

There are several proposals for a spin foam model of 4-dimensional Riemannian quantum gravity, the Barrett-Crane model, on 
which we focused our discussion, the Reisenberger model \cite{Re97b}, the Iwasaki model \cite{Iwa1}, the Gambini-Pullin 
model \cite{GambiniPullin}; there is also a spin foam model for the Lorentzian case, the Lorentzian Barrett-Crane model, that we have also
described in detail, and a general scheme of causal evolution of spin networks \cite{marsmo}, being also in the spin foam 
framework. 

The Barrett-Crane model is the most studied and the best understood. It exists, as we have just said, in both a Lorentzian
and a Riemannian formulation. We have several derivations of it from a lattice gauge
theory perspective and from a group field theory (that seems to be more general), and also its connection to a classical 
formulation of General Relativity (the Plebanski formulation or more generally a formulation of gravity as a constrained 
topological theory) seems to be clear if not rigorously extablished.
The model itself, on the other hand, is mathematically completely rigorous in its construction and definition, and much is 
known about it, as we have tried to show. 
We have its space of states, given by simple spin networks in the pure 
representation formalism or in the combined representations-plus-normals formulation, with an explicit link with canonical
approaches as the covariant loop quantum gravity one. We have a definition (given by the spin foam model itself) of 
transition amplitudes between these states and an interpretation of them as a canonical inner product. There is a good
understanding of the geometric meaning of the variables of the model and of the way classical simplicial geometry is encoded
 in its quantum amplitudes and partition function, with a clear link to a classical simplicial action. We know the boundary 
terms we need to add in the presence of boundaries and thus in the definition of these transition amplitudes. 
The quantum (or cosmological) deformation of the model using quantum groups is known and rigorous, in both the Riemannian 
and Lorentzian cases. We have at our disposal an asymptotic approximation of the partition function of the model.
There are several promising proposals for the coupling of matter and gauge field to pure gravity in this spin foam context.
The model itself, moreover, exists in different versions, some of them having striking finiteness properties interpretable 
in the group field theory context as perturbative finiteness to all orders (!), and 
including a modification of it that seems to incorporate a
dependence on the orientation of the geometric structures in the manifold, admits a causal interpretation in the Lorentzian 
case, and seems to be a crucial link  between loop quantum gravity, the spin foam or sum-over-histories approach, dynamical 
triangulations, quantum Regge calculus, and the quantum causal sets approach.
From a more conceptual point of view, what we have is a concrete implementation in an explicit model of the idea that a 
complete formulation of a theory of quantum spacetime should be background independent and fully relational, fundamentally 
discrete, non-perturbative, covariant in nature and based on symmetry considerations, purely algebraic and combinatorial,
with a notion of fundamental causality. This is not all, by any means, but it is a non-trivial achievement.

\section{...and perspectives}
Of course, the things that still need to be understood are many more than those we did understand.

The understanding of the Barrett-Crane model itself needs to be improved. Among the relevant issues we mention the problems 
that a description of simplicial geometry in terms of areas poses at both the classical and quantum level. We have seen in 
fact that the fundamental variables of the Barrett-Crane model, together with the
normal vectors in the first order formalism, the representations of the Lorentz group, have the interpretation as being the
areas of the triangles of the simplicial complex on which the model is based. The usual fundamental variables in the known 
descriptions of simplicial geometry are instead the edge lengths. Now, for a single 4-simplex the number of edges and of 
triangles match (10), and one can (almost) always invert any expression involving areas in terms of edge lengths, and 
vice versa, so that the two descriptions are in principle equivalent. For a generic triangulation, instead, the number of 
triangles is always higher than the number of edges, so assigning all the values of the areas of the triangles is giving 
more information than needed. This is true at the classical level, but it seems reasonable to think that the same holds 
at the quantum one. We have seen that assigning the values of the quantum areas of a tetrahedron characterizes completely
 its quantum state. One possibility to translate the classical overcompleteness of the area basis is that there exist 
several assignments of quantum areas giving the same state, and that we need to implement some form of constraints to have a
better description of quantum geometry. These constraints are already hard to find at the classical level, and in the 
quantum case may imply the need of a combination of Barrett-Crane partition functions for different assignments of 
areas to the same triangles, or an additional non-local constraint on the assignment of areas in the Barrett-Crane model. 
More work on this problem at both the classical and quantum level is certainly needed.
Another intriguing possibility is that one needs to refine the model inserting additional variables corresponding to the 
edge lengths, using them as true fundamental variables, decomposing then the new model in terms of all the possible (and 
compatible) assignments of area variables. A model of this kind was proposed by Crane and Yetter \cite{CYexp} and it is 
based on particular reducible representations of the Lorentz group, namely the so-called \lq\lq expansors" \cite{Dirac},
assigned in fact to the edges of a simplicial complex. The tensor product of expansors (with prescribed symmetry properties)
assigned to the edges of a given triangle decoposes into irreducible representations which are exactly the simple 
representations of the Barrett-Crane model, and thus a spin foam model based on expansors seems to be equivalent to a 
combination of Barrett-Crane models realized precisely in the way needed to have a unique and clear assignments of quantum 
lengths for all the edges of the triangulation, thus solving the problem of a purely area description. The expansor model, 
moreover, has very intriguing mathematical properties with interesting applications of ideas from the theory of 2-categories
 \cite{CT2cat,barr2cat}.
We have also seen that the asymptotic expression of the vertex amplitudes of the Barrett-Crane model is dominated by 
degenerate configurations of zero 4-volume. This can be a problem for the model if really its semiclassical limit is the 
asymptotic limit of the representations labelling the triangles; this is not to be taken for granted however, since how the 
classical theory should emerge from the quantum one is a non-trivial and subtle issue, about which much remains to be 
understood.
Another issue regarding the Barrett-Crane model that needs a clarification is the existence of several versions of it, as we
have seen, sharing symmetries and face and vertex amplitudes, but differing in the edge amplitude, i.e. the propagator 
of the theory, from the group field theory point of view. One possibility is that it does not really matter which version 
one works with, because of some universality argument, and all have the same continuum or classical limit; this may well be
 true, but even if it is, they may have drastically different quantum behaviours and this differences may be physically 
relevant. It may be that the question of which version is the right one does not make sense, because they all have to be 
used in some physically or mathematically motivated combination, e.g. for imposing a more exact correspondence with the 
classical gravity action. It may also be that conditions like the absence of quantum anomalies related to diffeomorphisms 
will fix the edge amplitude making it possible to choose uniquely what is the correct version (for some work on this see
\cite{perezbojowald}).
If the relation between the different versions of the Barrett-Crane model is important, so is its relation with the other 
existing spin foam models, in particular the Reisenberger model. We have seen how close is their origin from the classical 
Plebanski action, but at the quantum level their appearence is very different and the Reisenberger constraints are less easy
 to solve, and thus it is not easy to compare their quantum data.

The other model that deserves, we think, further study is the causal or orientation dependent model we have derived and 
discussed in chapter \Ref{sec:causal}. Not only does it seem to implement the orientation properties of the simplicial manifold
correctly and has an interpretation as defining causal transition amplitudes, but it also puts the spin foam model in the 
precise form of a path integral for a first order simplicial gravity action, and is suitable for the application of 
techniques from quantum Regge calculus, dynamical triangulations and causal sets, in addition to loop quantum gravity, being
a link between all these approaches. We need to know more about the representation theoretic properties of the model, to be 
able to generalise the construction leading to it to other cases, e.g. the 3-dimensional one, and to understand its 
description of quantum geometry. A very important step would be its derivation from a group field theory formalism, since
 this seems to represent the most complete formulation of spin foam models, implementing the needed sum over triangulations
and the wanted sum over topologies, and because in that context it may be easier to 
understand what kind of transition amplitudes the modified model defines. The idea would be that it provides a 
background independent definition, thus independent of any time variable, of a kind of \lq\lq time ordered product", in the 
sense that it furnishes the kind of amplitudes that in ordinary quantum field theory are defined by using a time 
ordered product of field operators. This may also require the use of a complex field instead of the usual real one. Work on
this is in progress. We mention the 
intriguing possibility that it may result from the extension of the symmetry group used from the Lorentz group $SO(3,1)$ to
 the full Lorentz group including discrete transformations, i.e. $O(3,1)$, or using a double covering of it, the so-called
$Pin$ groups, $Pin(3,1)$ or $Pin(1,3)$ \cite{PinRev,Pin2}. This sounds reasonable since what we are trying to do in the 
modified model is to implement non-trivial behaviours of the amplitudes under inversions of their argument, so its proper 
derivation and understanding may need a representation of parity or time reversal transformations. Also, we have mentioned, 
when discussing the possibility of developing a Fock space picture for the group field theory, that the presence of geometric 
anti-particles will probably turn out to be related to the properties under inversion of the orientation of the triangles, so 
that this again may require a group field theory based on the full Lorentz group $O(3,1)$; this is also to be expected since
the use of $O(3,1)$ would allow the defintion of an action of a charge conjugation operator related, via CPT theorem, to the
 total inversion $PT$.

There are then many issues deserving further work, which are not only related to the Barrett-Crane model, but to the spin 
foam approach in general.

The issue of causality, for example, that we mentioned concerning the modified Barrett-Crane model, is more general than 
that. How do we define spin foam transition amplitudes that reflect an ordering between their arguments? How consistent is the
interpretation of this ordering as causality in a background-independent context? If not in this way, how is a causal 
structure to emerge in a suitable limit from the full quantum theory formulated as a spin foam model? What is the correct 
notion of causality in the first place? How many different transition amplitudes can we define in this formalism? We have 
seen that in the relativistic particle 
case, and in formal path integral quantization of gravity, just as in quantum field theory, there is a number of different
2-point functions that can be defined, with different physical meanings and use. These questions may be best investigated 
in the context of the group field theory,
maybe within extended models based on $O(3,1)$, and maybe after a full quantization of them has been achieved, or directly
at the level of the spin foam amplitudes, using a class of causal models such as the modified Barrett-Crane model we discussed.

The development of the whole formalism of group field theories is another important point in the agenda, we think, with the 
construction of a particle picture as we have been arguing, coming from a more rigorous and complete quantization of them, 
but also with an improved understanding of their perturbative expansion, trying to extend to the 4-dimensional case the 
results obtained \cite{llsum} in the 3-dimensional one, and of the physical meaning of their coupling 
constant. An intriguing possibility is that this
parameter would turn out to be related to the cosmological constant,
so that the perturbation expansion of the field theory would have
similar nature to the approximation expansion used recently in loop
quantum gravity \cite{GamPul}. Even not considering this connection, a similar
result would be of much interest in itself and would shed considerable
light on the
nature of the field theory. Another related possibility is that the perturbation expansion, which is in increasing number 
of 4-simplices, is a kind of expansion on terms of increasing value of the proper time represented exactly by this number of
4-simplices \cite{Mikov}. In light of the importance of this expansion in the definition of the models, the understanding of
 it is crucial.

Then comes the question of the role of diffeomorphisms in spin foam models. Is a spin foam a diffeo-invariant encoding of 
the degrees of freedom of spacetime, or merely diffeo-covariant? How are diffeomorphisms to be defined in this context? 
How are these questions affected by the implementation of a sum over spin foams defining a spin foam model?  These questions
or, better, the answers to them, change the very way we deal with a spin foam model, may determine the choice of the 
amplitudes appearing in it, and influence our strategies in defining a continuum limit/approximation for it.
A point of view may be that the face, edge and vertex amplitudes are diffeomorphism invariant ways of encoding geometric 
information and therefore the only residual diffeomorphisms may be the automorphisms of the 2-complexes used in the sum over
spin foams, and these are taken into account in the definition of the perturbative expansion of the group field theory. 
Therefore, the resulting complete models are already diffeo-invariant. The situation would then be directly analogous to 
that in dynamical triangulations, except that now also the diffeo-invariant metric information would be dynamical. 
It is also possible however that diffeomorphisms still act non-trivially on the amplitudes, and that the study of the 
residual action of diffeomorphisms and the requirements of the absence of anomalies may fix the correct amplitudes 
helping in 
choosing among the various spin foam models available. The work already done in the context of quantum Regge
calculus are a warning against underestimating the subleties of this problem. 

A directly related issue is about the correct way to reconstruct the spacetime quantum geometry (and possibly topology) from
 the spin foam data. While the general idea is clear, and for several geometric
quantities, such as the areas of surfaces, it is known how to apply
it, a general reconstruction procedure is missing. Any such procedure would involve the embedding of the abstract labelled
2-complex in a topological manifold, with a consequent loss of diffeomorphism invariance, and a careful study of this 
embedding procedure and of the constraints implied by diffeomorphism invariance should be undertaken. Given the relationship
 mentioned between spin foam models and causal sets, the existing literature about this last approach will
be useful, since similar problems arise in that context. Also, an analysis of the properties (including the spectrum of
eigenvalues) of many other geometric operators, e.g tetrahedral volume, 4-simplex volume, curvature, etc. would be of much 
interest, and would help in the reconstruction problem. 
 
The crucial issue that spin foam models have still to solve, in a sence the true \lq\lq reality check" of any potential 
theory of quantum gravity, is the definition of a continumm limit/approximation and of a related semiclassical 
approximation (or low energy limit), with a consequent recovering of classical General Relativity. Unfortunately, this has 
not been achieved yet. If one succeeds in reformulating spin foam models in the language of
decoherent histories, then an interesting task would be to find a
suitable definition of a decoherence functional that may serve as the
technical tool to obtain such a semi-classical limit.
This involves studying a coarse graining procedure of the type used in
all sum-over-histories approaches, but formulated in a purely algebraic
and background independent manner, in order to be suitable for the
spin foam context.
 A new general approach to this issue has been developed by Markopoulou
\cite{Mark}, based on the mathematics of Hopf Algebras, and another interesting set of proposals and results was obtained in
\cite{robertrenorm}.  
 Also, the issue of perturbations around semiclassical
solutions in spin foam models, both in 3 and 4 dimensions has to
be studied; in particular, in the 4-dimensional case, this would
settle the question about the existence of gravitons, i.e. of
quantum gravitational waves or local propagating degrees of freedom, in the Barrett-Crane model; an
intriguing possibility is also that such a study would reveal a
link between spin foam models and string theory, as proposed by L.
Smolin.

As we have stressed, also the correct coupling of matter and gauge fields to pure gravity in a spin foam context is an 
important open issue. All the proposals we have described deserve to be better studied and developed, and in particular the 
coupled model for Yang-Mills and gravity can be first better etablished by further analysis of the classical reformulation 
of Yang-Mills theory as a constrained BF theory \cite{Diakonov}, and then extended to include scalar fields and fermions, 
using 
techniques from lattice gauge theory; then it may be possible to obtain predictions on the behaviour of gauge
fields in a quantum gravity context (i.e. propagation of light in quantum gravity backgrounds, modified dispersion 
relations, etc.). As a possible solution (among many) to the problem of including matter fields in the models, but also for 
its own interest and relevance, a supersymmetric extension of the Barrett-Crane model, and
of its lower-dimensional (Ponzano-Regge-Turaev-Viro) and
higher-dimensional analogues, would be a very nice development. This extension would amount to the replacement of the 
(quantum) group used for the assignment of labels to the spin foam with its
supersymmetric counterpart ($Osp(N,4)$ for $Spin(4)$ or $Osp(N,2)$ for
$SU(2)$, for example, in the Riemannian case).
In 3 dimensions a \lq\lq super-Turaev-Viro" model would be reliably related at the continuum level to a supersymmetric BF 
theory with $Osp(N,2)$ as gauge symmetry, and to 3-dimensional supergravity, but its relevance for the construction of a new
 topological invariant for 3-manifolds is not so clear. Such a model was constructed in \cite{eterarobert} and should be 
analysed further. 
In higher dimensions, the construction of supersymmetric spin foam 
models as a step in a quantization of supergravity is an intriguing possibility, but also no more than a speculation at 
present. 

This is as far as spin foam models alone are concerned. However, there is a vast amount of knowledge about quantum gravity
already developed in the context of other approaches, and the cross-fertilization between them can only be helpful. With
respect to this, spin foam models are in a very good position, since they are already a result of such a 
cross-fertilization, as we have stressed many times.  The obvious first step would be a clarification of the exact 
relationship between spin foam models, and the Barrett-Crane model in particular, and of their boundary states, with the 
states defined by loop quantum gravity, the details of which are still in part missing. Then one has to explore in greater 
details the aspects in common with dynamical triangulations, quantum Regge calculus and quantum causal sets (or histories),
looking for reciprocal improvements. The modified or causal Barrett-Crane model seems the best place to start this
exploration. In particular, the quantum causal histories framework deserve, we think, much attention and further work.

A separate word should be devoted to the possible relationship with the approach to quantum gravity represented by 
non-commutative geometries; here, the possible connections are less understood and more difficult to predict, so any 
suggestion would be extremely tentative; two possible places to look at are: the reformulation of causal sets (posets) as 
non-commutative lattices \cite{landi}, that could lead to a redefinition of the algebraic data on an oriented complex
as given by a spin foam model as a quantum theory on a non-commutative space; the role played by quantum groups and 
associated structures, which are at the very roots of the 
non-commutative geometry programme \cite{Majid2}, in spin foam models, the speculative idea being that like in the 
undeformed case a smooth geometry is expected to 
emerge from the spin foam, now in the q-deformed case the emerging spacetime would result in being endowed naturally with a
 non-commutative structure. 

The role played by quantum groups in spin foam models has to be understood more also with respect to another issue, which is 
the emergence and role of a cosmological constant. We have seen that the asymptotic expansion of the q-deformed spin foam 
models gives the action for gravity in the presence of a cosmological constant. This connection between quantum groups and
cosmological constant, and the mathematical and physical reasons for it, are not so well-understood as they should be.

A quantum
theory of gravity has to make contact with the physical world in order to
be considered seriously, and the first test for such a theory, like the
Barrett-Crane model, would probably be the correct prediction of black
hole entropy in the semiclassical regime. Such a prediction cen be obtained, also since previous results on this
in the context of loop quantum gravity \cite{Rovel}\cite{Ash} and
3-dimensional spin foam models \cite{SK} show that important results can
be achieved in spite of a not-yet complete formulation of the quantum
theory, maybe using a more \lq\lq principle-based" approach to the problem \cite{makebh}.
The issue of black hole entropy is at the same time linked to
several conceptual and technical problems within the spin foam
approach. For example, having a good understanding of the causal
structure of the Barrett-Crane model, one should try to define
causal horizons (like the black hole one, or cosmological
horizons) in a purely algebraic and combinatorial fashion and then
to study their properties. Probably, a local definition of a
horizon is what is to be looked for, making contact with current
work on locally defined horizons in classical general relativity
\cite{Ash2}. Another element that may enter in the study of black
hole entropy in spin foam models, but which is certainly of much
interest in itself, is the holographic principle. Background
independent formulations of it exist and some more suitable for
direct application to the spin foam context were also proposed
\cite{Bousso}\cite{MS}. Also, some study on holographic projections for a 3-dimensional
spin foam model were performed \cite{Arc} and should now be extended to the
4-dimensional case. 

Being ambitious, we would like also to ask our quantum gravity theory to make contact with experiment, i.e. to provide a 
(possibly) rich phenomenology that can be tested in one way or another. This would have been considered probably foolish 
until some years ago, but recent proposals for the possibility of a quantum gravity phenomenology \cite{AC} make not 
completely unmotivated, even if still rather optimistic, the hope of obtaining physically testable predictions (coming maybe 
from the discreteness or the foamy structure of spacetime) from spin foam models of quantum gravity. A route towards the 
experiments may be represented by the study of spin foam models using quantum deformations of the Lorentz group as symmetry 
group, since many of the current proposals for quantum gravity motivated modifications of low energy physics, including 
modifications of special relativity \cite{AC2,maguejosmolin}, arise from deformations of the Lorentz group to allow for an
invariant length scale (Planck length or cosmological constant?).  

We close by mentioning a few further lines of possible research, partly motivated by the work on spin 
foam models and quantum gravity, but even more broad and general in nature and aims.   

Work and thinking about quantum gravity has always led to questioning the foundations of our physical theories, and in 
particular the meaning and correct formulation of quantum mechanics. Starting from spin foam models, one should try to make 
contact with other formulations of quantum mechanics, and quantum gravity, in terms of histories. For
example, the possibility to formulate quantum causal histories in
terms of Hilbert spaces associated to whole histories of
spacetime, as in the consistent histories approach of
\cite{Isham2} should be studied. Moreover, in the causal histories
framework as well as in the consistent histories one, the very
logic that governs a quantum mechanical description of the world
is drastically different from the usual classical one. In both
cases, category theory, and in particular topos theory, plays a
major role in such a description, together with intuitionistic
logic (Heyting algebras), for example in the definition of the
observables and in the encoding of the spacetime causal structure
\cite{Ish}\cite{fotlogic}. These formulations represent a very promising framework for
applying ideas and tools from these areas of mathematics to the
theoretical physics problem of formulating a theory of quantum
gravity and of quantum cosmology.
Motivated partly by quantum gravity considerations, it has been repeatedly proposed that the same formalism or at least the 
interpretation of
quantum mechanics itself has to be suitably modified. The general
aim is to make it in a sense ``more relational'', to allow a
variety of descriptions of the same system, in the spirit of the
relationality of general relativity. A first step is to make
explicit the degree to which the notion of the state of a system
is already a relational construct in current formulations of
quantum mechanics and has to be understood in terms of the
information that a given system (observer) has of another
(observed system) \cite{carlo96}, but such an
information-theoretic formulation of quantum mechanics has still to be fully developed, using ideas and tools from 
quantum information theory \cite{Fuchs}. In such a
description, each system has associated with it a unique Hilbert
space (or its information-theoretic analogue), but the choice of a
state in this Hilbert space is relative to the observer, and has
to be characterized in terms of a flow of information from the
system to it. In a fully relational formulation of quantum
mechanics, representing a generalization of the current one, there
is no unique Hilbert space associated to a given system, but a
variety of them one for each pair system-observer, i.e. it is the
interaction between two systems that determines not only a state
but also an Hilbert space for the pair. Examples of this fully
relational quantum mechanics are the quantum causal histories
mentioned above, and the many-views approach to quantum theory,
based on histories, formalized and developed by Isham and
collaborators \cite{Ish}\cite{Isham2}. Also, some important work concerning a reformulation of quantum mechanics motivated 
by ideas from quantum gravity and topological field theories has been recently proposed in \cite{robertQM,robertQM2}, and 
should be developed further. 
These relational formulations may be applied to a description of spacetime geometry.
Applied in a cosmological context, this relational formulation of
quantum theory would realize the ``pluralistic view'' advocated by
L. Smolin \cite{Smolin} and L. Crane \cite{crane95}, for the solution
of technical and interpretative issues in quantum cosmology, also
with the use of category theoretic tools, in addition of
furnishing a framework for quantum gravity. 

Among the issues on which a complete formulation of quantum gravity is
expected to provide an answer is the so-called problem of time,
mentioned above. This is a major difficulty
in any canonical formulation of gravity, including canonical quantum
cosmology, and it can be expected that a sum-over-histories formulation, like spin
foam models, is more suited
to provide a clue for its solution. The knowledge developed in the context of decoherent
and consistent histories approaches to quantum mechanics
\cite{Halliwell2}\cite{Isham2} will certainly turn out to be essential in making progress on this difficult issue.
There are other approaches to this problem that deserve further
study: one is the idea that a correct description of time arises
from a statistical mechanical analysis of a quantum gravity
theory, so that a preferred physical time direction is singled out
by the settling of a gravitational system (the whole universe or a
part of it) in a particular statistical state, and by the
consequent definition of a preferred vector flow in terms of it;
this may be investigated in the context of spin foam models or at
the classical level of pure general relativity \cite{carloflowtime};  another route for
understanding the role of time in general relativistic quantum
theory, can be to take the opposite attitude and try to show how
conventional quantum theories can be rewritten in more generally
covariant terms, without reference to any preferred time variable,
and then how such a variables can be at the end singled out; work
on such a reformulation of classical mechanics in a way that
unifies both generally covariant and not-covariant theories has
been carried out by C. Rovelli \cite{carlomech}, and this line of research should be extended now to the quantum
domain.

\vspace{1cm}

As a conclusion, which is not a conclusion, we can just say that spin foam models in general, and the Barrett-Crane one in 
particular, represent a promising new approach to the construction of a quantum theory of gravity. It provides some answers 
and poses new questions, making use of interesting ideas and pointing to a rather radical conceptual change in the way we 
look at the Nature we are part of. The progress achieved in this research area in the last years is, to our opinion, quite 
impressive. At the same time, many aspects of these models are still obscure, and also their 
possible improvements and developments cannot be fully predicted, so that any opinion about them can just be provisional, 
and, while forced to be necessarily cautious, at the same time we have reasons to be optimistic. Much more work is certainly
 needed, but this is going to be interesting and worth doing, and we are surely going to have much fun doing it!

\end{document}